\documentclass[structabstract,longauth]{aa}
\usepackage[utf8]{inputenc}
\usepackage{threeparttable,pifont}
\usepackage{natbib}
\usepackage{pstricks}
\usepackage{multicol}
\usepackage{hyperref}
\usepackage{amsmath,amssymb,url}
\usepackage{graphicx}
\usepackage{xcolor}
\usepackage{subcaption}
\usepackage{txfonts}
\usepackage{longtable,dcolumn,verbatim}
\bibpunct{(}{)}{;}{a}{}{,}
\usepackage{array} % fixes the corners
\usepackage[normalem]{ulem}
\usepackage{threeparttable}
\usepackage{widetext}

\usepackage[squaren,textstyle]{SIunits}

\usepackage{hyperref}
\hypersetup{
    pdftitle={(216)~Kleopatra},
    pdfauthor={Yang et al.},
    colorlinks=true,        % false: boxed links; true: colored links
    linkcolor=gray,         % color of internal links
    citecolor=blue,         % color of links to bibliography
    filecolor=gray,         % color of file links
    urlcolor=gray           % color of external links
}

\usepackage{xspace}

\newcommand{\adam}{\texttt{ADAM}\xspace}

\newcommand{\mistral}{\texttt{Mistral}\xspace}

\newcommand{\DiamADAM}{119.3\,$\pm$\,2}

\begin{document}

%\title{ {First detailed physical model of large asteroid (704)~Interamnia}
%\title{ {High Precision Density Measurement of Nearly Spherical (216)~Kleopatra with SPHERE/VLT}
\title{(216) Kleopatra, a low density critically rotating M-type asteroid
%\title{Critically rotating asteroid (216)~Kleopatra
    \thanks{Based on observations made with 
      ESO Telescopes at the La Silla Paranal Observatory under program 199.C-0074 (PI Vernazza)},\thanks{The  reduced  images  are  available  at  the  CDS  via  anonymous   ftp  to \url{http://cdsarc.u-strasbg.fr/} or via \url{http://cdsarc.u-strasbg.fr/viz-bin/qcat?J/A+A/xxx/Axxx}}}

\titlerunning{Critically rotating asteroid (216)~Kleopatra}

\author{F.~Marchis\inst{\ref{seti},\ref{lam}} \and % LP
    L.~Jorda\inst{\ref{lam}}         \and % Moon   
    P.~Vernazza\inst{\ref{lam}}            \and % LP
    M.~Bro\v{z}\inst{\ref{prague}}         \and % orbital and SPH model
    J.~Hanu{\v s}\inst{\ref{prague}}       \and % 3D model
    M.~Ferrais\inst{\ref{lam}}           \and % LP
    F.~Vachier\inst{\ref{imcce}}           \and % Orbit
    N.~Rambaux\inst{\ref{imcce}}           \and % McLaurin ellipsoid
    M.~Marsset\inst{\ref{mit}}             \and % Satellite 
    M.~Viikinkoski\inst{\ref{tampere}}     \and % 3D model, ADAM
    E.~Jehin\inst{\ref{liege}}             \and % LP
    S.~Benseguane\inst{\ref{lam}}           \and % LP    
    E.~Podlewska-Gaca\inst{\ref{poznan}} \and % LP
    B.~Carry\inst{\ref{oca}}               \and % LP
    A.~Drouard\inst{\ref{lam}}             \and % LP
%Spectrum satellite
%    K.~Tazhenova\inst{\ref{lam}}           \and % Moon
%--- Observers/data contributors (mostly) by alphabetical order: No additional observers for Kleopatra, perhaps ONE from GaiaGOSA 
%    B.~Warner\inst{\ref{warner}}            % Observer
%
%    R.~Behrend\inst{\ref{behrend}}         \and % Obs, Behrend, raoul.behrend@unige.ch
    S.~Fauvaud\inst{\ref{fauvaud}}           \and % Obs, Behrend, sfauvaud@gmail.com
%    S.~Charbonnel\inst{\ref{charbonnel}}   \and % Obs, Behrend, scharbonnel949@gmail.com
%    J-F.~Coliac\inst{\ref{coliac}}         \and % Obs, Behrend, jfcoliac@free.fr
%    R.~Duffard\inst{\ref{duffard}}   \and % Obs, GaiaGOSSA
%    C.~Garcia\inst{\ref{gaiagosa}}   \and % Obs, GaiaGOSSA
%    A.~Jones\inst{\ref{gaiagosa}}   \and % Obs, GaiaGOSSA
%    A.~Leroy\inst{\ref{leroy}}             \and % Obs, Behrend, arnaud.leroy@club-internet.fr
%    D.~Molina\inst{\ref{gaiagosa}}   \and % Obs, GaiaGOSSA
%    O.~Pejcha\inst{\ref{teorka}}           \and % ASAS-SN
%    H.~Riemis\inst{\ref{riemis}}           \and % Obs, Behrend, ??? (Suys?)
%    B.~Shappee\inst{\ref{hawaii}}          \and % ASAS-SN 
%    F.~Sold\'an\inst{\ref{soldan}}         \and % Obs, Behrend, fsarrakis@gmail.com
%    D.~Suys\inst{\ref{riemis}}             \and % Obs, Behrend, ??? (Suys?)
%    R.~Szakats\inst{\ref{konkoly}}   \and % Obs, GaiaGOSSA
%    J.~Vantomme\inst{\ref{riemis}}         \and % Obs, Behrend, ??? (Suys?)
%
%--- Then by alphabetical order the other members of the LP
    M.~Birlan\inst{\ref{imcce}, \ref{aira}} \and % LP
%    Z. ~Benkhaldoun\inst{\ref{cau}} \and % TRAPIST
    J.~Berthier\inst{\ref{imcce}}          \and % Orbit
    P.~Bartczak\inst{\ref{poznan}}         \and % LP
    C.~Dumas\inst{\ref{tmt}}               \and % RAP
    G.~Dudzi\'{n}ski\inst{\ref{poznan}}    \and % LP
    J.~{\v D}urech\inst{\ref{prague}}      \and % LP
    J.~Castillo-Rogez\inst{\ref{jpl}}      \and % LP
    F.~Cipriani\inst{\ref{estec}}          \and % LP
    F.~Colas\inst{\ref{imcce}}             \and % LP
    R.~Fetick\inst{\ref{lam}}              \and % LP
    T.~Fusco\inst{\ref{lam},\ref{onera}}   \and % LP
    J.~Grice\inst{\ref{oca},\ref{ou}}      \and % LC
    A.~Kryszczynska\inst{\ref{poznan}}     \and % LP
    P.~Lamy\inst{\ref{lamos}}              \and % LP
    A.~Marciniak\inst{\ref{poznan}}        \and % LP
    T.~Michalowski\inst{\ref{poznan}}      \and % LP
    P.~Michel\inst{\ref{oca}}              \and % LP
    M.~Pajuelo\inst{\ref{imcce},\ref{puc}} \and % LP
    T.~Santana-Ros\inst{{\ref{uda},\ref{iccub}}}      \and % LP
    P.~Tanga\inst{\ref{oca}}               \and % LP
    A.~Vigan\inst{\ref{lam}}               \and % LP
    O.~Witasse\inst{\ref{estec}}           \and % LP
    B.~Yang\inst{\ref{eso}}             % LP
} 

   \institute{
      %---- Marchis
     SETI Institute, Carl Sagan Center, 189 Bernado Avenue, Suite 200, Mountain View CA 94043, USA %Marchis
     \label{seti}
      \and %---- Vernazza, Jorda, Ferrais, Fusco, Fetick, Drouard, Rambaux
     Aix Marseille Univ, CNRS, LAM, Laboratoire d'Astrophysique de Marseille, Marseille, France
     \label{lam}
    \and 
      %---- Hanus, Durech, Broz
     Institute of Astronomy, Faculty of Mathematics and Physics, Charles University, V~Hole{\v s}ovi{\v c}k{\'a}ch 2, 18000 Prague, Czech Republic
     \label{prague}
     \and %---- Vachier, Berthier, Colas, Birlan, Pajuelo
     IMCCE, Observatoire de Paris, PSL Research University, CNRS, Sorbonne Universit{\'e}s, UPMC Univ Paris 06, Univ. Lille, France
     \label{imcce}
     \and %---- Marsset
      Department of Earth, Atmospheric and Planetary Sciences, MIT, 77 Massachusetts Avenue, Cambridge, MA 02139, USA
     \label{mit}
     \and%---- Viikinkoski, Kaasalainen
     Mathematics \& Statistics, Tampere University,  PO Box 553, 33101, Tampere, Finland
     \label{tampere}
      \and %---- Jehin
     Space sciences, Technologies and Astrophysics Research Institute, Universit{\'e} de Li{\`e}ge, All{\'e}e du 6 Ao{\^u}t 17, 4000 Li{\`e}ge, Belgium
     \label{liege}
     %     %
%     \and %---- Benkhaldoun, 
%     Oukaimeden Observatory, High Energy Physics and Astrophysics Laboratory, Cadi Ayyad University, Marrakech, Morocco
%     \label{cau}
     %
     \and %---- Bartczak, Marciniak, Michalowski, Podlewska, 
     Faculty of Physics, Astronomical Observatory Institute, Adam Mickiewicz University, ul. S{\l}oneczna 36, 60-286 Pozna{\'n}, Poland
     \label{poznan}
     \and %----  Carry, Tanga, Michel
     Universit\'e C{\^o}te d'Azur, Observatoire de la C{\^o}te d'Azur, CNRS, Laboratoire Lagrange, France
     \label{oca}
     \and %---- Fauvaud
     Observatoire du Bois de Bardon, 16110 Taponnat, France % sfauvaud@gmail.com
     \label{fauvaud}
    \and %---- Birlan2 
     Astronomical Institute of Romanian Academy, 5, Cutitul de Argint Street, 040557 Bucharest, Romania
    \label{aira}
     \and %---- Dumas
     Thirty-Meter-Telescope, 100 West Walnut St, Suite 300, Pasadena, CA 91124, USA
     \label{tmt}
     \and %---- Castillo-Rogez
     Jet Propulsion Laboratory, California Institute of Technology, 4800 Oak Grove Drive, Pasadena, CA 91109, USA
     \label{jpl}
     \and %---- Cipriani
     European Space Agency, ESTEC - Scientific Support Office, Keplerlaan 1, Noordwijk 2200 AG, The Netherlands
     \label{estec}
    \and %---- Fusco2
     The French Aerospace Lab BP72, 29 avenue de la Division Leclerc, 92322 Chatillon Cedex, France
     \label{onera}
     \and %---- Grice
     Open University, School of Physical Sciences, The Open University, MK7 6AA, UK
     \label{ou}
     \and %---- Lamy
     Laboratoire Atmosph\`eres, Milieux et Observations Spatiales, CNRS \& 
    Universit\'e de Versailles Saint-Quentin-en-Yvelines, Guyancourt, France
    \label{lamos}
     \and %---- Pajuelo
     Secci{\'o}n F{\'i}sica, Departamento de Ciencias, Pontificia Universidad Cat{\'o}lica del Per{\'u}, Apartado 1761, Lima, Per{\'u}
     \label{puc}
     %
%     \and %---- Edyta only
%     Institute of Physics, University of Szczecin, Wielkopolska 15, 70-453 Szczecin, Poland
%     \label{edyta}
     %
     \and %---- Santana-Ros
     Departamento de Fisica, Ingenier\'ia de Sistemas y Teor\'ia de la Señal, Universidad de Alicante, Alicante, Spain
    \label{uda}    
     \and %---- Santana-Ros
    Institut de Ci\'encies del Cosmos (ICCUB), Universitat de Barcelona (IEEC-UB), Martí Franqu\'es 1, E08028 Barcelona, Spain
    \label{iccub}
     \and     %---- Yang
     European Southern Observatory (ESO), Alonso de Cordova 3107, 1900 Casilla Vitacura, Santiago, Chile
     \label{eso}
}

   \date{Received x-x-2020 / Accepted x-x-2019}
% \abstract{}{}{}{}{} 
% 5 {} token are mandatory
 
  \abstract
   {The recent estimates of the 3D shape of the M/Xe-type triple asteroid system (216)~Kleopatra indicated a density of $\sim$5 g.cm$^{-3}$, which is by far the highest for a small Solar System body. Such a high density implies a high metal content as well as a low porosity which is not easy to reconcile with its peculiar ``dumbbell'' shape. 
   }
  {Given the unprecedented angular resolution of the VLT/SPHERE/ZIMPOL camera, here, we aim to constrain the mass (via the characterization of the orbits of the moons) and the shape of (216)~Kleopatra with high accuracy, hence its density.
   } 
   {We combined our new VLT/SPHERE observations of (216)~Kleopatra recorded during two apparitions in 2017 and 2018 with archival data from the W.M. Keck Observatory, as well as lightcurve, occultation, and delay-Doppler images, to derive a model of its 3D shape using two different algorithms (\adam{}, MPCD). Furthermore, an N-body dynamical model allowed us to retrieve the orbital elements of the two moons as explained in the accompanying paper.}
   {The shape of (216)~Kleopatra is very close to an equilibrium dumbbell figure with two lobes and a thick neck. Its volume equivalent diameter $(118.75\pm 1.40)\,{\rm km}$ and mass $(2.97\pm 0.32)\cdot 10^{18}\,{\rm kg}$ (i.e., 56\% lower than previously reported) imply a bulk density of $(3.38\pm 0.50)\,{\rm g}\,{\rm cm}^{-3}$. Such a low density for a supposedly metal-rich body indicates a substantial porosity within the primary. This porous structure along with its near equilibrium shape is compatible with a formation scenario including a giant impact followed by reaccumulation. 
   (216)~Kleopatra's current rotation period and dumbbell shape imply that it is in a critically rotating state. The low effective gravity along the equator of the body, together with the equatorial orbits of the moons and possibly rubble-pile structure, opens the possibility that the moons formed via mass shedding.
   }
   %A calculation of the effective potential with the current period shows that the system could have formed via an oblique collision. The moons could be the by-product of gravitational instabilities. %Unique shape of (216)~Kleopatra- lobes are ellipsoidal, neck is thick, not a bilobated-shape density  is low. Can’t be explained with the formation. Formation is unclear. Is that a trait of M-type asteroids? Atypique objects of the solar system (radiogenic heat). Moons orbits are not constrained with well-known orbital model. why?
    {    
    (216) Kleopatra is a puzzling multiple system due to the unique characteristics of the primary. This system certainly deserves particular attention in the future, with the { Extremely Large Telescopes} and possibly a dedicated space mission, to decipher its entire formation history. %Now that we have laid out a good estimate of its shape and moons' orbits, future work should be dedicated to characterizing the largest critically-rotating asteroid known so far.
    }
 

  \maketitle

%%%%%%%%%%%%%%% INTRO %%%%%%%%%%%%%%%%%%%%%%%%%%%%%%%%%%%%%%%%%%%%%%%%%%%%%%%%%%%%%
\section{Introduction}\label{sec:introduction}
Moons around main-belt asteroids have been known to exist since the discovery in 1993 of Dactyl, the companion of (243)~Ida \citep{Binzel1995}. Since then, {using mostly lightcurve inversion, but also} with the use of adaptive optics (AO) on 8-10m class telescopes and the Hubble Space Telescope (HST), $\sim$190 multiple systems have been discovered \citep{Johnston2020}, starting in 1998 with Petit-Prince, around (45) Eugenia \citep{Merline1999}, followed by the discovery of the first triple asteroid (87)~Sylvia a few
years later \citep{Marchis2005}. Today, about 30 of them have been observed by direct imaging (AO on 8-10m class telescopes, HST), providing insights into their formation and evolution \citep{Yang2016}. { The images bring indirect information about the interior based on direct information about mass and volume,} and hence density \citep{Margot2015,Scheeres2015}.\\

The arrival of a second generation of AO, such as the Spectro-Polarimetric High-contrast Exoplanet Research instrument (SPHERE) at the Very Large Telescope \citep[VLT,][]{Beuzit2019} and the Gemini Planet Imager (GPI) at GEMINI-South \citep{Macintosh2014}, offers a great opportunity to constrain, via direct imaging, the 3D shape and mass of large multiple asteroid systems where the primary's diameter exceeds 100 km. In 2017, we started a survey of about forty large (D$\geq$100km) main-belt asteroids through a European Southern Observatory (ESO) large programme \citep[id: 199.C-0074,][]{Vernazza2018}, including the following six known multiple systems: (22)~Kalliope, (41)~Daphne \citep{Carry2019}, (45)~Eugenia, (87)~Sylvia \citep{Carry2021arXiv}, (130)~Elektra, and (216)~Kleopatra. In addition, our programme allowed the discovery of a new binary asteroid \citep[(31)~Euphrosyne,][]{Yang2020a, Yang2020b}.\\ 

Among the six known systems, (216)~Kleopatra is of particular interest because of the various density estimates reported for this object, ranging from $\sim$3.6 g.cm$^{-3}$ to $\sim$5 g.cm$^{-3}$ \citep{Descamps2011,Hanus2017b,Shepard2018}, and for its unique dumbbell shape so far {\citep{Ostro2000,Descamps2015,Shepard2018}}. (216)~Kleopatra is a {M/Xe-type \citep{DeMeo2009,Hardersen2011} triple asteroid system \citep{Descamps2011}
with a high radar albedo \citep{Shepard2018},} likely implying the presence of a substantial fraction of metal at its surface.

Here, we present new AO  observations of (216)~Kleopatra with VLT/SPHERE/ZIMPOL (Zurich Imaging Polarimeter), which were obtained as part of our ESO large programme (Sect.~\ref{sec:data}). Combining these new observations, with disk-integrated photometry, stellar occultations, and delay-Doppler images, we derived two shape models with the \adam{} and MPCD reconstruction methods \citep{Viikinkoski2015,Capanna2012} (Sect.~\ref{sec:shape}). The images were further used to constrain the orbital properties of the two moons and thus constrain the mass of (216)~Kleopatra, and hence its density (see Sect.~\ref{sec:moon} and accompanying paper by Bro\v{z} et al.). In Sect.~\ref{sec:analysis}, we perform a thorough analysis of Kleopatra's shape and propose a formation scenario of this peculiar triple system in Sect.~\ref{sec:discussion}.
%%% global picture

%%%%%%%%%%%%%%% DATA %%%%%%%%%%%%%%%%%%%%%%%%%%%%%%%%%%%%%%%%%%%%%%%%%%%%%%%%%%%%%%%%%%%%%
\section{Observations \& data reduction}\label{sec:data}

\subsection{Disk-resolved data with SPHERE}\label{sec:ao}

Asteroid (216)~Kleopatra was observed at two different epochs in July--August 2017 and December 2018{ -- January 2019}, { using ZIMPOL of SPHERE} \citep{Thalmann2008} in the classical imaging mode with the narrow band filter (N$\_$R filter; filter central wavelength = 645.9 nm, width = 56.7 nm). The angular size of Kleopatra was in the range of 0.09--0.11$\arcsec$. At the time of the observations, the asteroid was close to an equator-on geometry. Therefore, the SPHERE images of Kleopatra obtained from seven epochs allowed us to reconstruct a reliable 3D shape model with well defined dimensions. The reduced images were further deconvolved with the \mistral algorithm \citep{Fusco2003}, using a parametric point-spread function \citep{Fetick2019}.
Table~\ref{tab:ao} lists information about the images, while  Figs.~\ref{fig:Deconv1} and~\ref{fig:Deconv2} display all obtained images with SPHERE.

We complemented our dataset with 14 disk-resolved images obtained by the NIRC2 camera mounted on the W. M. Keck II telescope {(Table~\ref{tab:aoKeck} and Fig.~\ref{fig:comparisonKeck})}. These data were already compiled and used for Kleopatra's shape modeling in the study of \citet{Hanus2017b}.

{The pixel scale of the Zimpol instrument is 3.6 mas, which is almost a factor of three improvement compared to the Keck's NIRC2 camera with a pixel scale of 9.942 mas. We also note that the pixel scale of the VLT/NACO instrument, the decommissioned predecessor of SPHERE, was 13.24 mas.}

\subsection{Disk-integrated optical photometry}\label{sec:lcs}

We compiled a rich dataset of Kleopatra's disk-integrated optical photometry (180 lightcurves from 15 different apparitions). The oldest data were obtained in 1977 \citep{Scaltriti1978} and the most recent ones in 2015 (St\'{e}phane Fauvaud). A large fraction of the data spans seven different apparitions in the 1980s \citep[e.g.,][]{Pilcher1982, Weidenschilling1987}. Additionally, we used data from apparitions in 1994 \citep{Fauvaud2001} and 2006 \citep{Warner2006c}. Finally, the largest dataset covering apparitions in 2008, 2010, 2011, and 2013 comes from the SuperWASP survey \citep{Grice2017}. The list of lightcurves is summarized in Table~\ref{tab:lcs}{; a subset is then shown in Figs.~\ref{fig:comparisonLC1},~\ref{fig:comparisonLC2}. We did not use any sparse data for their redundancy. Each lightcurve densely sampled the brightness variations for several hours. We treated the lightcurves as relative only, so we normalized the fluxes to unity. We corrected the epochs on light-time effect. Because each lightcurve is only several hours long, it was not necessary to correct for the phase angle effect. }

\subsection{Stellar occultations}\label{sec:occ}

We utilized five stellar occultations of Kleopatra {(Table~\ref{tab:occ}, \citealt{Herald2020})}. While three (from 2009, 2015, and 2016) are of a sufficient quality to be utilized for the shape modeling (i.e., multiple chords sampling the object's projection well), the remaining two (from 1980 and 1991) served as validity checks. We note that the occultation in 1980 is of particular historical interest as two observers independently spotted a 0.9-second star disappearance too far from the primary to be related with it. At that time, the scientific community was not yet ready to accept the existence of tiny moons around asteroids. This, however, changed in 1994 when the Galileo probe sent images of asteroid (243)~Ida with its moon Dactyl. Fortunately, the awareness about these peculiar data persisted. The evident explanation of the data is the occultation by one of Kleopatra's two moons. We list the suspected moon position in Table~\ref{tab:positions}. 

\subsection{Delay-Doppler images}\label{sec:radar}

Delay-Doppler images of Kleopatra were obtained in 2008 and 2013 using the 2380 MHz radar at the Arecibo observatory \citep{Shepard2018}. The 2008 observations were almost equatorial, but with a weak signal-to-noise ratio. {Therefore we did not include them in the modeling}. Higher quality observations in 2013 had an aspect angle of --50$^{\circ}$ from the equatorial plane. The nominal range resolution is {5.25} km in range and 10 Hz in frequency.

%%%%%%%%%%%%%%% ADAM & MPCD %%%%%%%%%%%%%%%%%%%%%%%%%%%%%%%%%%%%%%%%%%%%%%%%%%%%%%%%%%%%%%%%%%%%%
\section{3D shape modeling}\label{sec:shape}

\subsection{ADAM shape model}\label{sec:ADAM}
%\setkeys{Gin}{draft=false}
\begin{figure*}%[!t]
\begin{center}
\resizebox{0.81\hsize}{!}{\includegraphics{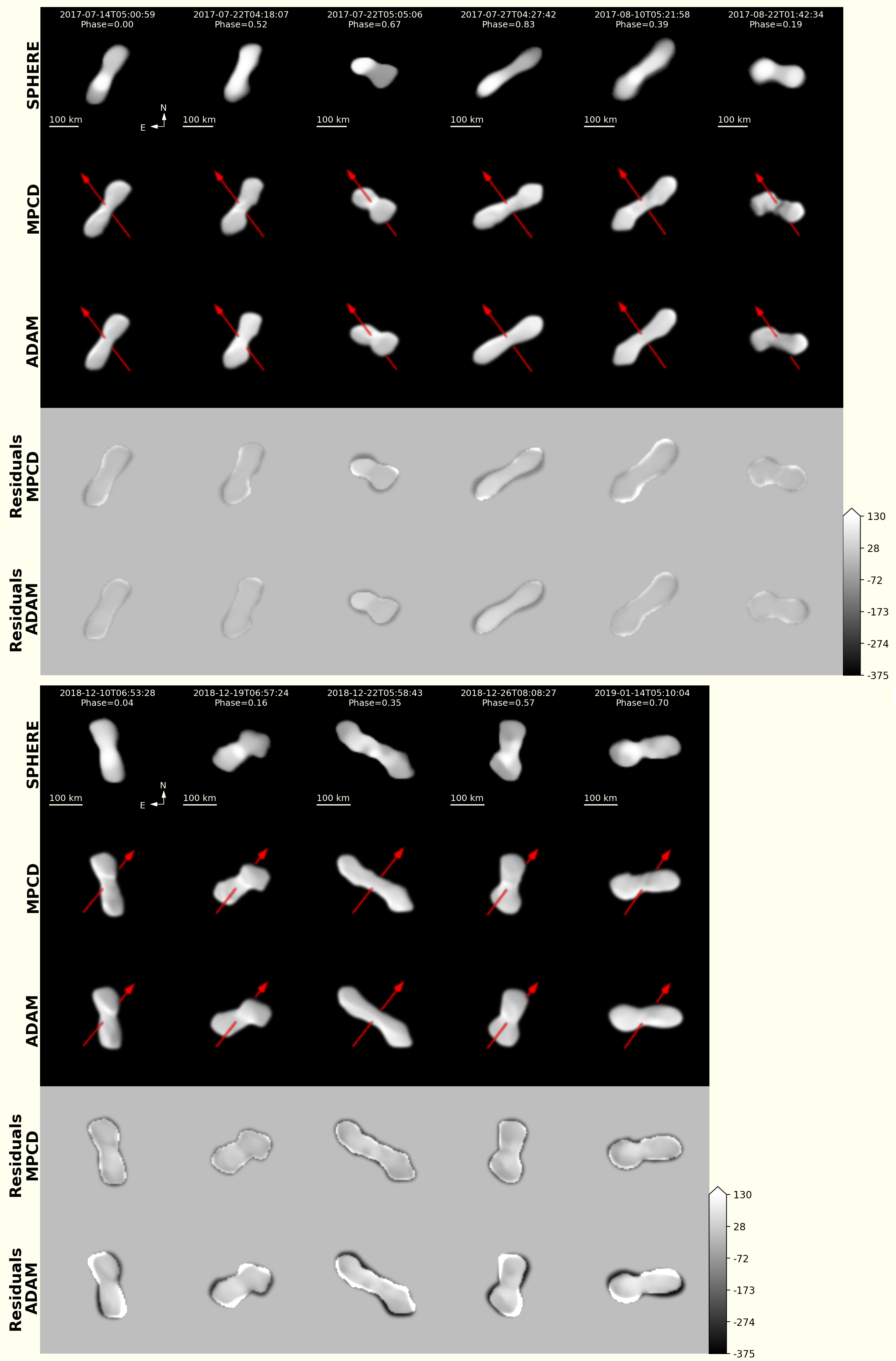}}
%\resizebox{0.99\hsize}{!}{\includegraphics{./figs/octdec_de1Ph2_1.png}}\\
\end{center}
\caption{\label{fig:compAO}Comparison between the VLT/SPHERE/ZIMPOL deconvolved images of Kleopatra and the corresponding projections of our MPCD and \adam{} shape models. The red arrow indicates the orientation of the spin axis. We used a realistic illumination to highlight the local topography of the model using the OASIS software \citep{Jorda2010}. The residuals of both models are shown in the two bottom rows, more specifically those are chi-square pixel residuals in units of the instrumental noise associated to each pixel (photon and readout noise).}
\end{figure*}
%\setkeys{Gin}{draft=true}

All-Data Asteroid Modeling \citep[\adam{},][and references therein]{Viikinkoski2015} is an inversion algorithm commonly used for the reconstruction of shape models of asteroids from their combined disk-integrated and disk-resolved data \citep{Viikinkoski2015b, Viikinkoski2018, Hanus2017b}. The key elements of \adam{} are the a priori knowledge of the rotation state (i.e., sidereal rotation period and spin vector orientation) and the existence of disk-resolved data. The former is usually available as convex shape models have been derived for the majority of the largest asteroids \citep[see the Database of Asteroid Models from Inversion Techniques, DAMIT\footnote{ \url{https://astro.troja.mff.cuni.cz/projects/damit/}},][]{Durech2010}. The most common disk-resolved data are the {high-resolution angular} images obtained with the 8--10m class telescopes equipped with AO systems (Keck, VLT, Gemini), but also the more scarce delay-Doppler images \citep{Shepard2018} or the ALMA interferometry \citep{Viikinkoski2015b}. Finally, stellar occultations can also be considered as disk-resolved data; however, only those with multiple chords with proper timings, sampling the asteroid's on-sky projection well, are useful for constraining the shape.    

We applied ADAM to our dataset of 180 optical lightcurves, 14 disk-resolved images from Keck, 55 disk-resolved images from SPHERE, three stellar occultations, and {15} delay-Doppler images from the Arecibo {Observatory} \citep{Shepard2018}. We used the rotation state \citep{Hanus2017b} as an initial value for the \adam{} modeling with a low shape model resolution {(1152 facets)} and the octantoid shape parametrization \citep{Viikinkoski2015}. Then, we increased the shape model resolution {(2048 facets)} and used the low-resolution model as a starting point for the shape model improvement. We also increased the relative weight of the SPHERE data with respect to other datasets. We show the comparison of the shape model projections with the corresponding SPHERE and Keck/NIRC2 images in Figs.~\ref{fig:compAO} { and \ref{fig:comparisonKeck},} and with the stellar occultations in Fig.~\ref{fig:comparisonOCC}. { The fit to a subset of optical lightcurves is shown in Figs.~\ref{fig:comparisonLC1},~\ref{fig:comparisonLC2}}. Our solution is robust against variations in data weighting and \adam{} regularization functions. We generated several models both with and without delay-Doppler images. {The optimization method used for radar images is described in detail in \citet{Viikinkoski2015}. We used the cosine scattering law with constant albedo.
By increasing the relative weight of radar data with respect to AO images, the shape solution approaches the shape presented in \citet{Shepard2018}. The choice of weights between different data sources is always a somewhat subjective matter. However, in this case, both the coverage and the resolution of AO images is clearly superior compared to radar images, so it seems prudent that the AO observations from SPHERE are given predominance.}  The comparison between radar data and the shape model is shown in Fig.~\ref{fig:comparisonRD}.

{Our final \adam{} shape model fits all datasets sufficiently well. Specifically, we have not identified any substantial disagreement between the model and the data. Considering the superior quality, the resolution, and the coverage of the SPHERE data, the shape model is already well constrained by them. The other data (Keck, occultations, radar) are usually fitted naturally and are mostly complementary. }

\subsection{MPCD shape model}\label{sec:MPCD}
\begin{figure*}%[!t]
\begin{center}
\begin{tabular}{ccc}
Radar & \adam{} & MPCD \\
\includegraphics[width=0.3\linewidth]{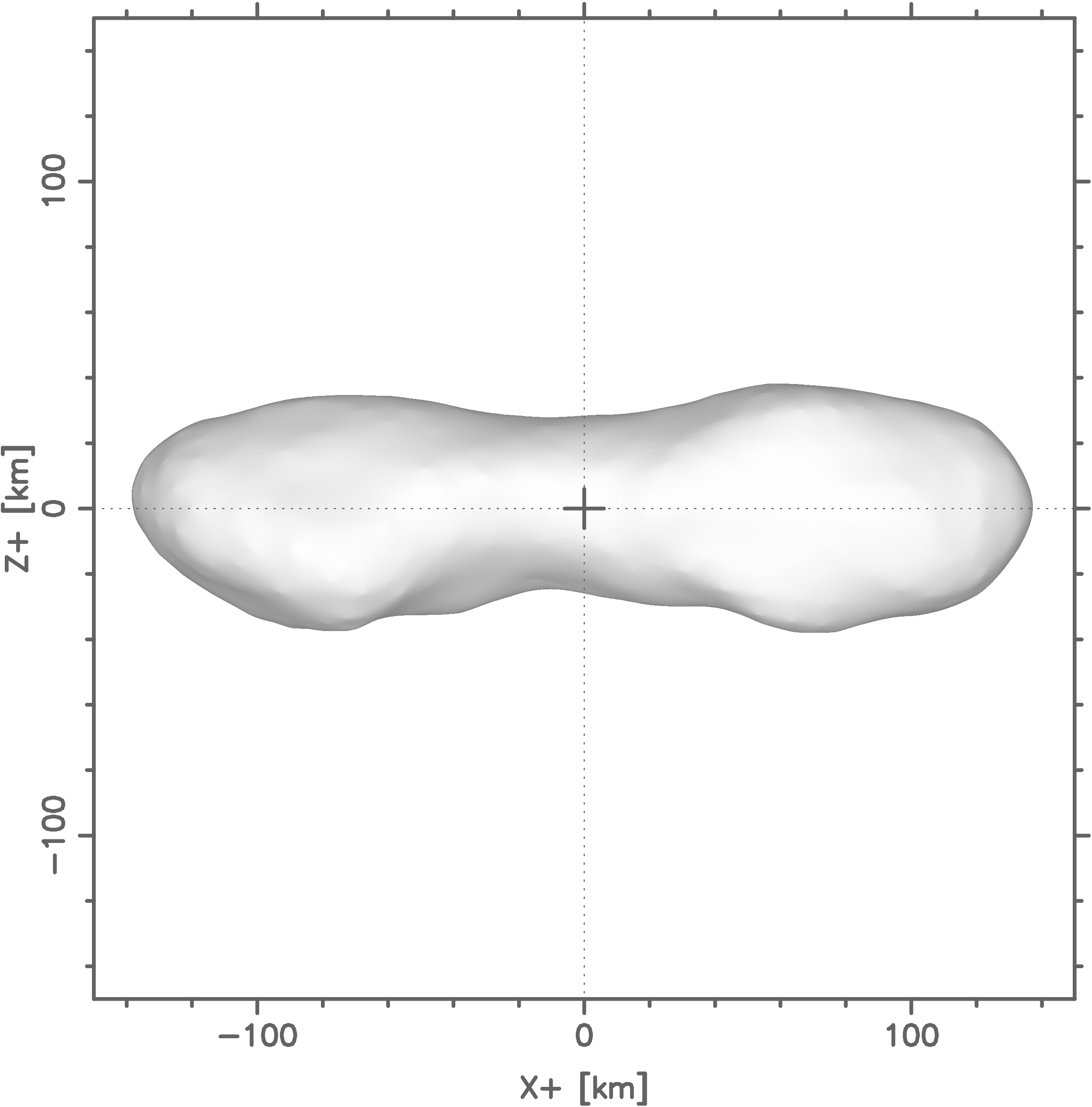} &
\includegraphics[width=0.3\linewidth]{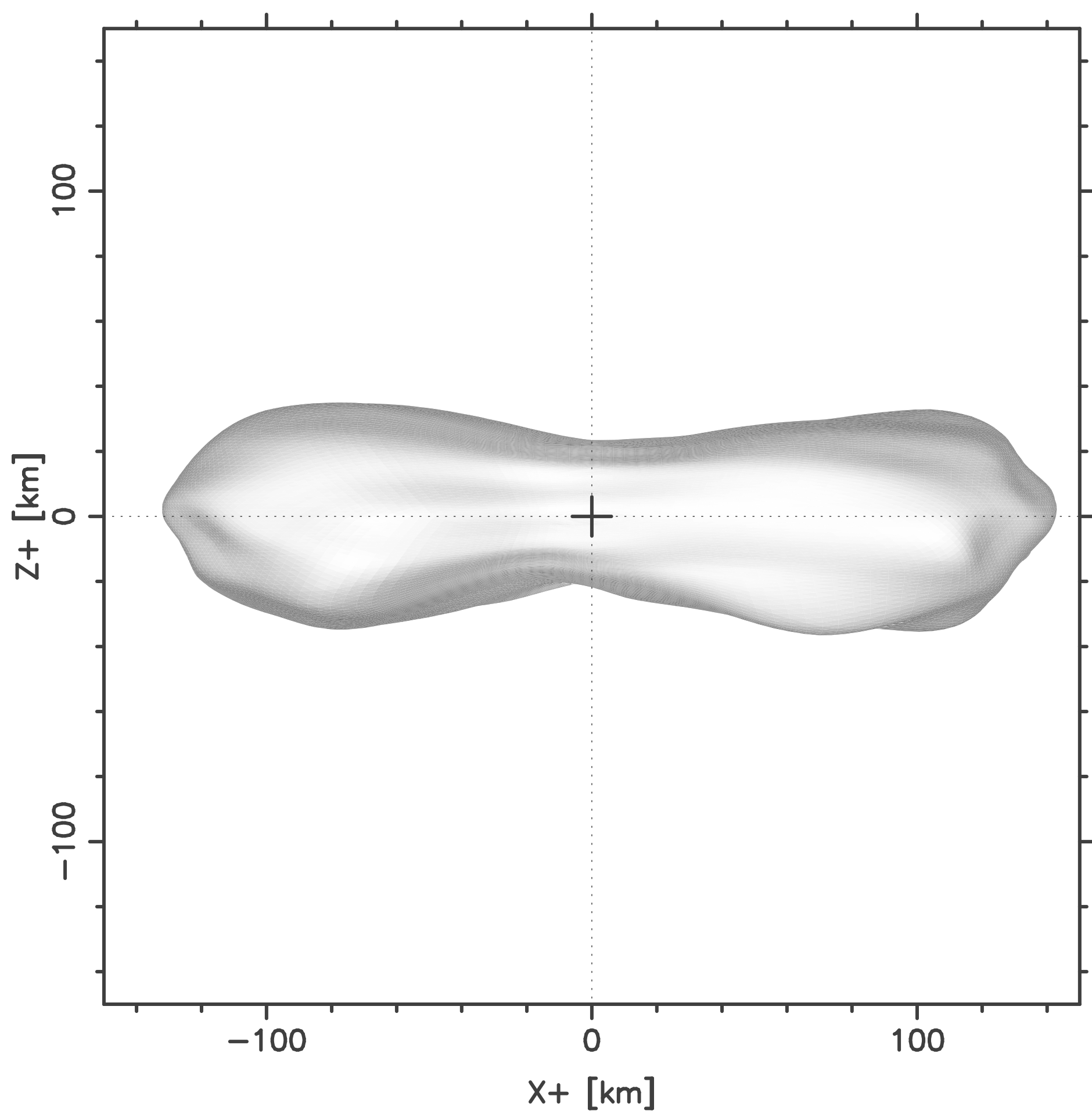} &
\includegraphics[width=0.3\linewidth]{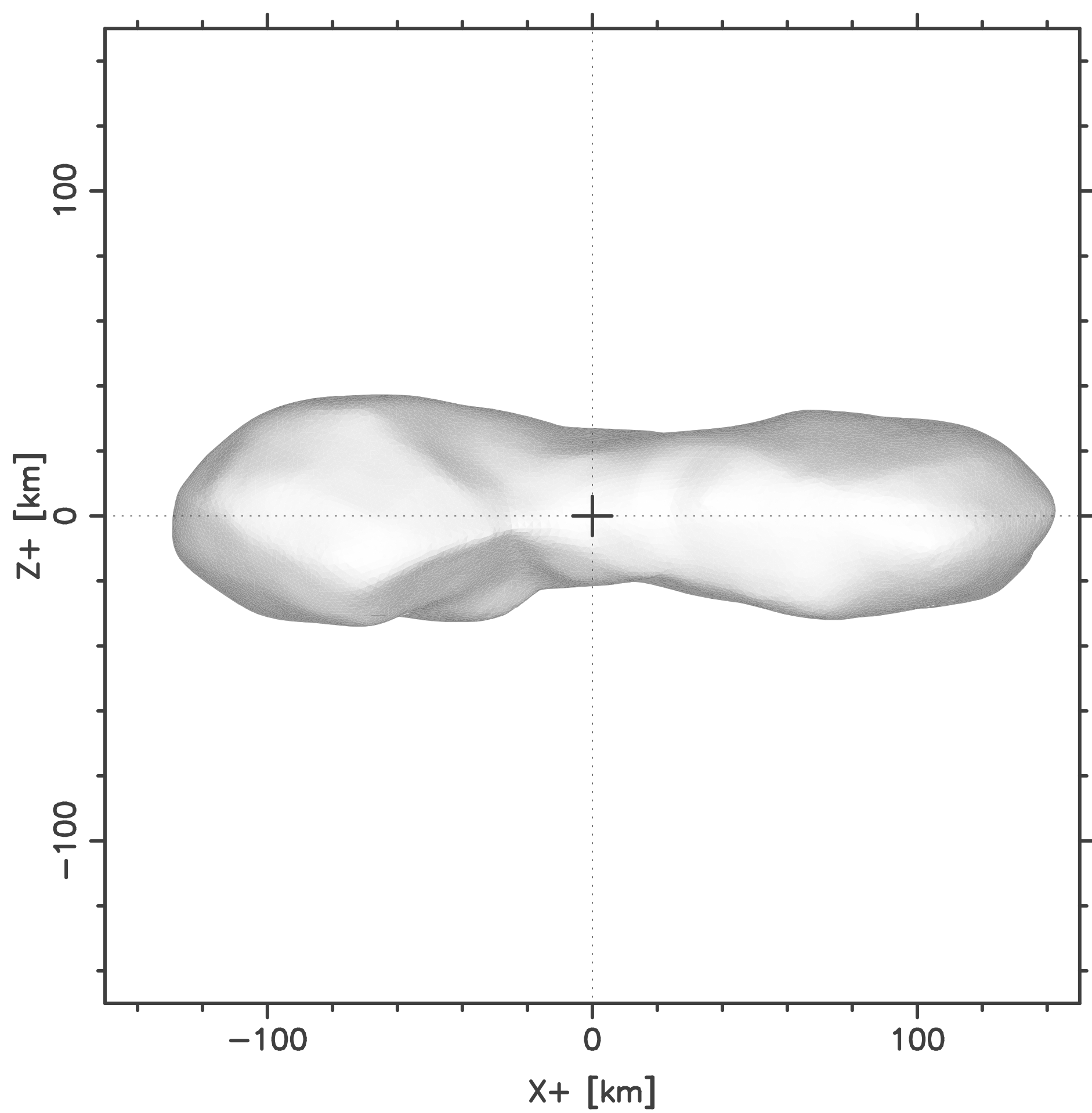} \\
\includegraphics[width=0.3\linewidth]{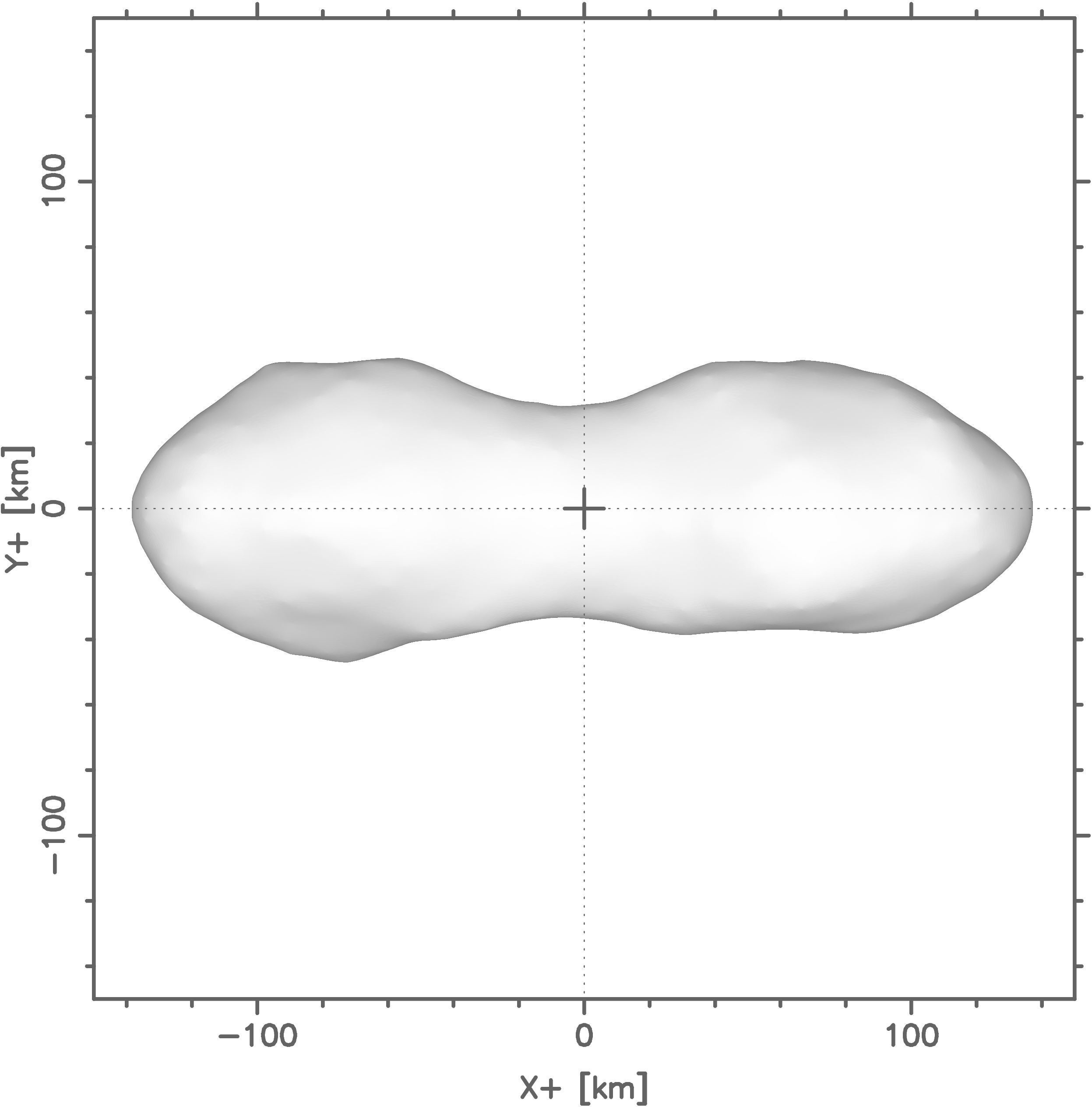} &
\includegraphics[width=0.3\linewidth]{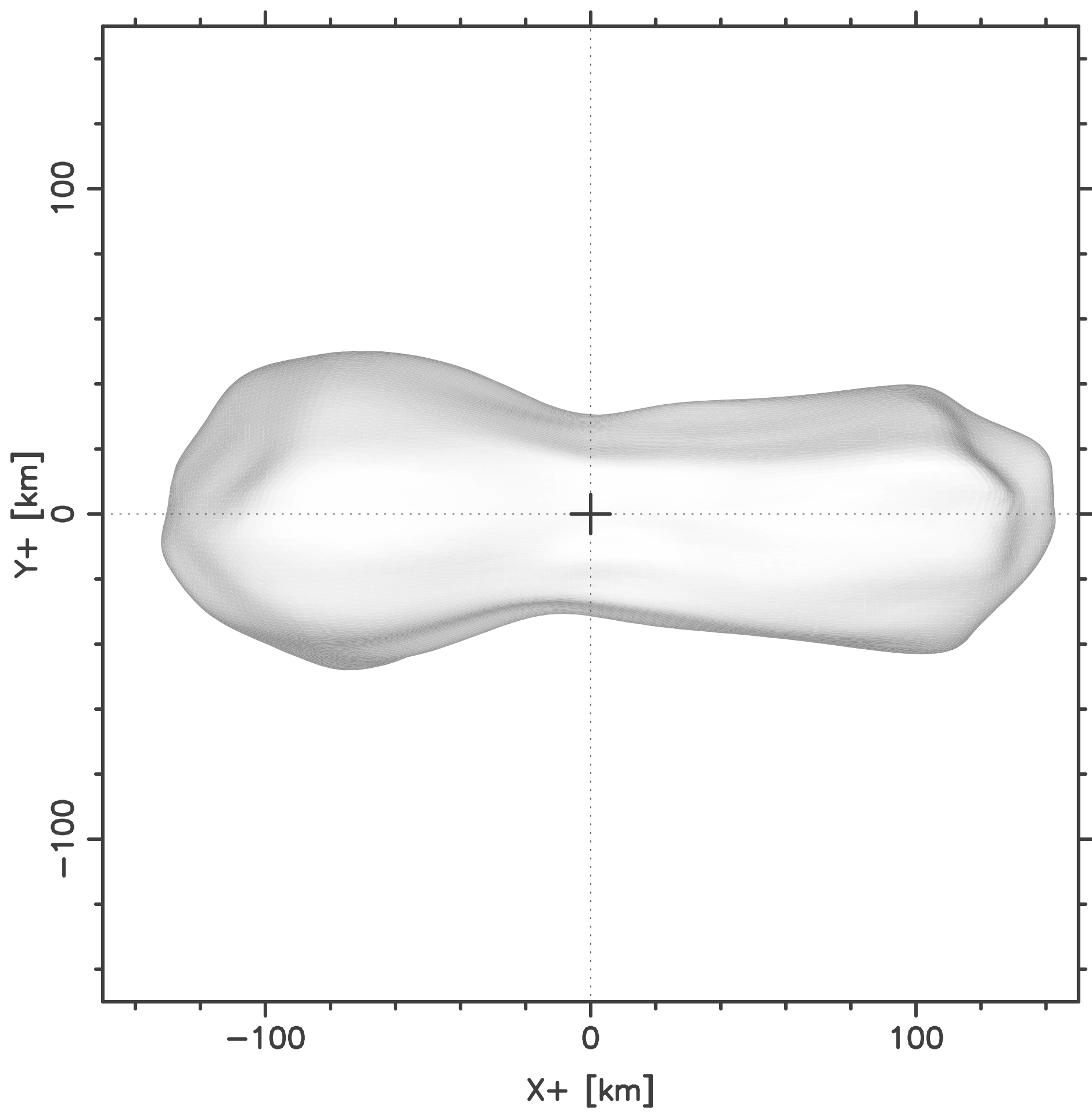} &
\includegraphics[width=0.3\linewidth]{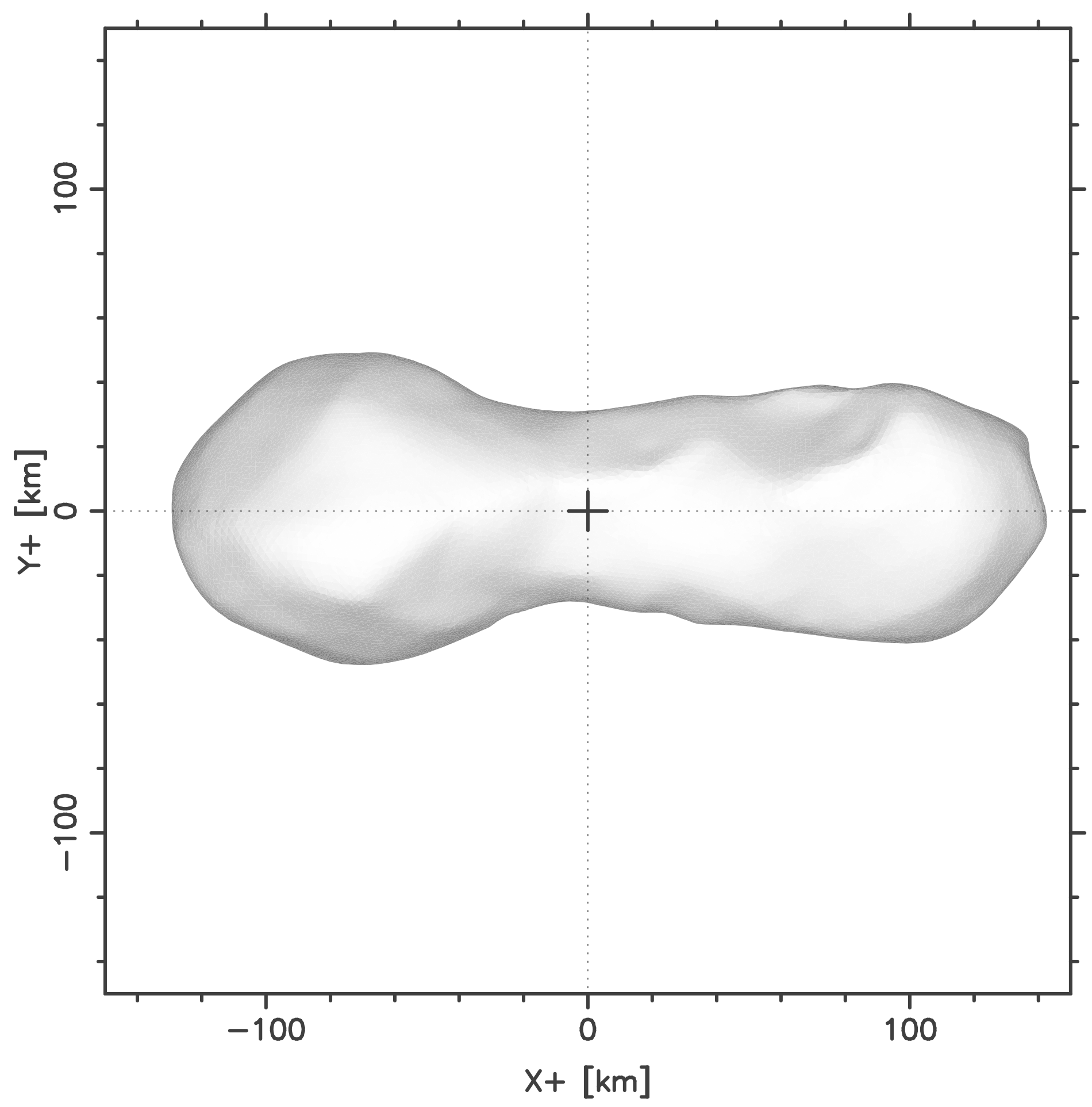} \\
\end{tabular}
\end{center}
\caption{\label{fig:compshapes}Rendered views of the radar shape model of \cite{Shepard2018} (left panels), and \adam{} and MPCD models (middle and right panels) as seen from the $Y-$ (top panels) and $Z+$ (bottom panels) body axes. The mass deficit or crater is visible on the small lobe of the MPCD model.}
%\caption{\label{fig:comparison3}Rendered views of the radar shape model (left panels) of \cite{Shepard2017}, \adam{} model (middle panels} and MPCD model (right panels) as seen from the $Y-$ (top panels) and $Z+$ (bottom panels) body axes.}
\end{figure*}

%{\bf LAURENT}
The Multiresolution PhotoClinometry by Deformation \citep[MPCD,][]{Capanna2012,Jorda2016} is a 3D shape reconstruction method that utilizes an initial shape model to give a better fit to disk-resolved images. Therefore, our \adam{} shape model is further modified by MPCD by fitting solely the high-resolution SPHERE AO data. 
The MPCD algorithm minimizes the chi-square pixel-to-pixel differences between a set of observed images and the synthetic images built from the shape model for optimization.
The reflectance function is Hapke's five-parameters function with the parameters listed in \citet{Descamps2011}.
The shape is optimized through shifts of the vertices along the local normal vector.
The method goes through several increasing steps of resolutions of both the observed images and shape before converging toward the final optimized model.
In the case of (216)~Kleopatra, we used the sample of 33 SPHERE images obtained during 11 visits.
The shape was reconstructed starting from a decimated \adam{} model with only 196 facets and ending with a final model of 3136 facets, after optimization in three levels.
Furthermore, we also optimized the Euler angles describing the orientation of the spin pole after noticing unusual systematic residuals between our sets of observed and synthetic images.
The shape was reoptimized with this new pole orientation in the same manner to produce the final MPCD model of (216)~Kleopatra.
As expected, the chi-square between observed and synthetic AO images decreases from 135 (\adam{} model) to 50 (MPCD model).
A comparison between synthetic SPHERE images generated from the \adam{} and MPCD models and an observed image for each visit is shown in Fig.~\ref{fig:compAO}.
The final MPCD model is shown in Fig.~\ref{fig:compshapes} alongside the \adam{} model and the radar model of \cite{Shepard2018}. {The physical properties of those three models are listed in Tab. \ref{tab:param}.}
The MPCD method also provides an albedo map calculated together with the slope and height errors of each facet from their corresponding average residual pixel values.
However, the only significant albedo features are found near limbs and a careful inspection shows that they likely correspond to faint artifacts introduced by the deconvolution process \citep{Fetick2019} at the edges of the object.

\begin{table}
%\resizebox{0.99\hsize}{!}
 \caption{\label{tab:param}
 Physical properties of (216)~Kleopatra based on \adam{} and MPCD shape modeling of our VLT/SPHERE images: Sidereal rotation period $P$, spin-axis ecliptic J2000 coordinates $\lambda$ and $\beta$, volume-equivalent diameter $D$, dimensions along the major axis $a$, $b$, $c$, their ratios $a/b$ and $b/c$, mass $M$ from Bro\v z et al. (accompanying paper), volume $V$, and bulk density $\rho$. Uncertainties correspond to 1\,$\sigma$ values. The values based on radar data \citep{Shepard2018} are also reported. { The $b$ and $c$ extents reported for the radar model are the maximum extents in those directions.}}
 \centering
% \begin{threeparttable}
% \centering
 \begin{tabular}{lcccc}
\hline
  Parameter            & S18              & \adam{}      & MPCD\\  \hline
   $P$ (h)             & 5.385280(1)      & 5.385282(1)  & 5.385282(1) \\
   $\lambda$ ($\degr$) & 74$\pm${ 2}   & 73.5$\pm$0.5 & 74.1$\pm$0.5 \\
   $\beta$ ($\degr$)   & 20$\pm${ 2}   & 20.8$\pm$0.5 &  21.6$\pm$0.5 \\
   $D$ (km)            & 122$\pm${ 10} & \DiamADAM    & 118.2$\pm$0.8 \\
   $a$ (km)            & 276$\pm${ 14} & 270$\pm$4    & 267$\pm$6 \\
   $b$ (km)            & 94$\pm${ 5}   & 62$\pm$4     & 61$\pm$6 \\
   $c$ (km)            & 78$\pm${ 4}   & 38$\pm$4     & 48$\pm$6 \\
   $a/b$               &   { 2.9}      & 4.35$\pm$0.3 & 4.4$\pm$0.4 \\
   $b/c$               &   1.20           & 1.63$\pm$0.2 & 1.3$\pm$0.2 \\
   $V$ ($10^5$~km$^3$) &     { 9.56}   &  8.90$\pm$0.45 & 8.65$\pm$0.17 \\
%%   $R_{eq}$ (km)       &      ???     &    59.6      & 58.2$\pm$XX \\
   $M$ ($10^{18}$~kg)  & 4.64$\pm$0.02    & 2.97$\pm$0.32  & 2.97$\pm$0.32 \\
   $\rho$ (g cm$^{-3}$)& 4.9$\pm$0.5      & 3.34$\pm$0.53  & 3.43$\pm$0.38 \\
  \hline 
$\chi^2$               &      {209}        & 135          & 50  \\ 
\hline
 \end{tabular}
% \begin{tablenotes}[para,flushleft]
%     \centering $^a$ \citet{Michalak2001}. 
% \end{tablenotes}
% \end{threeparttable}
\end{table}

\section{Mass and bulk density}\label{sec:moon}
%\section{Positions of the satellites}\label{sec:moon}

\begin{figure*}
\includegraphics[width=\textwidth]{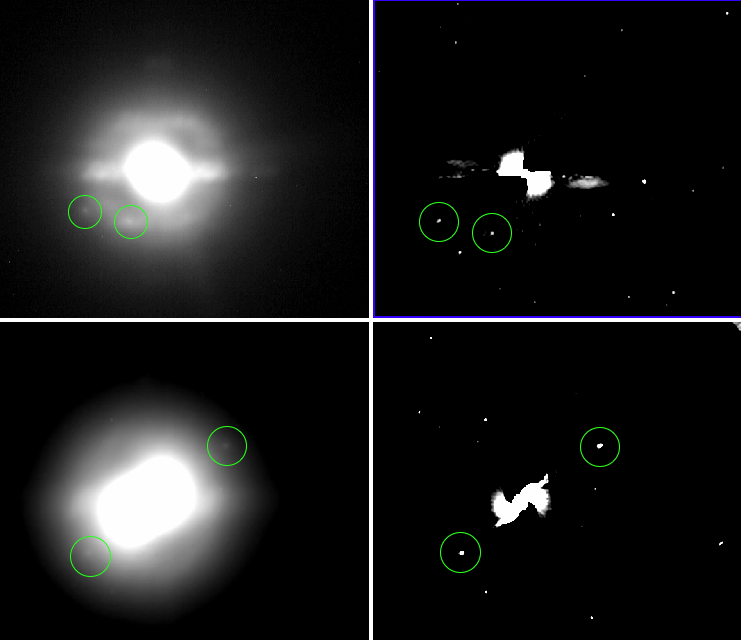}
\caption{Processed ZIMPOL images on the left, revealing the presence of the satellites, CleoSelene and AlexHelios around (216)~Kleopatra at two epochs (bottom: 2017-07-14, top: 2017-08-22). To reveal the moons which are as faint as the halo (due to imperfect AO correction), we subtracted a rotational average of the image centered on the primary (right image).  The circle points to the location of the satellites in the images. {The other dots in the images are bad pixels.}}
\label{fig:figsat}
\end{figure*}

In the first step, each image obtained with SPHERE/ZIMPOL was further processed to remove the bright halo surrounding Kleopatra, following the procedure described in detail in \cite{Yang2016} and \cite{Pajuelo2018}, and {shown in Fig.~\ref{fig:figsat}}. The residual structures after the halo removal were minimized using the processing techniques introduced in \cite{Wahhaj_etal_2013ApJ...779...80W}, where the radial structures were removed using a running median in a $\sim\,$30-pixel box in the radial direction and the images were smoothed by convolving a Gaussian function with a FWHM of $\sim\,$8~pixels. %In five of the seven epochs, a faint non-resolved source was clearly detected in the vicinity of Kleopatra (Fig~\ref{figsat}). %The discovery of the satellite, S/2019 (216) 1, was reported once we {\bf had} confirmed its genuine link with Kleopatra (CBET 4627, 2019).

In the second step, we measured the relative positions on the plane of the sky between Kleopatra
and its satellites. We used the unmodified images (i.e., without halo removal) and we fit both photocenters by a suitable 2D Gaussian (see \citealt{Carry2019}).
The dispersion of the Gaussian function for the moons was chosen
conservatively as comparable to the residual (AO-corrected) PSF.
The astrometric positions are reported in Table~\ref{tab:positions}.
Sometimes, the identification of the two satellites was ambiguous.
Nevertheless, it was possible to recover the correct identification
later  (see Bro\v z et al.). We report the corrected data here.

Furthermore, we estimated the offsets between the photocenter
and the center of mass for Kleopatra. Because the central body
is so extended and irregular, the offset may reach up to a few
milli-arcseconds, as reported in Table~\ref{tab:offset}.
We use these offset adjustments for further analysis of orbits
because our dynamical model requires the centers of mass,
not the centers of light. Alternatively, one can use
a relative astrometry of the two moons (second with respect to the first),
which is unaffected by these photocenter motions.

Uncertainties in the measurements are approximately 0.01\,arcsec,
based on repeated and/or close-in-time measurements. As of now,
we do not account for the orbital motion of the satellites during
five consecutive exposures and we take their average position,
although in principle that motion could be detected. It would however require
a fitting by an asymmetric PSF, elongated along the orbital motion.

The dynamical model required to interpret the motion of the moons is more complex due to the irregular shape of
Kleopatra, and the mutual interactions of the moons and the solar tides.
We thus used an advanced n-body model with the multipole expansion
up to the order of $\ell = 10$,
which is described in detail in Bro\v z et al.,
in order to determine the orbital elements. 
Our best solution fits the observed positions
with a root mean square (RMS) residuals of 17\,mas.
Most importantly, the phase coverage of new VLT/SPHERE observations
allowed us to derive the true periods
$P_1 = (1.822359\pm0.004156)\,{\rm d}$ and
$P_2 = (2.745820\pm0.004820)\,{\rm d}$,
which results in the revised mass
$m_1 = (2.97\pm0.32)\cdot10^{18}\,{\rm kg}$ for Kleopatra.
This is significantly lower than the previously reported value of $4.64\cdot10^{18}\,{\rm kg}$ \citep{Descamps2011}.
The orbits of both satellites are circular, prograde, and equatorial, similar to most known satellites around large main belt asteroids.
(e.g., \citealt{Marchis2008a,Berthier2014,Margot2015,Carry2019,Yang2020a}).

Taking the average volume of the \adam{} and MPCD models (Table 1), the density of Kleopatra amounts to $(3.38 \pm 0.50)\,{\rm g}\,{\rm cm}^{-3}$.
A comparison with the previous estimate (3.6 ${\rm g}\,{\rm cm}^{-3}$,  \citealt{Descamps2011}) is not pertinent given that both the mass and the volume were revised.
This density has important implications for the interpretation
of the shape in Sec.~\ref{sec:analysis}.

%\setkeys{Gin}{draft=false}
%\setkeys{Gin}{draft=true}  %%%%%%%% FIGURES ARE OFF %%%%%%%%%%

\begin{table}
\centering
\caption{\label{tab:positions}Positions of Kleopatra's satellites  with respect to its photocenter. Uncertainties are approximately 0.01\,arcsec. The position from 1980 is based on the stellar occultation and those from 2008 are taken from \citet{Descamps2011}.}
\begin{tabular}{ccrr}
\multicolumn{4}{c} {AlexHelios}\\\hline
Date & UT & $u$ [arcsec] & $v$ [arcsec] \\\hline 
1980-10-10 & 07:00        & $0.2711$  & $0.4564$ \\
2008-09-19 & 11:38:00     & $-0.18\phantom{00}$   & $0.35\phantom{00}$ \\
2008-09-19 & 11:51:00     & $-0.20\phantom{00}$   & $0.36\phantom{00}$ \\
2008-10-05 & 09:13:00     & $-0.27\phantom{00}$   & $0.37\phantom{00}$ \\
2008-10-05 & 09:49:00     & $-0.29\phantom{00}$   & $0.39\phantom{00}$ \\
2008-10-05 & 10:03:00     & $-0.32\phantom{00}$   & $0.39\phantom{00}$ \\
2008-10-09 & 05:46:00     & $-0.32\phantom{00}$   & $0.29\phantom{00}$ \\
2017-07-14 & 05:00:59     & $-0.4158$ & $ 0.2952$ \\
2017-07-22 & 04:18:07     & $-0.4262$ & $ 0.2444$ \\
2017-07-22 & 05:00:55     & $-0.4291$ & $ 0.2614$ \\
2017-08-22 & 01:42:34     & $ 0.1843$ & $-0.3150$ \\ 
2018-12-10 & 06:47:17     & $-0.4894$ & $-0.0958$ \\
2018-12-19 & 06:45:02     & $ 0.1871$ & $-0.2869$ \\
2018-12-22 & 05:58:43     & $-0.2540$ & $-0.3438$ \\
2018-12-26 & 08:14:42     & $-0.4973$ & $-0.0428$ \\
2019-01-14 & 04:57:43     & $-0.5319$ & $-0.1177$ \\
\multicolumn{4}{c} {CleoSelene}\\\hline
2008-09-19  & 06:17:00     & $-0.25\phantom{00}$   & $ 0.50\phantom{00}$ \\
2008-09-19  & 08:44:00     & $-0.34\phantom{00}$   & $ 0.54\phantom{00}$ \\
2008-09-19  & 11:38:00     & $-0.44\phantom{00}$   & $ 0.57\phantom{00}$ \\
2008-09-19  & 11:51:00     & $-0.44\phantom{00}$   & $ 0.57\phantom{00}$ \\
2008-09-19  & 12:02:00     & $-0.44\phantom{00}$   & $ 0.58\phantom{00}$ \\
2008-10-05  & 09:13:00     & $-0.29\phantom{00}$   & $ 0.46\phantom{00}$ \\
2008-10-05  & 09:49:00     & $-0.31\phantom{00}$   & $ 0.47\phantom{00}$ \\
2008-10-05  & 10:03:00     & $-0.32\phantom{00}$   & $ 0.46\phantom{00}$ \\
2008-10-09  & 09:36:00     & $ 0.28\phantom{00}$   & $-0.44\phantom{00}$ \\
2017-07-14  & 05:01:00     & $ 0.3070$ & $-0.2600$ \\
2017-07-22  & 04:18:07     & $-0.2714$ & $ 0.2941$ \\
2017-07-22  & 05:00:55     & $-0.2621$ & $ 0.2963$ \\
2017-08-22  & 01:42:34     & $ 0.4423$ & $-0.2401$ \\
2018-12-10  & 06:47:17     & $ 0.1949$ & $-0.1505$ \\
2018-12-19  & 06:45:02     & $ 0.3763$ & $ 0.0929$ \\
2018-12-22  & 05:58:43     & $ 0.4555$ & $-0.0119$ \\
2018-12-26  & 08:14:42     & $ 0.3937$ & $ 0.1271$ \\
2019-01-14  & 04:57:43     & $-0.1008$ & $-0.3298$ \\
\end{tabular}
\end{table}

\begin{table}
\caption{{Positions of the photocenter minus the center of mass for Kleopatra.}}
\label{tab:offset}
\centering
\begin{tabular}{ccrr}
JD & UT & $u$ [mas] & $v$ [mas] \\\hline
2008-09-19  & 06:17:00 & $ 0.762$ & $ 0.789$ \\
2008-09-19  & 08:44:00 & $-0.948$ & $-1.395$ \\
2008-09-19  & 11:38:00 & $ 1.178$ & $ 1.069$ \\
2008-09-19  & 11:51:00 & $-1.462$ & $-0.737$ \\
2008-09-19  & 12:02:00 & $-3.723$ & $-1.784$ \\
2008-10-05  & 09:13:00 & $ 4.586$ & $ 2.911$ \\
2008-10-05  & 09:49:00 & $ 4.059$ & $ 1.363$ \\
2008-10-05  & 10:03:00 & $ 2.212$ & $-0.395$ \\
2008-10-09  & 05:46:00 & $ 0.548$ & $-2.190$ \\
2008-10-09  & 09:36:00 & $ 6.923$ & $ 5.164$ \\
2017-07-14  & 05:01:00 & $ 0.223$ & $-1.194$ \\
2017-07-22  & 04:18:07 & $ 0.609$ & $ 1.934$ \\
2017-07-22  & 05:00:55 & $ 0.666$ & $ 1.929$ \\
2017-08-22  & 01:42:34 & $ 0.288$ & $ 0.901$ \\
2018-12-10  & 06:47:17 & $-1.674$ & $-2.402$ \\
2018-12-19  & 06:45:02 & $ 0.391$ & $-3.132$ \\
2018-12-22  & 05:58:43 & $ 2.014$ & $-1.430$ \\
2018-12-26  & 08:14:42 & $-0.848$ & $ 0.791$ \\
2019-01-14  & 04:57:43 & $-0.455$ & $-1.266$ \\
\end{tabular}
\end{table}

%%%%%%%%%%%%%%%%%%%%%%%%%%%%%%%%%%%%%%%%%%%%%%%%%%%%%%%%%%%%%%%%%%%%%%

\section{Shape analysis}\label{sec:analysis}

\subsection{ Lobes interpretation}

\begin{figure*}%[!ht]
\begin{center}
\resizebox{0.49\hsize}{!}{\includegraphics{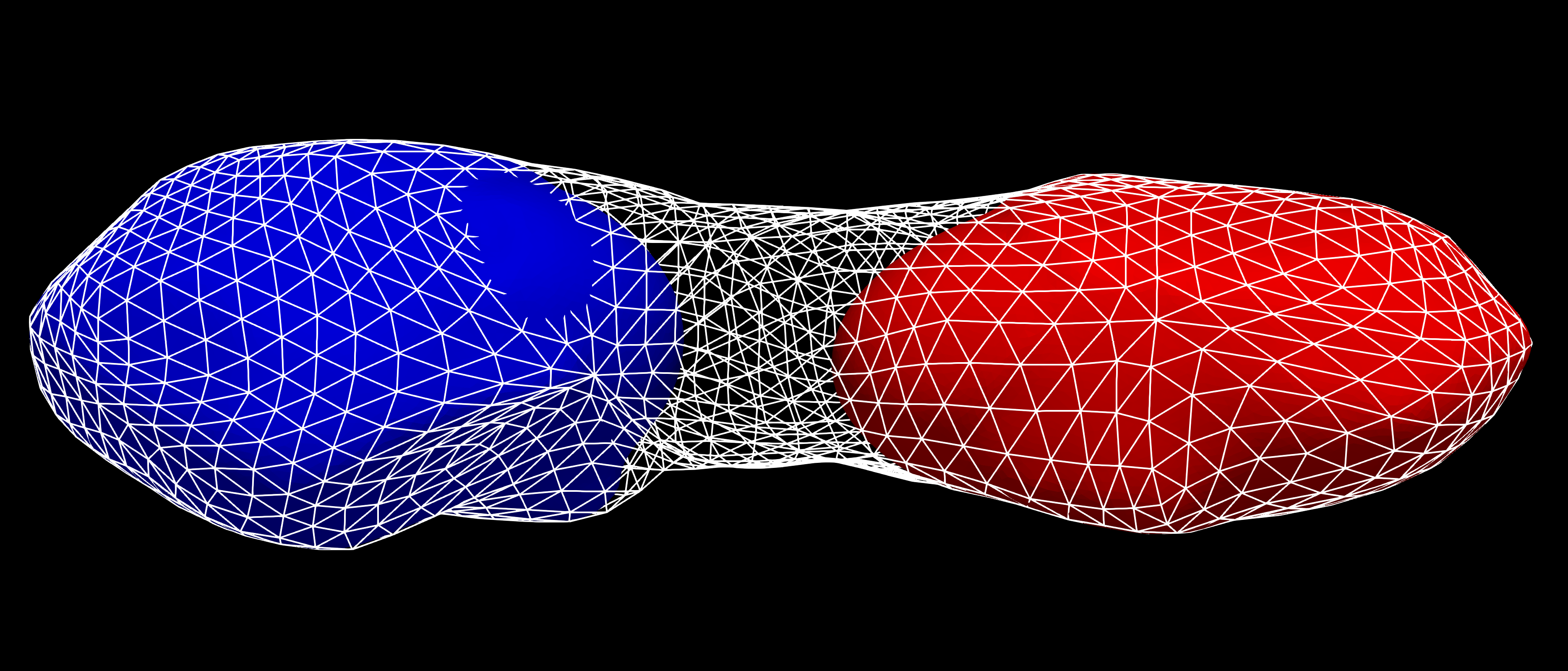}}
\resizebox{0.49\hsize}{!}{\includegraphics{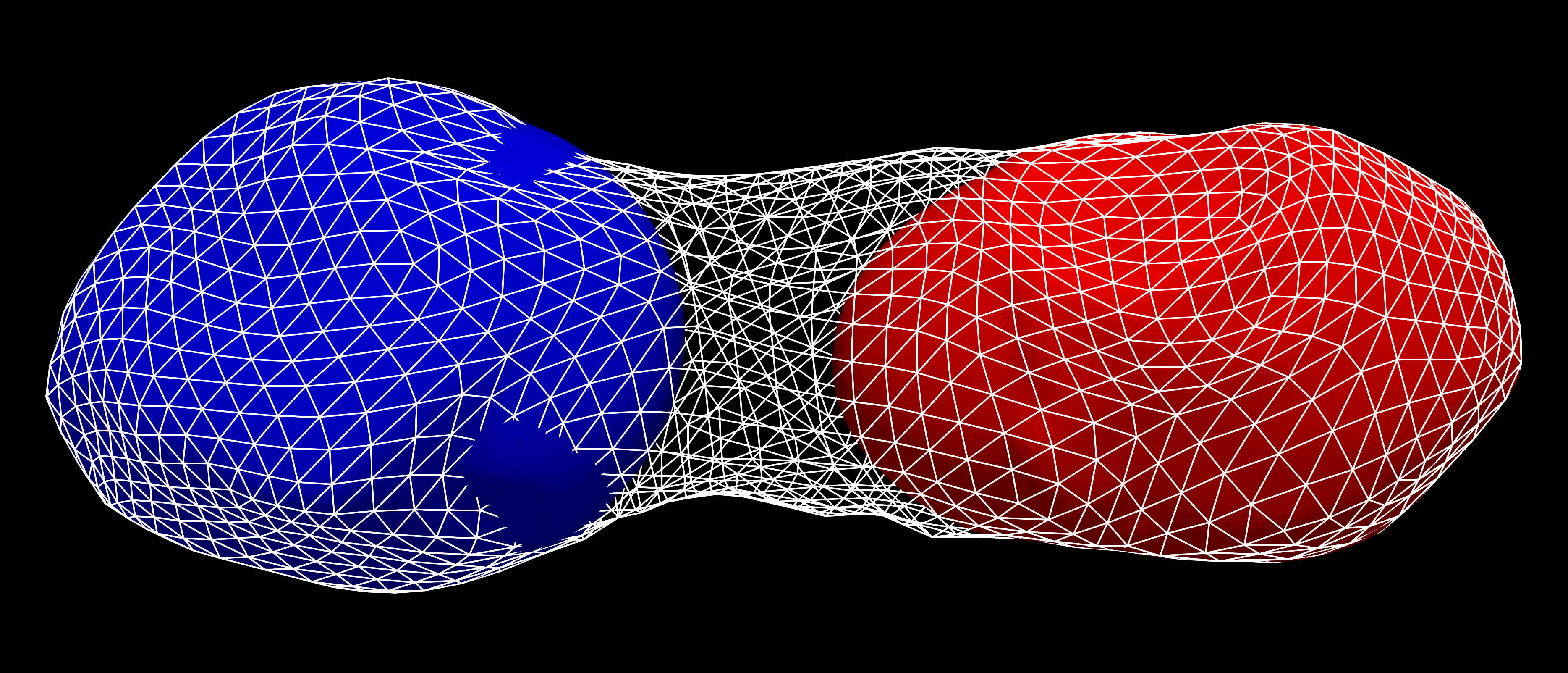}}
%\resizebox{0.3\hsize}{!}{\includegraphics{figs/snapshot02_crop.png}}
\end{center}
\caption{\label{fig:lobes}
{Views of the two lobes A (blue) and B (red) extracted from the shape model as seen from the Y+ (left) and Z+ (right) axes. The initial shape model is displayed as a wireframe for comparison, allowing one to visualize the neck between the two lobes.}
}
\end{figure*}

A visual inspection of the global shape confirms the presence of two lobes separated by a neck, at the first order similar to the shape derived in \cite{Ostro2000}.
For this reason, we extracted the individual  shapes of the two lobes from the MPCD global shape model in order to characterize their physical
properties.
We used an approach similar to that applied to comet 67P/C-G \citep{Jorda2016}.
In this approach, the facets belonging to each lobe are manually selected in the ``Meshlab software'' \citep{Cignoni2008}.
The best-fit ellipsoid of each lobe is then computed by fitting the coordinates of the extracted
vertices.
Finally, the lobe models are merged with those of the best-fit ellipsoids to compute their closed shapes and
volumes (for details on the method, see \citealt{Jorda2016}).
This leads to the individual shapes of the two lobes shown in Fig.~\ref{fig:lobes}.

\begin{table}
 \caption{\label{tab:lobes}
 Physical parameters (volume and diameters) of the two lobes computed from their reconstructed shape models,
 as well as geometric parameters (center coordinates, Euler angles, and tilt of the lobes' Z-axis with respect to that of the object), resulting from the best-fit ellipsoid of the selected facets of the MPCD shape model belonging to each lobe.}
 \centering
\begin{tabular}{lcc}
\hline
{\bf Parameter}   & {\bf Lobe A} & {\bf Lobe B} \\
\hline
Diameter $a$ (km) &    118 &  126 \\
Diameter $b$ (km) &     94 &   79 \\
Diameter $c$ (km) &     66 &   61 \\
Volume ($10^5$ km$^3$)   & 4.1 &  3.5 \\
Volume (\%)       &     47 &      40 \\
Volume-equiv. diameter $D$ (km) &  91.9  &  87.1 \\
\hline
Center coordinate $X$ (km) & --69.9 & 79.8 \\
Center coordinate $Y$ (km) &   0.0 &  --0.5 \\
Center coordinate $Z$ (km) &   0.2 &  --0.9 \\
Euler angle $\psi$($\degr$)  &  39.0 &  --21.7 \\
Euler angle $\theta$($\degr$)  &  --5.6 &    5.0 \\
Euler angle $\phi$($\degr$)  & --35.2 &   26.3 \\
Z-axis tilt($\degr$)   & 5.6 & 5.0 \\
Ellipsoid fit residuals (km) & 2.2 & 1.8 \\
Ellipsoid fit residuals (\%) & 3.1 & 2.4 \\
 \hline
\end{tabular}
\end{table}

% lobe analysis
We determined the diameters of the two lobes along the X-, Y-, and Z-axes and along their
principal axes of inertia, as well as their volume  (summarized in Table~\ref{tab:lobes}).
It appears that the X- and Z- axes have the same diameters within their error bars.
However, the Y-axis is significantly different between the two lobes, leading to a volume difference of 16\,\% in favor of lobe~A.
Furthermore, both lobes appear highly ellipsoidal, with a deviation between the lobes
and their best-fit ellipsoids of only $\sim\!2.5-3\,\%$, a value comparable to those found for large asteroids
with equilibrium shapes, such as 10~Hygiea \citep{Vernazza2020} and 4~Vesta \citep{Ferrais2020}.
The size difference between the two lobes along the Y-axis can possibly be explained by a depression
observed on lobe B (see Fig.~\ref{fig:lobes}), possibly formed by an impact.

% neck size
As a next step, we computed the length and mean radii of the neck as well as its volume from the parameters of Table~\ref{tab:lobes}.
Its volume appears to be 13~\% of the total volume of the object, whereas the minimum length of the neck along the X-axis is $\sim 25$ km.

\subsection{{\bf Dumbbell interpretation}}

\begin{figure}%[!ht]
\begin{center}
\resizebox{0.49\hsize}{!}{\includegraphics{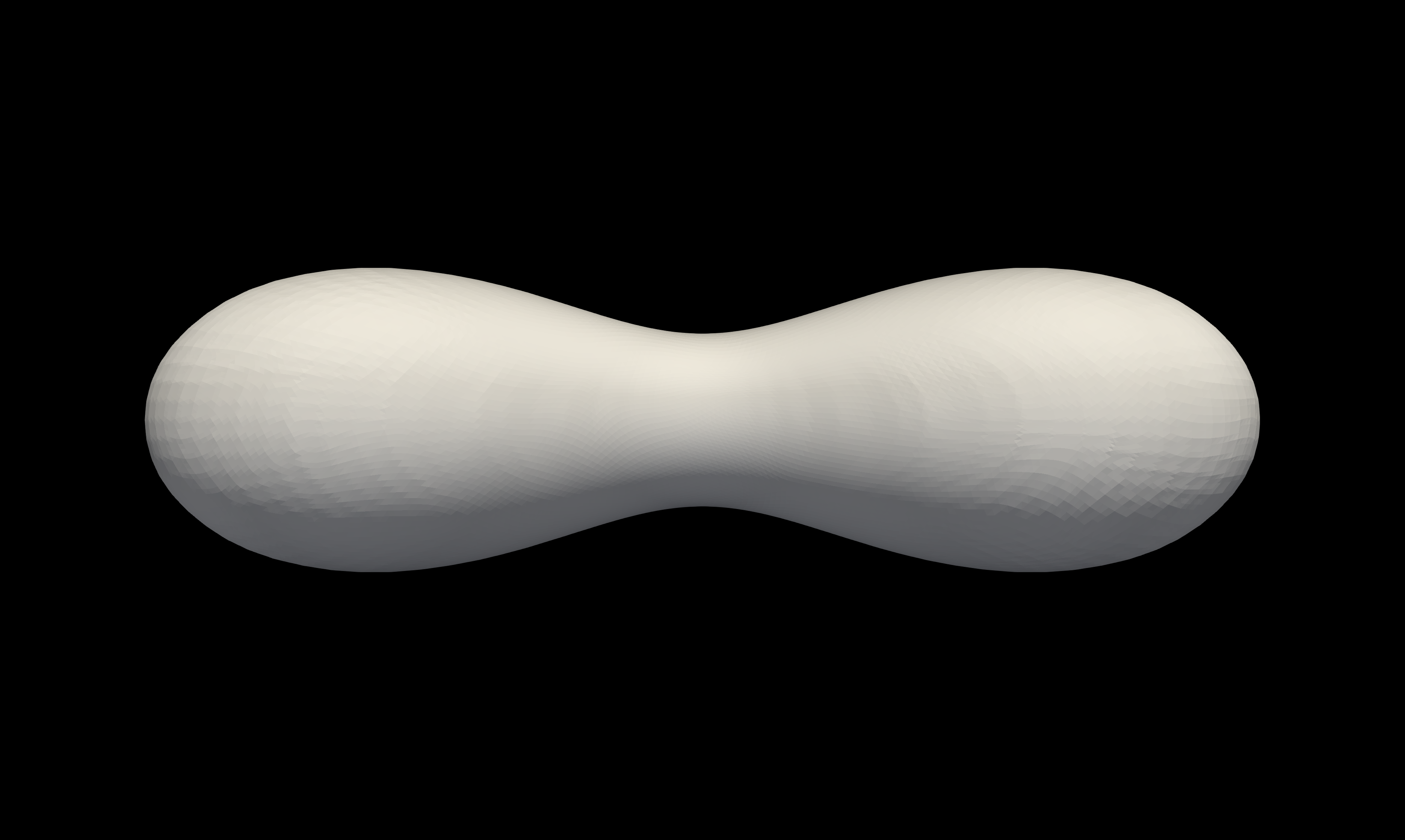}}
\resizebox{0.49\hsize}{!}{\includegraphics{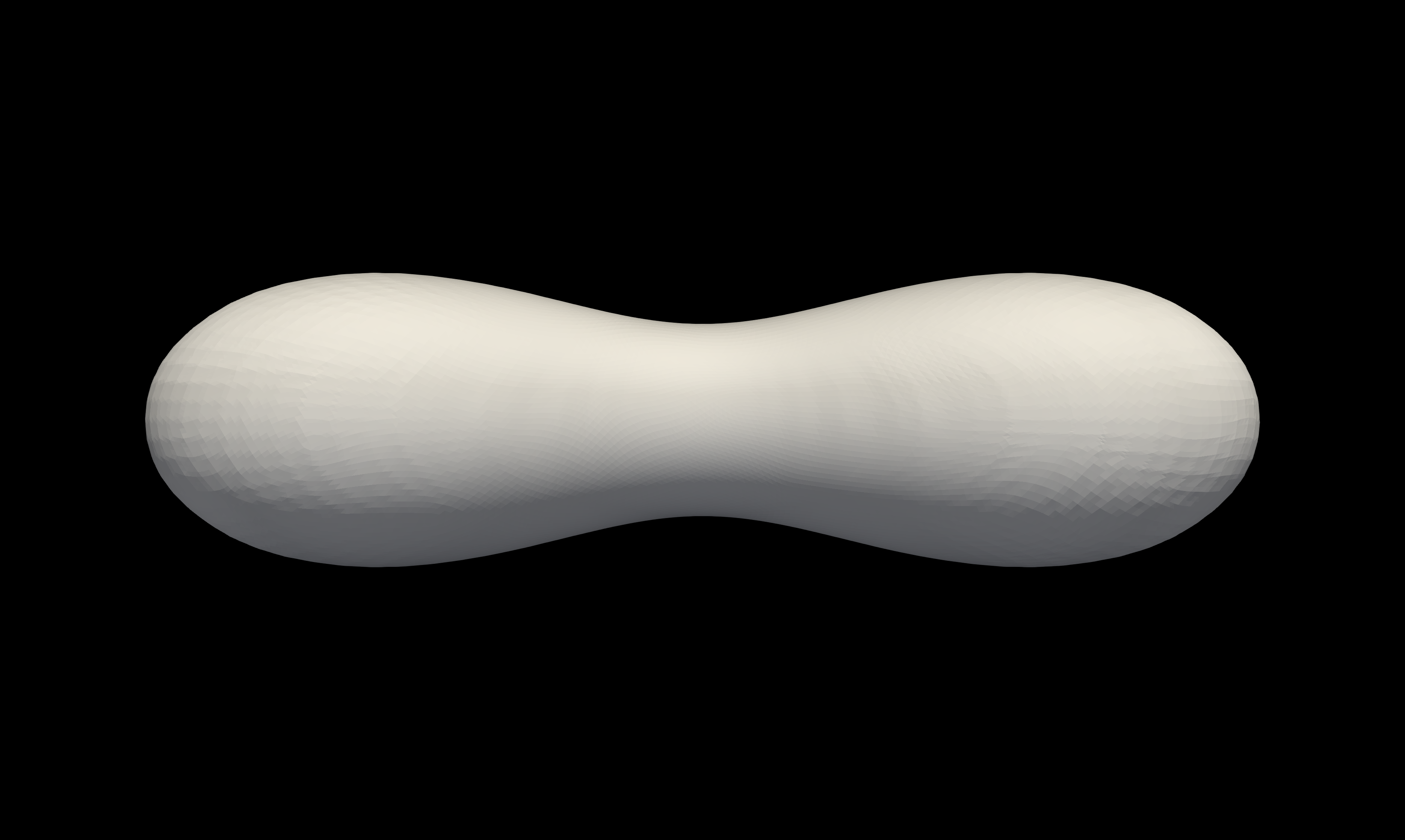}}
\resizebox{0.49\hsize}{!}{\includegraphics{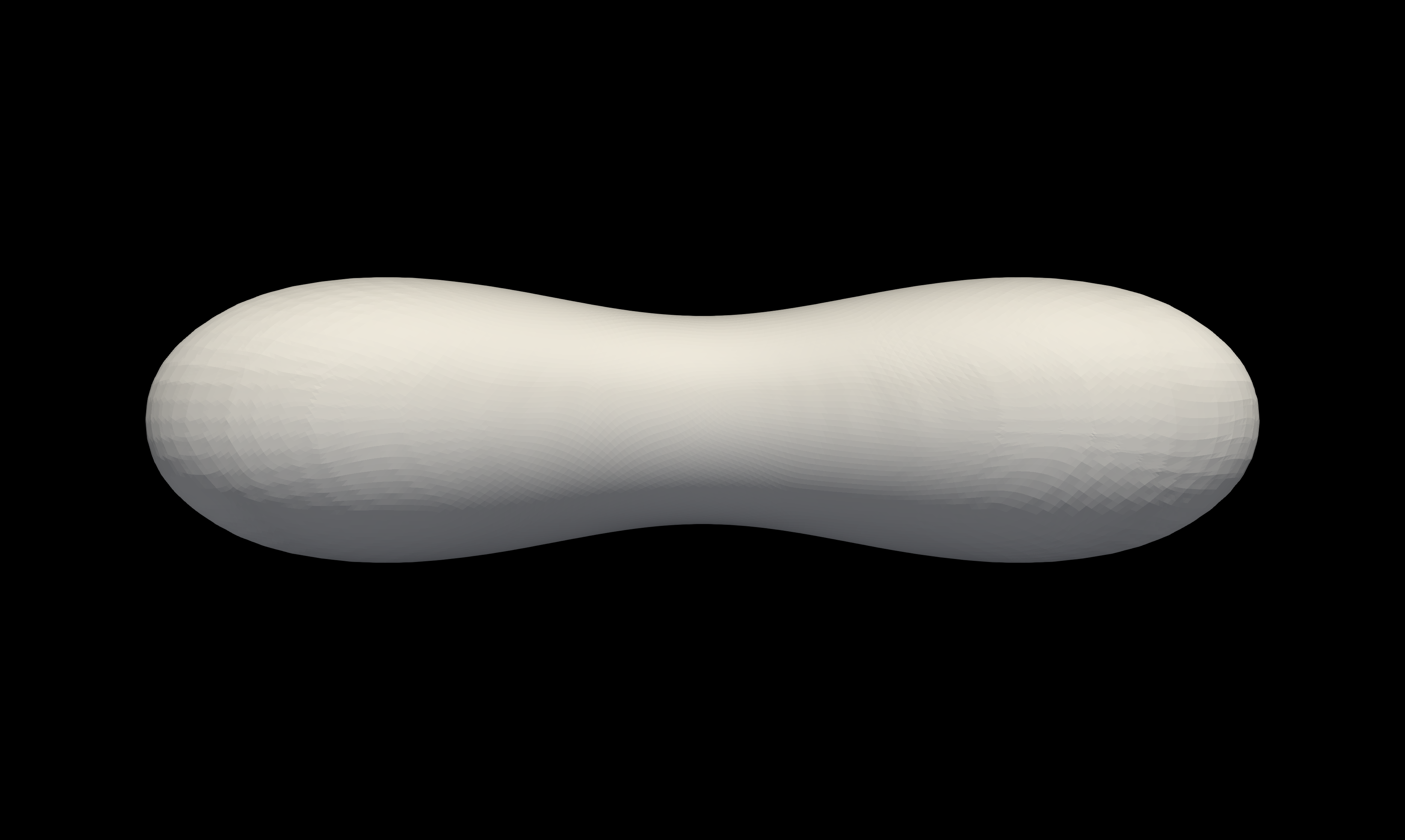}}
\resizebox{0.49\hsize}{!}{\includegraphics{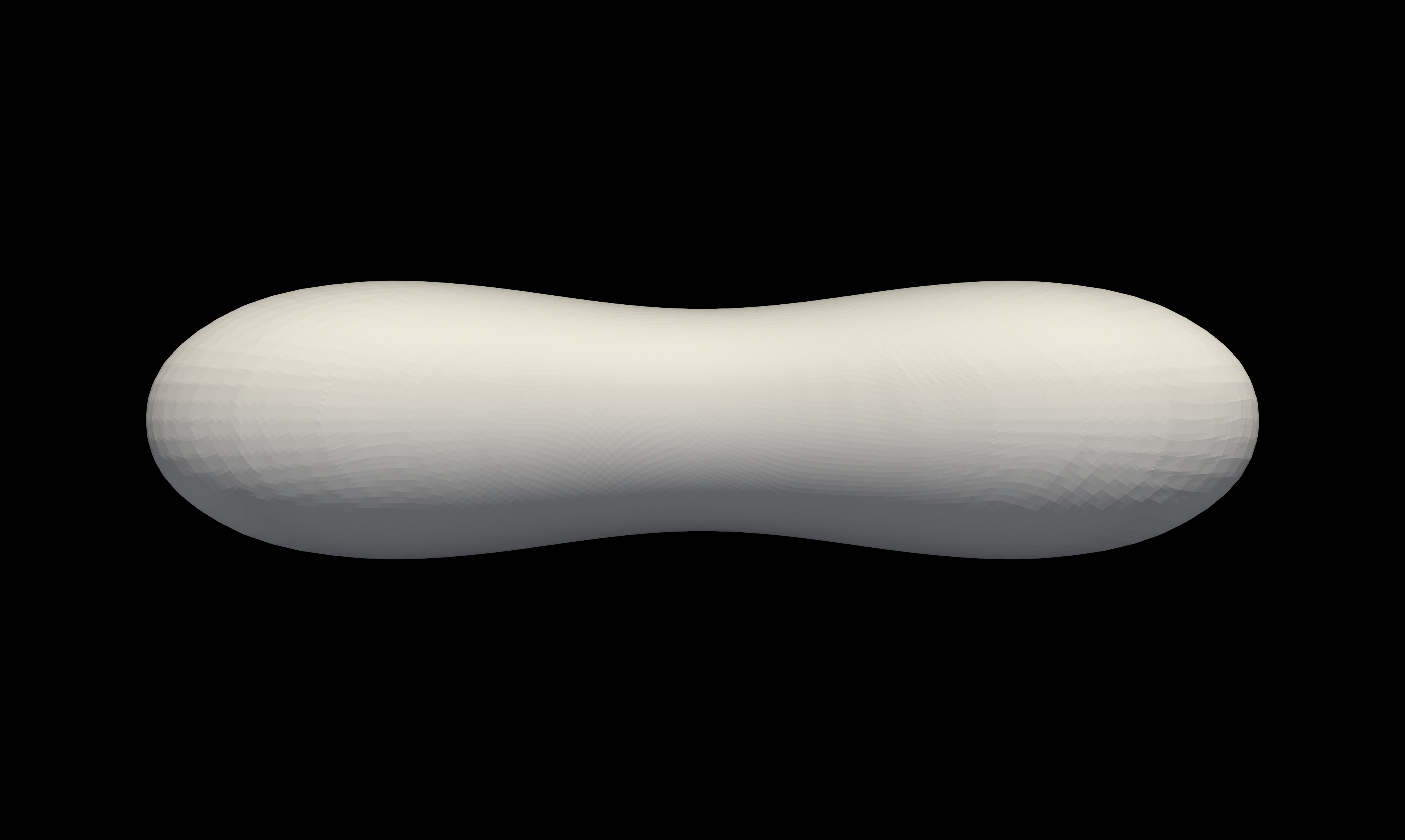}}
\resizebox{0.49\hsize}{!}{\includegraphics{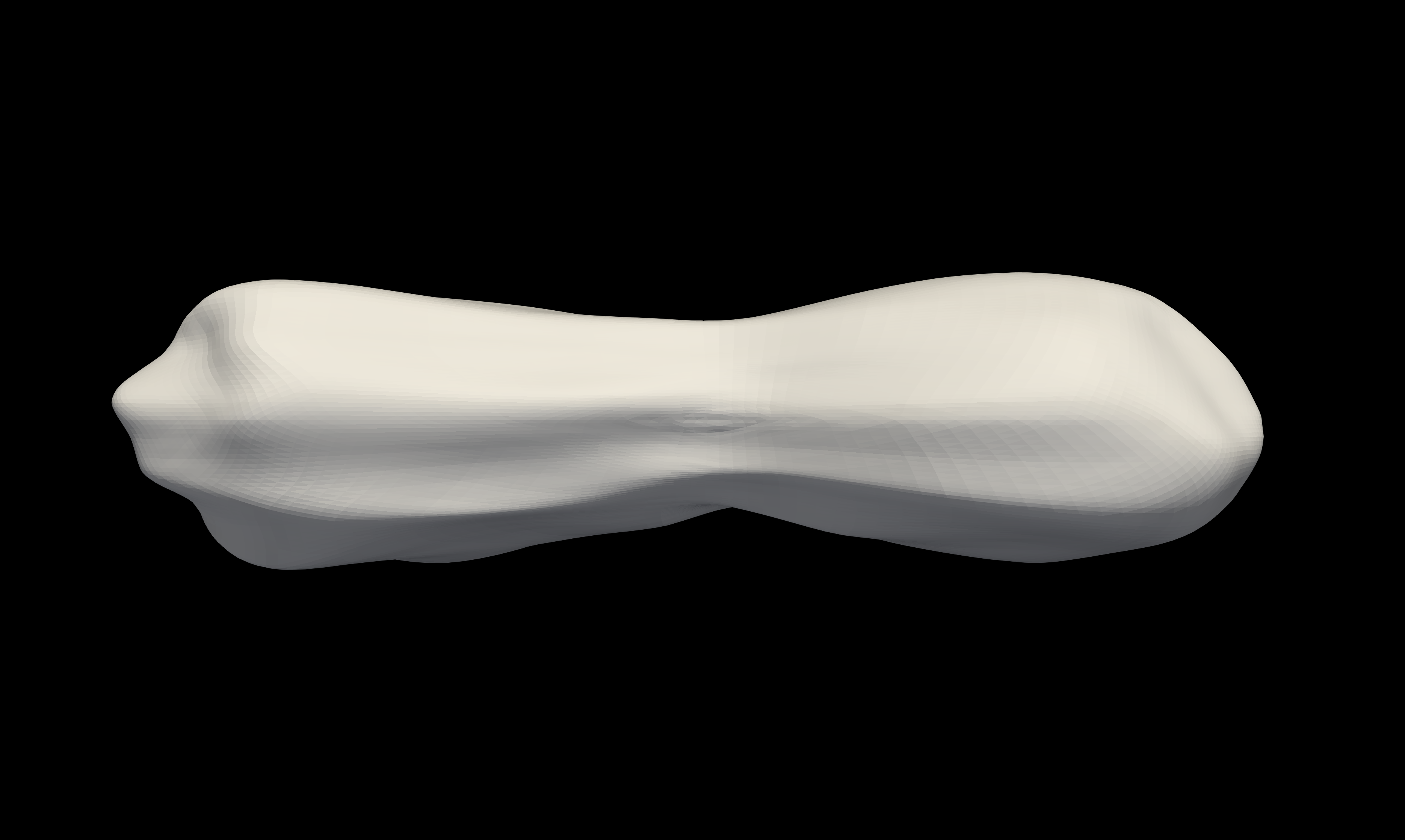}}
\resizebox{0.49\hsize}{!}{\includegraphics{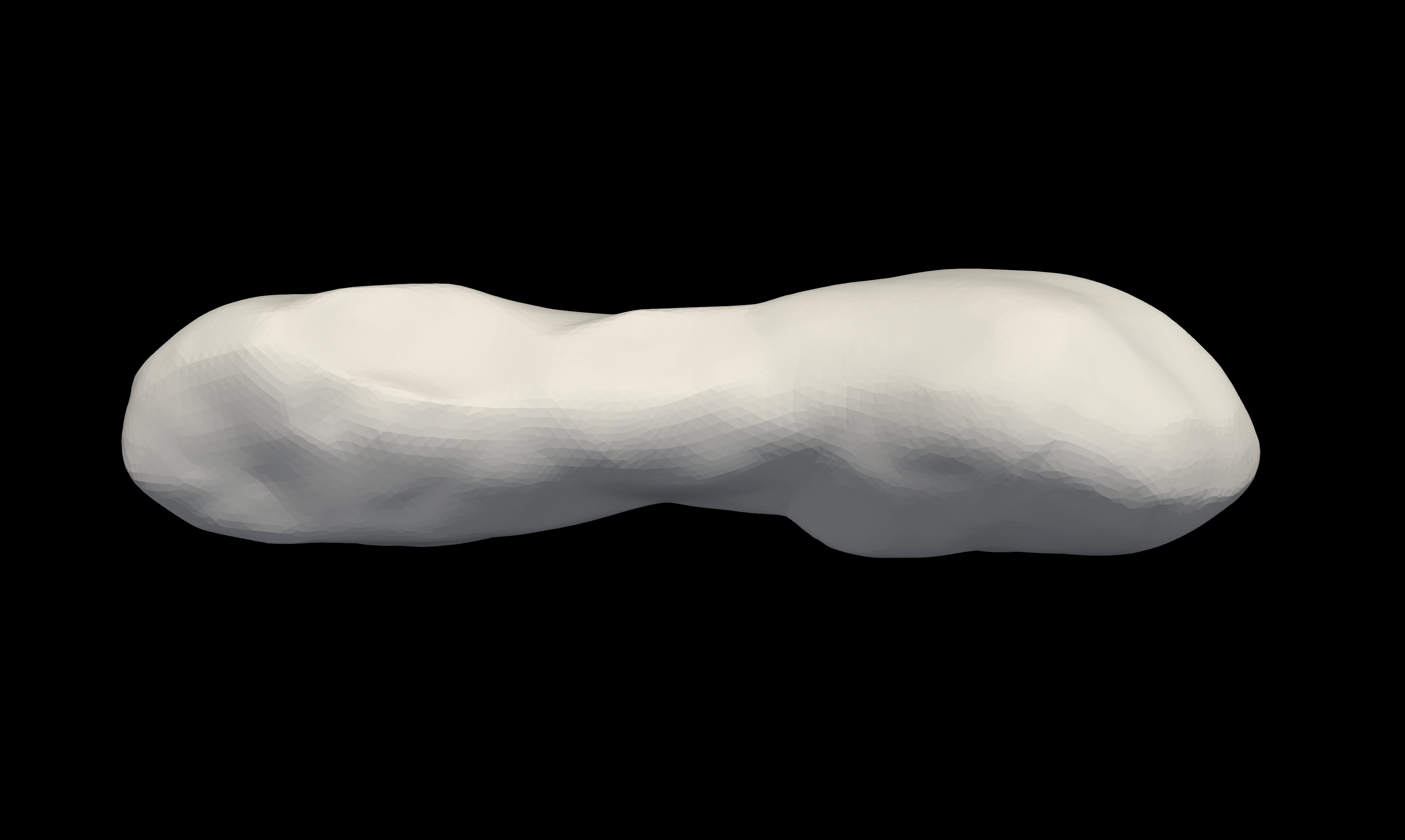}}
\end{center}
\caption{\label{fig:dumbbell}
Views of the dumbbell equilibrium shapes for values of the normalized angular velocity $\Omega$ equal to
$0.31$ (top left), $0.32$ (top right), $0.33$ (middle left), and $0.34$ (middle right).
The two bottom panels show the reconstructed ADAM (left) and MPCD (right) shapes.
All shape models are seen from the $Y-$ axis.}
\end{figure}

% comparison with equilibrium figs (P Descamps)
It is striking that in Fig.~\ref{fig:compshapes}, the shape of Kleopatra resembles the ``dumb-bell'' equilibrium shapes studied by \citet{Descamps2015}.
To test whether Kleopatra formed at equilibrium, we computed {the rms of the deviation} between the MPCD shape model of Kleopatra and several shape models of dumbbell equilibrium figures corresponding to different values of
the normalized angular velocity~$\Omega$ defined by \citet{Descamps2015} as $\omega\sqrt{3/(4\pi G\rho)}$,
where $\rho$ denotes the bulk density.
In order to perform this comparison, we rescaled the equilibrium shapes so that their lengths match that of Kleopatra and we shifted the center of the figure by $-3.8\,{\rm km}$ along the X-axis.
This shift leads to a displacement of the center of mass toward the larger lobe (lobe A), which is coherent if we assume that the two lobes have the same density. 
The lowest final rms of the distance between the models is equal to $\sim3.5$~km for a normalized angular velocity $\Omega=0.32-0.33$, but figures with values of $\Omega$ in the range from $0.31-0.34$ remain compatible with the shape of Kleopatra (rms below 4~km), as illustrated { in Fig.~\ref{fig:dumbbell}.}
The $\Omega$ value calculated from Kleopatra's current rotation
period $5.385\,{\rm h}$ and density $3.38 \pm 0.50\,{\rm g}\,{\rm cm}^{-3}$
is $0.334$, which is in striking agreement with the values reported above.
We thus confirm that Kleopatra is an equilibrium shape.

\subsection{{\bf Surface acceleration}}\label{sec:acc}

\begin{figure*}
\centering
\includegraphics[height=4.5cm]{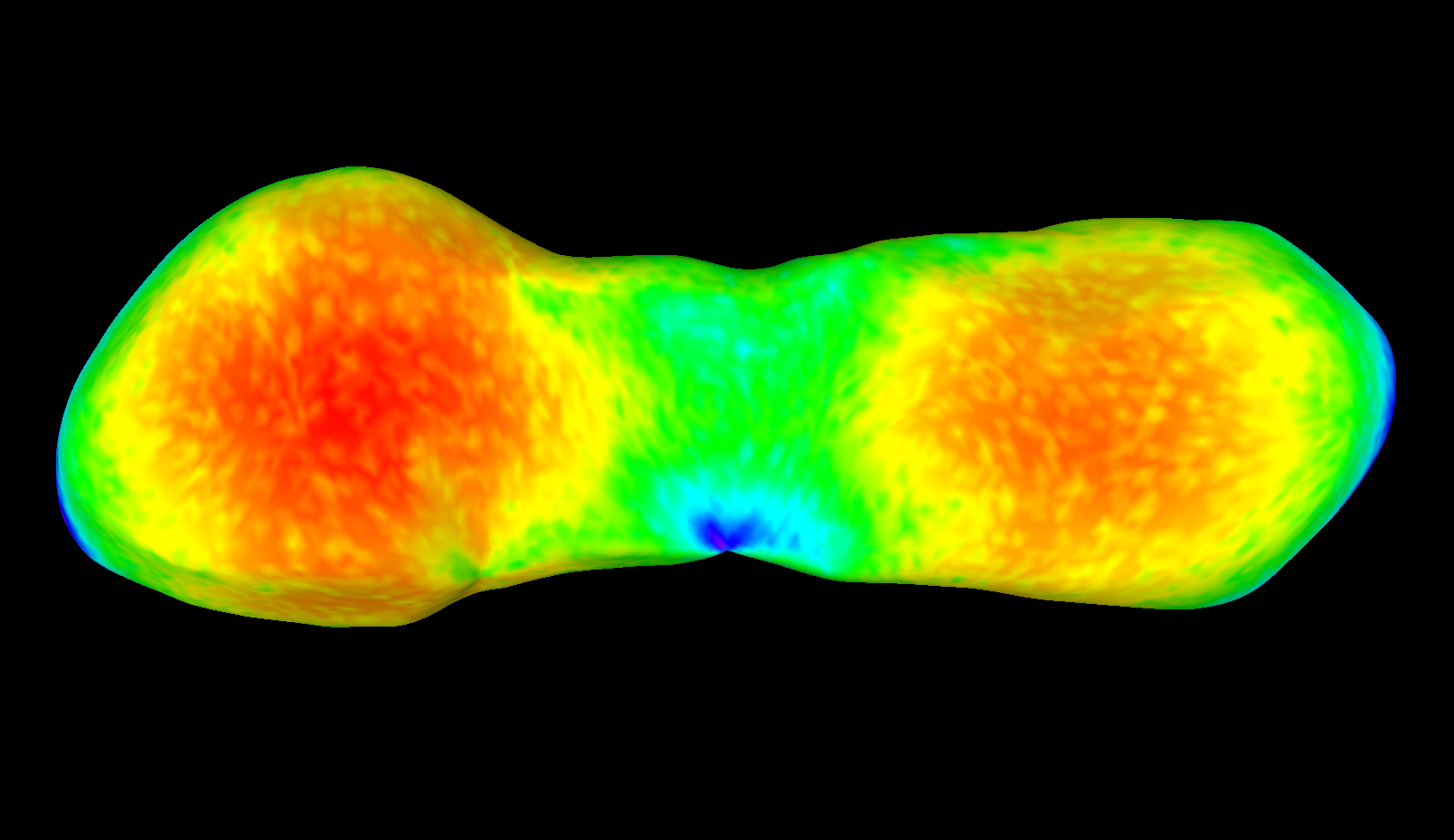}
\includegraphics[height=4.5cm]{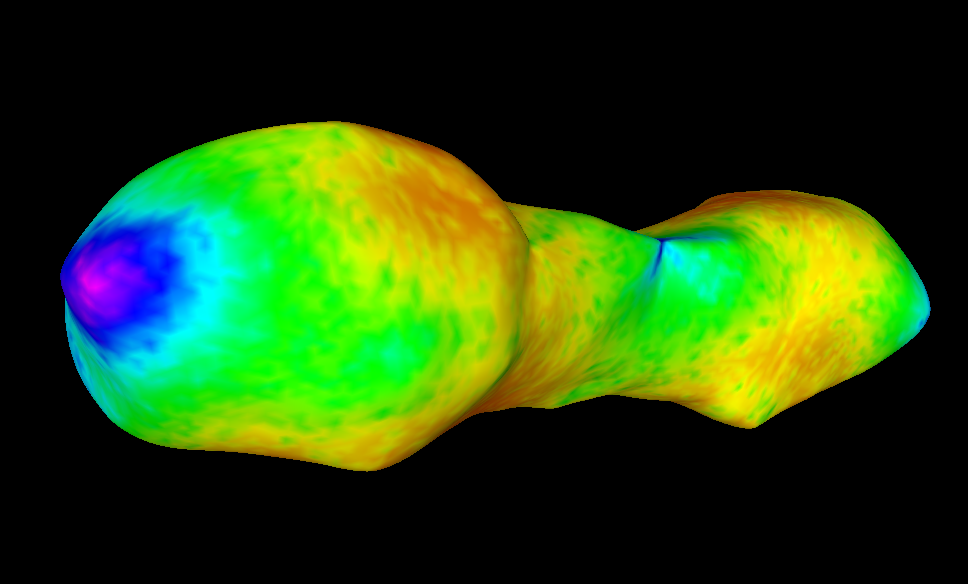}
\includegraphics[height=4.5cm]{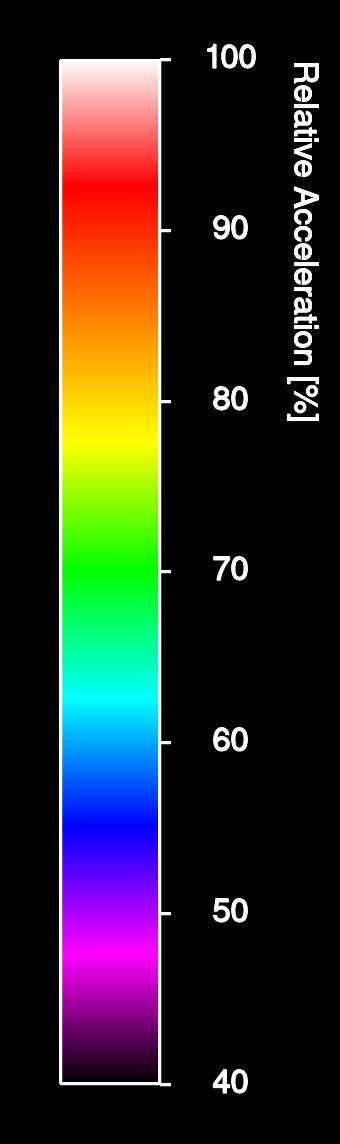}
\caption{
Local acceleration at the surface of Kleopatra normalized by that of a sphere of an equivalent radius and
spin period. View from the $Z+$ (left) axis and oblique view along the equator (right). We note the very low value at the edges and
higher values around the two lobes.
}
\label{fig:216_acceleration}
\end{figure*}

To probe the effect of the shape on the internal compaction, we computed the gravitational plus the centrifugal acceleration at each point of the surface and compared it with that of a sphere of equivalent mass (Fig.~\ref{fig:216_acceleration}). It appears that the mean acceleration at the surface of Kleopatra amounts to 76\% of that of a sphere of equivalent mass with values below 50\% locally. It follows that Kleopatra's highly elongated shape does not favor compaction and it supports a higher macroprosity compared to more spherical or ellipsoidal bodies of an equivalent mass.

\begin{figure*}
\centering
\resizebox{1.0\hsize}{!}{\includegraphics{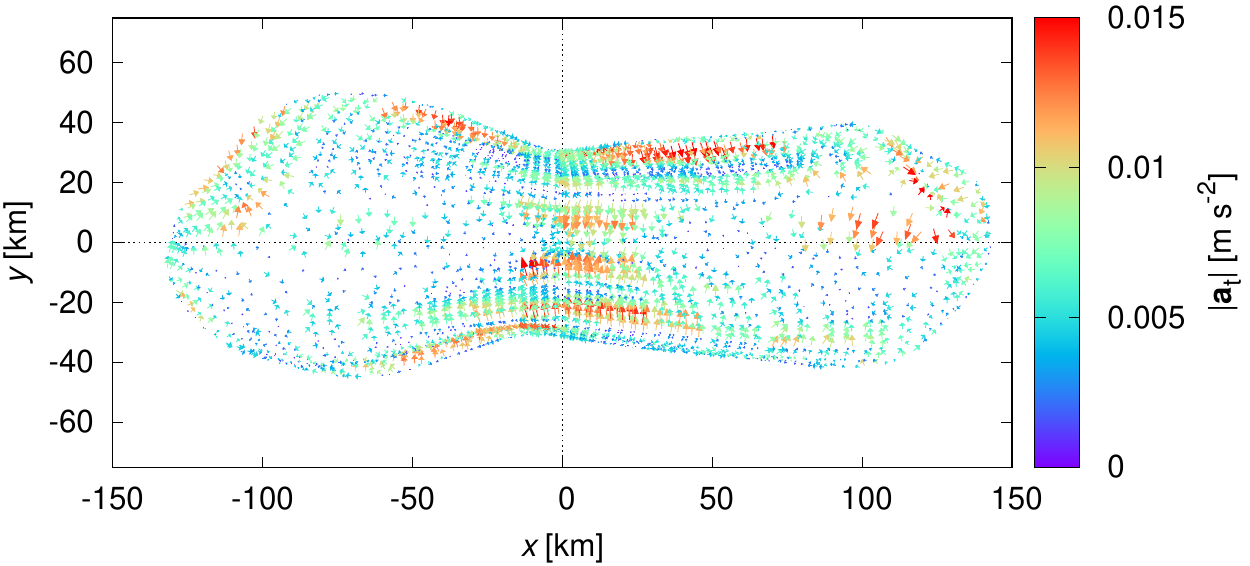}\includegraphics{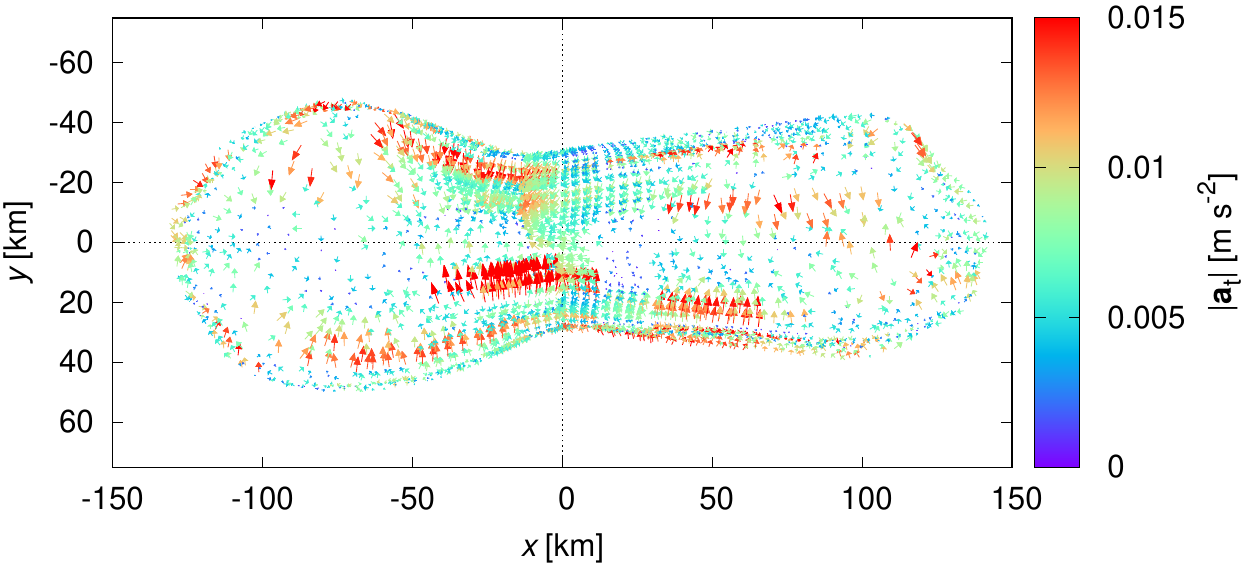}}

\resizebox{1.0\hsize}{!}{\includegraphics{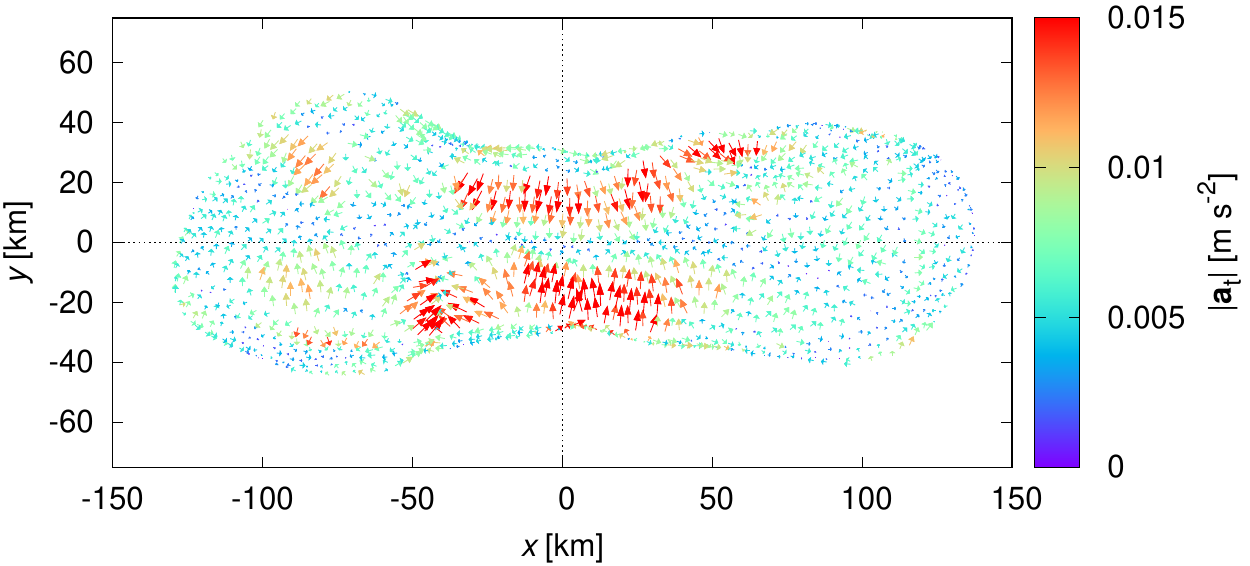}\includegraphics{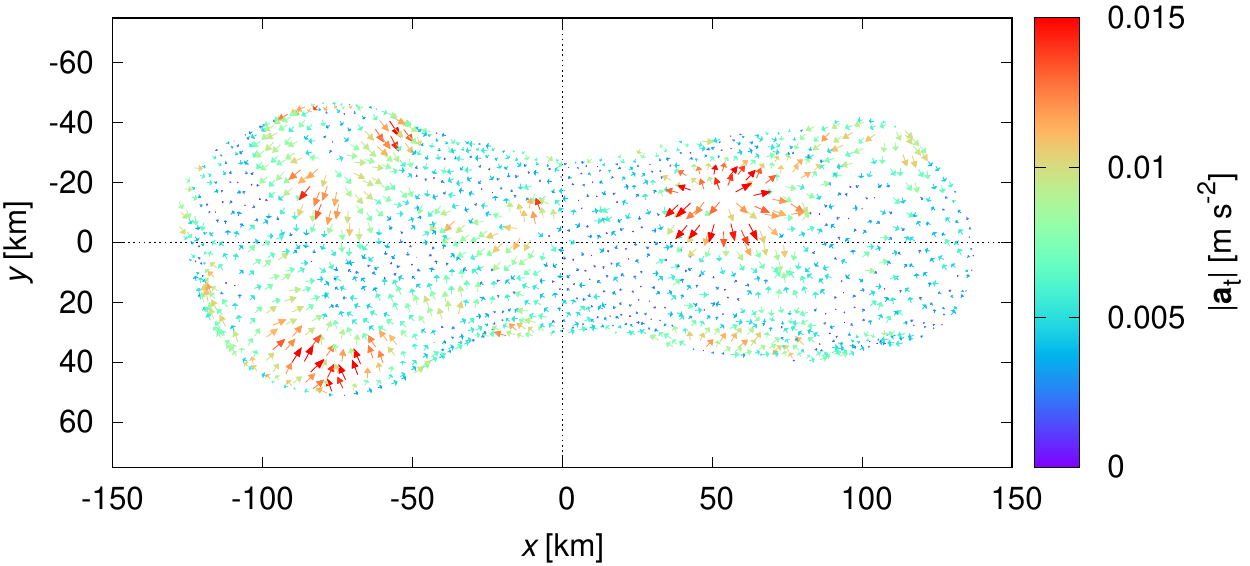}}
\caption{{ Tangential surface accelerations $|a_\mathrm{t}|$ computed for ADAM (top) and MPCD (bottom) models. Views on the left are north pole-on ($+z$), while those on the right are south pole-on ($-z$).}
}
\label{fig:tangential}
\end{figure*}

{ We also computed tangential surface accelerations $|a_\mathrm{t}|$, which indicate possible material motion (Fig.~\ref{fig:tangential}). It seems that a convergence is located on the left lobe, at (x, y, z) = ($-$120, 20, 20) km, or on the neck ($\pm$20, 0, 20) km. On the other hand,
a divergence is on the "hill" at (50, 15, $-$35) km. These locations are
common to both ADAM and MPCD models. We do not report on the locations with
possible shape artifacts. The maximum accelerations reach $\sim$1 cm/s$^2$.
Whether this is sufficient to sustain global motion is uncertain
because it depends on local topography, regolith structure, roughness,
friction, impacts, seismicity, etc. Alternatively, cratering impacts
can initiate ballistic transport, with complex near-surface dynamics,
determined by the proximity of the critical L$_{1}$ equipotential.}

\subsection{{\bf Critical equipotentials}}

\begin{figure}
\centering
\includegraphics[width=7.5cm]{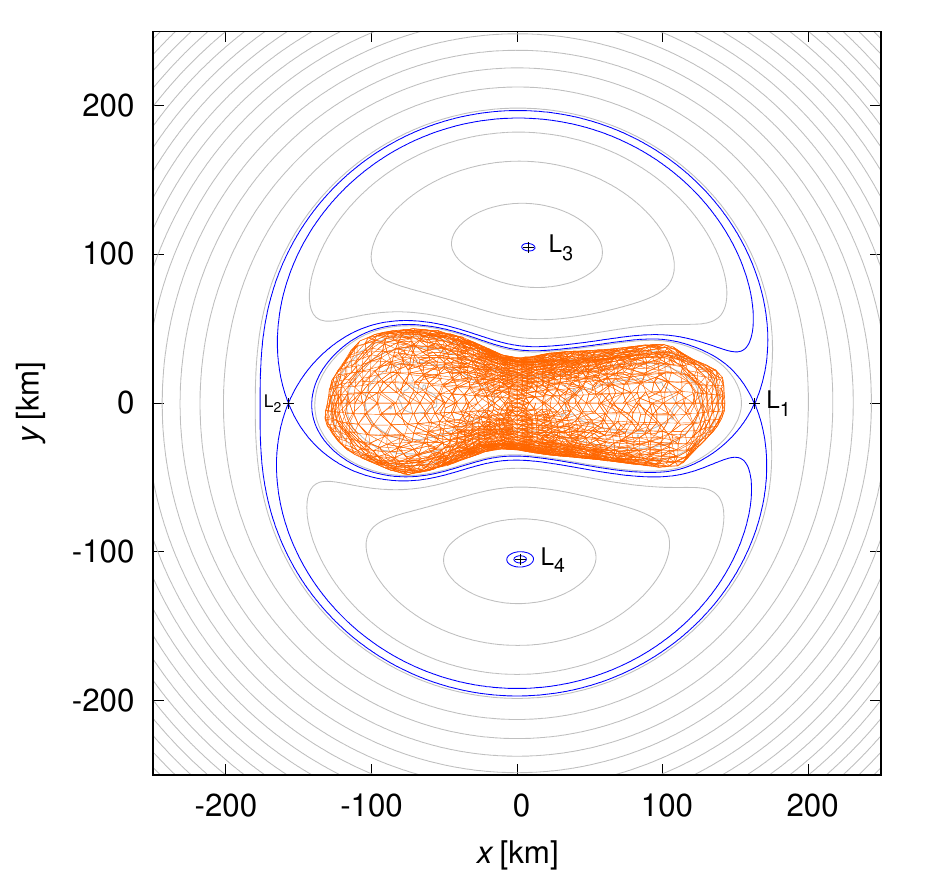}
\caption{
Effective potential $U_{\rm eff}$ in the $(x,y)$ plane (gray lines),
critical equipotentials (blue lines),
and the \adam{} shape model (orange).
The density is $\rho = 3.34\,{\rm g}\,{\rm cm}^{-3}$
and the rotation period $P = 5.385\,{\rm h}$.
Four critical points are denoted: L$_1$, L$_2$, L$_3$, and L$_4$.
The L$_1$ critical point is on the right.
The L$_1$ equipotential touches the surface of Kleopatra.
}
\label{216_bf_xy_rho3.3_adam_gravity3__5.385}
\end{figure}

\begin{figure}
\centering
\includegraphics[width=7.5cm]{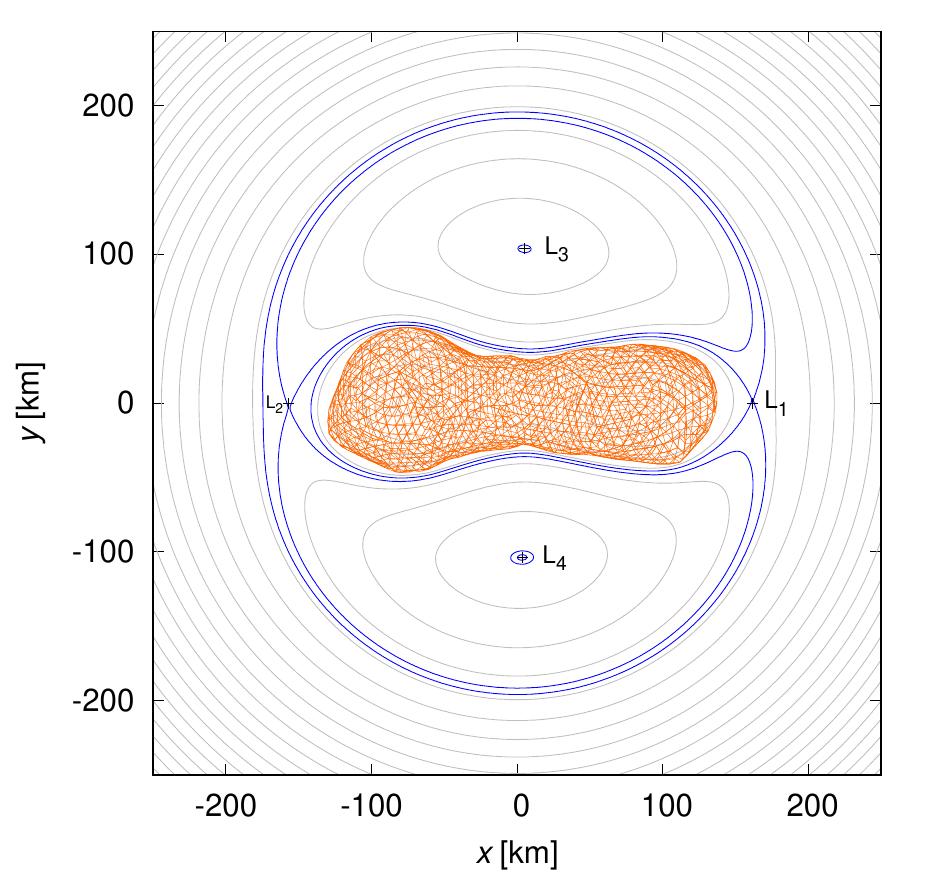}
\caption{Effective potential $U_{\rm eff}$ in the $(x,y)$ plane (gray lines),
critical equipotentials (blue lines),
for the MPCD model (orange) and density $3.43\,{\rm g}\,{\rm cm}^{-3}$.
}
\label{216_bf_xy_rho3.5_mpcd_gravity3__5.385}
\end{figure}

\begin{figure}
\centering
\includegraphics[width=7.5cm]{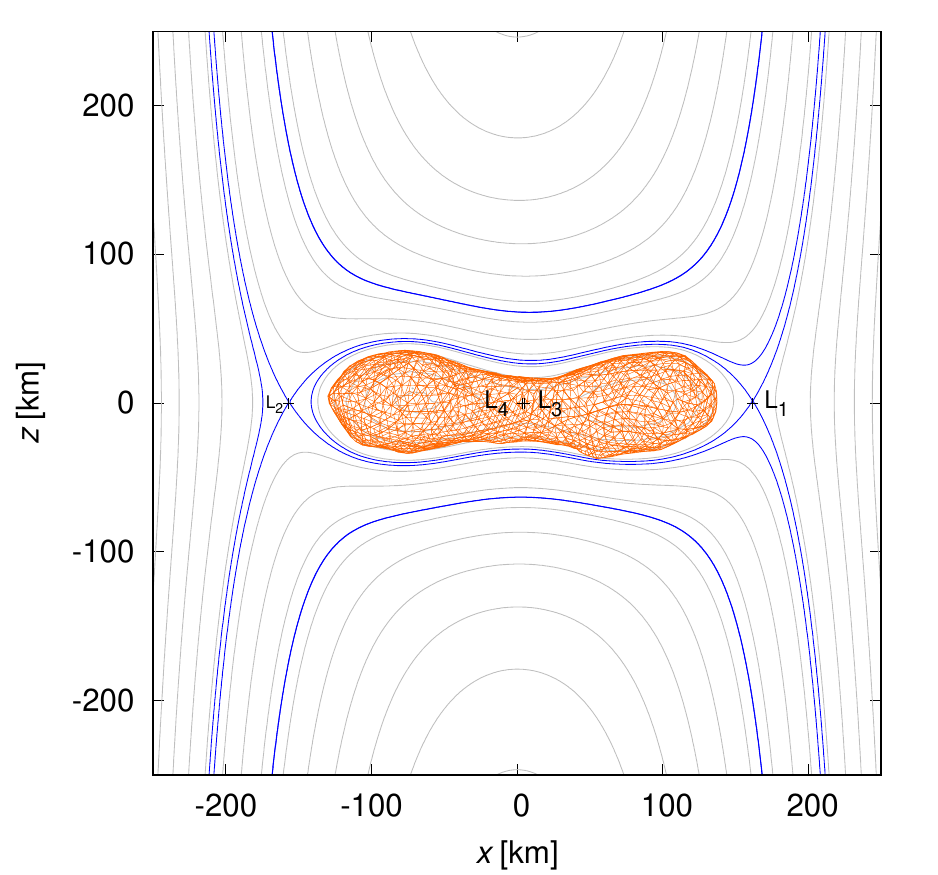}
\caption{
Effective potential $U_{\rm eff}$ in the $(x,z)$ plane (gray lines),
critical equipotentials (blue lines),
for the MPCD model (orange) and density $3.43\,{\rm g}\,{\rm cm}^{-3}$.
}
\label{216_bf_xz_rho3.5_mpcd_gravity3__5.385}
\end{figure}

The effective potential
$U_{\rm eff} = U_{\rm g} - \frac{1}{2}\omega(x^2+y^2)$
was computed using the same algorithm as in Appendix~\ref{sec:slopes}.
We plotted its equipotentials together with the four critical points
in Fig.~\ref{216_bf_xy_rho3.3_adam_gravity3__5.385}.
The major result is that the shape extends to a distance that is very close to the
critical L$_1$ equipotential. In fact, the \adam{} shape model
almost touches it at one point in the $(x,y)$ plane.
At the same time, they are separated by several (or more) kilometers
in the $(x,z)$ plane. An analogous analysis of the MPCD shape shows a very similar result
(see Figs.~\ref{216_bf_xy_rho3.5_mpcd_gravity3__5.385} and
\ref{216_bf_xz_rho3.5_mpcd_gravity3__5.385}).
A minor difference is that the shape touches the equipotential
at two different points along its $(x,y)$  and $(x,z)$
circumference.  The closest distance is less than a kilometer.

Our work includes three major differences with respect to 
\cite{Hirabayashi_Scheeres_2014ApJ...780..160H}.
(i)~We used the currently observed $P = 5.385\,{\rm h,}$
together with $\rho = 3\,380\,{\rm kg}\,{\rm m}^{-3}$.
 Consequently, we did not need any mechanism for a spin-down
(e.g., from $2.8\,{\rm h}$)
to explain why the shape is critical.
(ii)~We did not use any scaling. Instead, the absolute volume is
constrained by the AO observations.
(iii)~Our L$_1$ point is on the right ($+x$) and not on the left, implying that the shape model presented here is different to the previous one used by \cite{Hirabayashi_Scheeres_2014ApJ...780..160H} .
This possibly affects near-surface dynamics and
surface locations from which material is more likely to escape.

%%%%%%%%%%%%%%%%%%%%%%%%%%%%%%%%%%%%%%%%%%%%%%%%%%%%%%%%%%%%%%%%%%%%%%
\section{Discussion}\label{sec:discussion}

Our density estimate of  $(3.38 \pm 0.50)\,{\rm g}\,{\rm cm}^{-3}$ for Kleopatra is surprising, considering its high radar albedo of 0.43 $\pm$ 0.10 \citep{Shepard2018} that implies a high surface bulk density and a large metal content. In comparison, the density of metal-rich asteroid (16) Psyche is $(4.2 \pm 0.6)\,{\rm g}\,{\rm cm}^{-3}$ \citep{Ferrais2020}. One possible explanation for such a low density for Kleopatra is the presence of substantial porosity within the body. \citet{Wilson1999} showed that gravitationally reaccreted asteroids should have porosities of $\sim$20--40\%. It is of great interest that the highly elongated shape of Kleopatra actually supports a higher macroporosity than that expected for a spherical or ellipsoidal body. Accordingly, acknowledging a rubble-pile structure of Kleopatra from its specific angular momentum, this range of porosities implies a grain density in the 4.2--5.8 ${\rm g}\,{\rm cm}^{-3}$ range, suggestive of a mixture of NiFe metal-rich with the inclusion of silicates \citep{Marchis2003}. The presence of silicates is supported by the presence of a 0.9 micron band in Kleopatra's spectrum \citep{Hardersen2011}.

This rubble-pile structure along with its near equilibrium shape is compatible with a formation scenario including a giant impact \citep{Sugiura2018}, followed by reaccumulation during which Kleopatra behaved as a fluid as suggested by \citet{Descamps2010}.  The dumbbell equilibrium shape also explains the very unusual long neck between the two lobes.
The small volume deficit {due to the depression
observed on lobe B (see Figures ~\ref{fig:compshapes} and ~\ref{fig:lobes}}) is coherent with a later smaller impact.

{ The critical rotation would be 5.250 h (compared to 5.385 h) if we require the $L_1$ equipotential to be in contact with the surface. Similarly, the critical density should be 3.2 ${\rm g}\,{\rm cm}^{-3}$. 
 The whole surface actually does not follow the $L_1$ equipotential exactly, since even very small impacts could eject fragments beyond the equipotential surface. Moreover, the critical value of $P$ (as well as $\rho$) is sensitive to small variations in the topography.}

Finally, the low gravity at the edges and along the equator of the body, together with its rubble-pile structure and the equatorial orbits of the moons, opens the possibility that the latter formed via mass shedding. 
The large offset between the L1 equipotential and the surface at the edges of the asteroid supports this interpretation.

%%%%%%%%%%%%%%%%%%%%%%%%%%%%%%%%%%%%%%%%%%%%%%%%%%%%%%%%%%%%%%%%%%%%%%
\section{Summary \& conclusion}\label{sec:conclusions}

New AO observations of the triple system (216) Kleopatra with VLT/SPHERE allowed us to constrain the 3D shape and mass of the primary with high accuracy (see Bro\v{z} et al., companion paper for the mass estimate). Both mass and volume estimates of (216)~Kleopatra imply a low density of  $(3.38 \pm 0.50)\,{\rm g}\,{\rm cm}^{-3}$. Such low density for a metallic asteroid suggests the presence of substantial porosity {within the metal-rich primary or a significant content of silicate in the composition of the asteroid.}

This rubble-pile structure along with its near equilibrium shape is compatible with a formation scenario including a giant impact followed by reaccumulation. (216)~Kleopatra’s rotation period and dumbbell shape imply that it is in a critically rotating state. The low gravity at the edges and along the equator of the body, together with its rubble-pile structure and the equatorial orbits of the moons, opens the possibility that the latter formed via mass shedding as suggested by \cite{Descamps2008}.

Future observations of (216)~Kleopatra with current AO systems such as SPHERE/ZIMPOL could reveal long-term perturbations in the moon orbits related to the shape of the primary. Similar observations of the primary, but at low phase angles (less than 5 deg), could also provide more accurate contours and thus help refine its shape. 

%Kleopatra's rotation period ($P = 5.385 h$) and dumb-bell 3D shape imply that the L1 equipotential is very close to the surface of Kleopatra. This suggests that (216)~Kleopatra is in critically rotating state and that material from the primary can escape the surface. As suggested by \cite{Descamps2008}, the moons CleoSelene and AlexHelios, may have formed by mass shedding. 

%Now we have an accurate model of the moon orbits, we can focus our attention in characterizing them in details, specifically to derive their size and shape. The mobilization of network of small telescopes to characterize moons by stellar occultation at the right place and time to capture a short event is difficult but possible if the moon orbits are well constrained similarly to what was done for (87) Sylvia \citep{Berthier2014}. 

We can speculate that high resolution images of Kleopatra's surface could help to truly understand the origin of the moons by revealing the presence of surface heterogeneities (e.g., albedo variegations), or anomalies such as concavities, that would help to link the moons directly to their parent body. Future observations with high angular resolution imaging data provided by the next generation of extremely large telescopes could help marginally by providing color and spectroscopic constraints on the moons and refining the shape model of the primary. A future space mission to (216)~Kleopatra and its two moons CleoSelene and AlexHelios would definitively shed light on the origin and current dynamics of this complex system.  In situ measurements could, for instance, reveal the ejection of particles from Kleopatra similar to what was recently seen on the near-Earth asteroid (101955) Bennu \citep{Lauretta2019}.

\begin{acknowledgements}
This material is partially based upon work supported by the National Science Foundation under Grant No. 1743015. This work has been supported by the Czech Science Foundation through grant 20-08218S (J. Hanu\v s, J. \v Durech), 21-11058S (M. Bro\v z) and by the Charles University Research program No. UNCE/SCI/023. P.~Vernazza, A.~Drouard, M. Ferrais and B.~Carry were supported by CNRS/INSU/PNP.  M.M. was supported by the National Aeronautics and Space Administration under grant No. 80NSSC18K0849 issued through the Planetary Astronomy Program. The work of TSR was carried out through grant APOSTD/2019/046 by Generalitat Valenciana (Spain). This work was supported by the MINECO (Spanish Ministry of Economy) through grant RTI2018-095076-B-C21 (MINECO/FEDER, UE). The research leading to these results has received funding from the ARC grant for Concerted Research Actions, financed by the Wallonia-Brussels Federation. TRAPPIST is a project funded by the Belgian Fonds (National) de la Recherche Scientifique (F.R.S.-FNRS) under grant FRFC 2.5.594.09.F. TRAPPIST-North is a project funded by the University of Liège, and performed in collaboration with Cadi Ayyad University of Marrakesh. E.~Jehin is a FNRS Senior Research Associate.\\
The data presented herein were obtained partially at the W. M. Keck Observatory, which is operated as a scientific partnership among the California Institute of Technology, the University of California and the National Aeronautics and Space Administration. The Observatory was made possible by the generous financial support of the W. M. Keck Foundation. The authors wish to recognize and acknowledge the very significant cultural role and reverence that the summit of Maunakea has always had within the indigenous Hawaiian community.  We are most fortunate to have the opportunity to conduct observations from this mountain.
\end{acknowledgements}

\bibliography{mybib,references}
\bibliographystyle{aa}

%\newpage
\begin{appendix}

%%%%%%%%%%%%%%%%%%%%%%%%%%%%%%%%%%%%%%%%%%%%%%%%%%%%%%%%%%%%%%%%%%%%%%%%

\section{{\bf Surface slopes}}\label{sec:slopes}

\begin{figure}
\centering
\includegraphics[width=8cm]{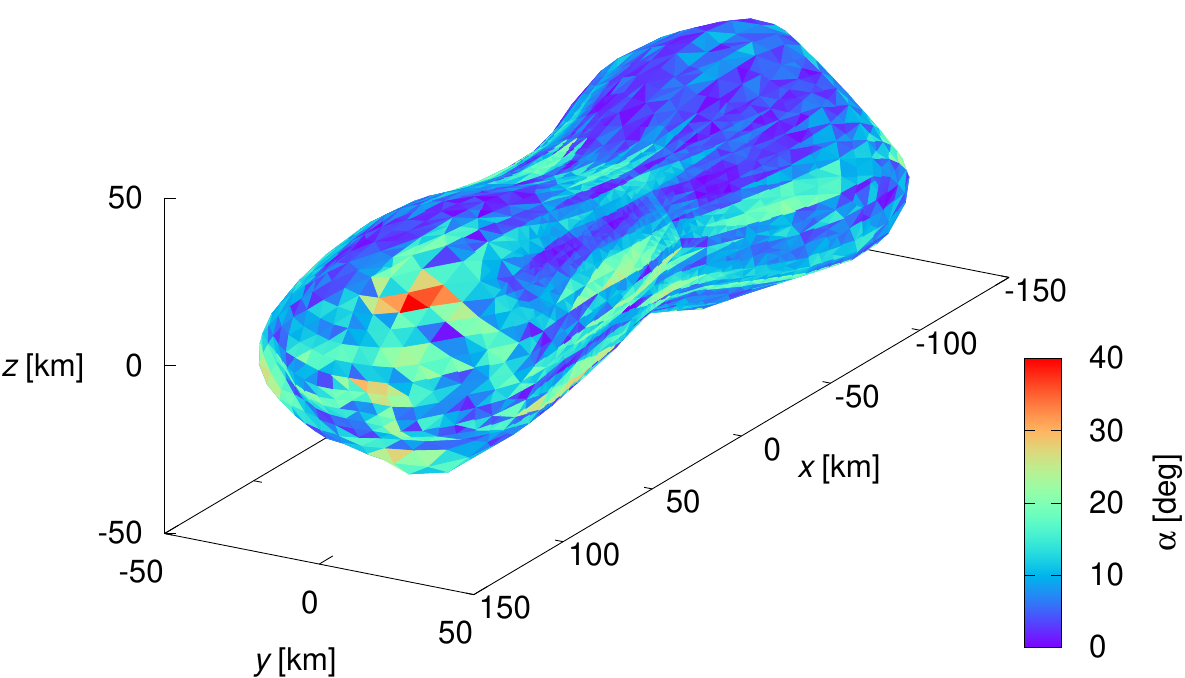}
\caption{
Surface slopes~$\alpha$ (shown as color)
for the \adam{} shape model.
They were computed for centers of faces
and gravitational plus centrifugal accelerations.
The density is $\rho = 3\,340\,{\rm kg}\,{\rm m}^{-3}$
and the rotation period $P = 5.385\,{\rm h}$.
Higher slopes are present on the lobes.
}
\label{216_slope_rho3.3_gravity4__70_120}
\end{figure}

Another useful characteristics is the statistics of surface slopes.
We computed the slope~$\alpha$ as the angle between the local normal
and the total acceleration in the corotating frame.
To evaluate the respective volumetric integrals,
we triangulated the volume via the Tetgen program \citep{Si_2006},
with 24099 elements,
and the brute-force algorithm.
The result is shown in Fig.~\ref{216_slope_rho3.3_gravity4__70_120}.
Most of the slopes are between $2^\circ$ and $15^\circ$.
Steeper slopes (up to $20^\circ$) are located on the lobes,
at the largest distances from the $z$-axis,
and some of them are on the neck.
A few outliers (over $25^\circ$) are related to individual facets,
which might be elongated or degenerate.

\begin{figure}
\centering
\includegraphics[width=8cm]{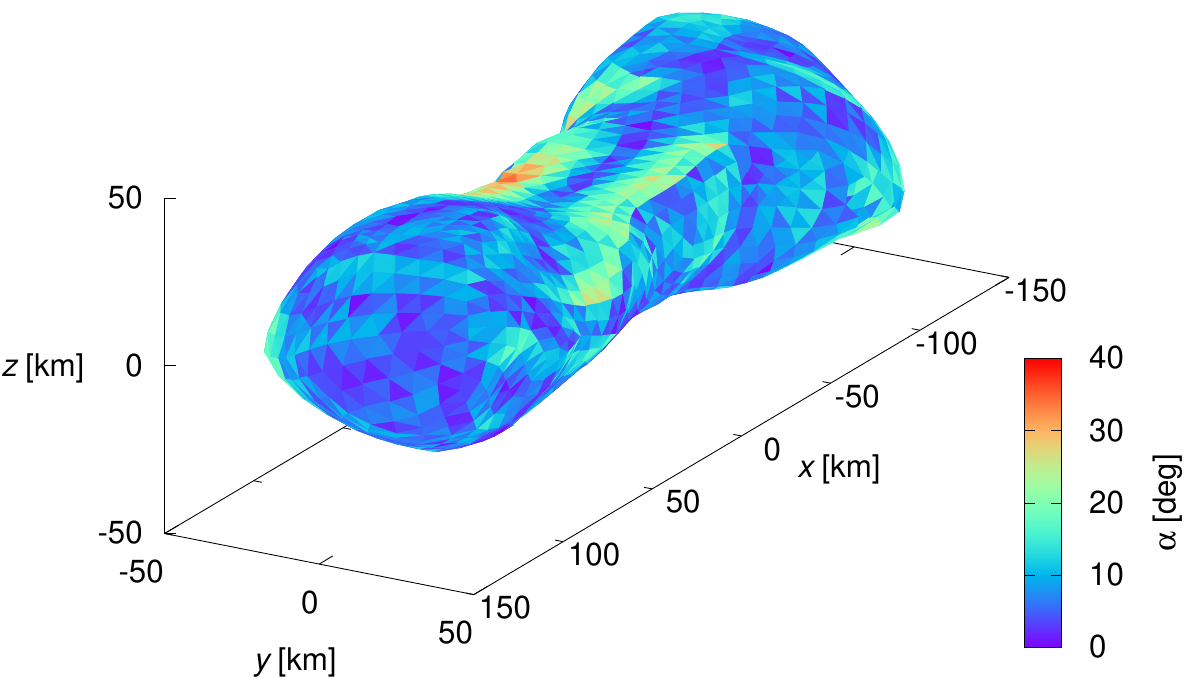}
\caption{
Same as Fig.~\ref{216_slope_rho3.3_gravity4__70_120}, but for the MPCD model
and density $\rho = 3\,430\,{\rm kg}\,{\rm m}^{-3}$.
Higher slopes are on the neck.
}
\label{216_slope_rho3.5_mpcd_gravity4__70_120}
\end{figure}

\begin{figure}
\centering
\includegraphics[width=7.5cm]{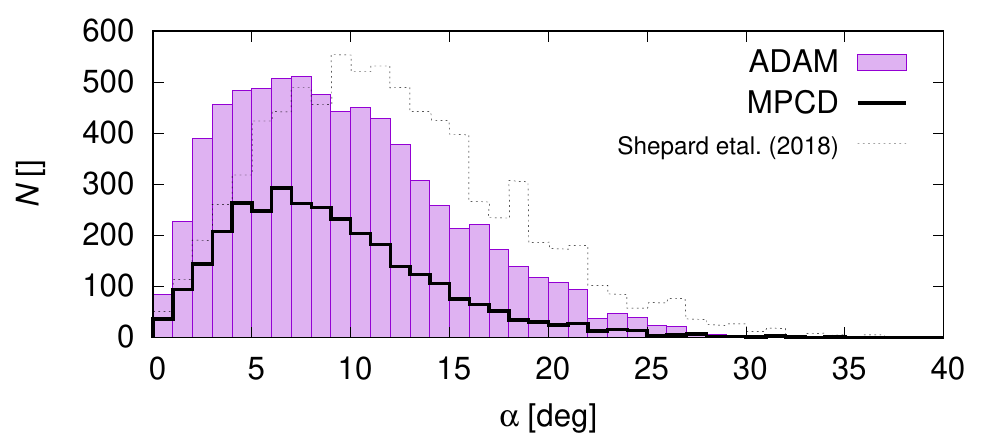}
\caption{
Histograms of slopes~$\alpha$ for the \adam{} and MPCD models.
They are most commonly between $2^\circ$ and $15^\circ$.
The distribution is a bit wider for the \adam{} model,
with a few outliers.
For comparison, we also plotted a histogram from \cite{Shepard2018}.
}
\label{216_slope_rho3.5_mpcd_hist}
\end{figure}

For comparison, we also show  the slopes computed on the basis of the MPCD model
(Fig.~\ref{216_slope_rho3.5_mpcd_gravity4__70_120}).
This shape model has a lower number of faces
which are more regular.
Interestingly, the steeper slopes are located elsewhere,
mostly on the neck.
The slopes on the $+x$ lobe are shallower (around $10^\circ$).
Given the nature of the shape models (\adam{} versus MPCD),
we think that local normals are better constrained
by the MPCD model.
The overall statistics (Fig.~\ref{216_slope_rho3.5_mpcd_hist})
from both models is comparable, though.
A comparison to \cite{Shepard2018} shows
that our mean slope is significantly lower ($9^\circ$ versus $12^\circ$).
Let us recall, however, that this difference might be partly "enforced"
by our/their regularization techniques.

%%%%%%%%%%%%%%%%%%%%%%%%%%%%%%%%%%%%%%%%%%%%%%%%%%%%%%%%%%%%%%%%%%%%%%%%

\section{Additional figures and tables}

\setkeys{Gin}{draft=false}
\begin{figure*}%[!h]
\begin{center}
\resizebox{1.0\hsize}{!}{\includegraphics{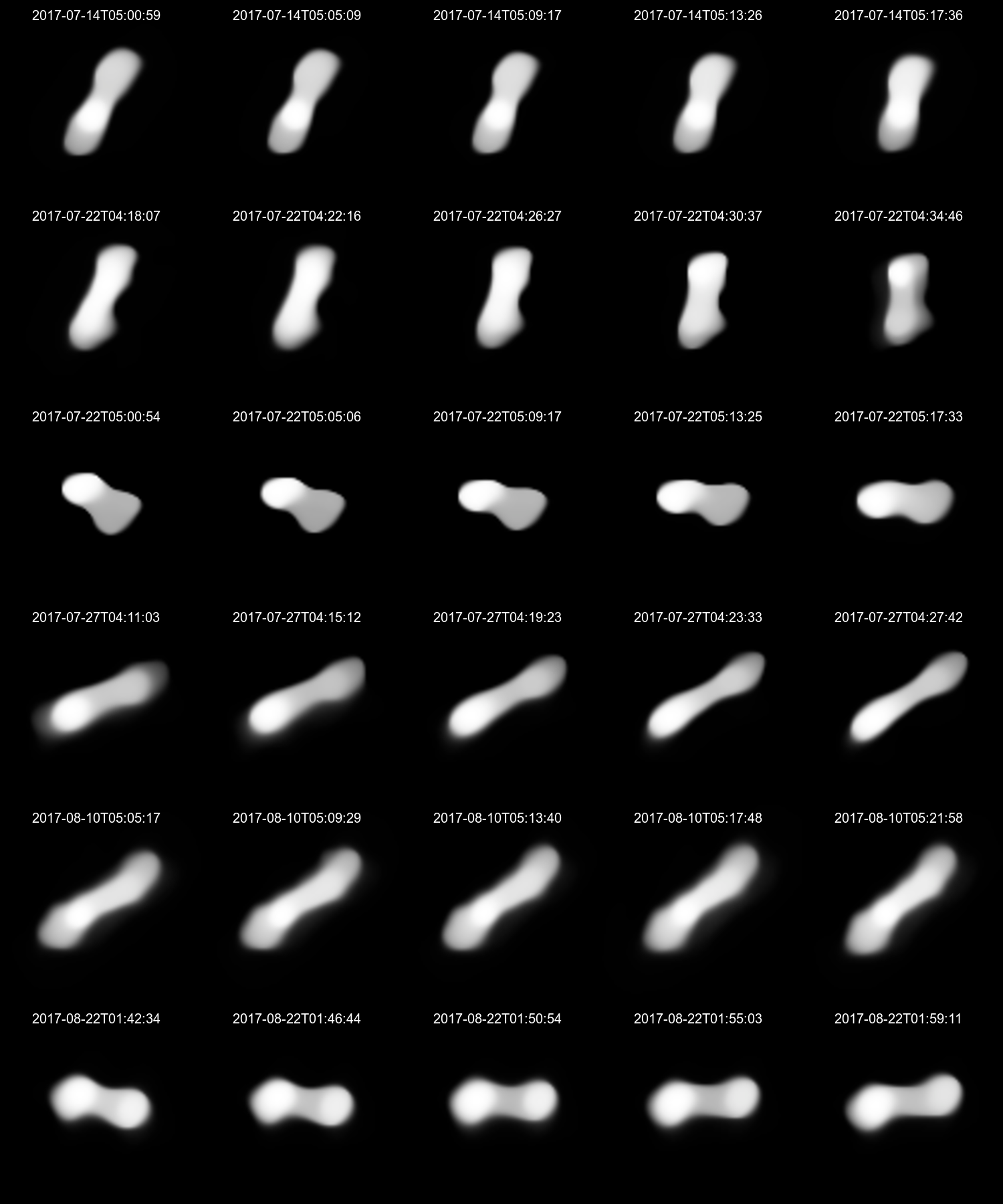}}
\end{center}
\caption{\label{fig:Deconv1}Full set of VLT/SPHERE/ZIMPOL images of (216)~Kleopatra from the apparition in 2017. The images were deconvolved by the \mistral~algorithm. { The pixel scale is 3.6 mas.} Additional information about the data can be found in Table~\ref{tab:ao}.}
\end{figure*}
\setkeys{Gin}{draft=true}

\setkeys{Gin}{draft=false}
\begin{figure*}%[!h]
\begin{center}
\resizebox{1.0\hsize}{!}{\includegraphics{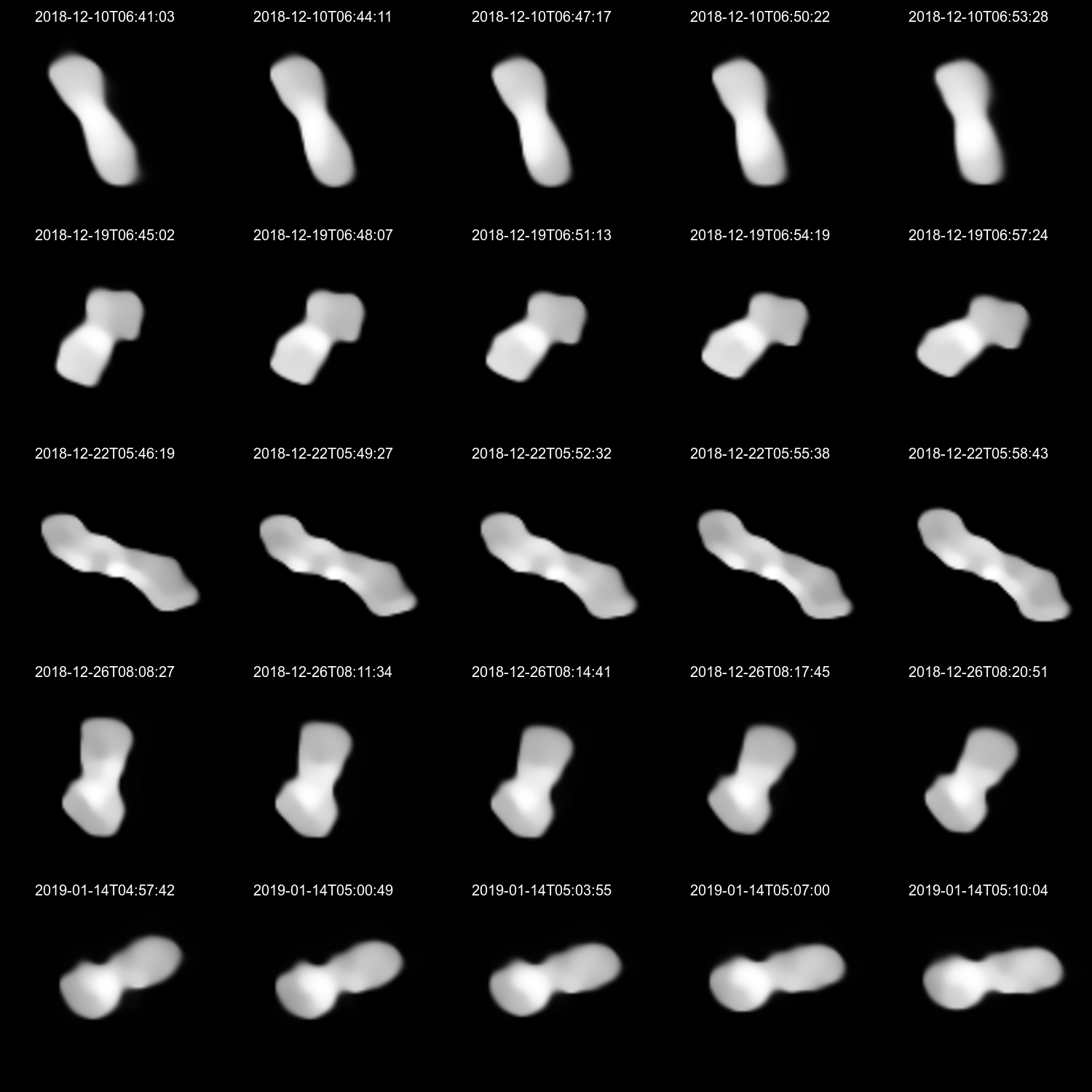}}
\end{center}
\caption{\label{fig:Deconv2}Full set of VLT/SPHERE/ZIMPOL images of (216)~Kleopatra from the apparition in {2018/2019}. The images were deconvolved by the \mistral~algorithm. { The pixel scale is 3.6 mas.} Additional information about the data can be found in Table~\ref{tab:ao}.}
\end{figure*}
\setkeys{Gin}{draft=true}

\setkeys{Gin}{draft=false}
\begin{figure*}%[!t]
\begin{center}
\resizebox{0.99\hsize}{!}{\includegraphics{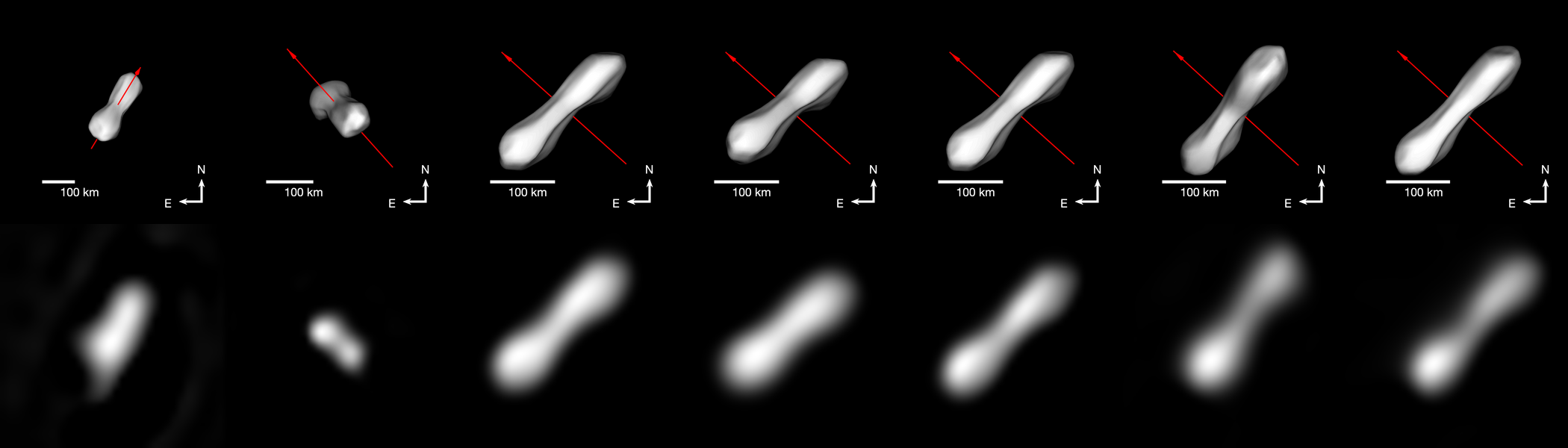}}

\resizebox{0.99\hsize}{!}{\includegraphics{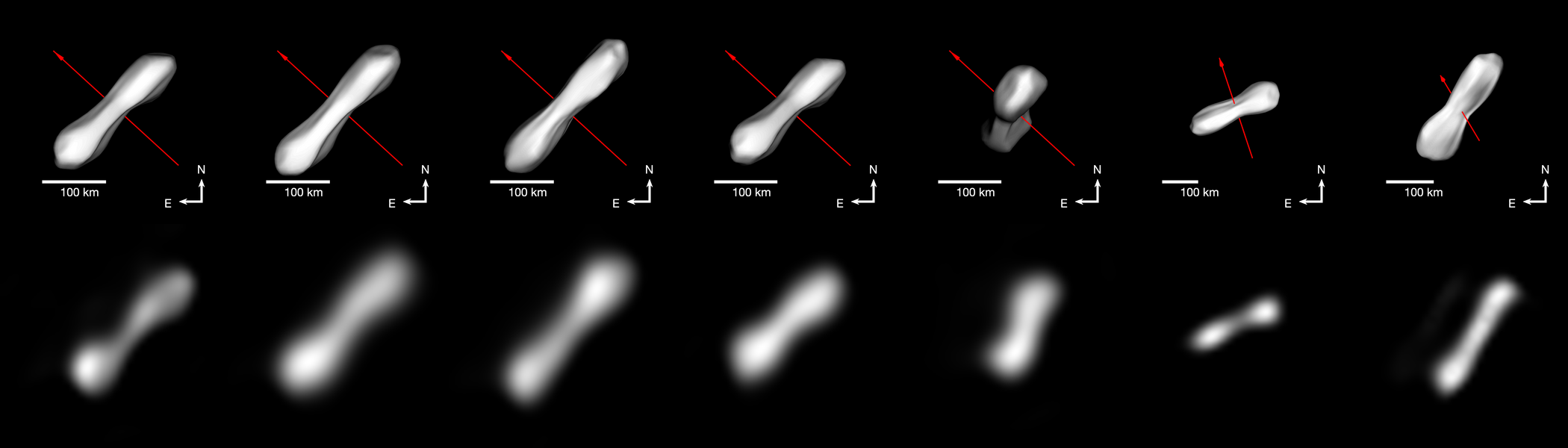}}
\end{center}
\caption{\label{fig:comparisonKeck}{Comparison between the disk-resolved images from the Keck/NIRC2 camera and the corresponding projections of our \adam{} shape model. The red arrow indicates the orientation of the spin axis. The ordering of the images follows the chronological order in Table~\ref{tab:aoKeck}. Given the lower resolution of the Keck data (pixel scale 9.942 mas) compared to those of the SPHERE (pixel scale 3.6 mas), the agreement between the images and the model projections is quite reasonable and sufficient. }}
\end{figure*}
\setkeys{Gin}{draft=true}

\setkeys{Gin}{draft=false}
\begin{figure*}%[!t]
\begin{center}
\resizebox{0.99\hsize}{!}{\includegraphics{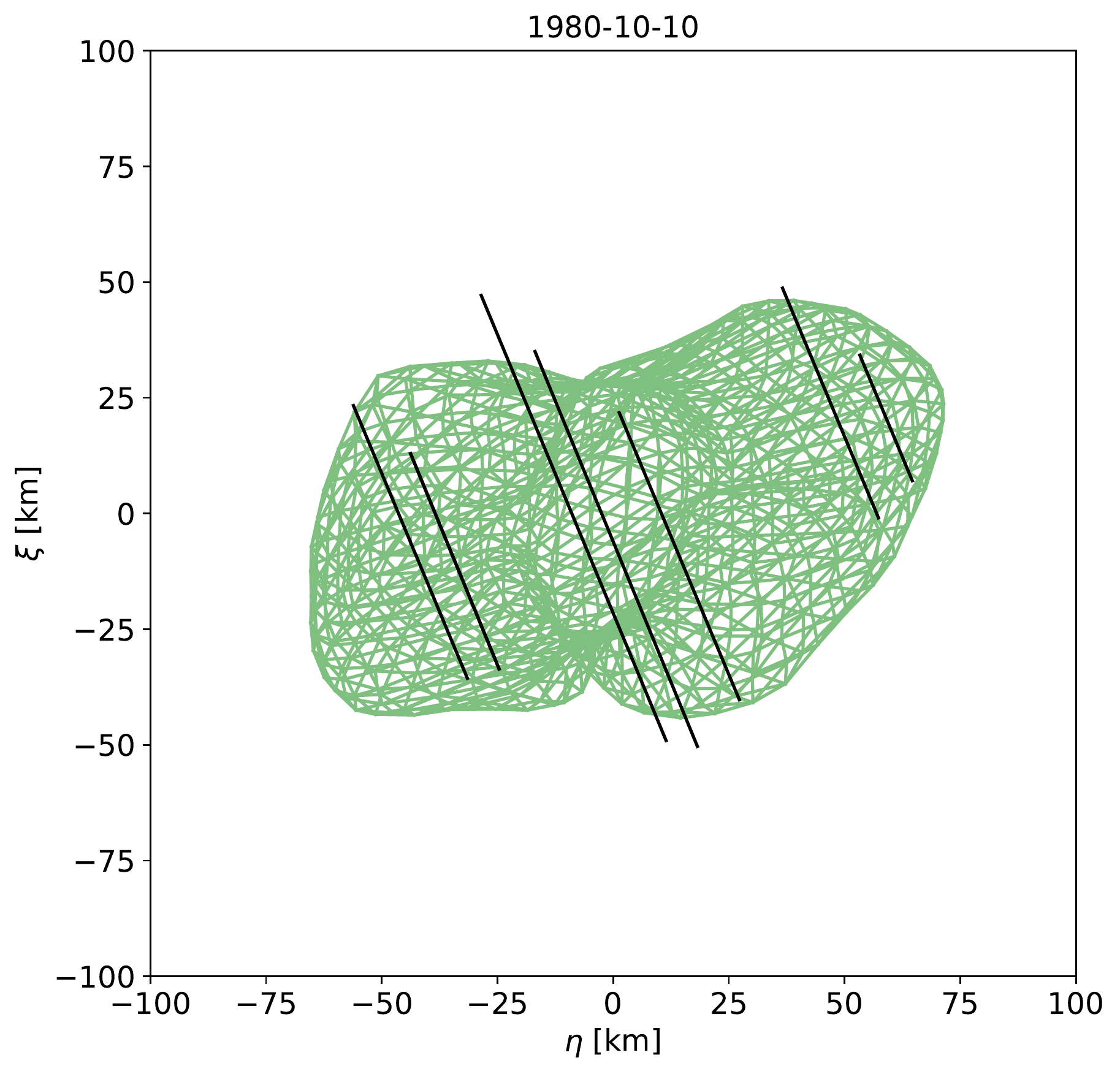}\includegraphics{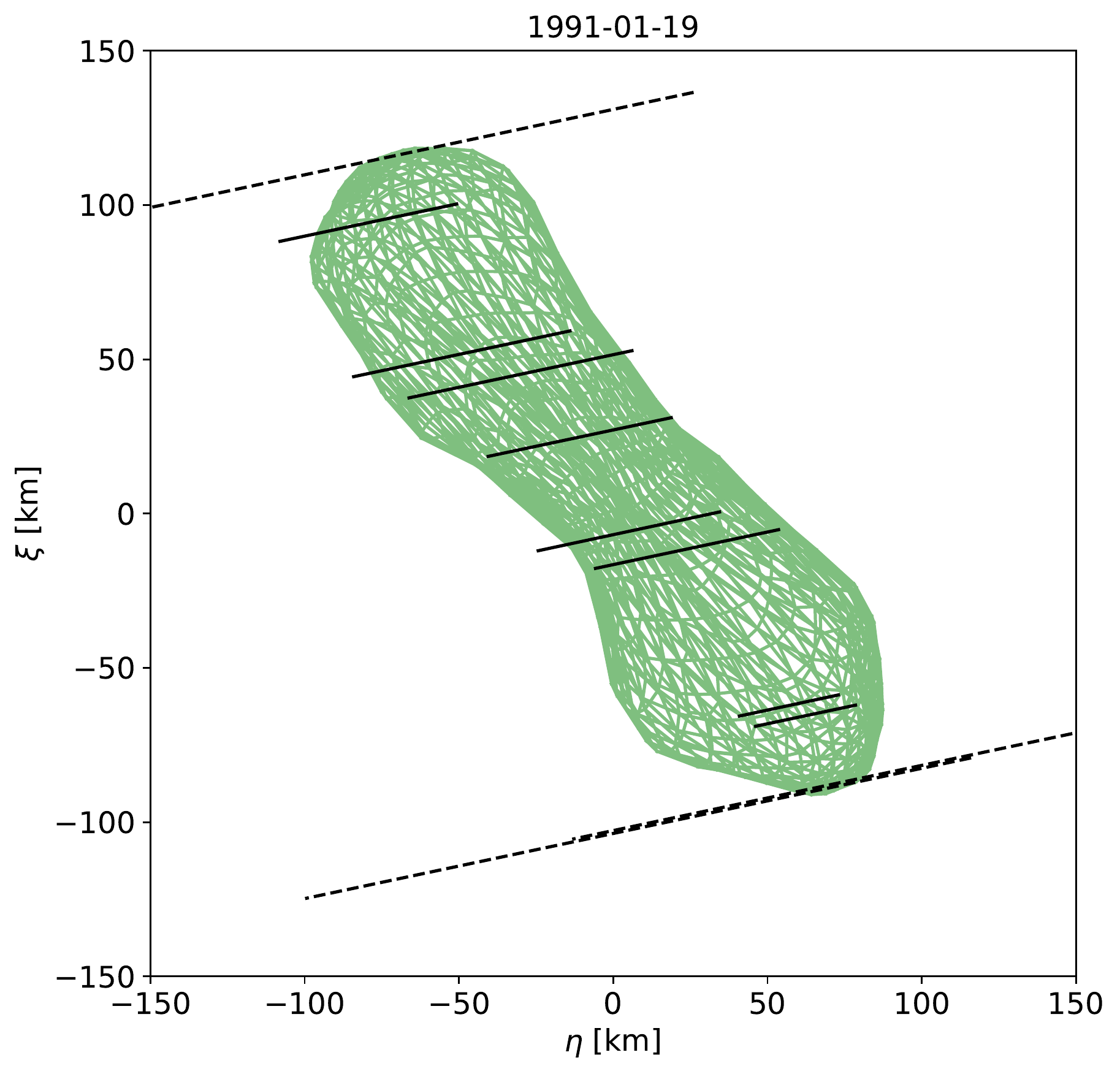}\includegraphics{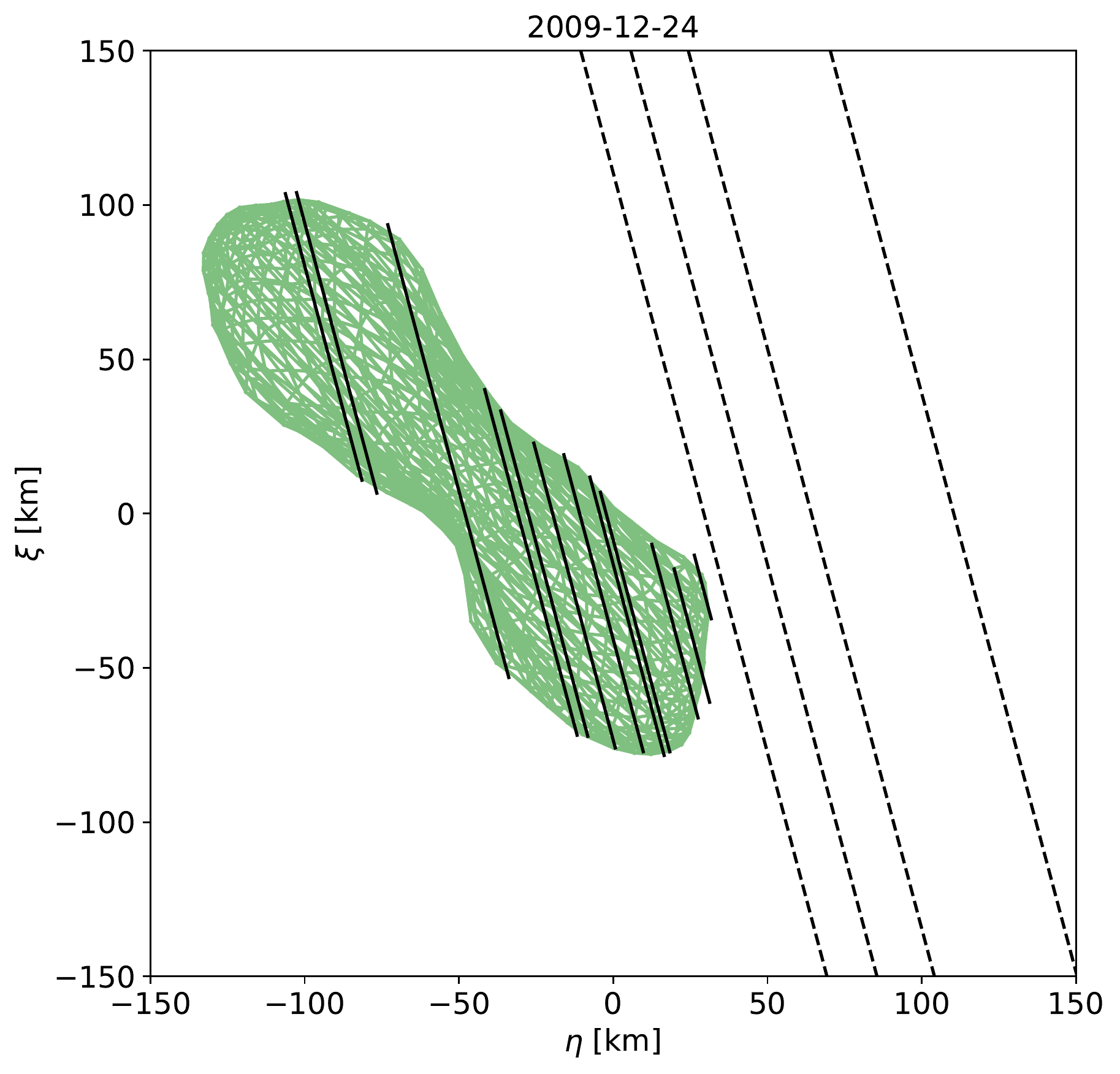}}

\resizebox{0.66\hsize}{!}{\includegraphics{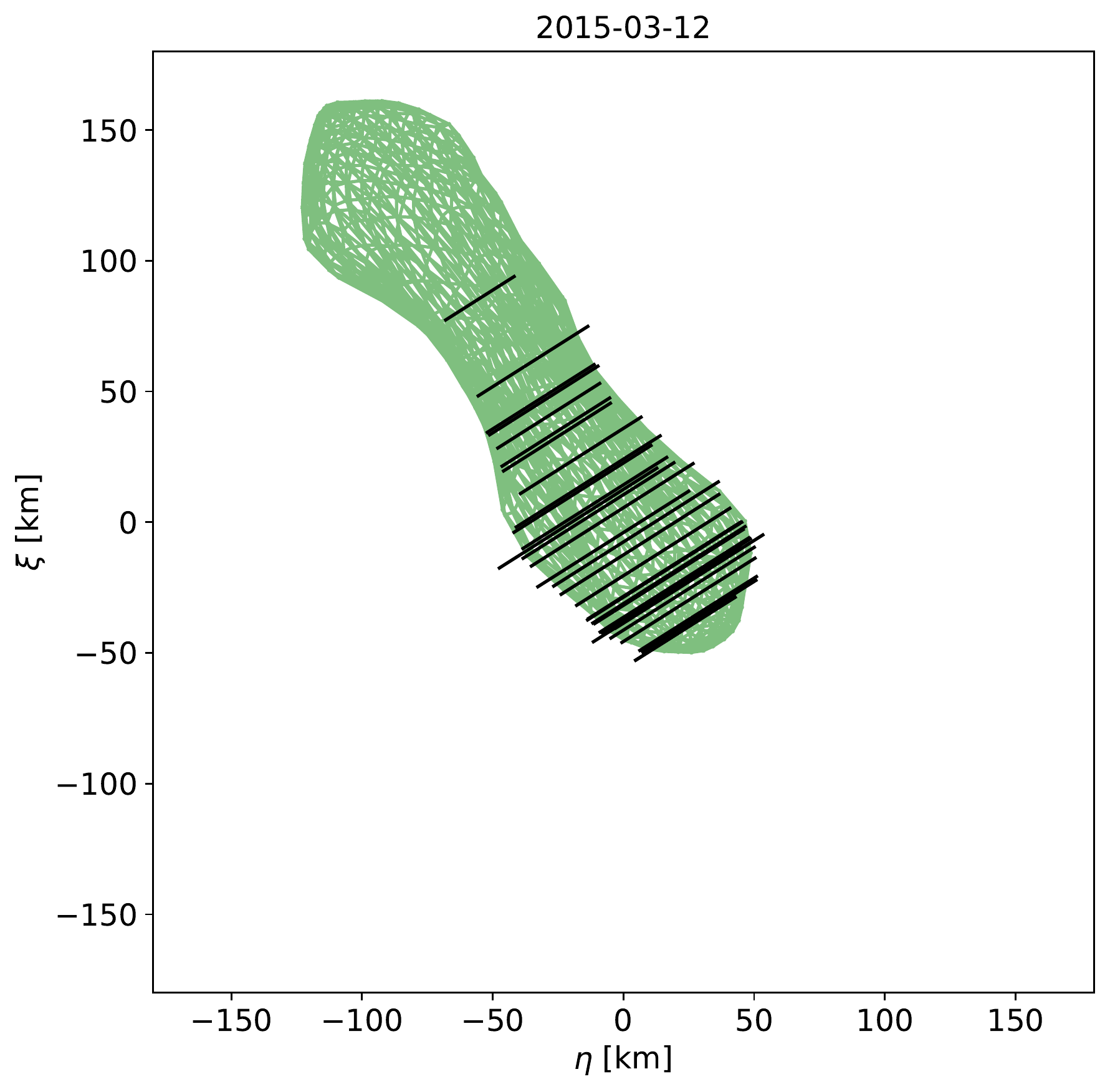}\includegraphics{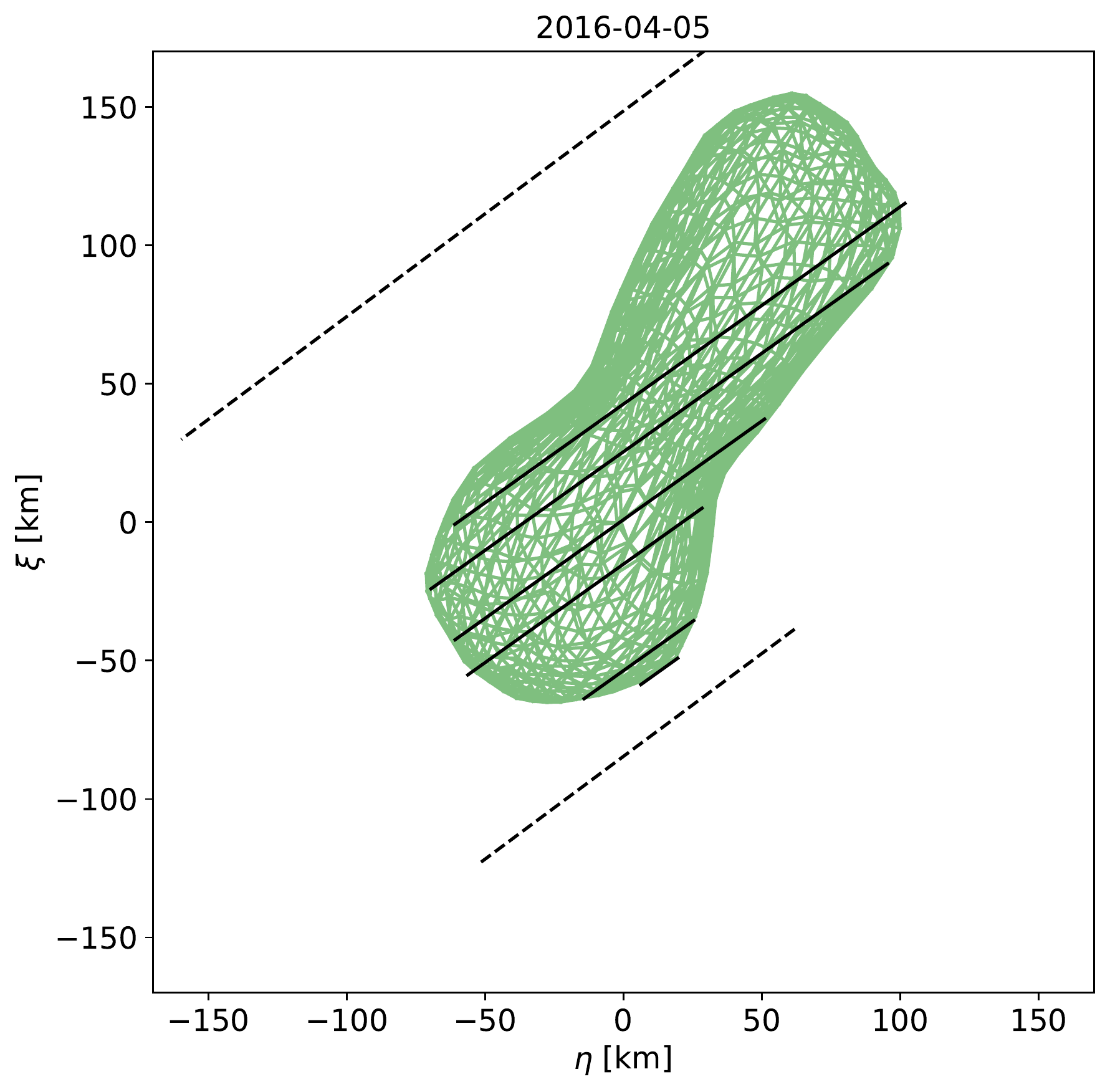}}
\end{center}
\caption{\label{fig:comparisonOCC}Comparison between the stellar occultations of Kleopatra { \citep{Herald2020}} and the corresponding projections of our \adam{} shape model. Dashed lines are the negative observations. Occultations are ordered chronologically; the first two were not included in the \adam{} shape modeling and are shown here for consistency reasons.}
\end{figure*}
\setkeys{Gin}{draft=true}

%\setkeys{Gin}{draft=false}
%\begin{figure*}%[!t]
%\begin{center}
%\resizebox{0.99\hsize}{!}{\includegraphics{./figs3/lcs_fit_3.png}\includegraphics{./figs3/lcs_fit_10.png}\includegraphics{./figs3/lcs_fit_17.png}}\\
%\resizebox{0.99\hsize}{!}{\includegraphics{./figs3/lcs_fit_18.png}\includegraphics{./figs3/lcs_fit_29.png}\includegraphics{./figs3/lcs_fit_45.png}}\\
%\resizebox{0.99\hsize}{!}{\includegraphics{./figs3/lcs_fit_48.png}\includegraphics{./figs3/lcs_fit_56.png}\includegraphics{./figs3/lcs_fit_62.png}}\\
%\resizebox{0.99\hsize}{!}{\includegraphics{./figs3/lcs_fit_68.png}\includegraphics{./figs3/lcs_fit_172.png}\includegraphics{./figs3/lcs_fit_175.png}}\\
%\end{center}
%\caption{\label{fig:comparisonLC}Comparison between a subset of optical lightcurves of Kleopatra and the corresponding modeled lightcurves based on our \adam{} shape model. The fit to all lightcurves from our dataset is available in DAMIT database.}
%\end{figure*}
%\setkeys{Gin}{draft=true}

\setkeys{Gin}{draft=false}
\begin{figure*}%[!t]
\begin{center}
\resizebox{0.99\hsize}{!}{\includegraphics{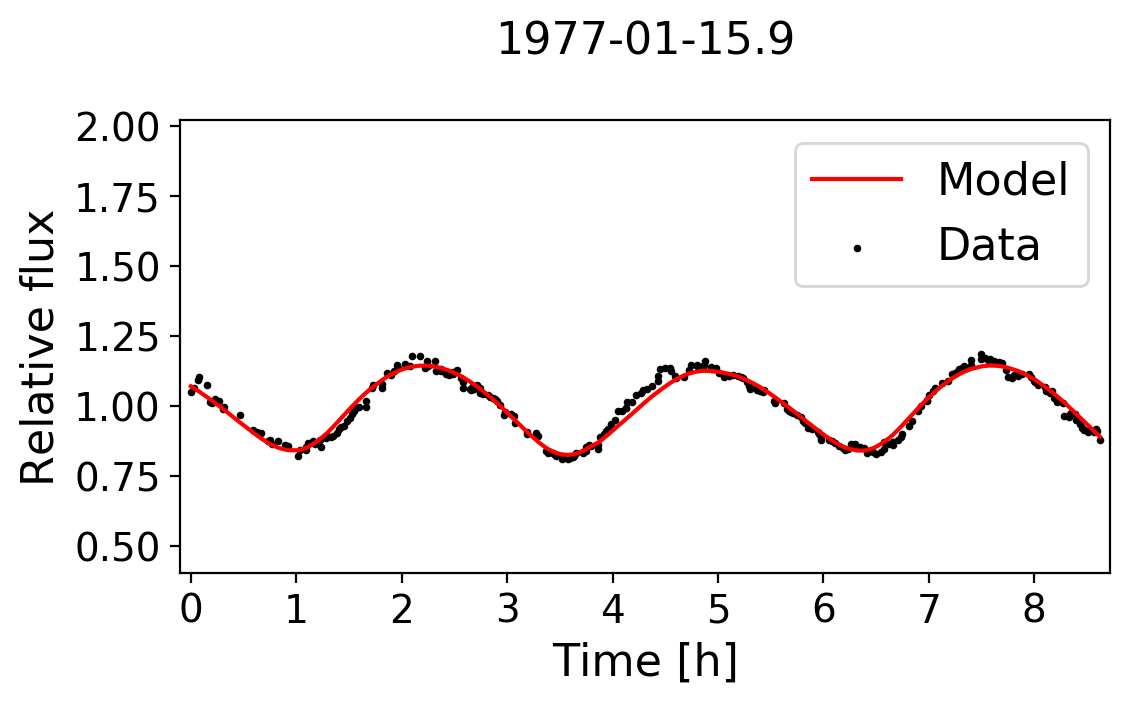}\includegraphics{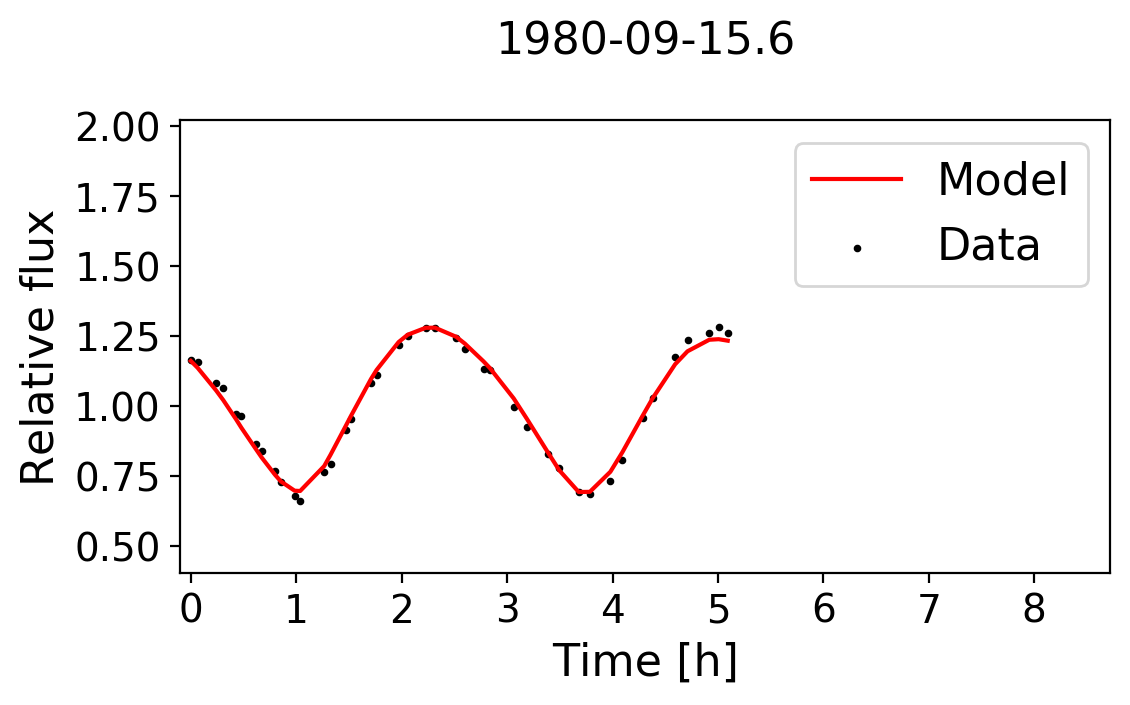}}

\resizebox{0.99\hsize}{!}{\includegraphics{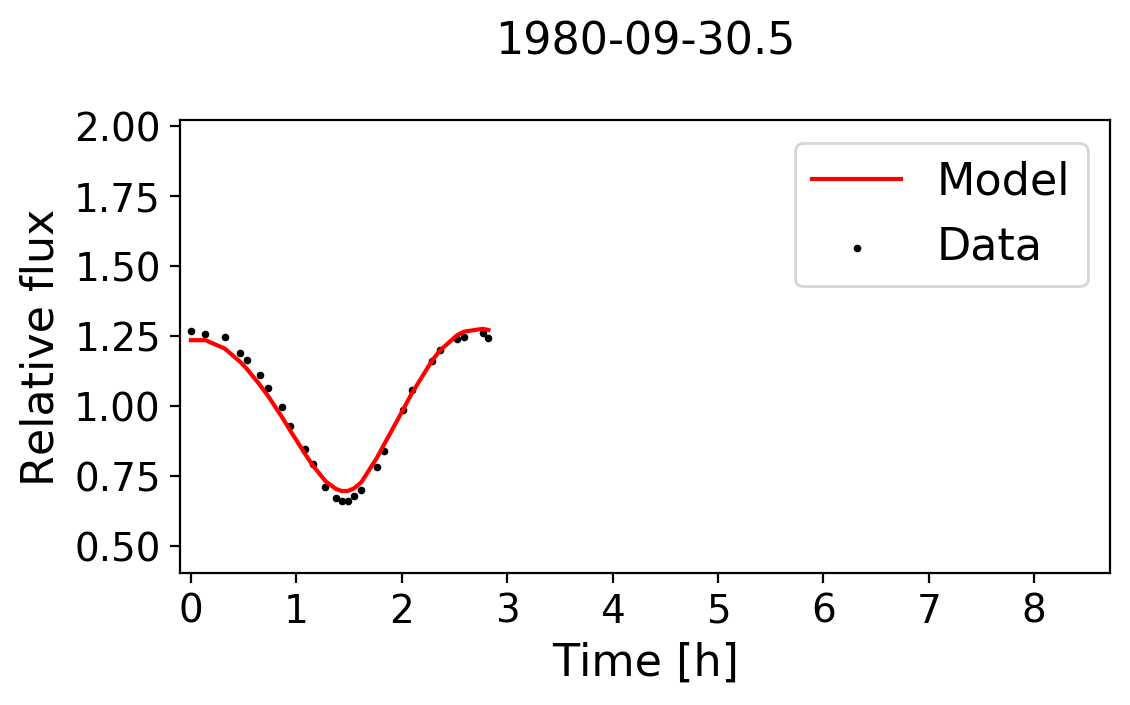}\includegraphics{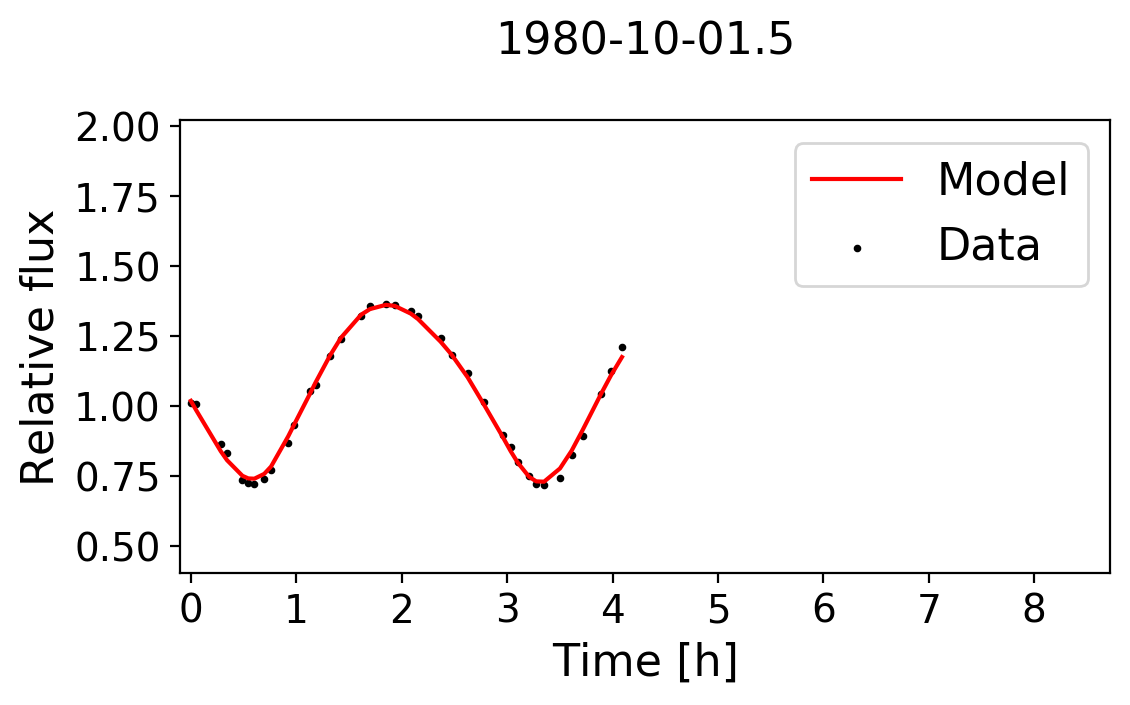}}

\resizebox{0.99\hsize}{!}{\includegraphics{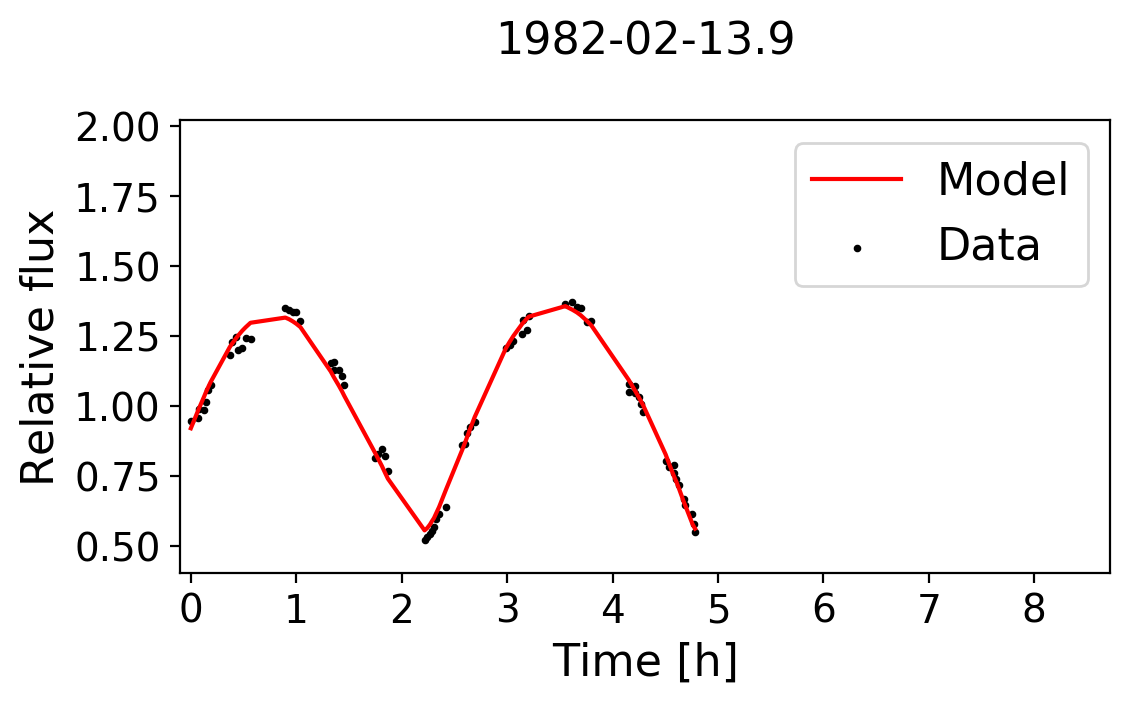}\includegraphics{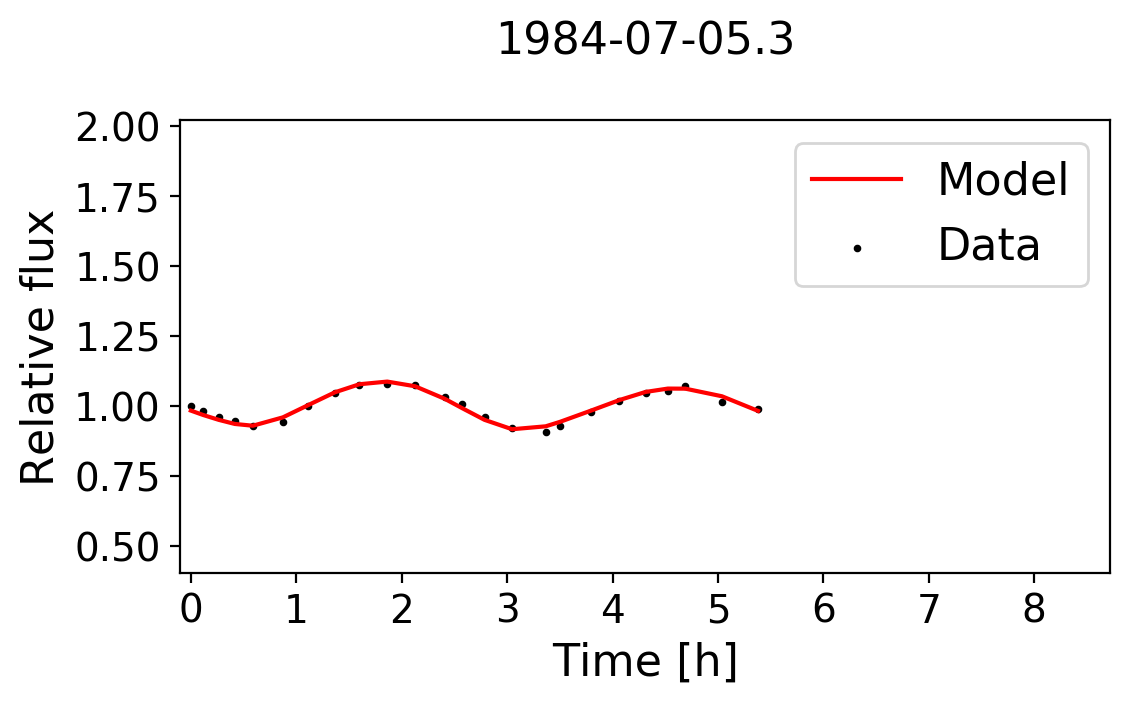}}
\end{center}
\caption{\label{fig:comparisonLC1}Comparison between a subset of optical lightcurves of Kleopatra and the corresponding modeled lightcurves based on our \adam{} shape model. The fit to all lightcurves from our dataset is available in the DAMIT database.}
\end{figure*}
\setkeys{Gin}{draft=true}

\setkeys{Gin}{draft=false}
\begin{figure*}%[!t]
\begin{center}
\resizebox{0.99\hsize}{!}{\includegraphics{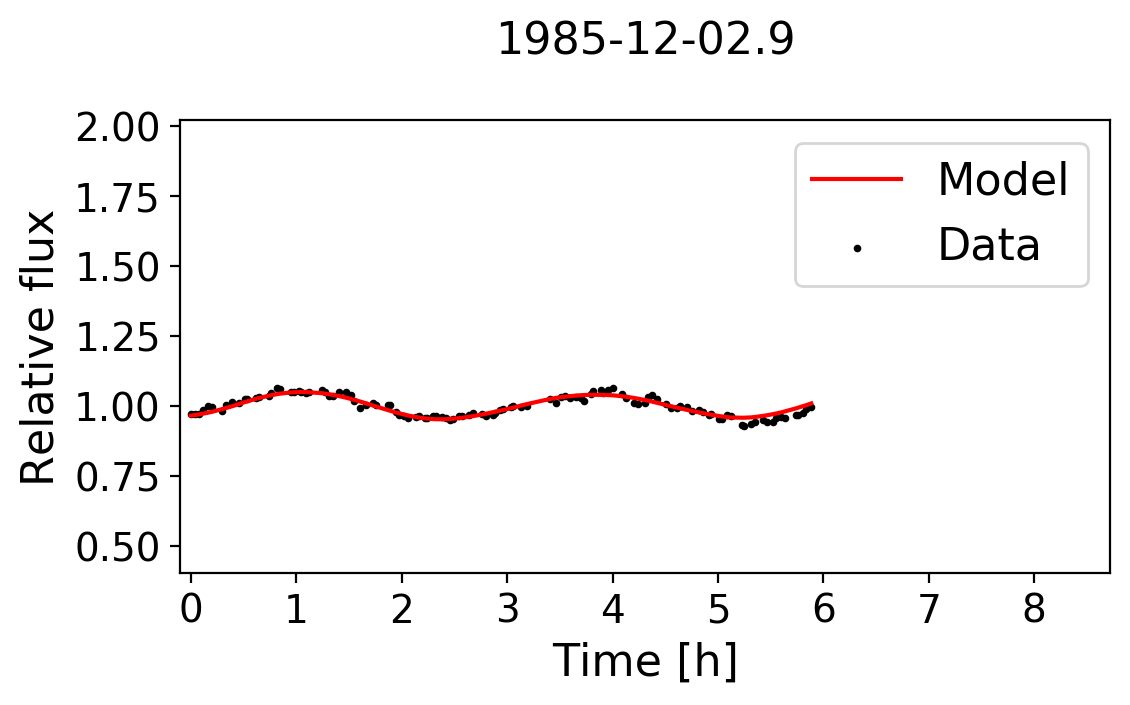}\includegraphics{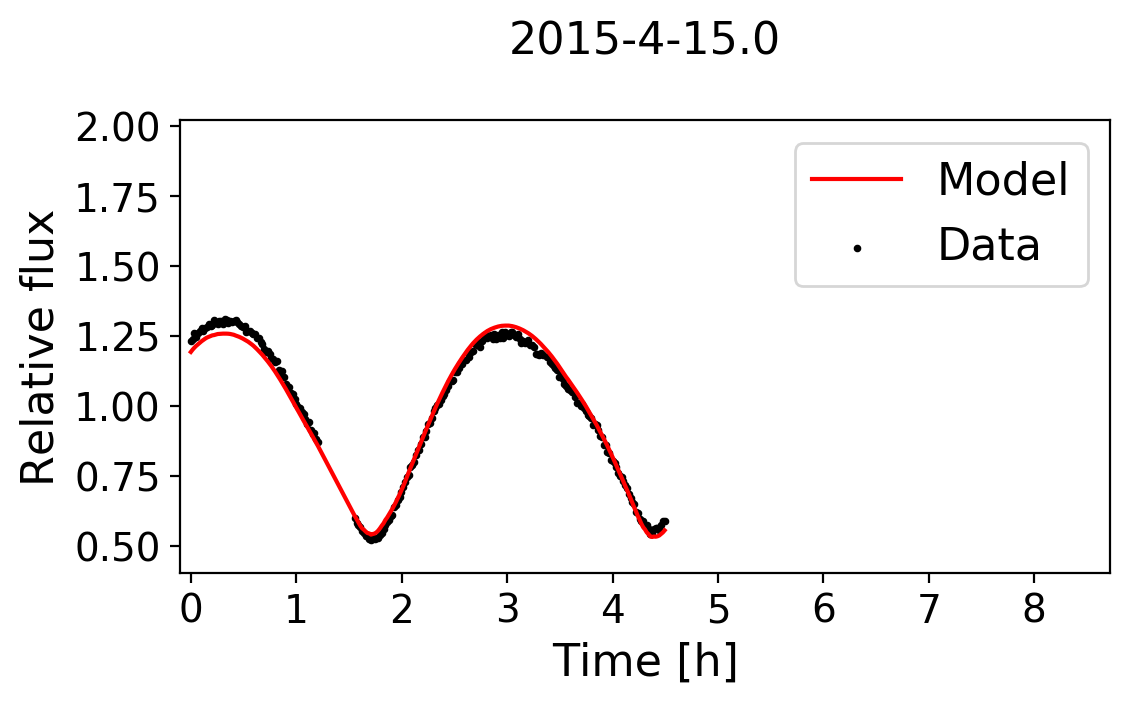}}

\resizebox{0.99\hsize}{!}{\includegraphics{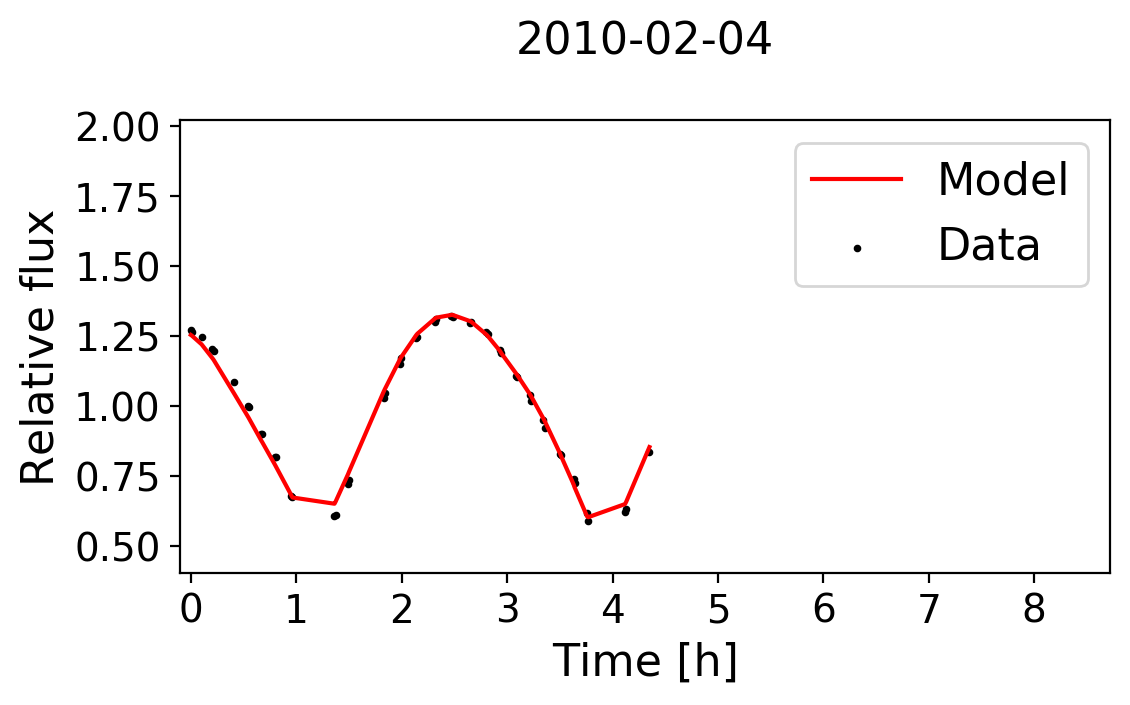}\includegraphics{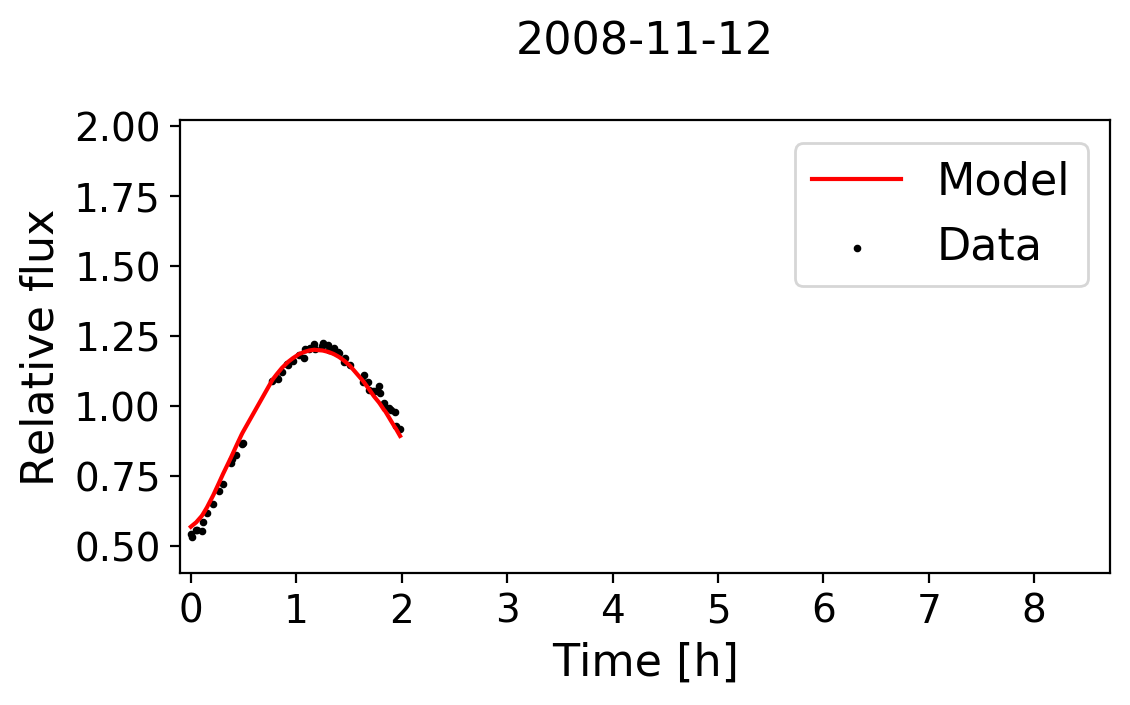}}

\resizebox{0.99\hsize}{!}{\includegraphics{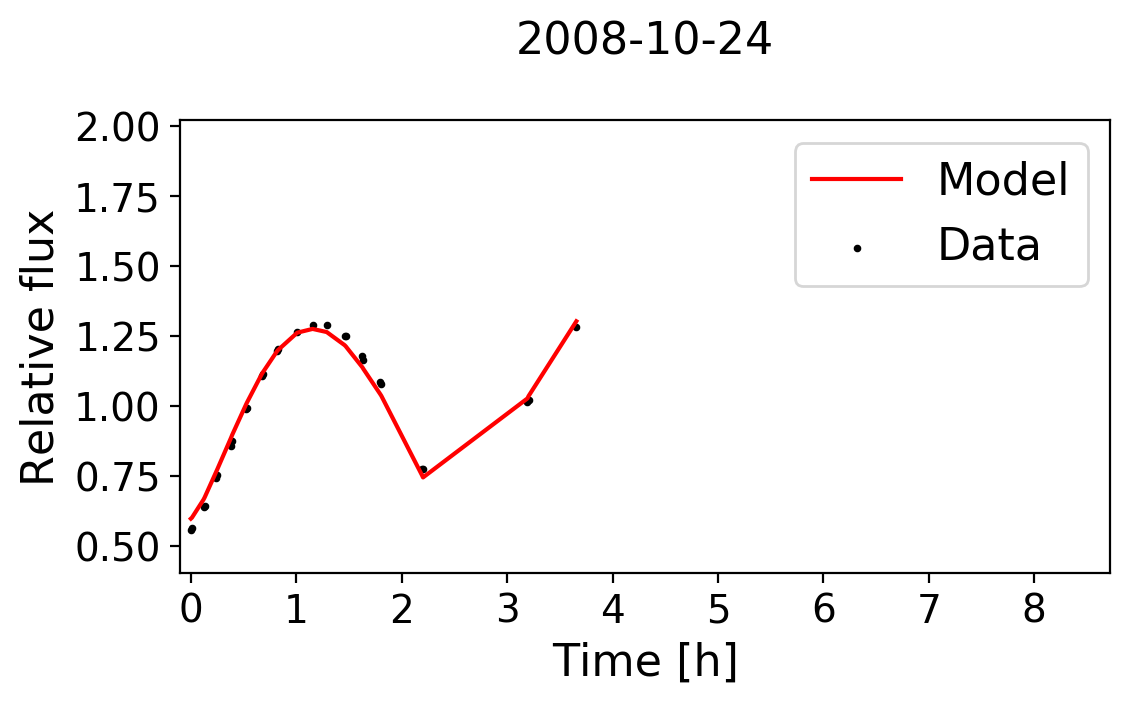}\includegraphics{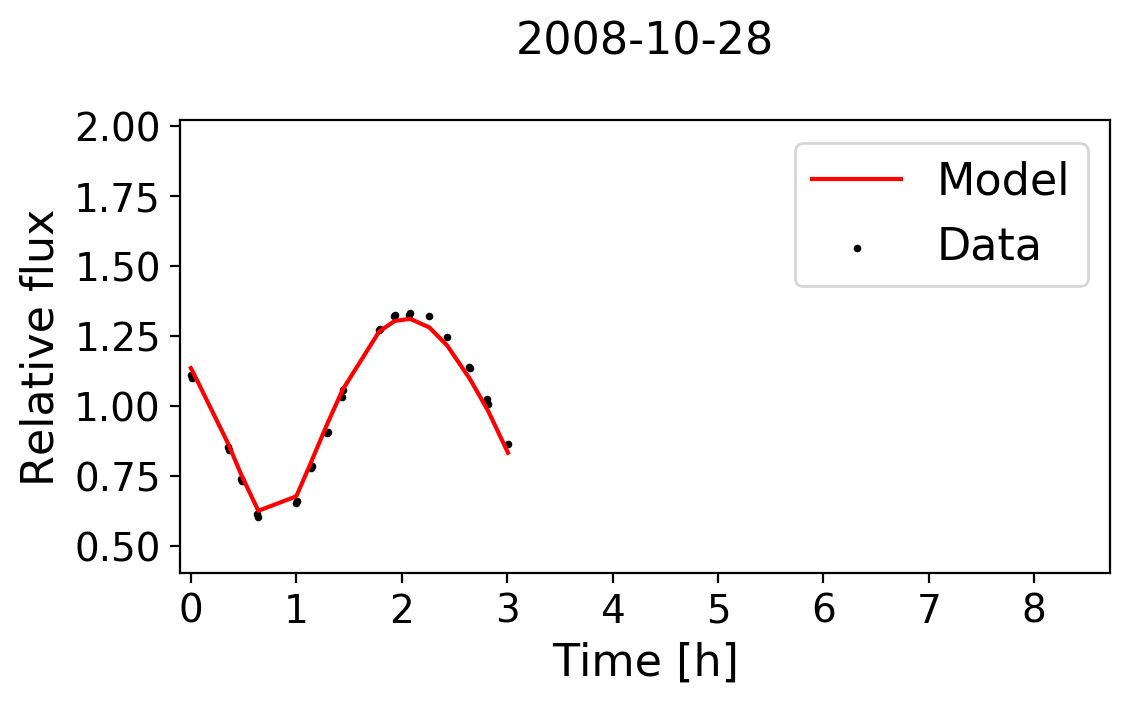}}
\end{center}
\caption{\label{fig:comparisonLC2}Comparison between a subset of optical lightcurves of Kleopatra and the corresponding modeled lightcurves based on our \adam{} shape model. The fit to all lightcurves from our dataset is available in the DAMIT database.}
\end{figure*}
\setkeys{Gin}{draft=true}

\setkeys{Gin}{draft=false}
\begin{figure*}%[!t]
%\centering
    \begin{subfigure}[b]{0.16\textwidth}
    \centering
      \includegraphics[clip=true,scale=0.25,trim=50 50 50 50]{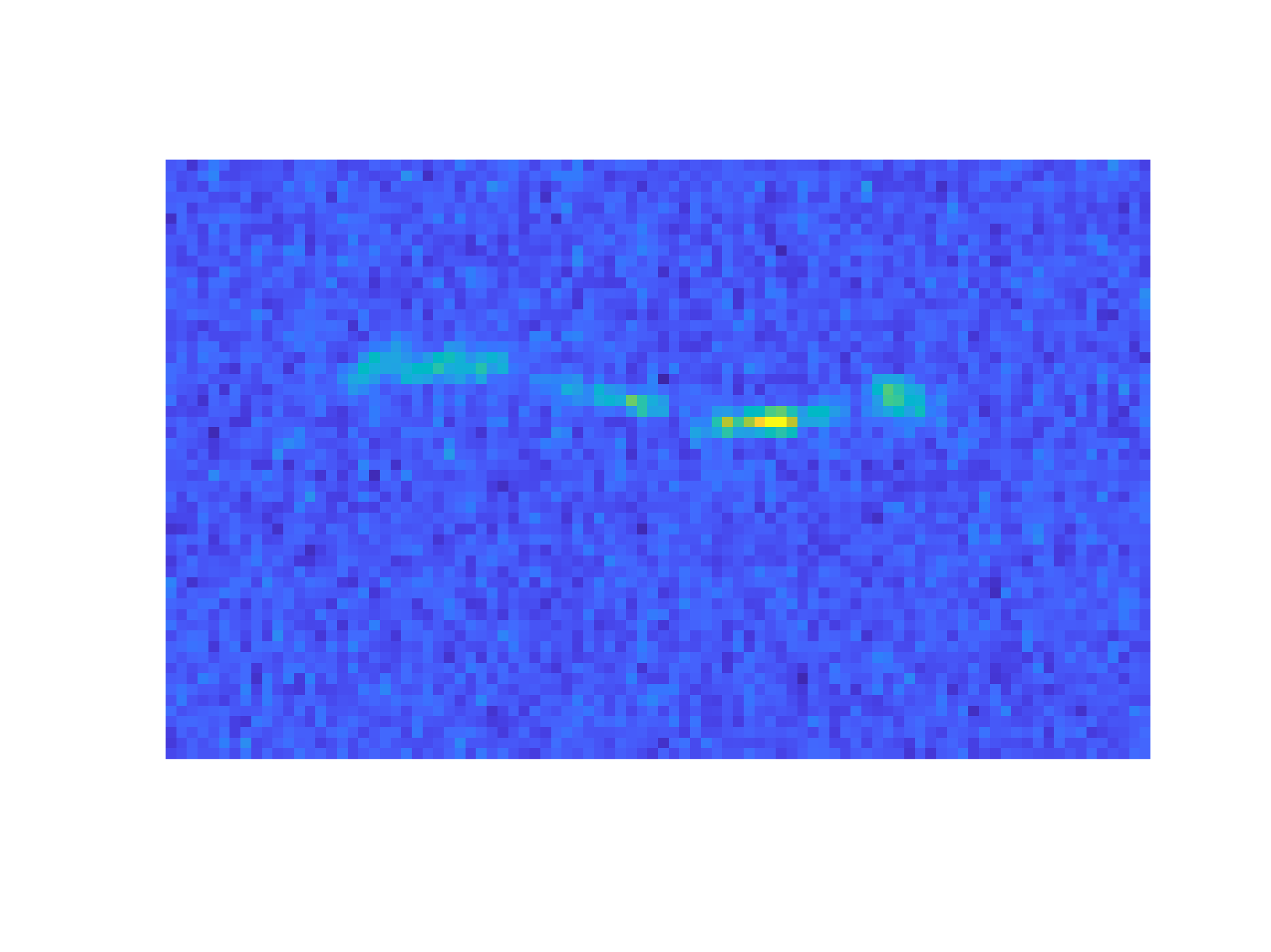}
      \end{subfigure}
      \begin{subfigure}[b]{0.16\textwidth}
      \includegraphics[clip=true,scale=0.25,trim=50 50 50 50]{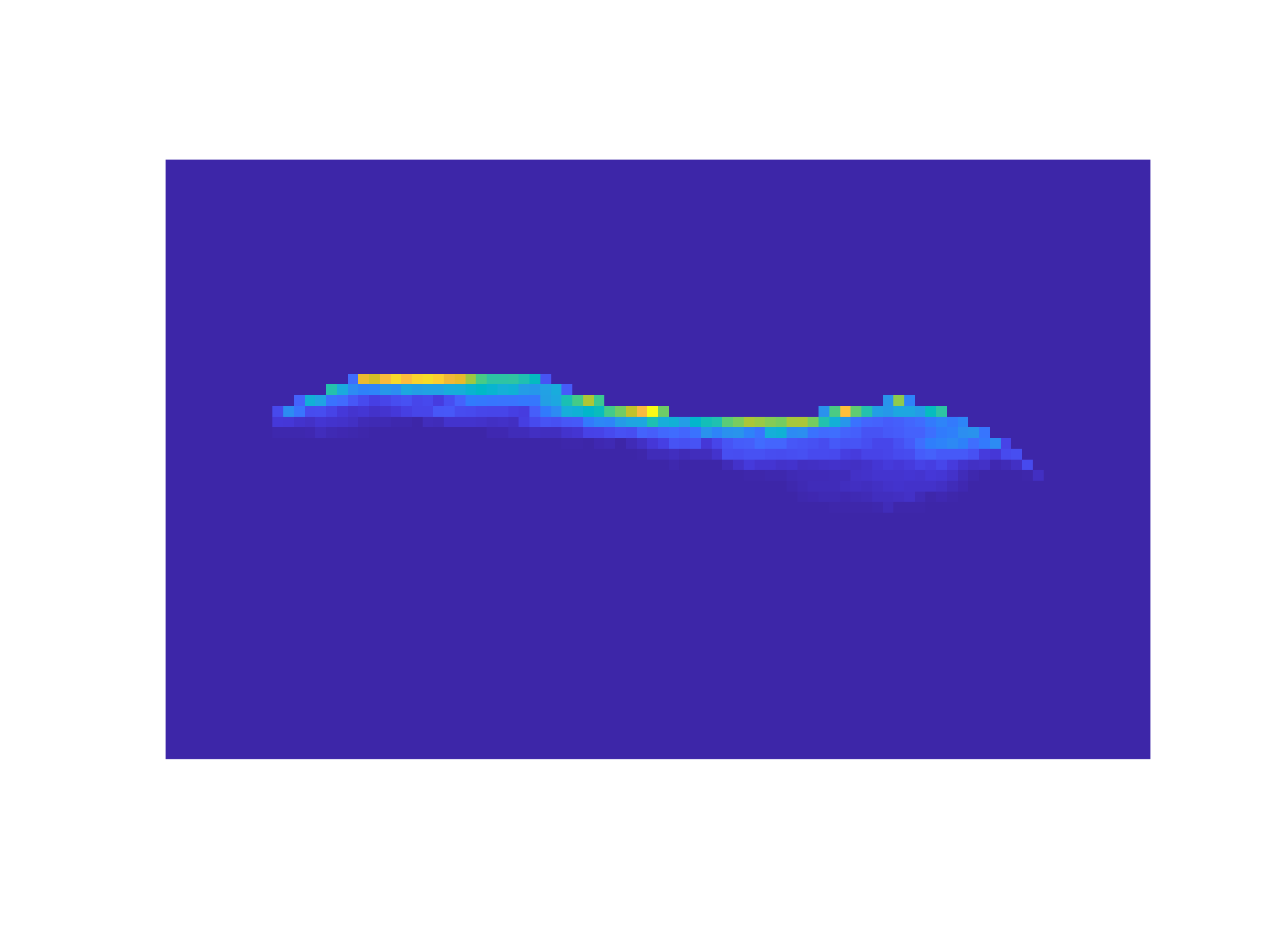}
      \centering
      \end{subfigure}
      \begin{subfigure}[b]{0.16\textwidth}
      \includegraphics[clip=true,scale=0.2,trim=15 30 15 30]{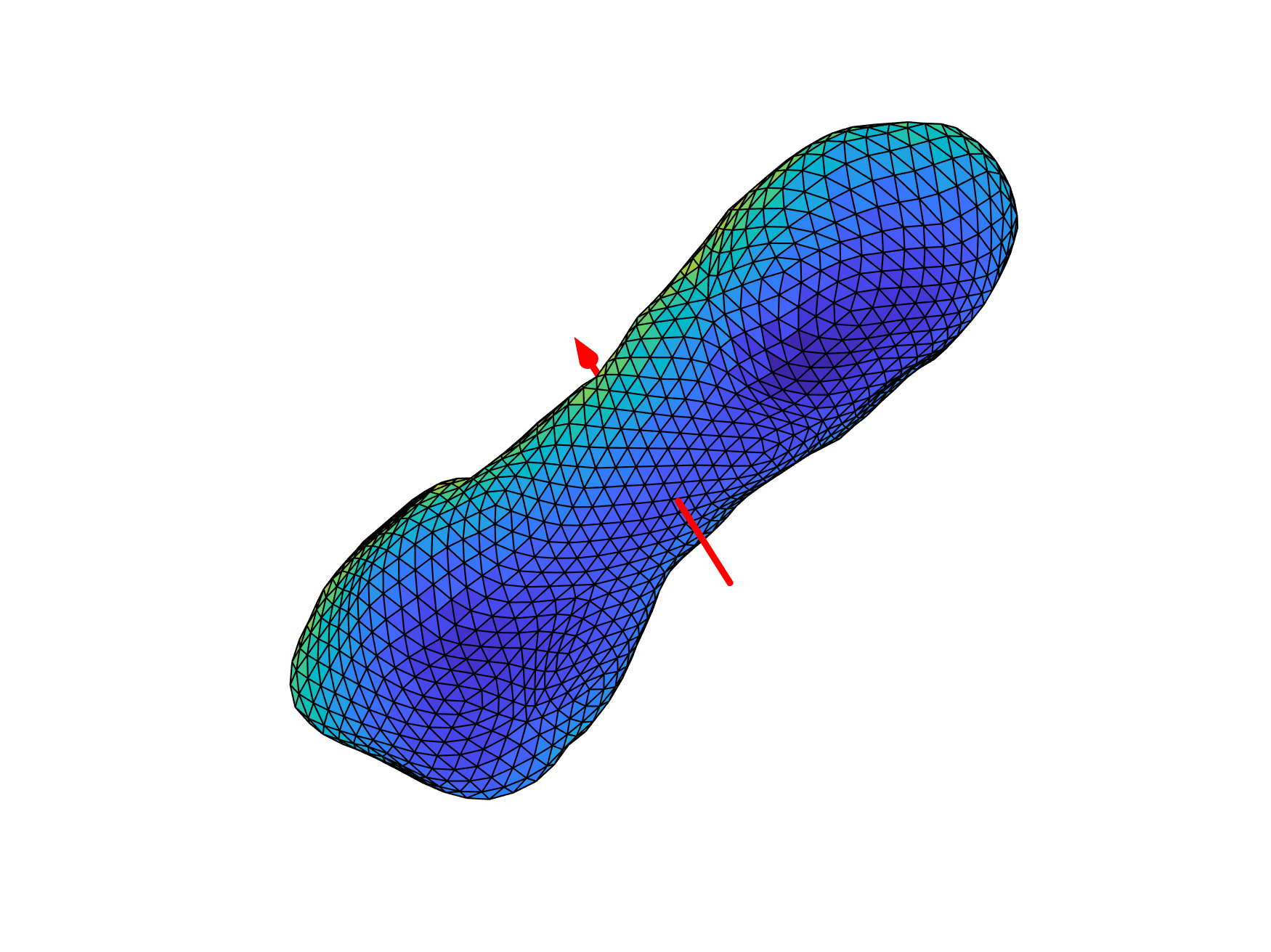}
      \centering
      \end{subfigure}
      \begin{subfigure}[b]{0.16\textwidth}
    \centering
      \includegraphics[clip=true,scale=0.25,trim=50 50 50 50]{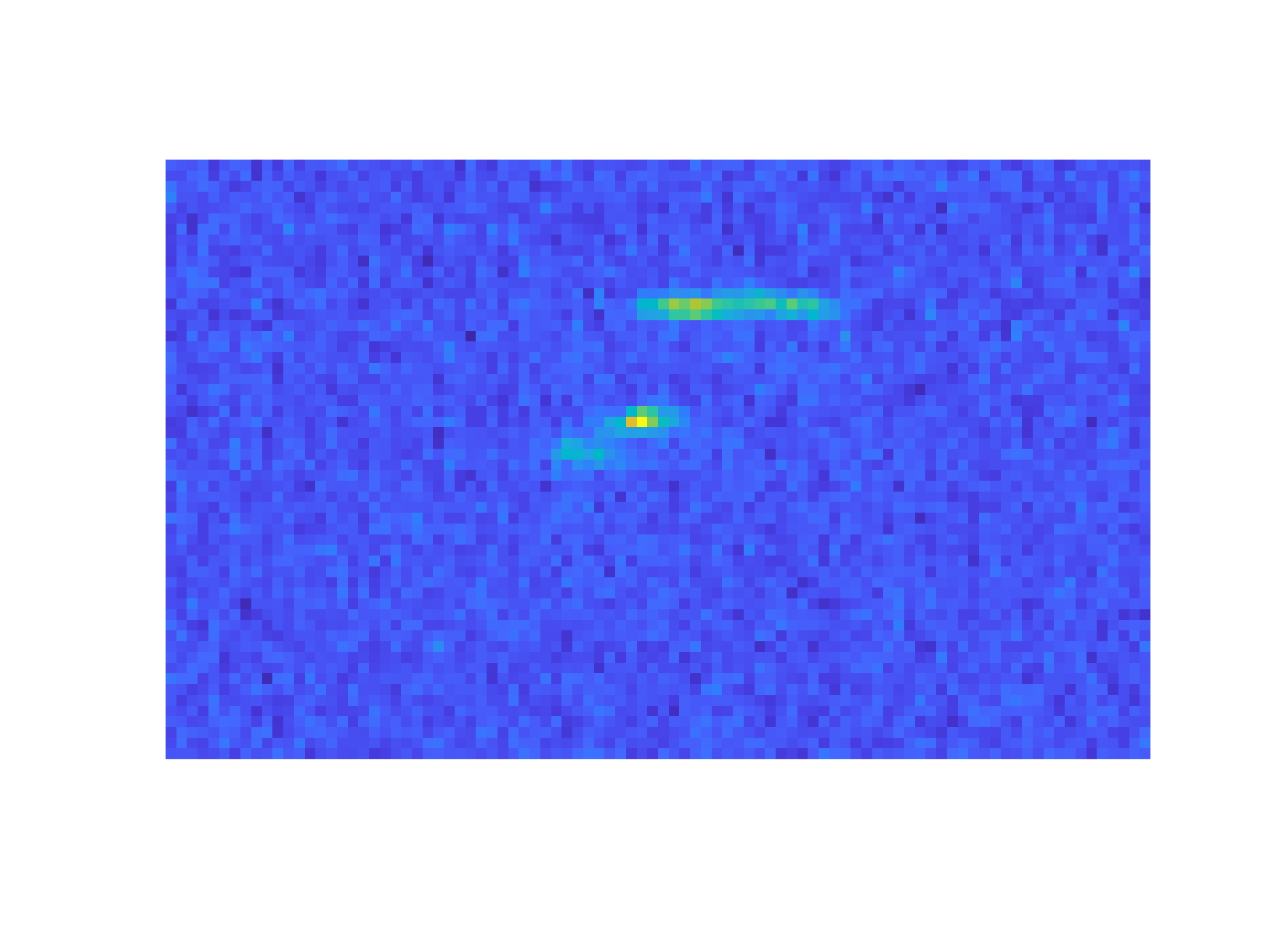}
      \end{subfigure}
      \begin{subfigure}[b]{0.16\textwidth}
      \includegraphics[clip=true,scale=0.25,trim=50 50 50 50]{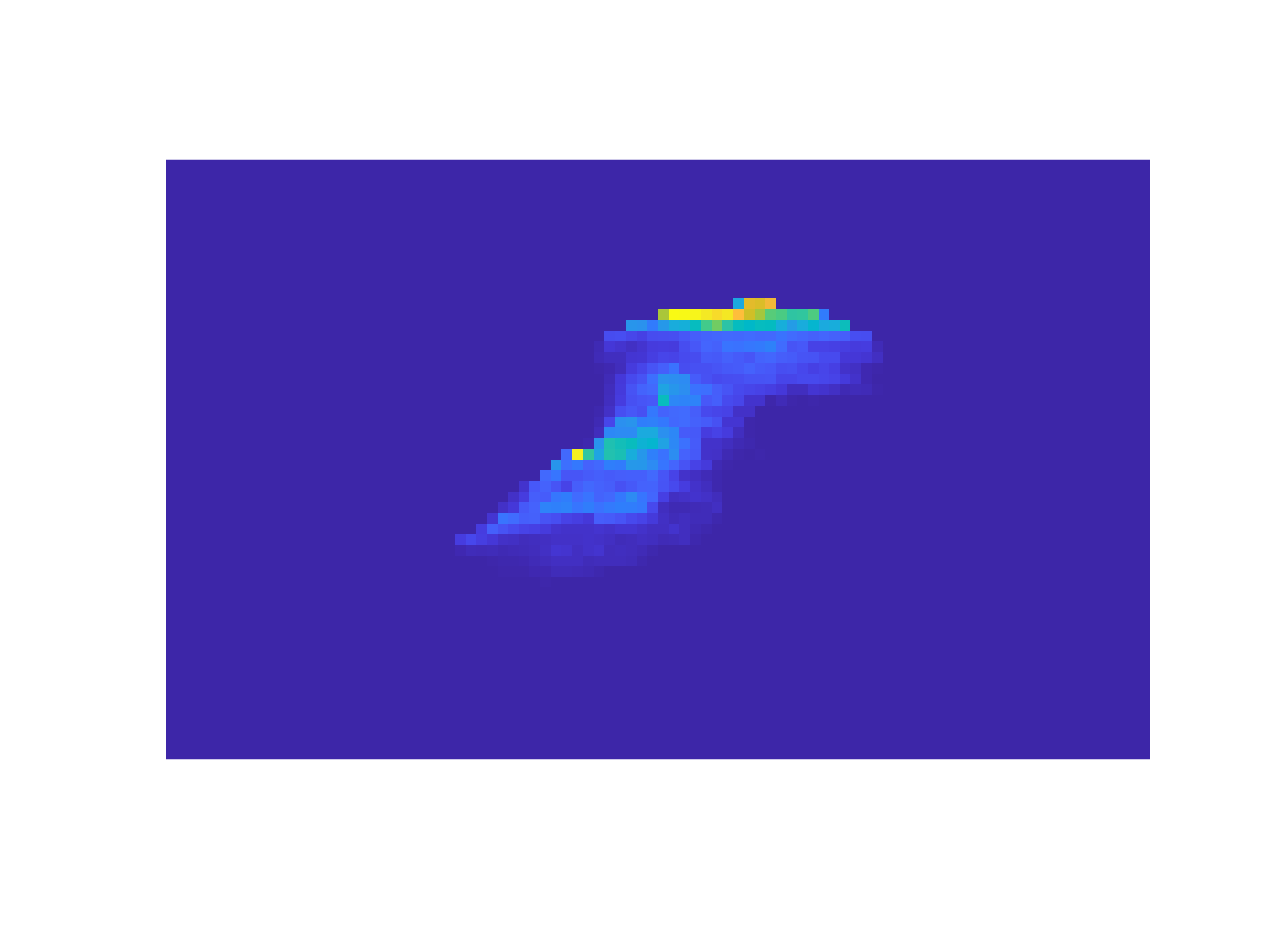}
      \centering
      \end{subfigure}
      \begin{subfigure}[b]{0.16\textwidth}
      \includegraphics[clip=true,scale=0.2,trim=15 30 15 30]{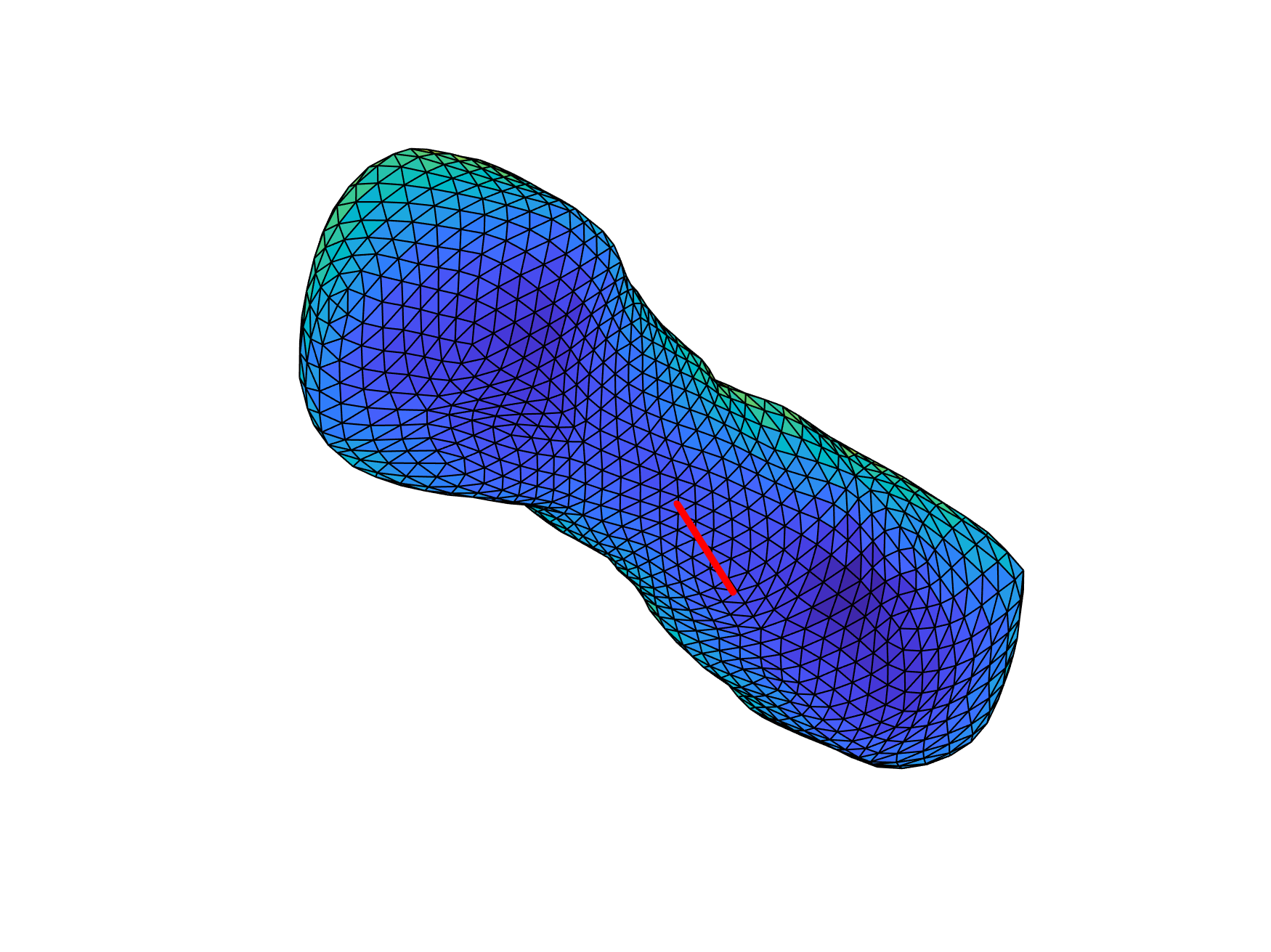}
      \centering
      \end{subfigure}
      \begin{subfigure}[b]{0.16\textwidth}
    \centering
      \includegraphics[clip=true,scale=0.25,trim=50 50 50 50]{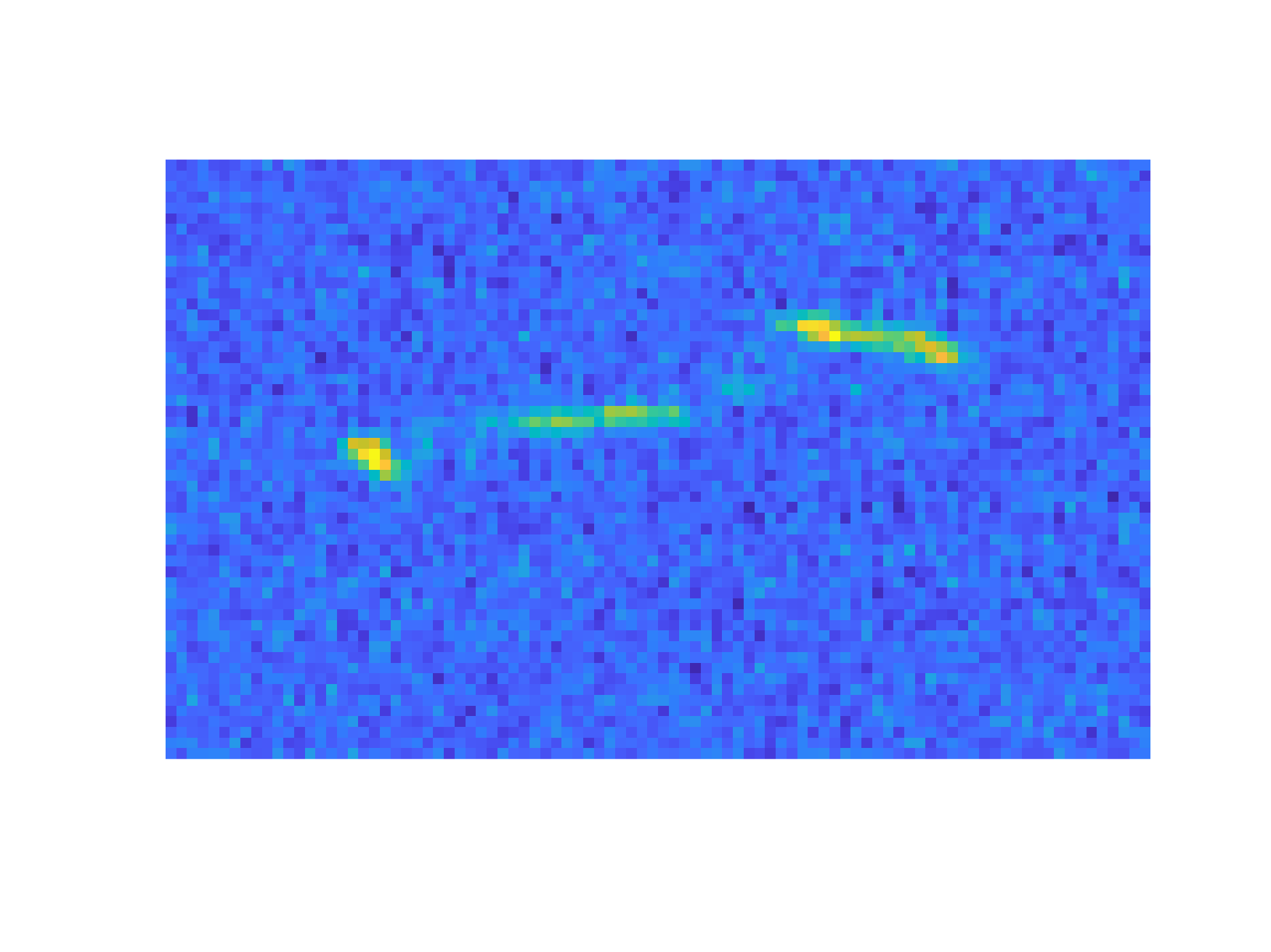}
      \end{subfigure}
      \begin{subfigure}[b]{0.16\textwidth}
      \includegraphics[clip=true,scale=0.25,trim=50 50 50 50]{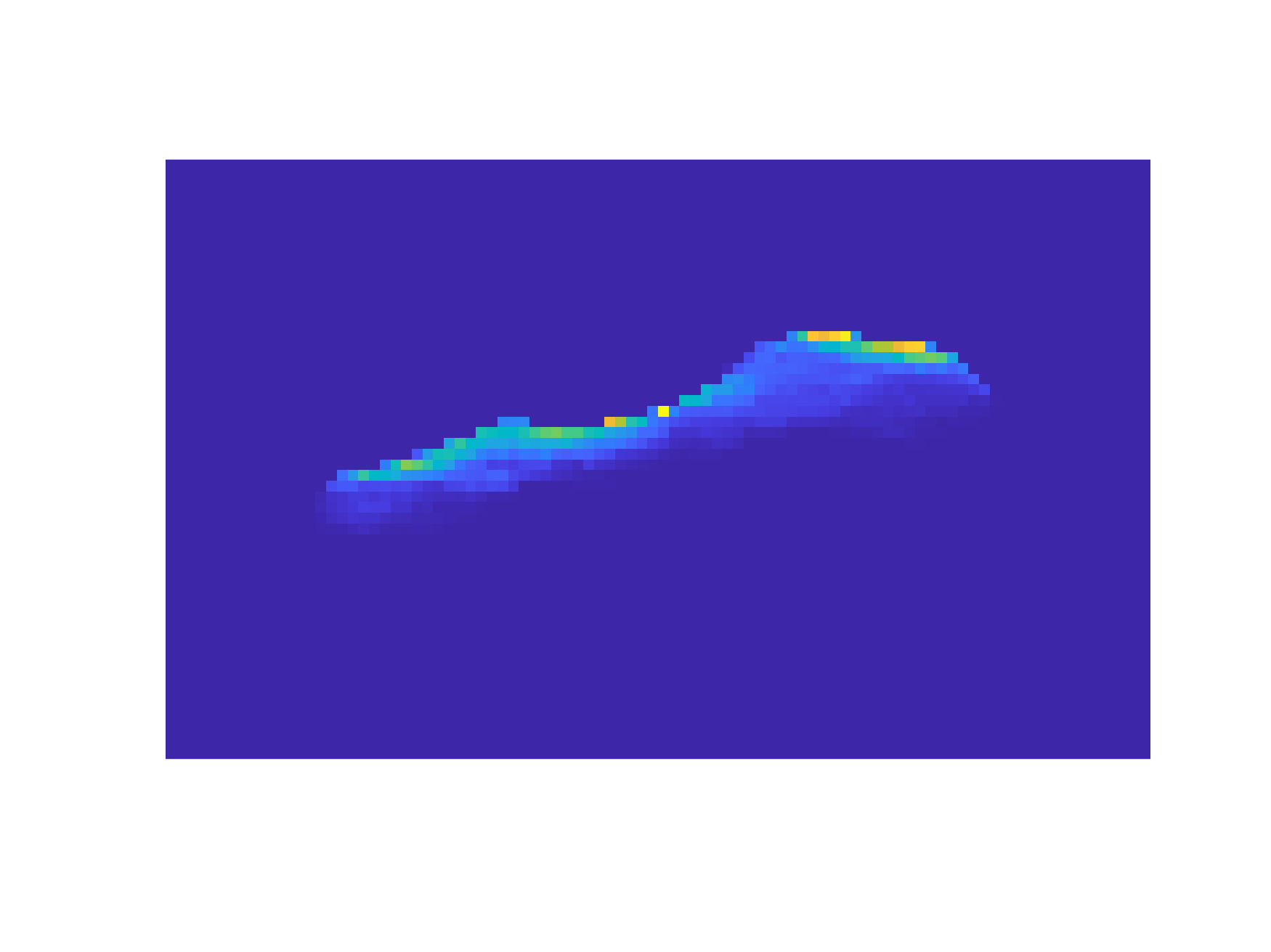}
      \centering
      \end{subfigure}
      \begin{subfigure}[b]{0.16\textwidth}
      \includegraphics[clip=true,scale=0.2,trim=15 30 15 30]{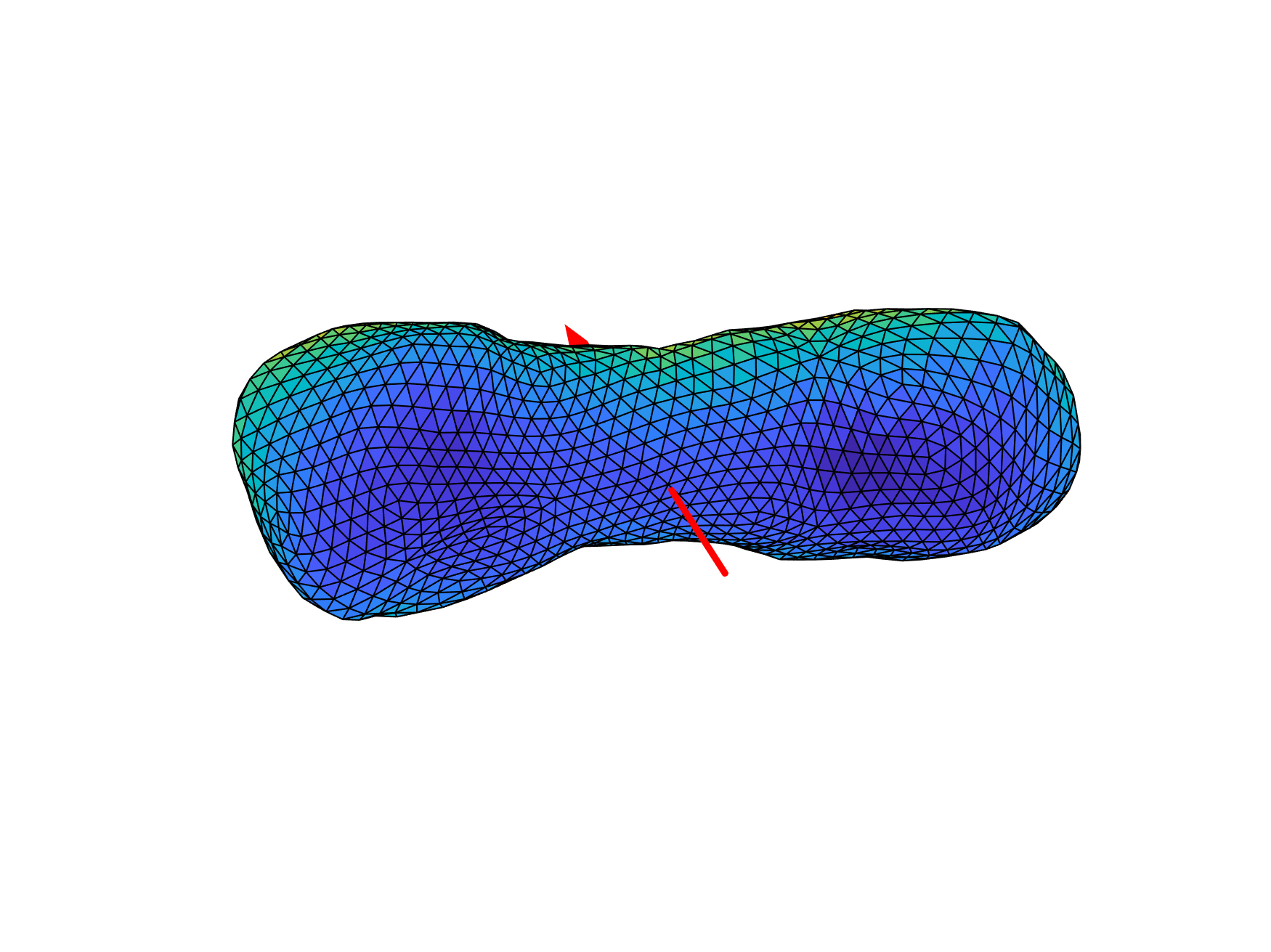}
      \centering
      \end{subfigure}
      \begin{subfigure}[b]{0.16\textwidth}
    \centering
      \includegraphics[clip=true,scale=0.25,trim=50 50 50 50]{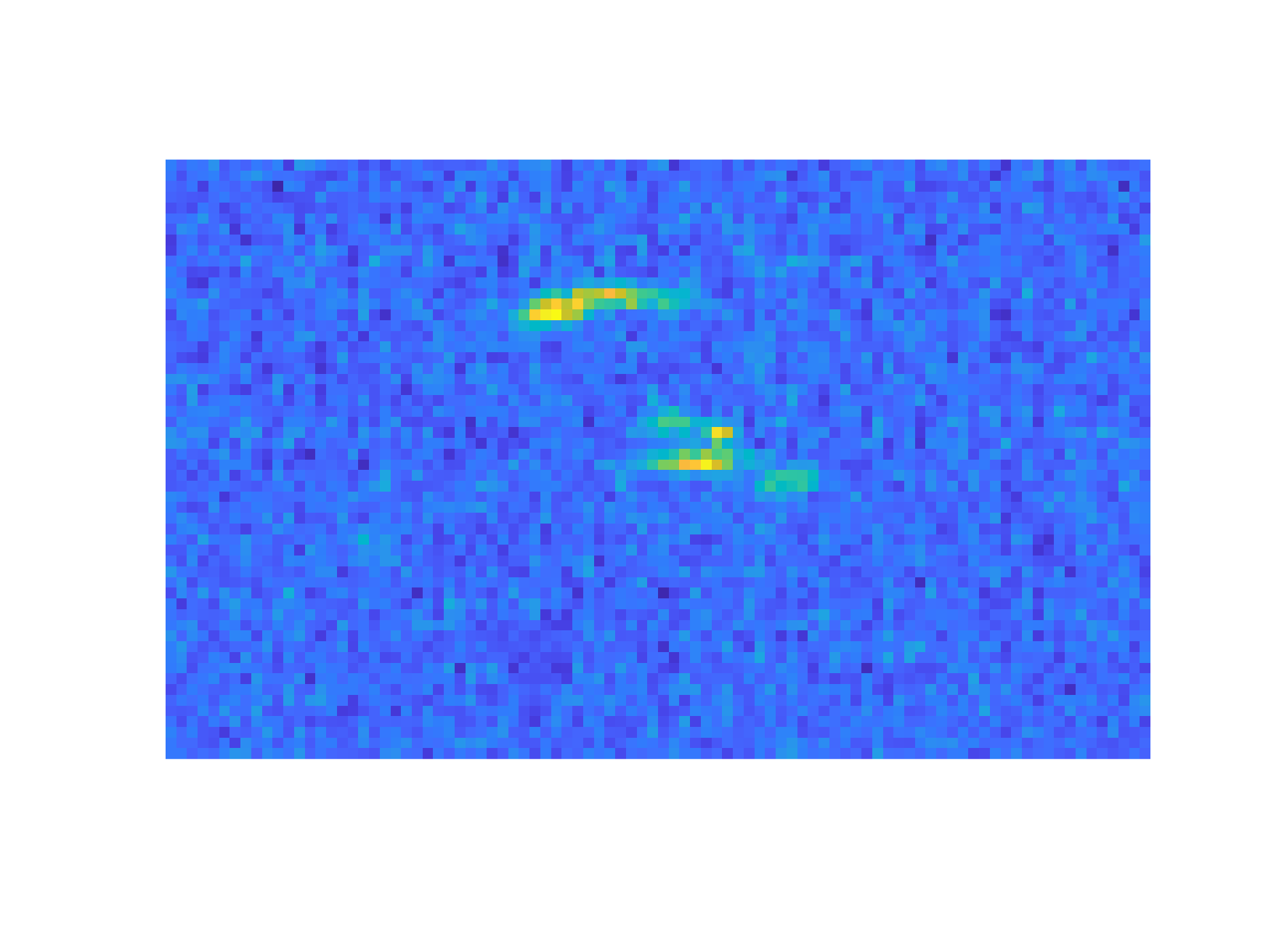}
      \end{subfigure}
      \begin{subfigure}[b]{0.16\textwidth}
      \includegraphics[clip=true,scale=0.25,trim=50 50 50 50]{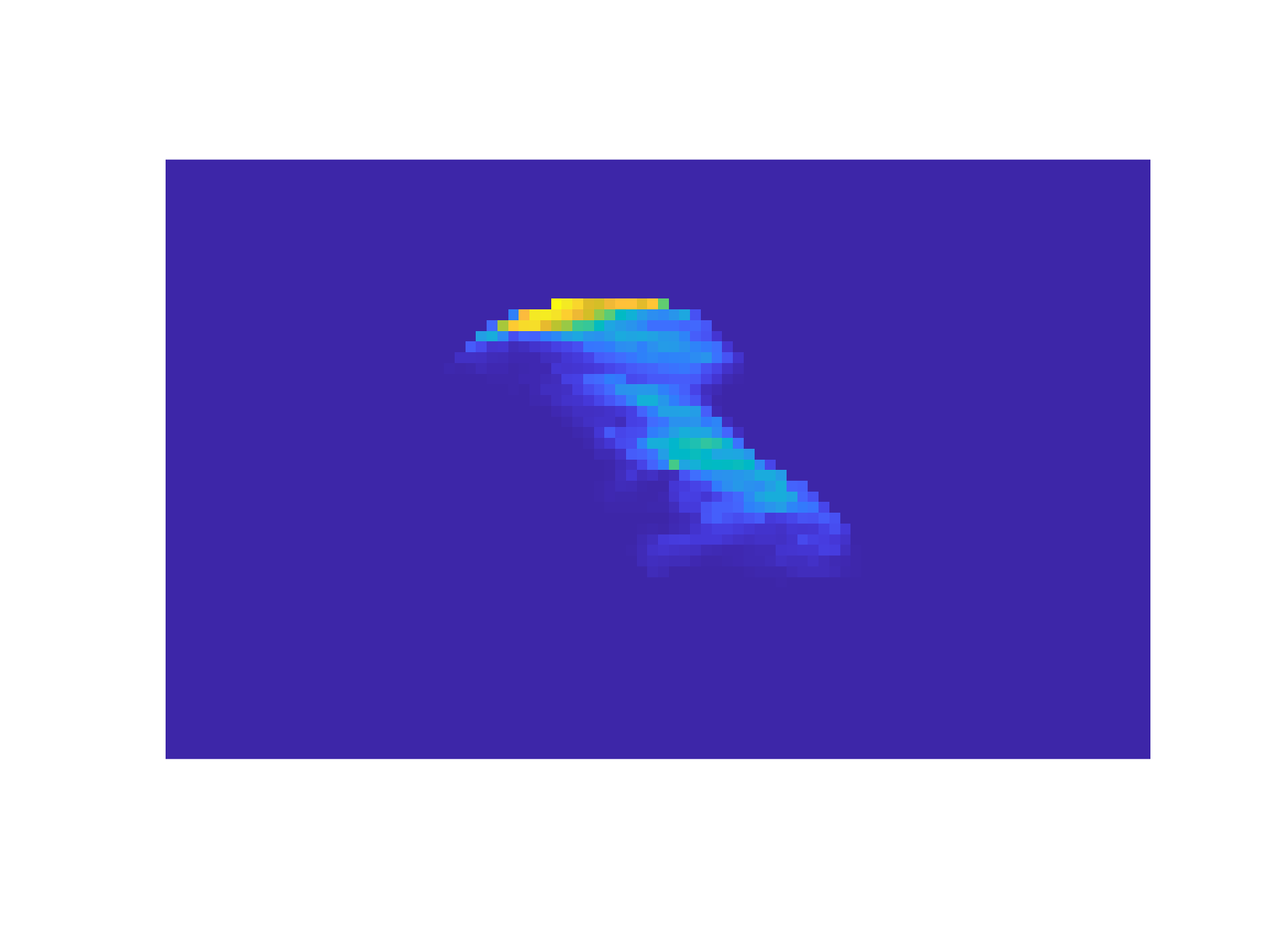}
      \centering
      \end{subfigure}
      \begin{subfigure}[b]{0.16\textwidth}
      \includegraphics[clip=true,scale=0.2,trim=15 30 15 30]{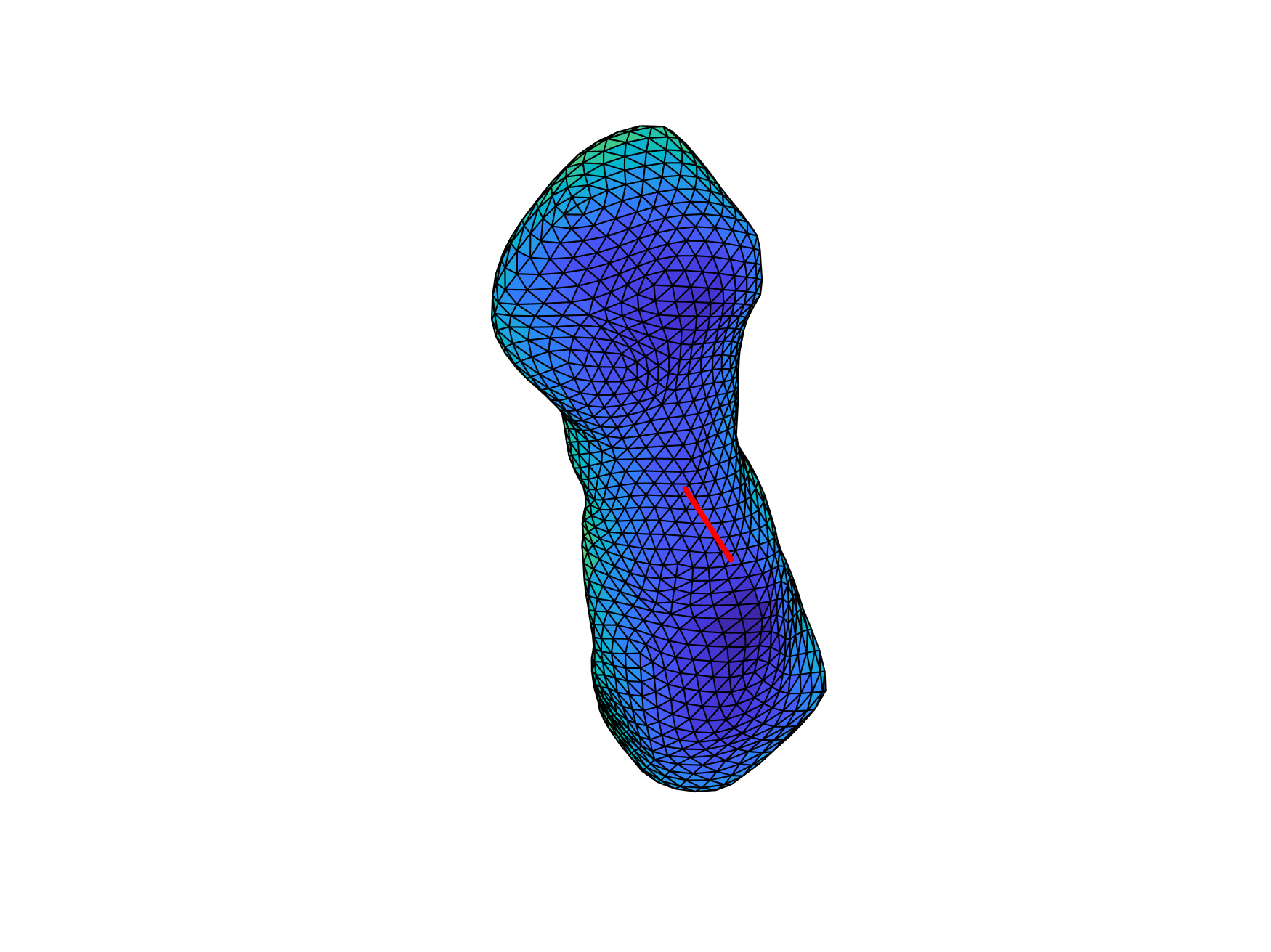}
      \centering
      \end{subfigure}
      \begin{subfigure}[b]{0.16\textwidth}
    \centering
      \includegraphics[clip=true,scale=0.25,trim=50 50 50 50]{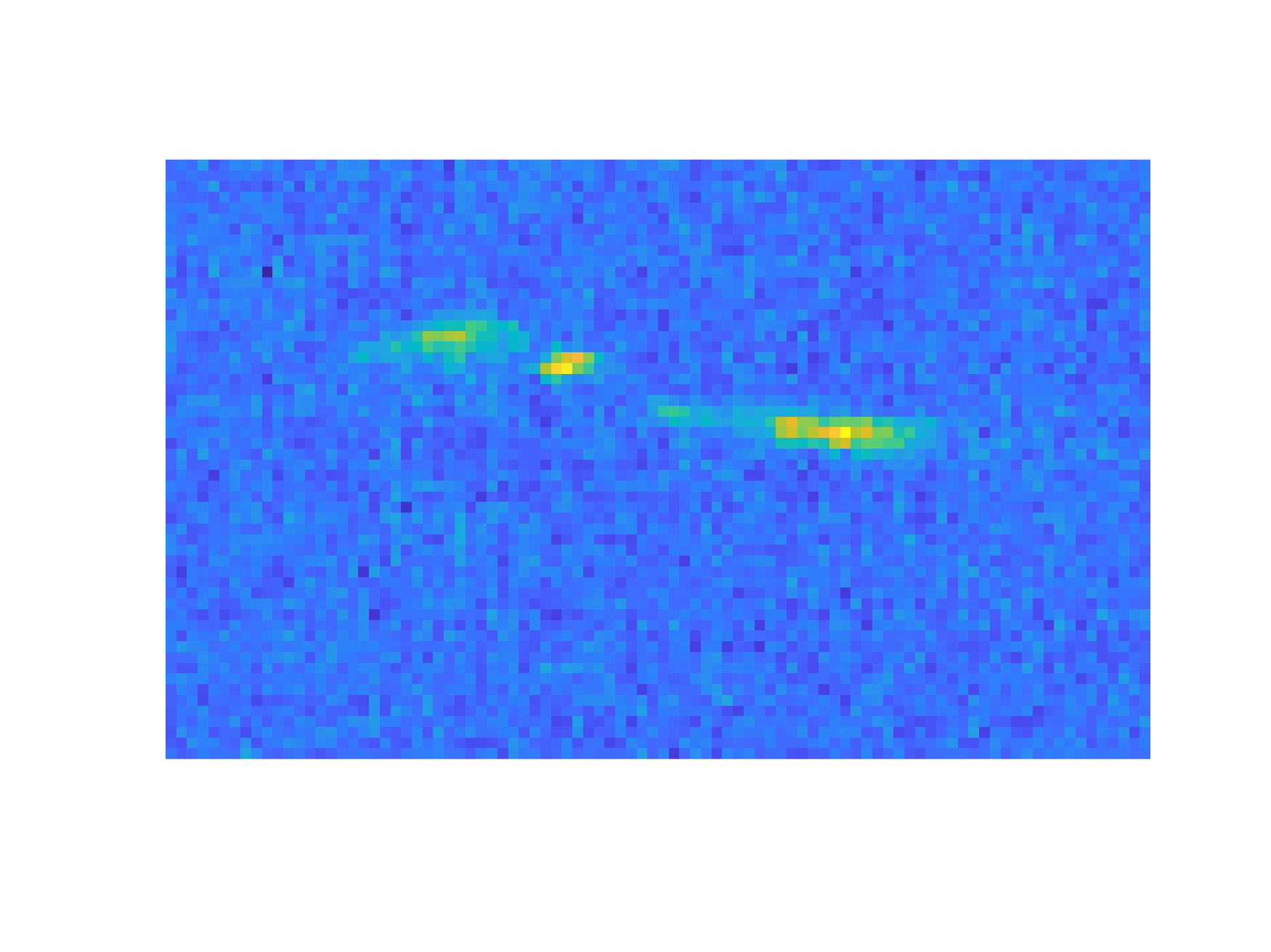}
      \end{subfigure}
      \begin{subfigure}[b]{0.16\textwidth}
      \includegraphics[clip=true,scale=0.25,trim=50 50 50 50]{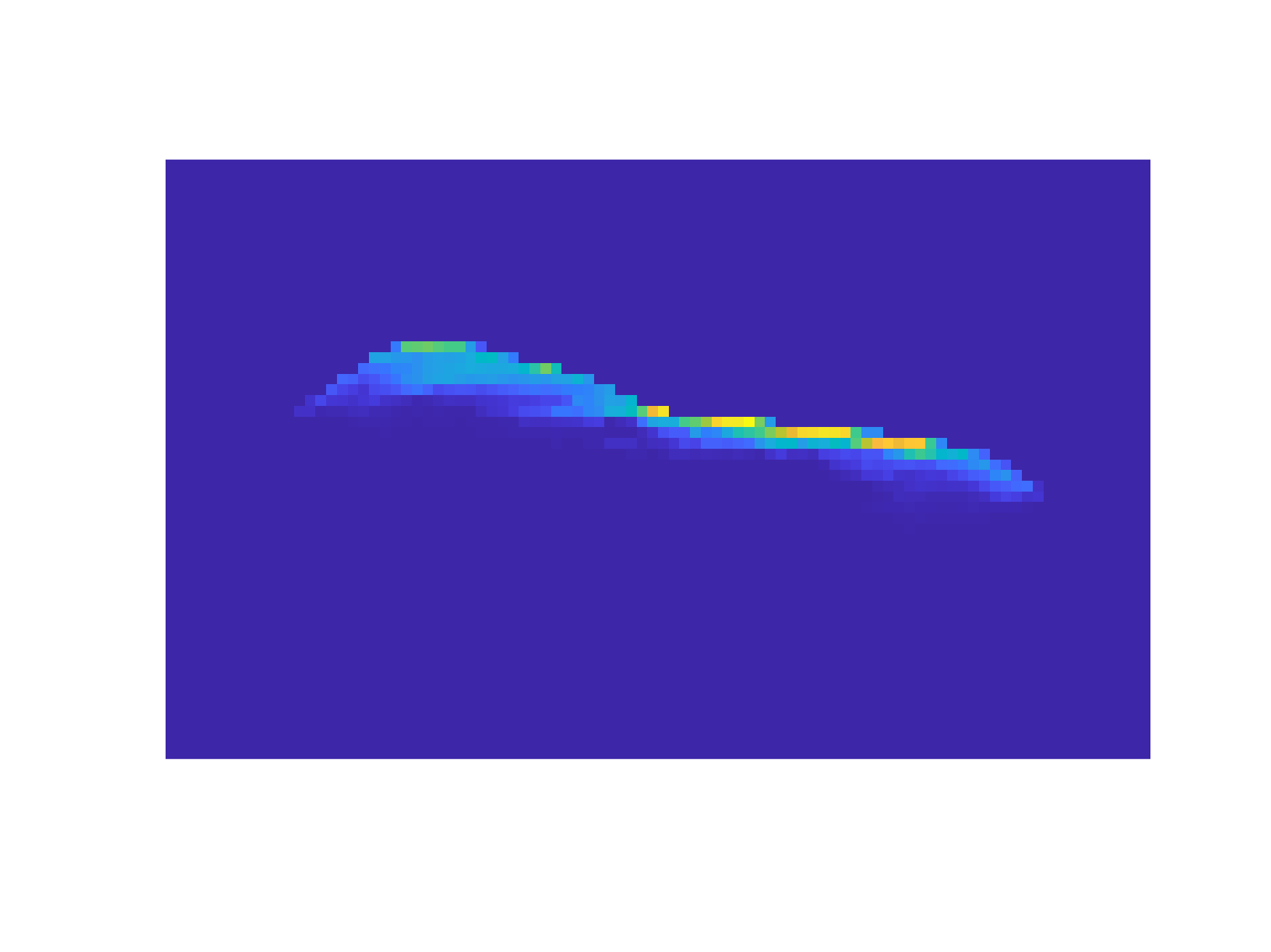}
      \centering
      \end{subfigure}
      \begin{subfigure}[b]{0.16\textwidth}
      \includegraphics[clip=true,scale=0.2,trim=15 30 15 30]{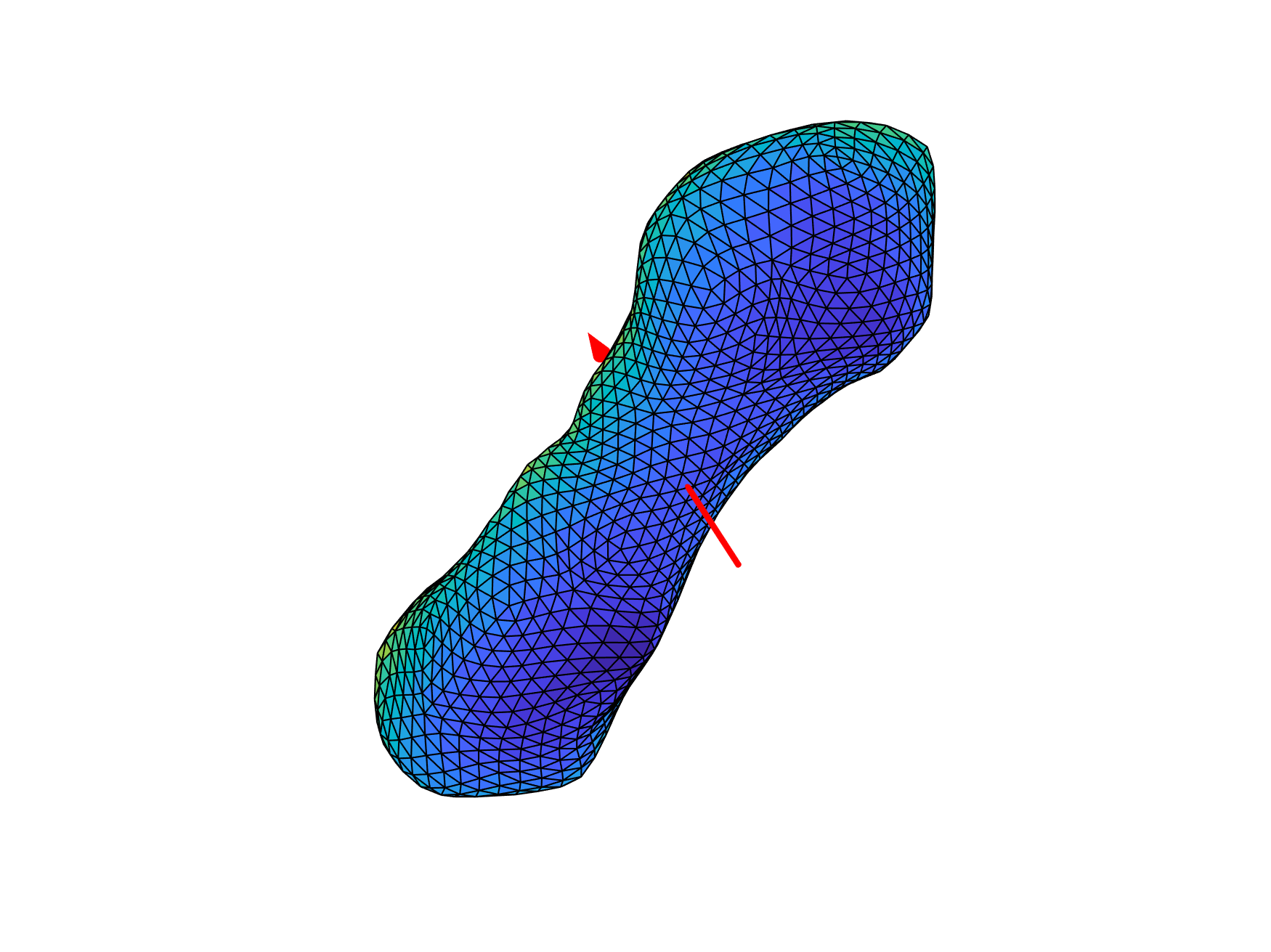}
      \centering
      \end{subfigure}
      \begin{subfigure}[b]{0.16\textwidth}
    \centering
      \includegraphics[clip=true,scale=0.25,trim=50 50 50 50]{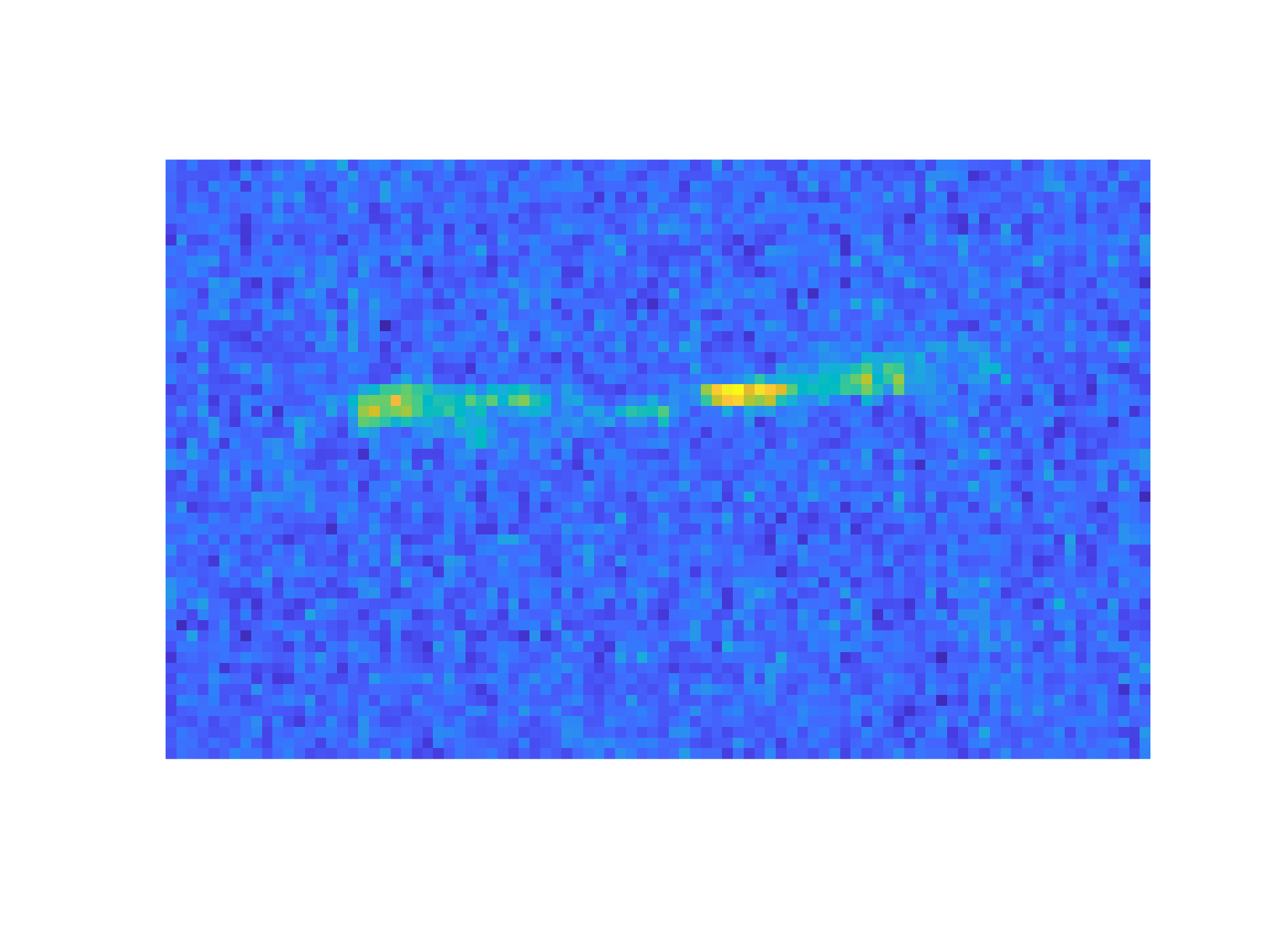}
      \end{subfigure}
      \begin{subfigure}[b]{0.16\textwidth}
      \includegraphics[clip=true,scale=0.25,trim=50 50 50 50]{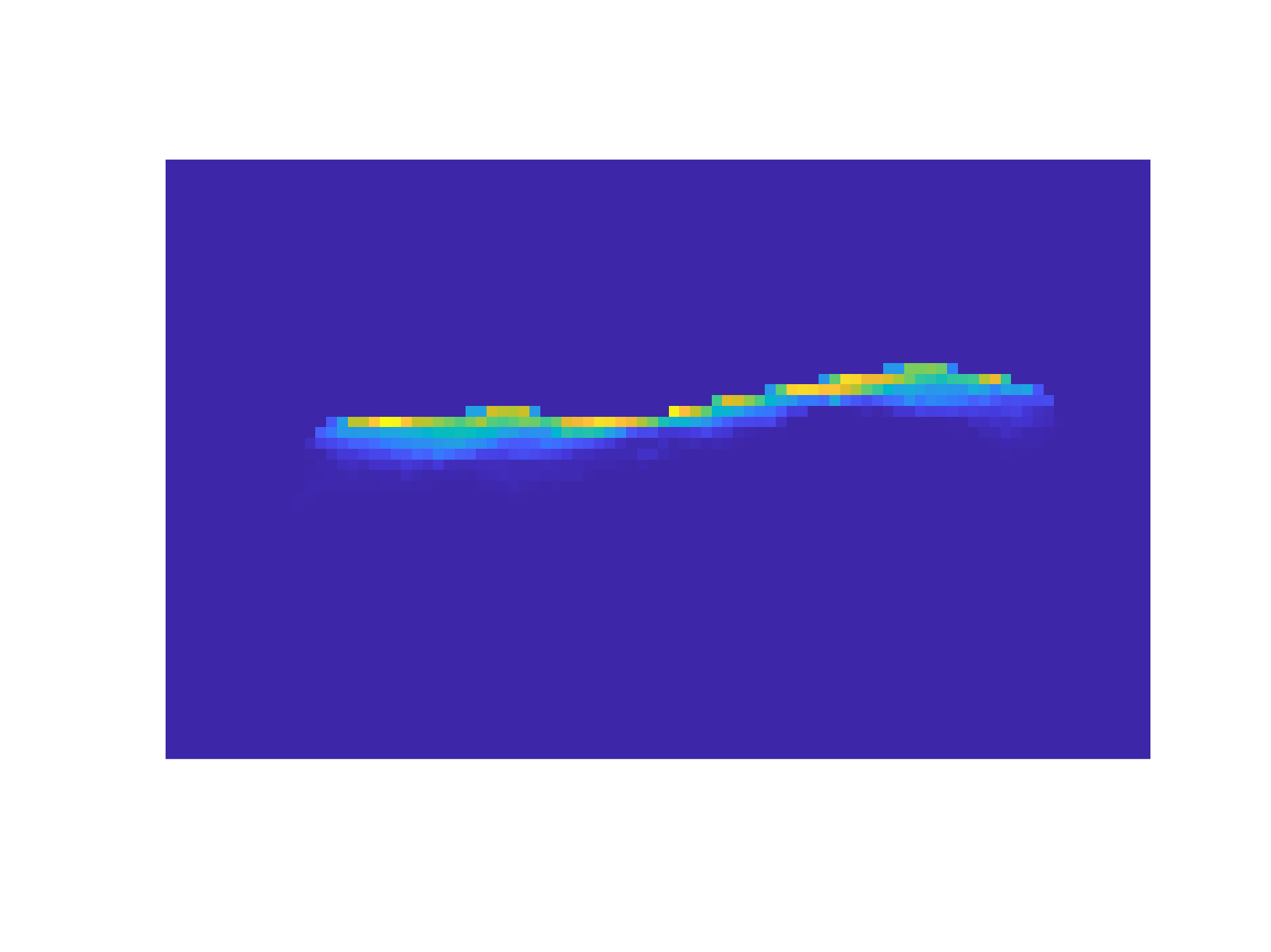}
      \centering
      \end{subfigure}
      \begin{subfigure}[b]{0.16\textwidth}
      \includegraphics[clip=true,scale=0.2,trim=15 30 15 30]{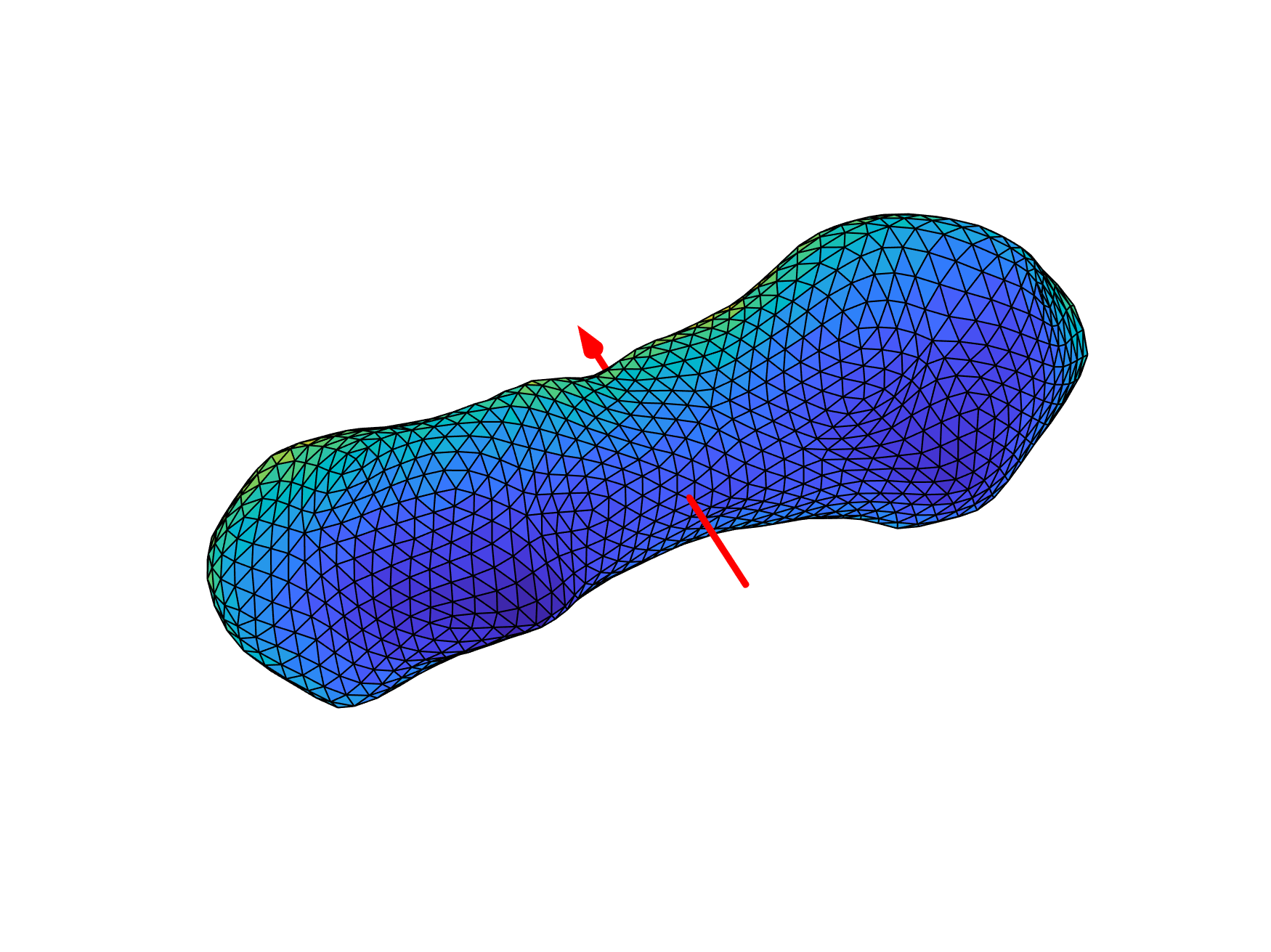}
      \centering
      \end{subfigure}
      \begin{subfigure}[b]{0.16\textwidth}
    \centering
      \includegraphics[clip=true,scale=0.25,trim=50 50 50 50]{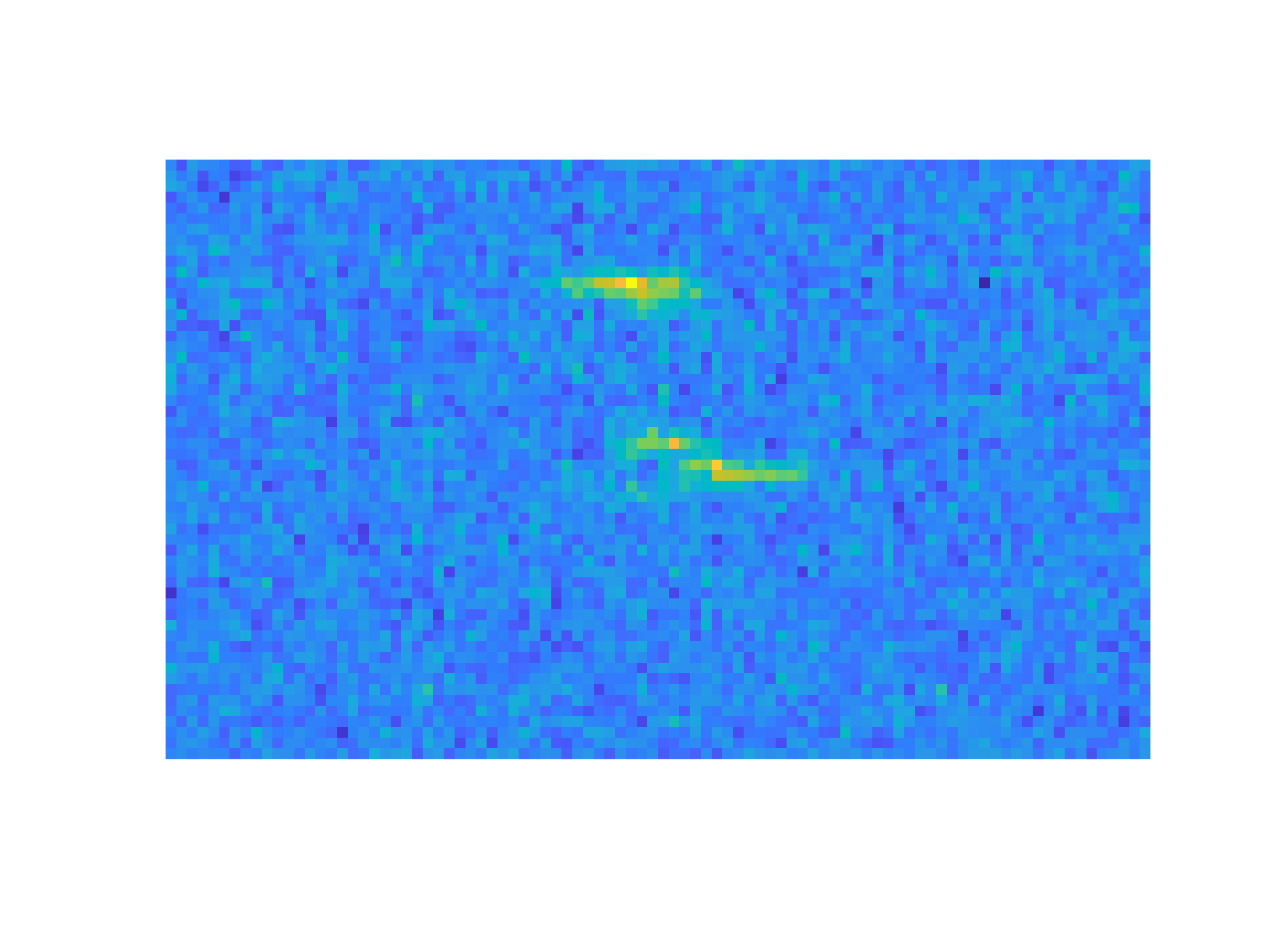}
      \end{subfigure}
      \begin{subfigure}[b]{0.16\textwidth}
      \includegraphics[clip=true,scale=0.25,trim=50 50 50 50]{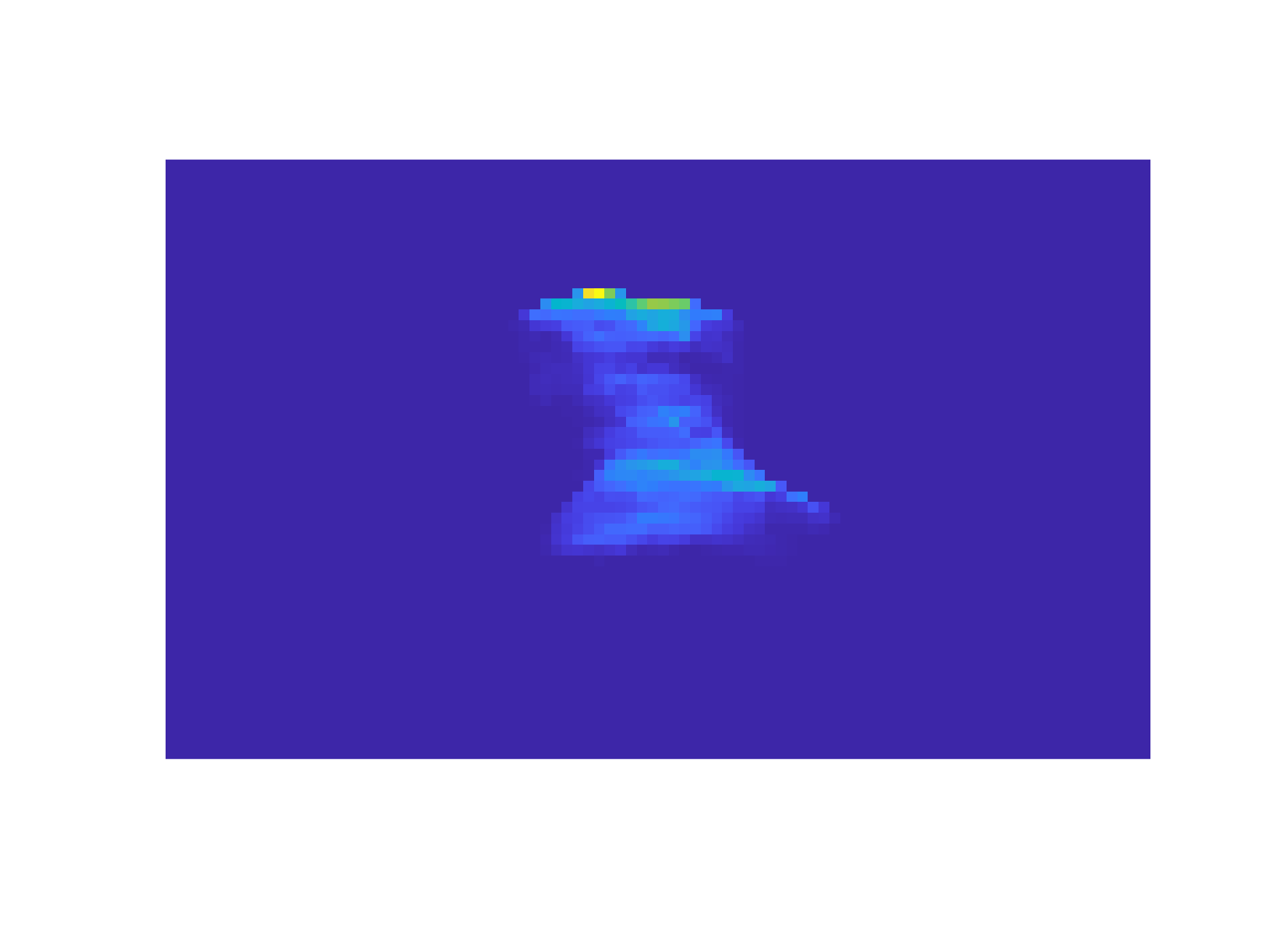}
      \centering
      \end{subfigure}
      \begin{subfigure}[b]{0.16\textwidth}
      \includegraphics[clip=true,scale=0.25,trim=15 30 15 30]{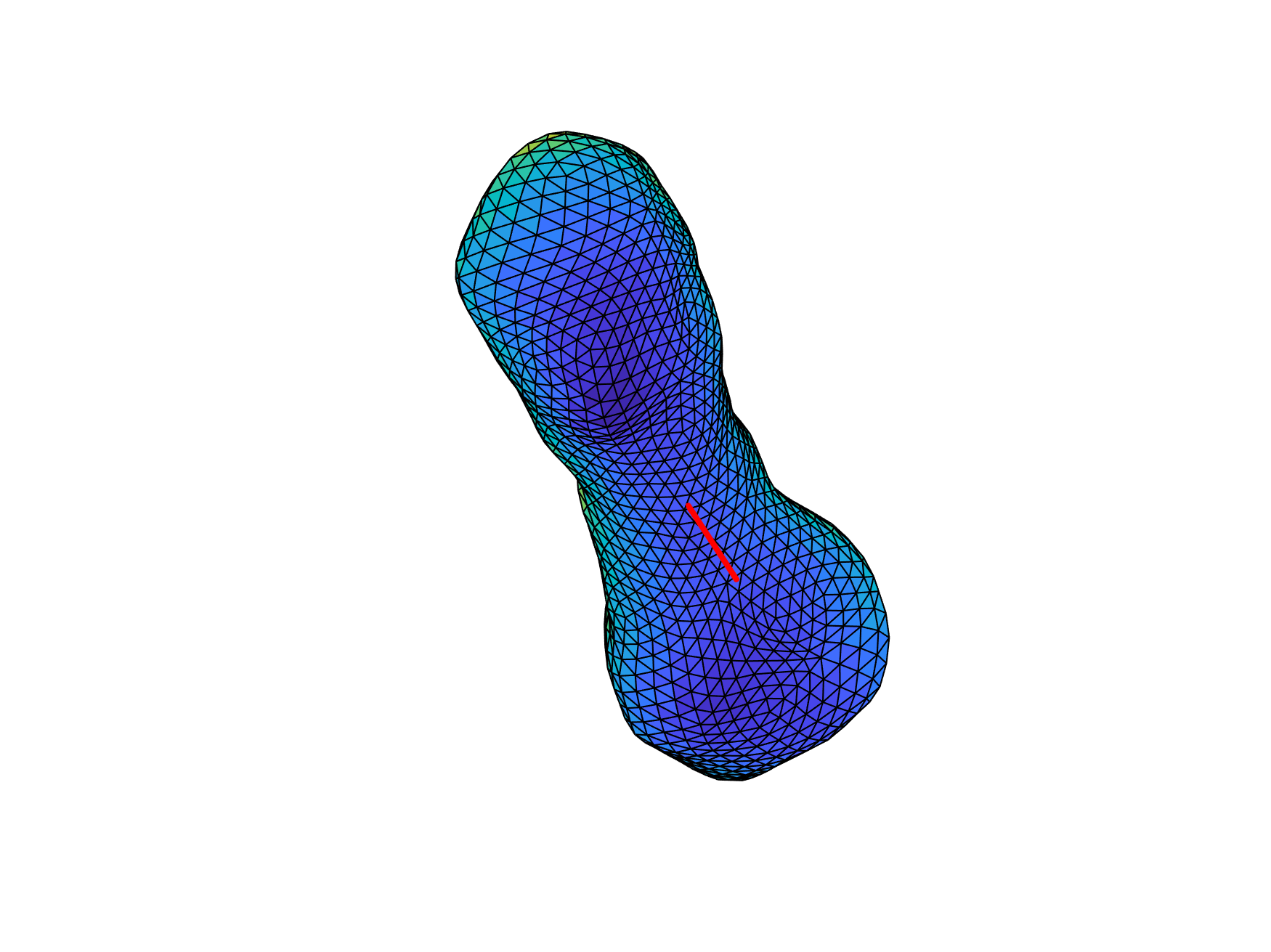}
      \centering
      \end{subfigure}
      \begin{subfigure}[b]{0.16\textwidth}
    \centering
      \includegraphics[clip=true,scale=0.25,trim=50 50 50 50]{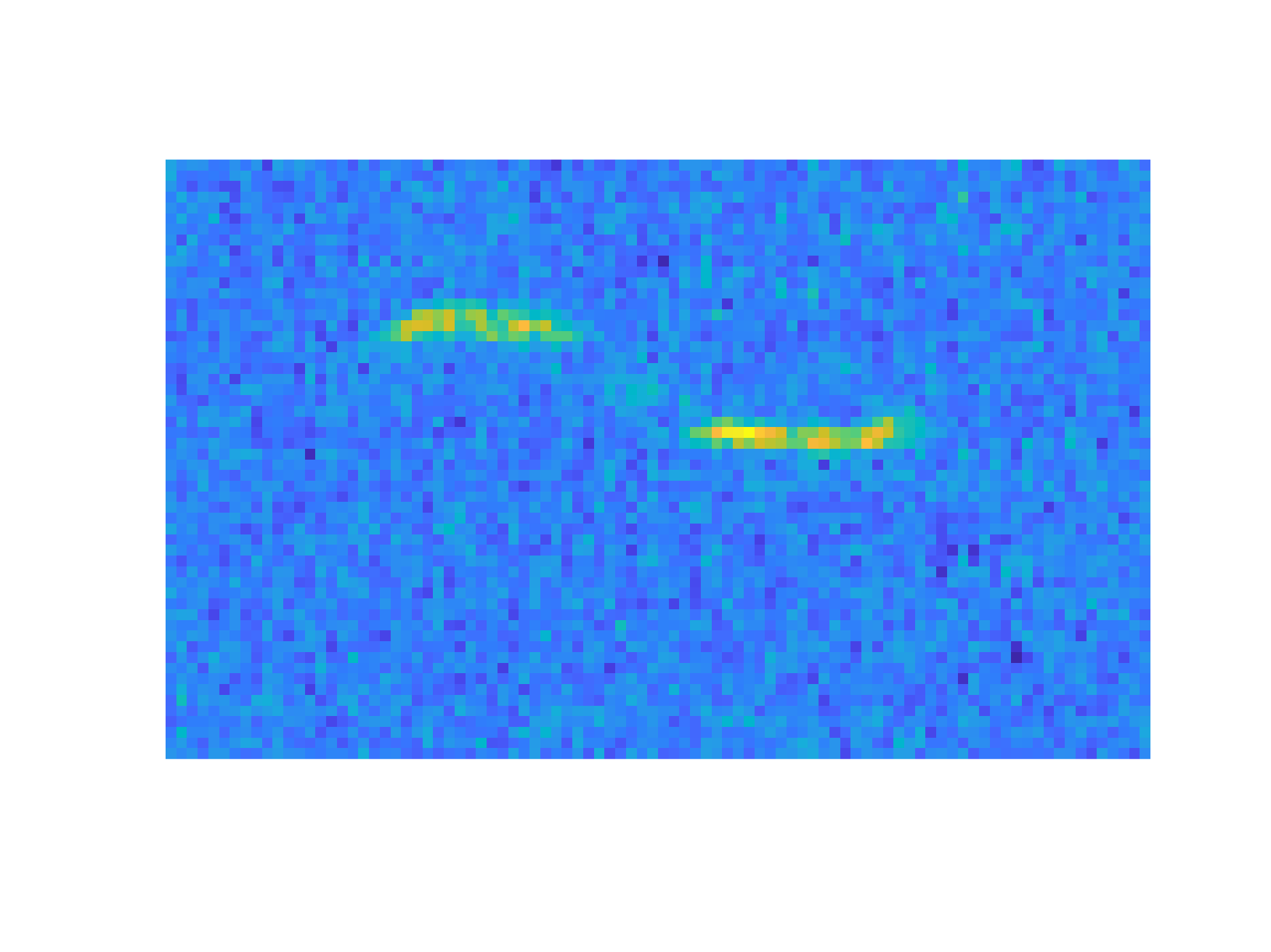}
      \end{subfigure}
      \begin{subfigure}[b]{0.16\textwidth}
      \includegraphics[clip=true,scale=0.25,trim=50 50 50 50]{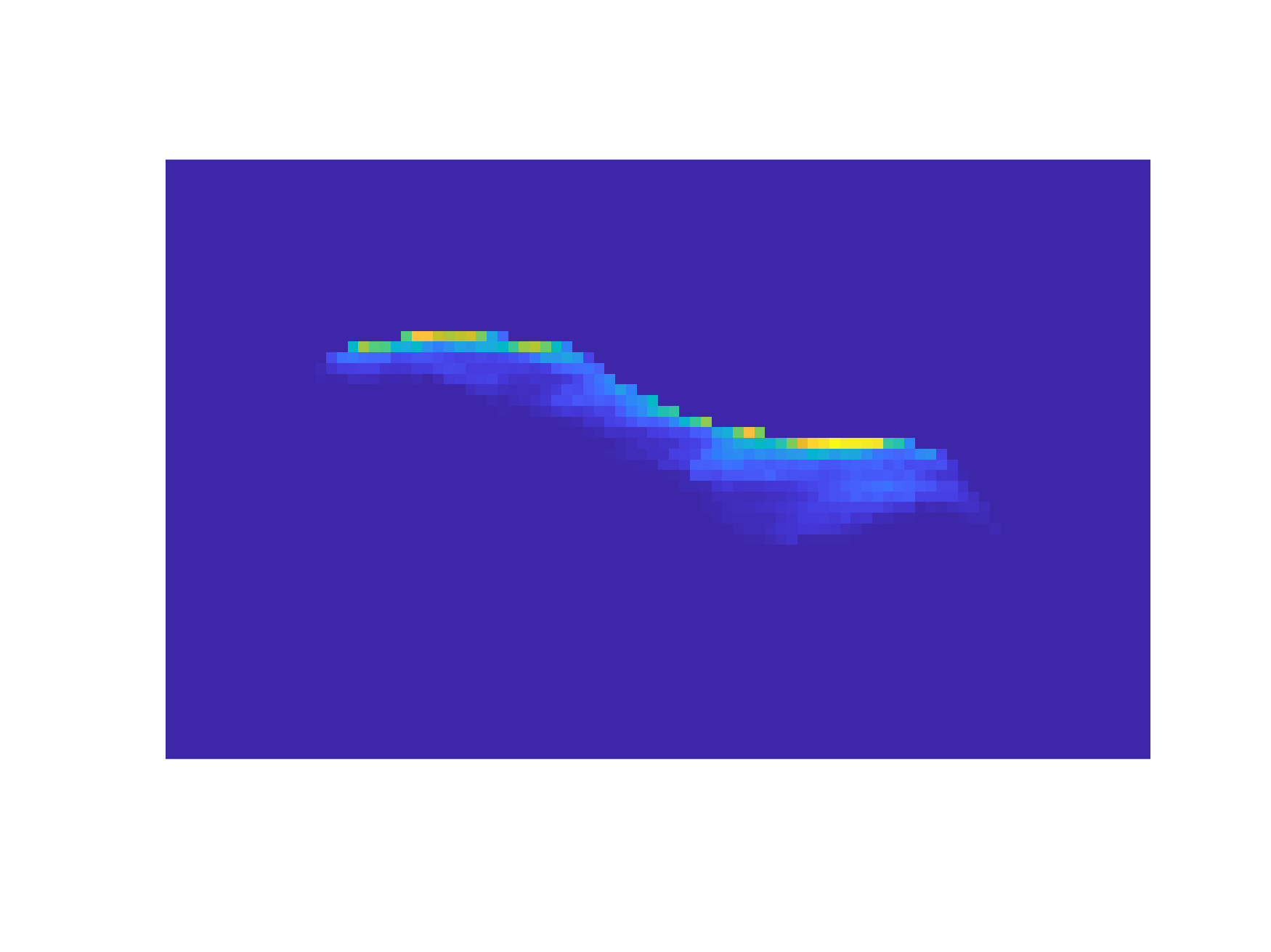}
      \centering
      \end{subfigure}
      \begin{subfigure}[b]{0.16\textwidth}
      \includegraphics[clip=true,scale=0.2,trim=15 30 15 30]{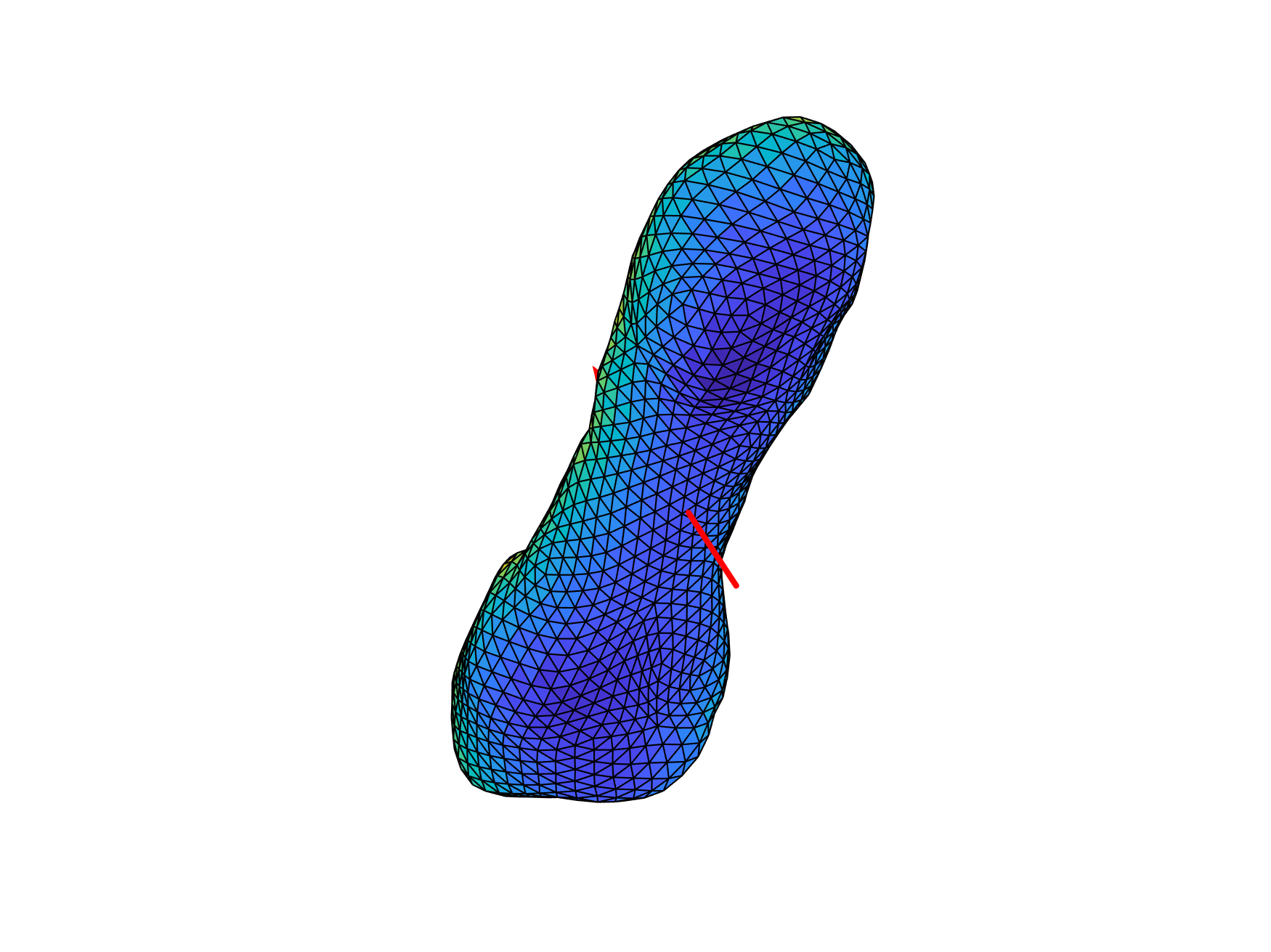}
      \centering
      \end{subfigure}
      \begin{subfigure}[b]{0.16\textwidth}
    \centering
      \includegraphics[clip=true,scale=0.25,trim=50 50 50 50]{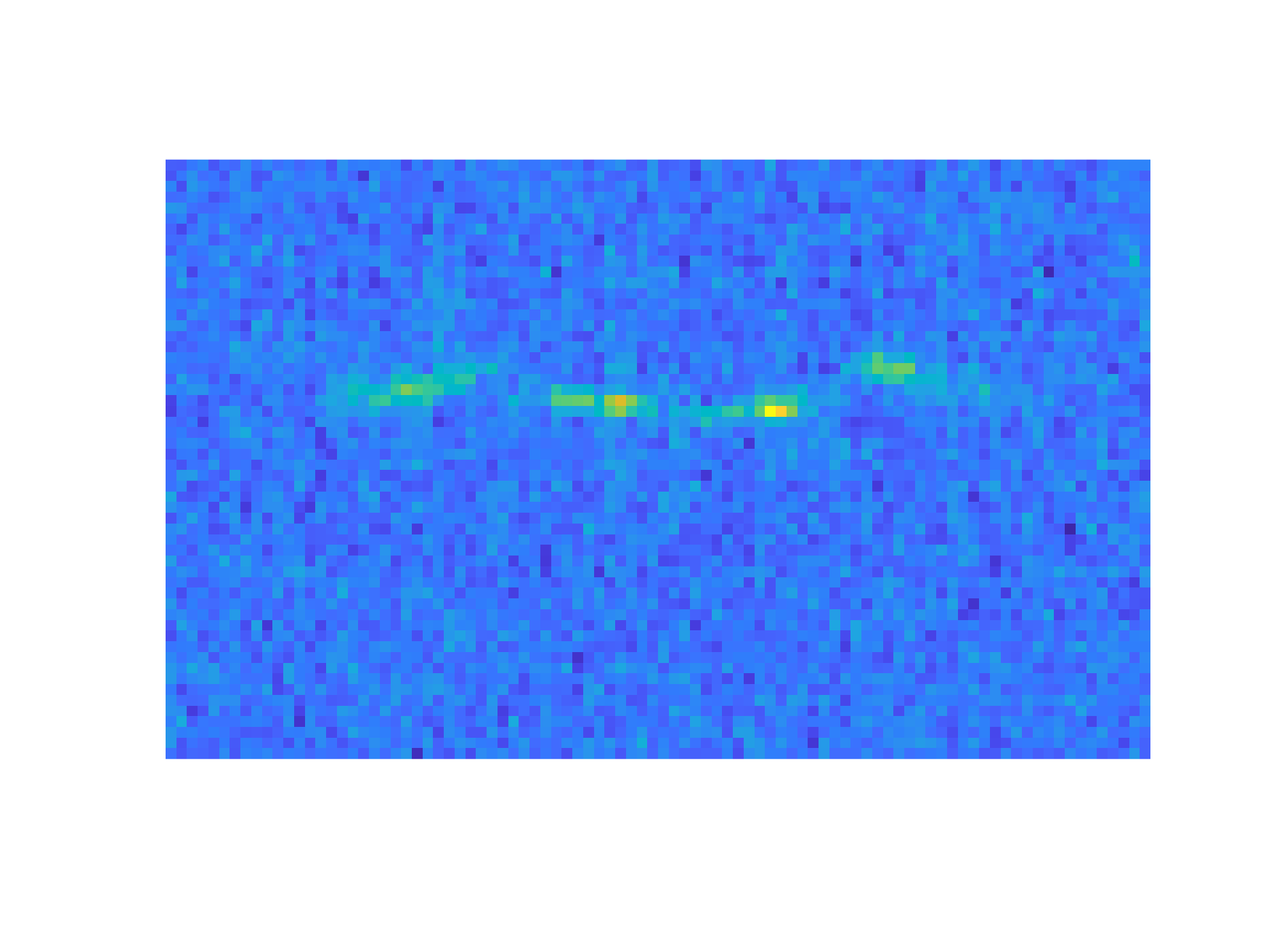}
      \end{subfigure}
      \begin{subfigure}[b]{0.16\textwidth}
      \includegraphics[clip=true,scale=0.25,trim=50 50 50 50]{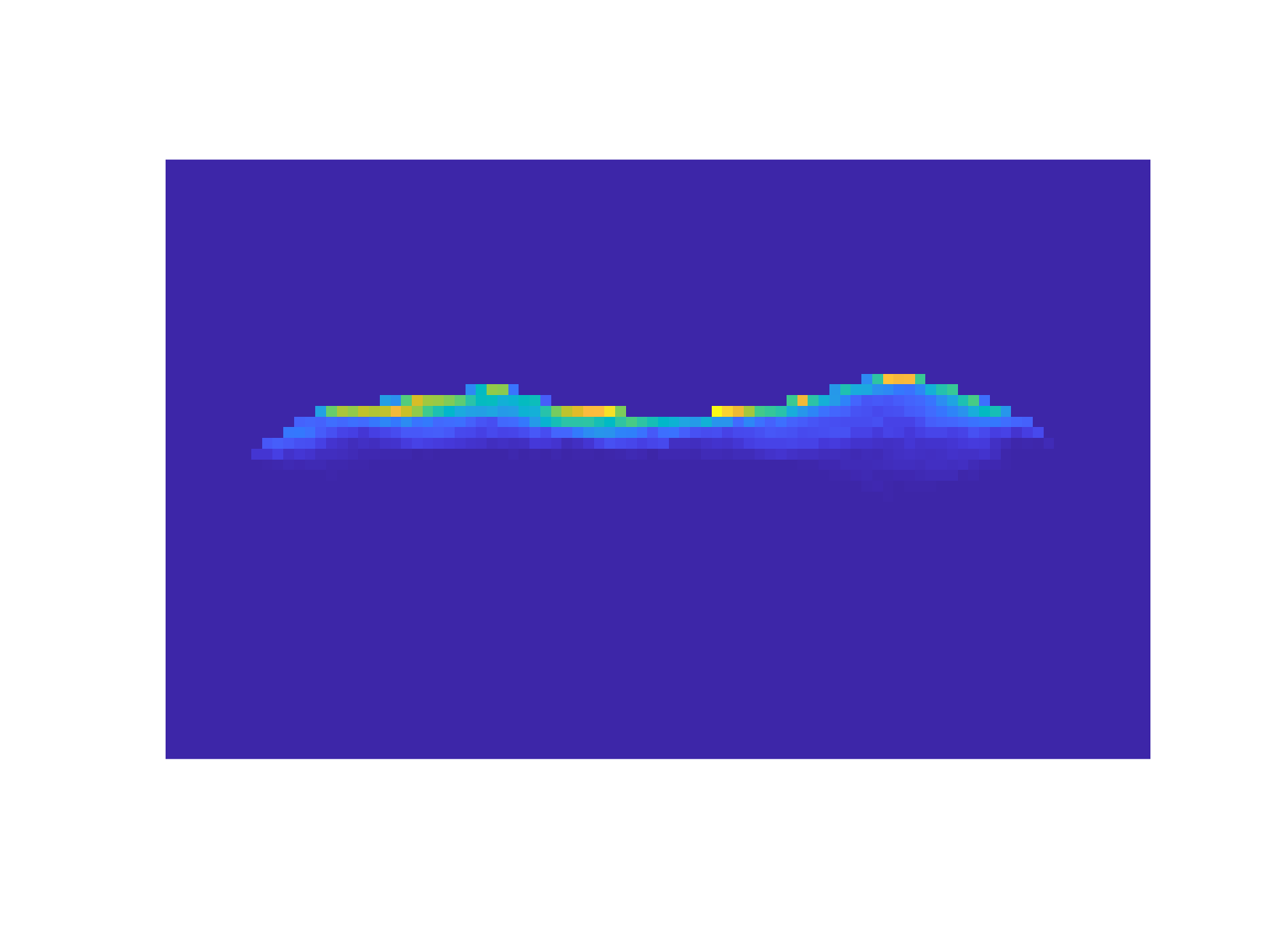}
      \centering
      \end{subfigure}
      \begin{subfigure}[b]{0.16\textwidth}
      \includegraphics[clip=true,scale=0.2,trim=15 30 15 30]{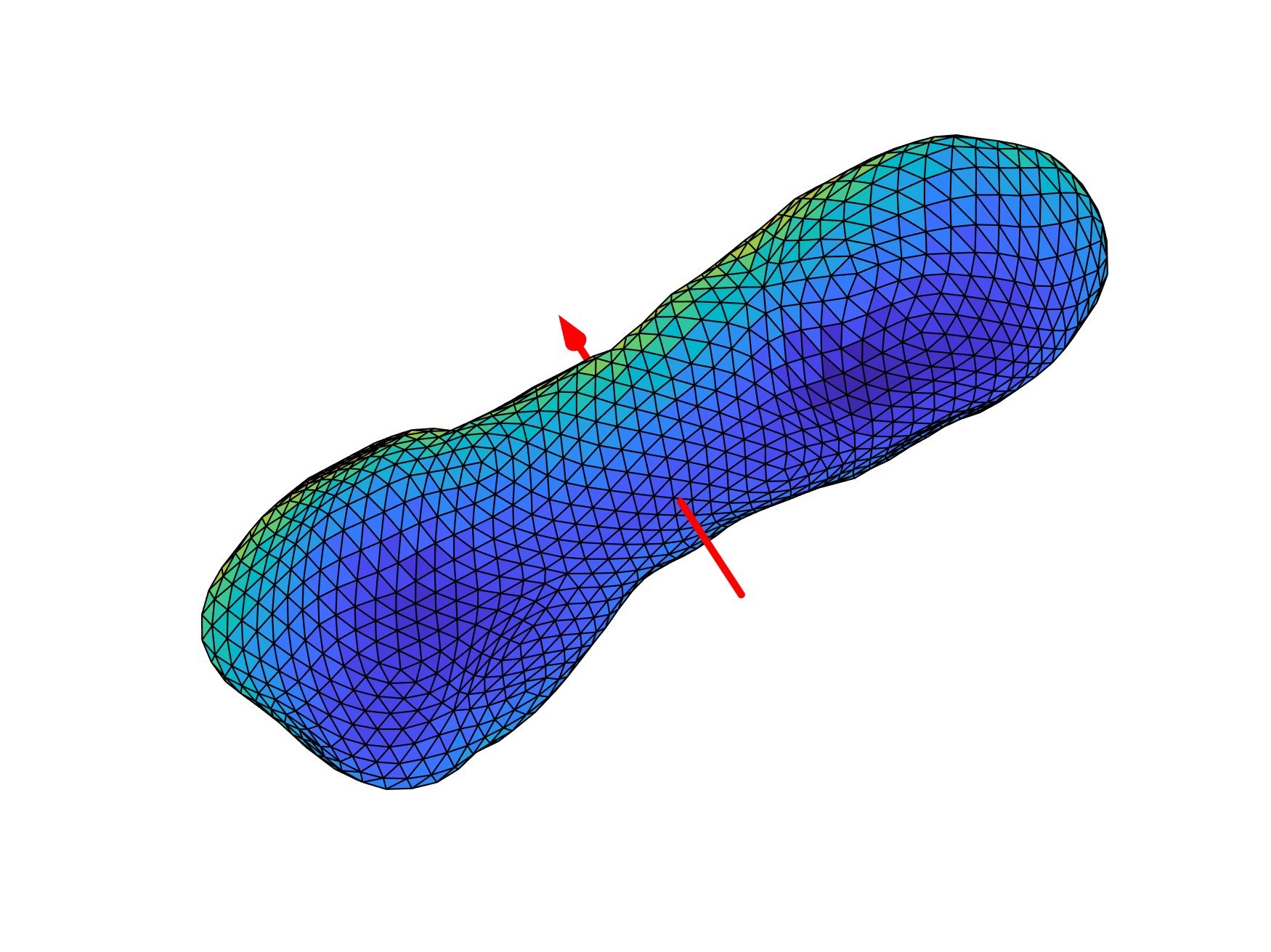}
      \centering
      \end{subfigure}
      \begin{subfigure}[b]{0.16\textwidth}
    \centering
      \includegraphics[clip=true,scale=0.25,trim=50 50 50 50]{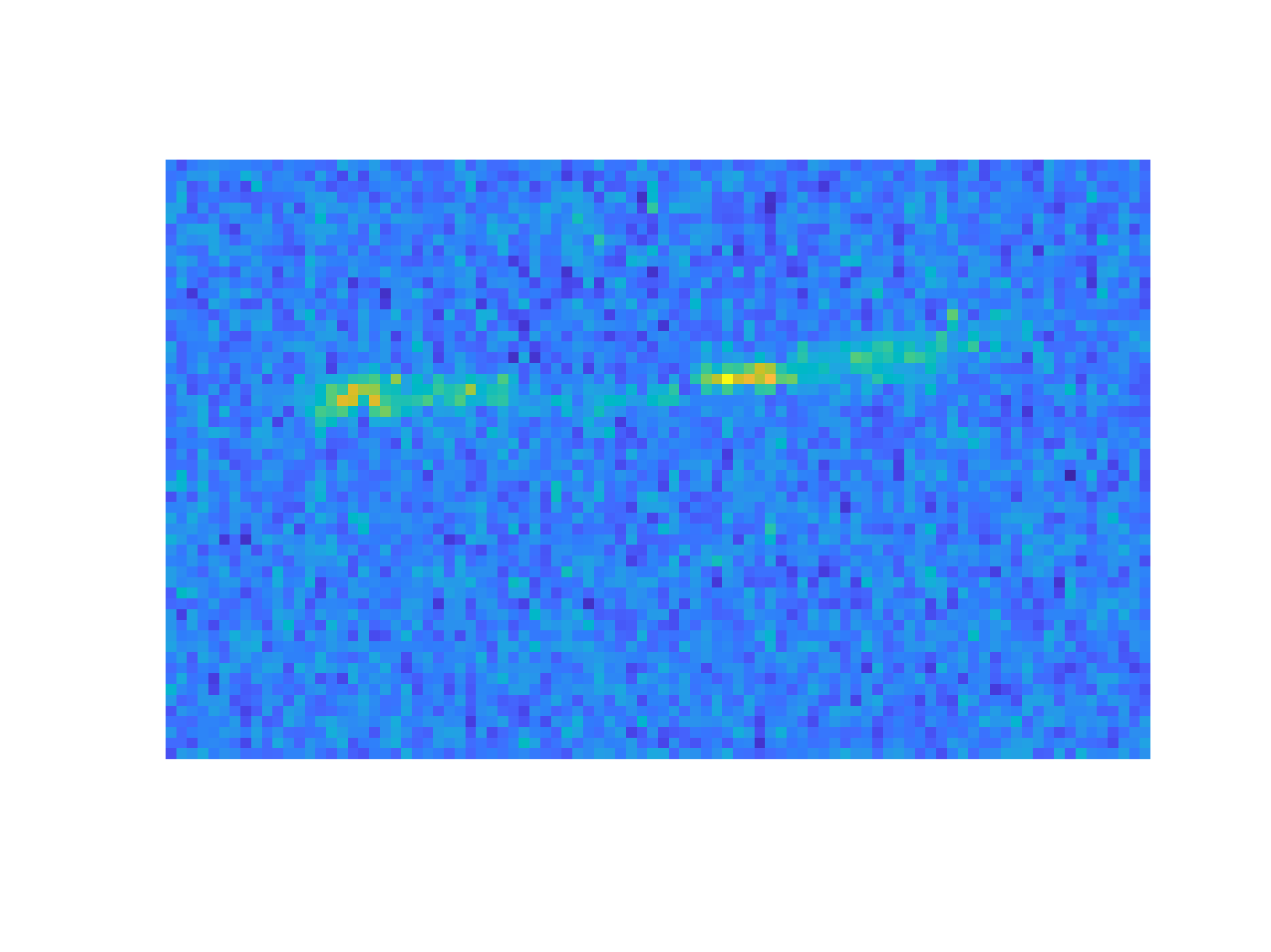}
      \end{subfigure}
      \begin{subfigure}[b]{0.16\textwidth}
      \includegraphics[clip=true,scale=0.25,trim=50 50 50 50]{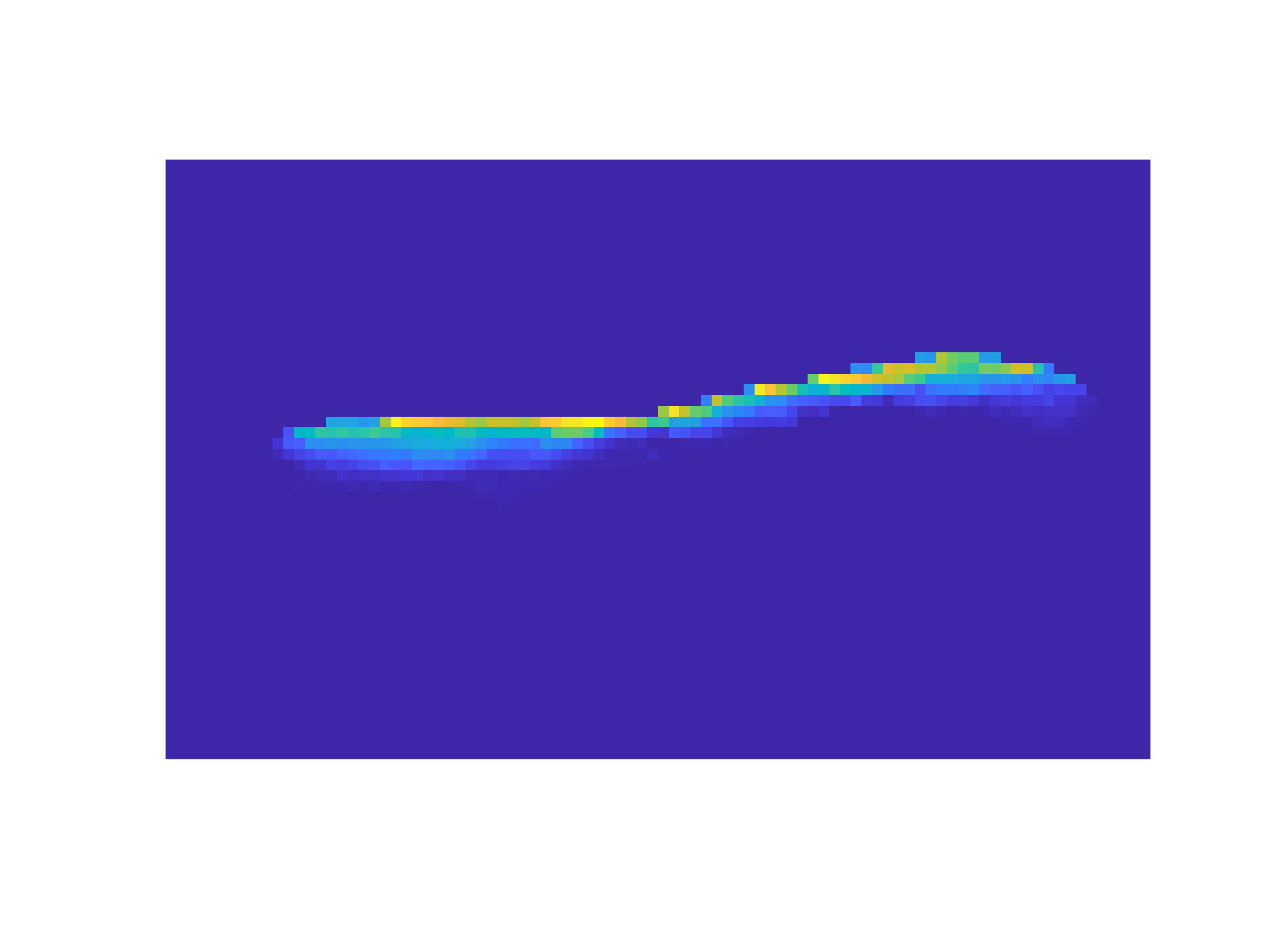}
      \centering
      \end{subfigure}
      \begin{subfigure}[b]{0.16\textwidth}
      \includegraphics[clip=true,scale=0.2,trim=15 30 15 30]{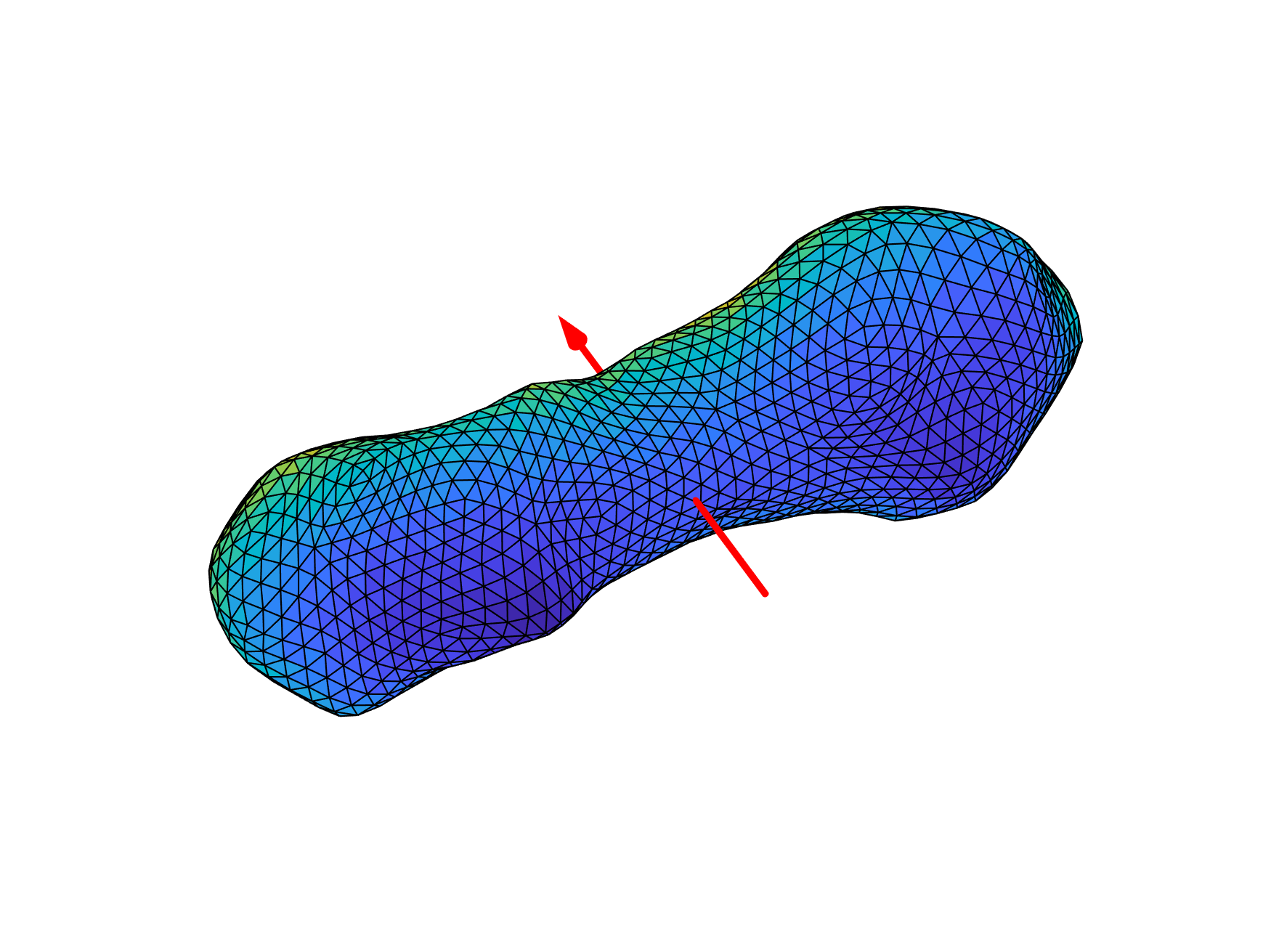}
      \centering
      \end{subfigure}
      \begin{subfigure}[b]{0.16\textwidth}
    \centering
      \includegraphics[clip=true,scale=0.25,trim=50 50 50 50]{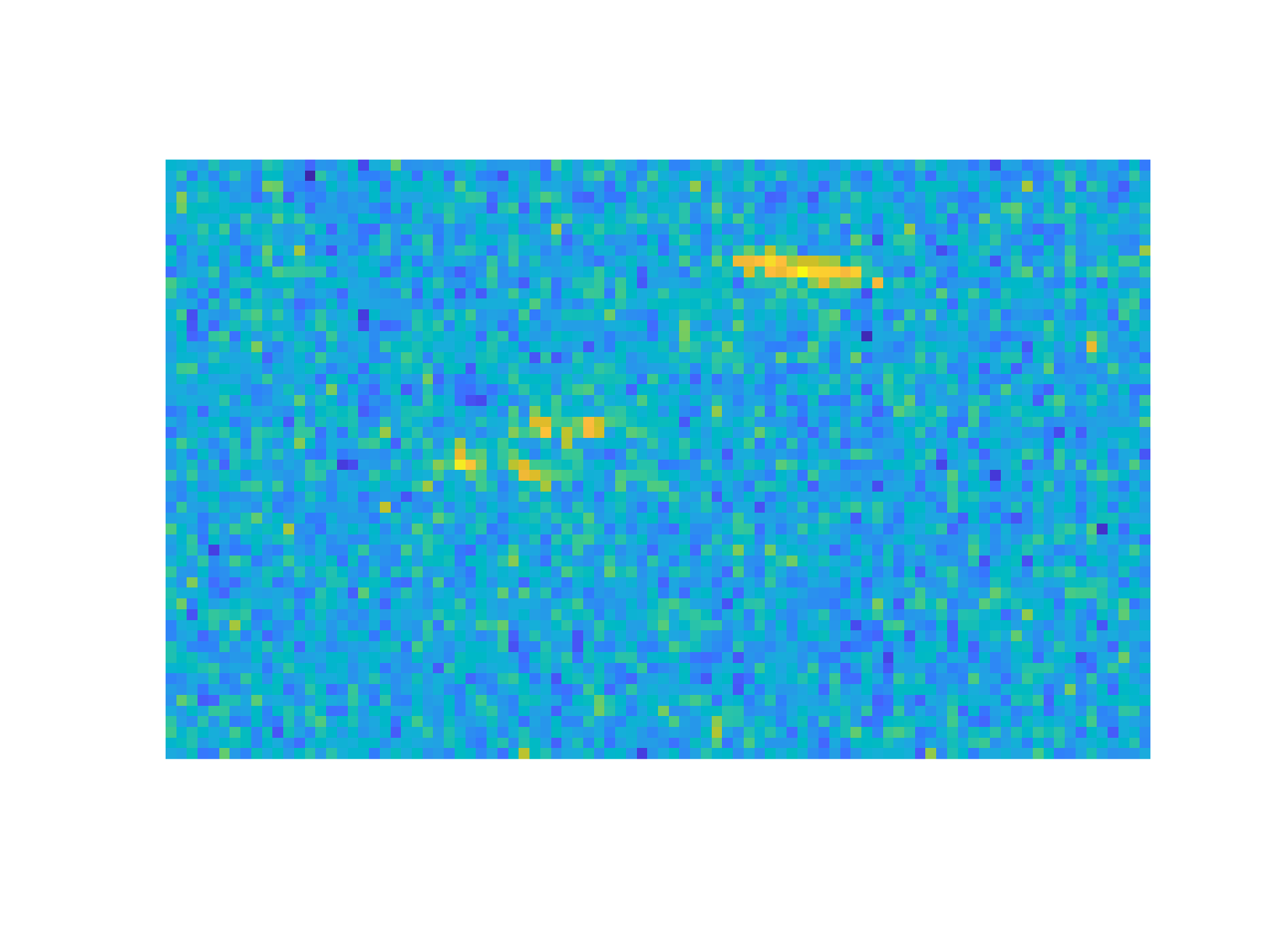}
      \end{subfigure}
      \begin{subfigure}[b]{0.16\textwidth}
      \includegraphics[clip=true,scale=0.25,trim=50 50 50 50]{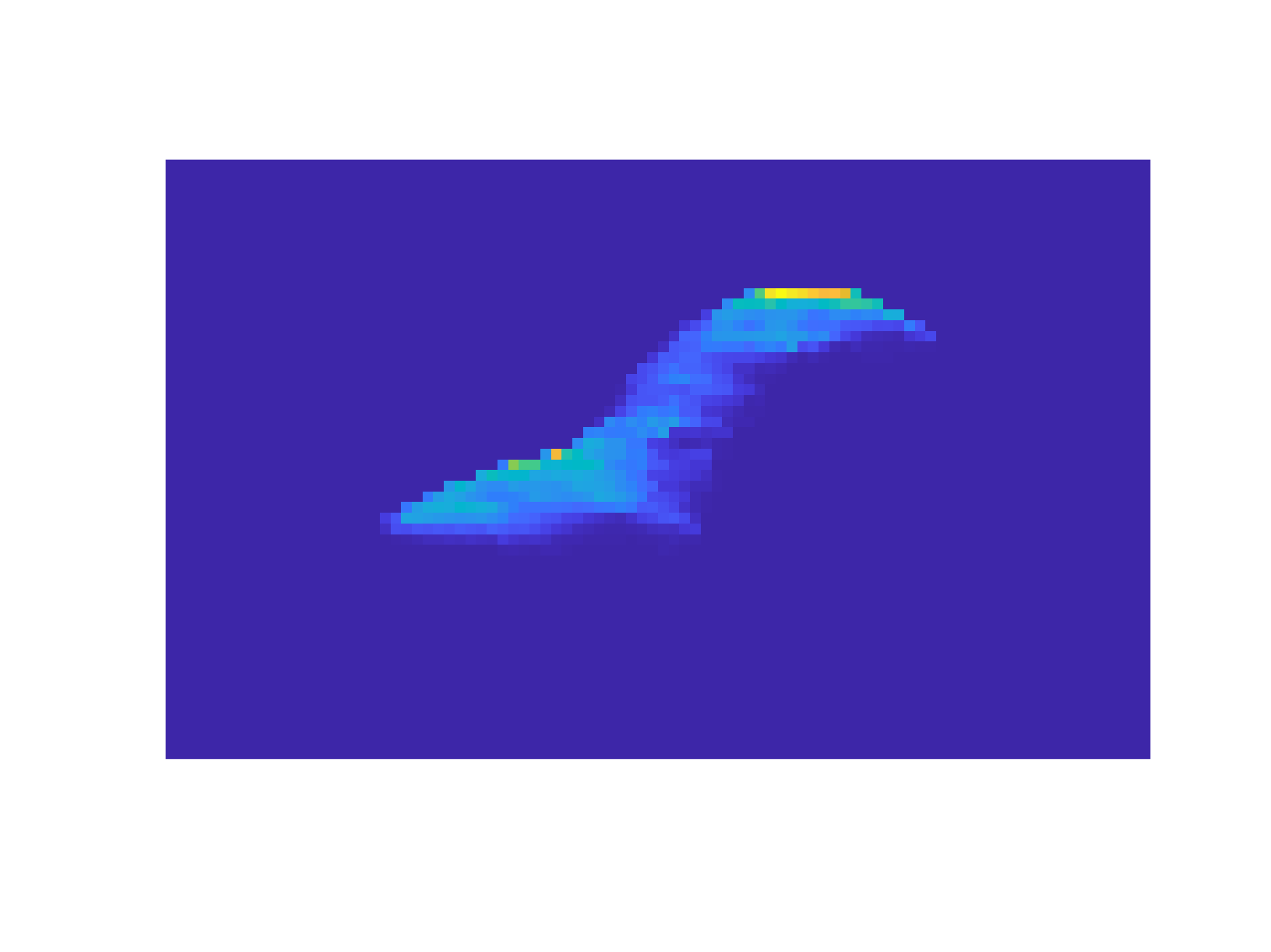}
      \centering
      \end{subfigure}
      \begin{subfigure}[b]{0.16\textwidth}
      \includegraphics[clip=true,scale=0.2,trim=15 15 15 15]{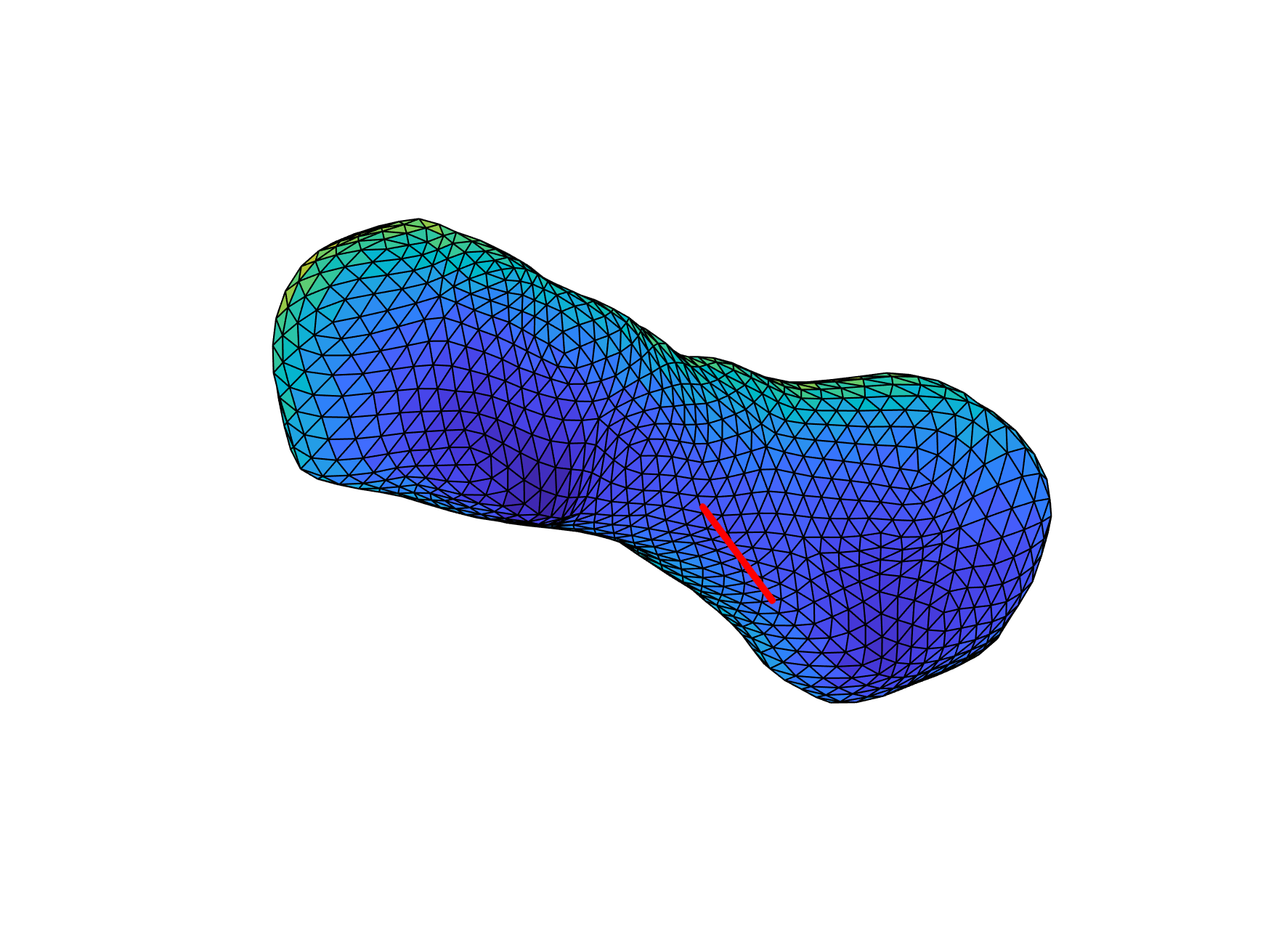}
      \centering
      \end{subfigure}
      \begin{subfigure}[b]{0.16\textwidth}
    \centering
      \includegraphics[clip=true,scale=0.25,trim=50 50 50 50]{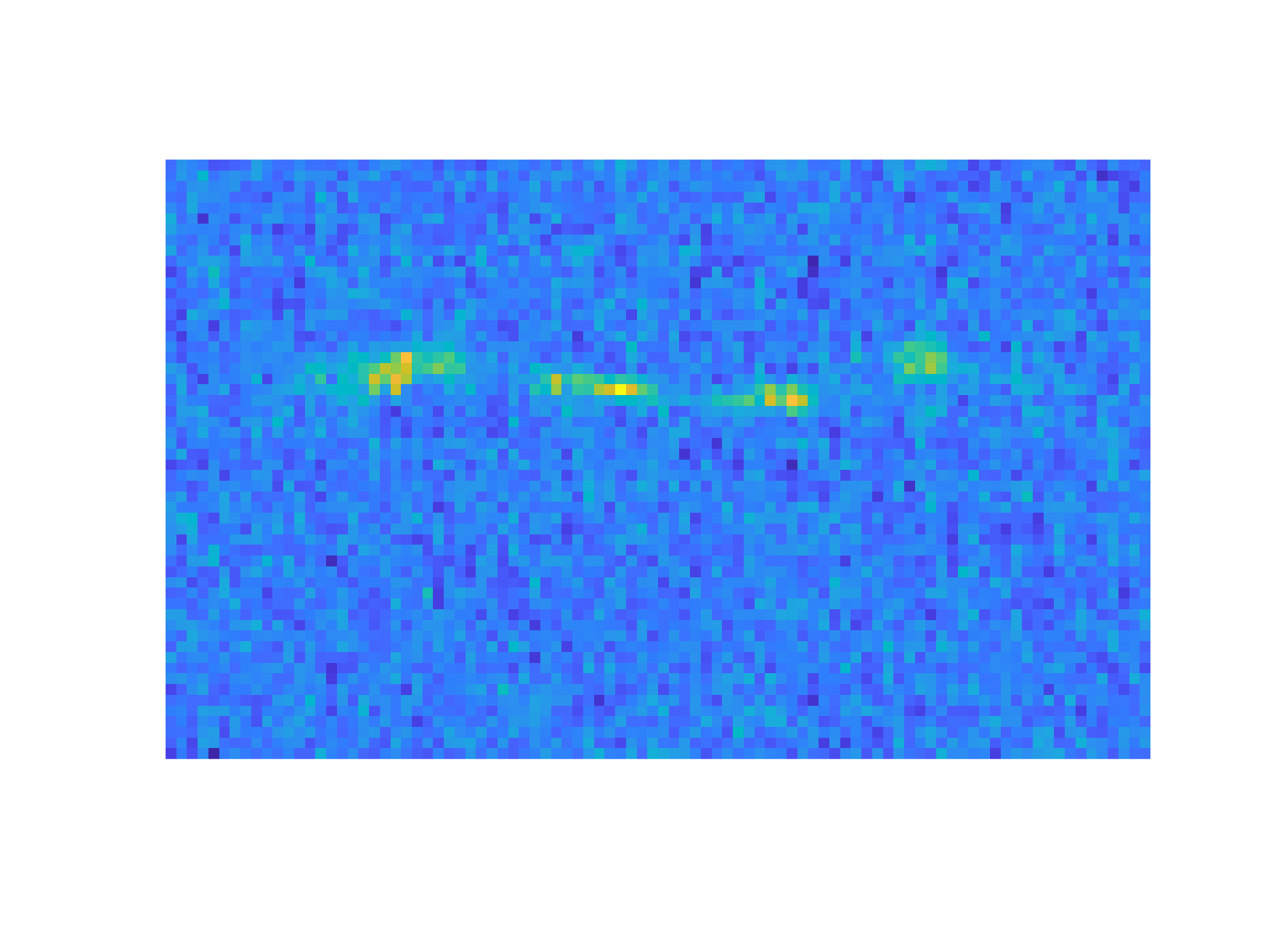}
      \end{subfigure}
      \begin{subfigure}[b]{0.16\textwidth}
      \includegraphics[clip=true,scale=0.25,trim=50 50 50 50]{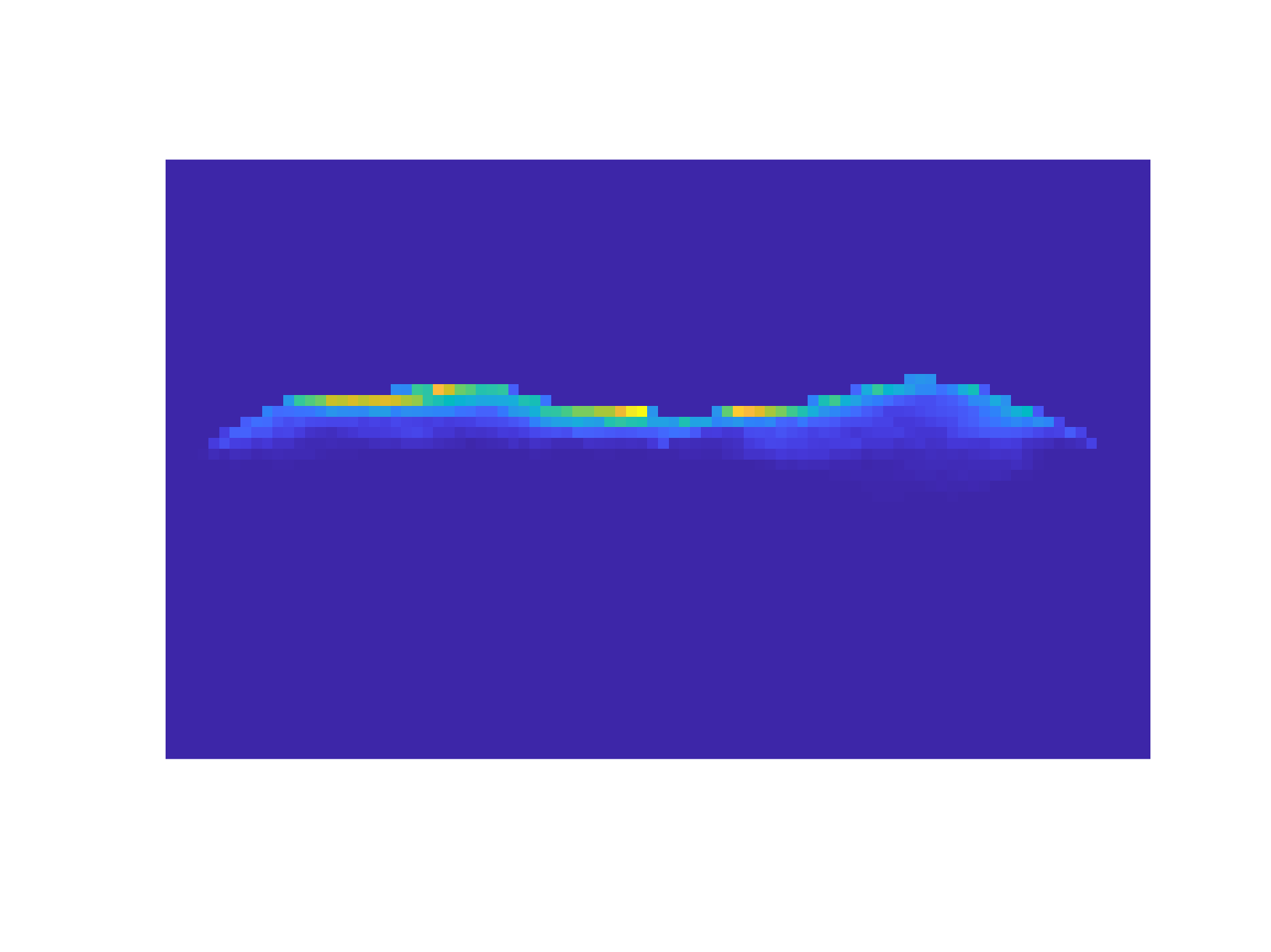}
      \centering
      \end{subfigure}
      \begin{subfigure}[b]{0.16\textwidth}
      \includegraphics[clip=true,scale=0.2,trim=15 30 15 30]{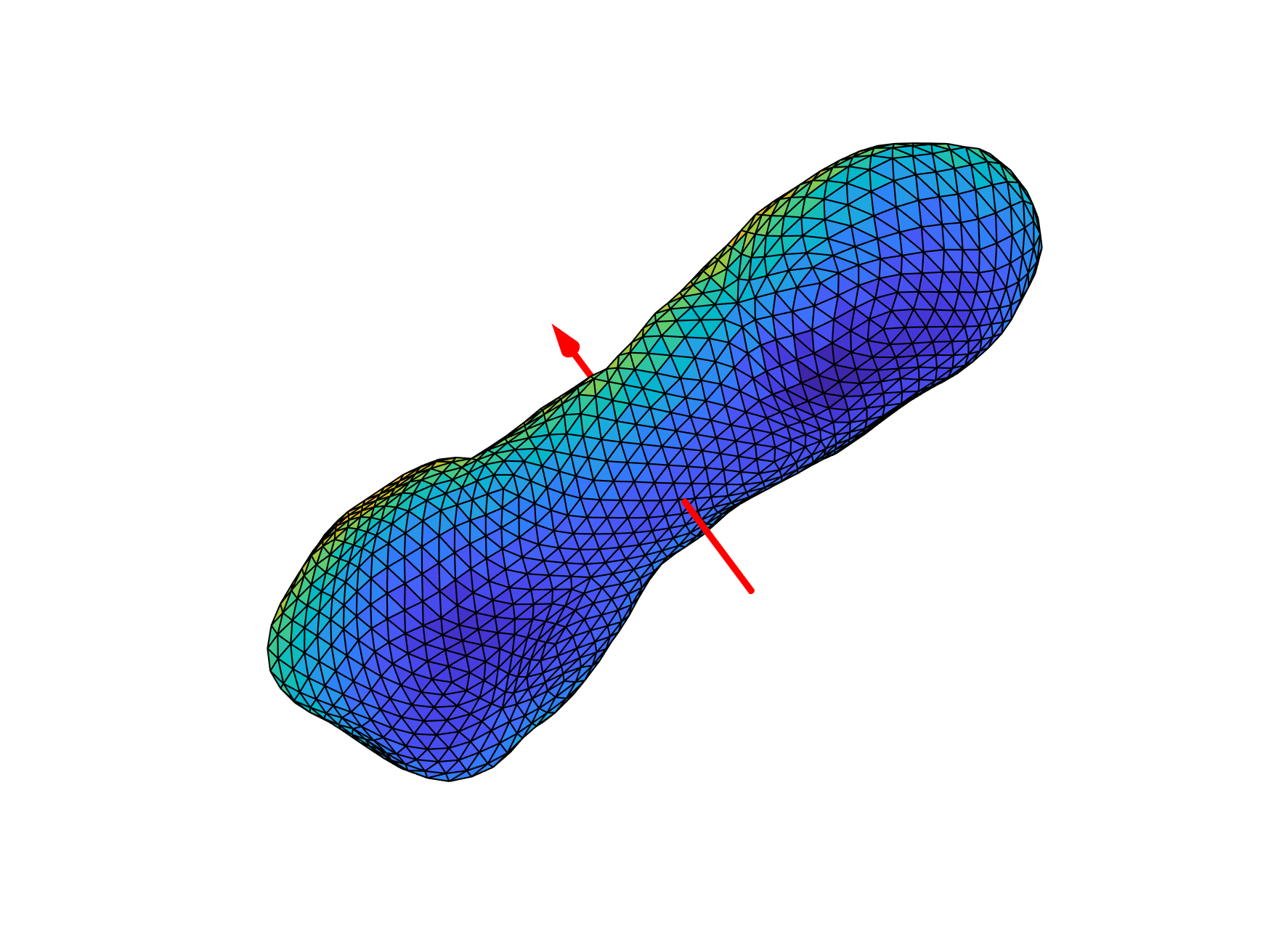}
      \centering
      \end{subfigure}
      \begin{subfigure}[b]{0.16\textwidth}
      \centering
      \includegraphics[clip=true,scale=0.25,trim=50 50 50 50]{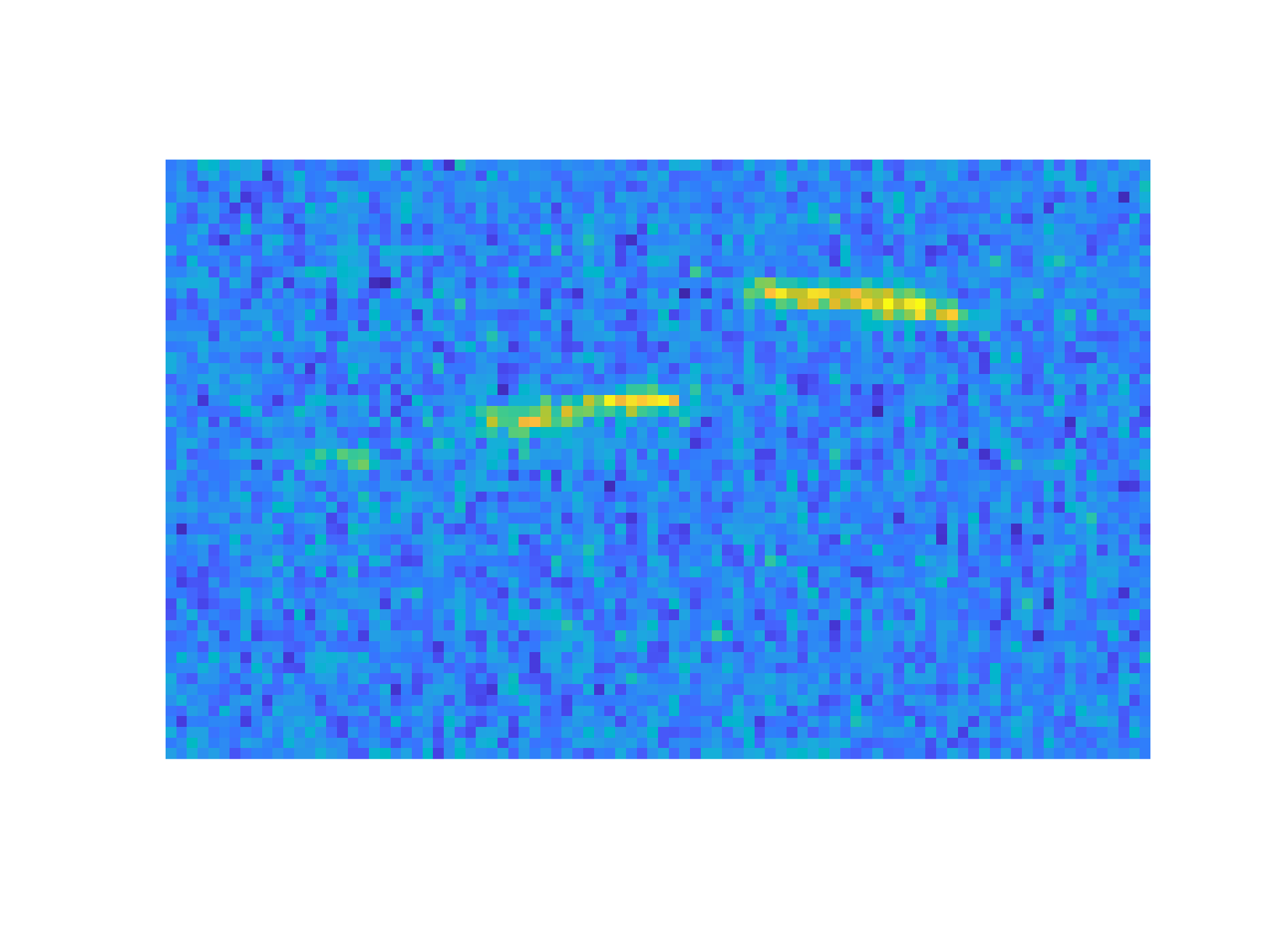}
      \end{subfigure}
      \begin{subfigure}[b]{0.16\textwidth}
      \includegraphics[clip=true,scale=0.25,trim=50 50 50 50]{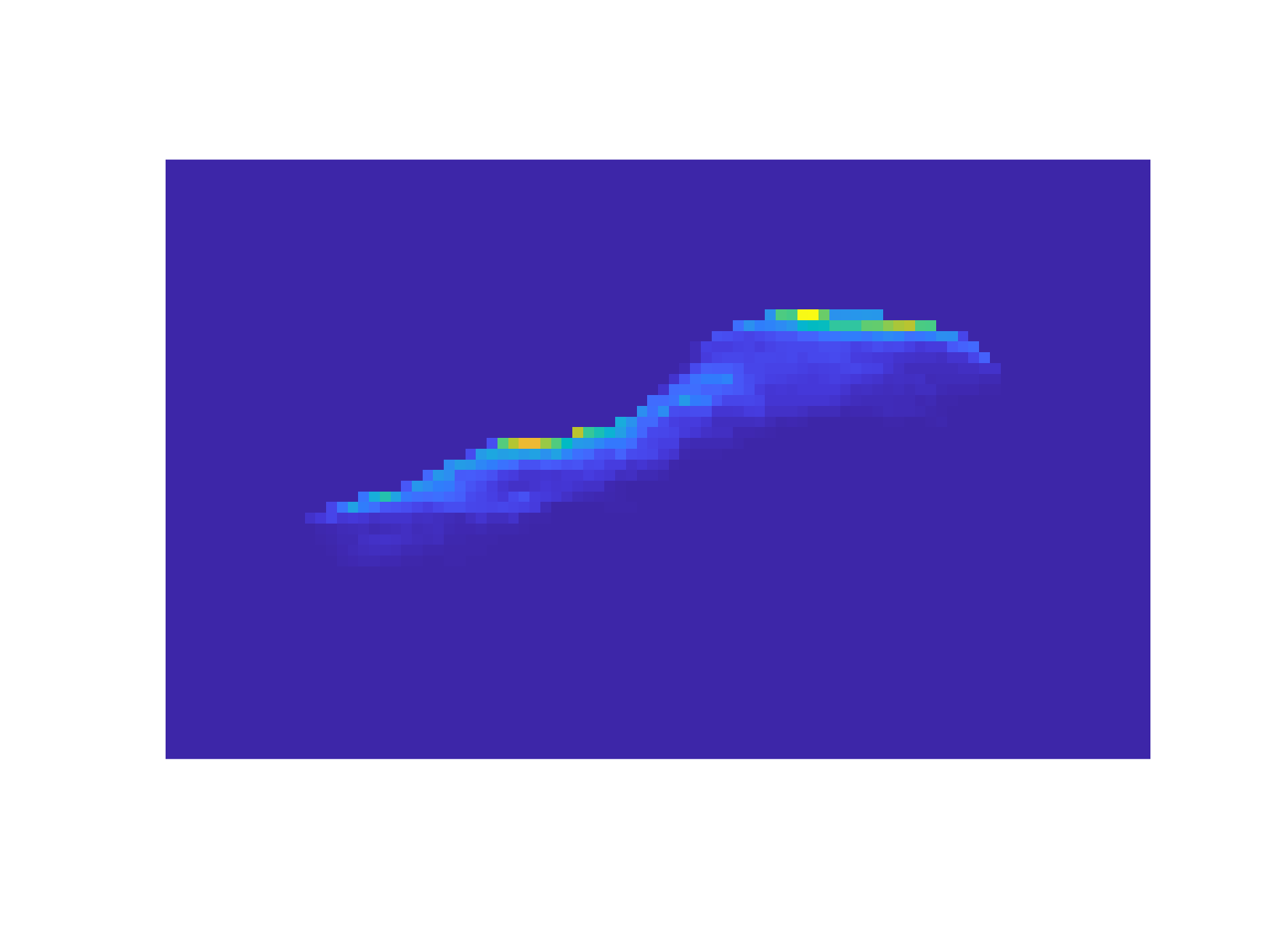}
      \centering
      \end{subfigure}
      \begin{subfigure}[b]{0.16\textwidth}
      \includegraphics[clip=true,scale=0.2,trim=15 30 15 30]{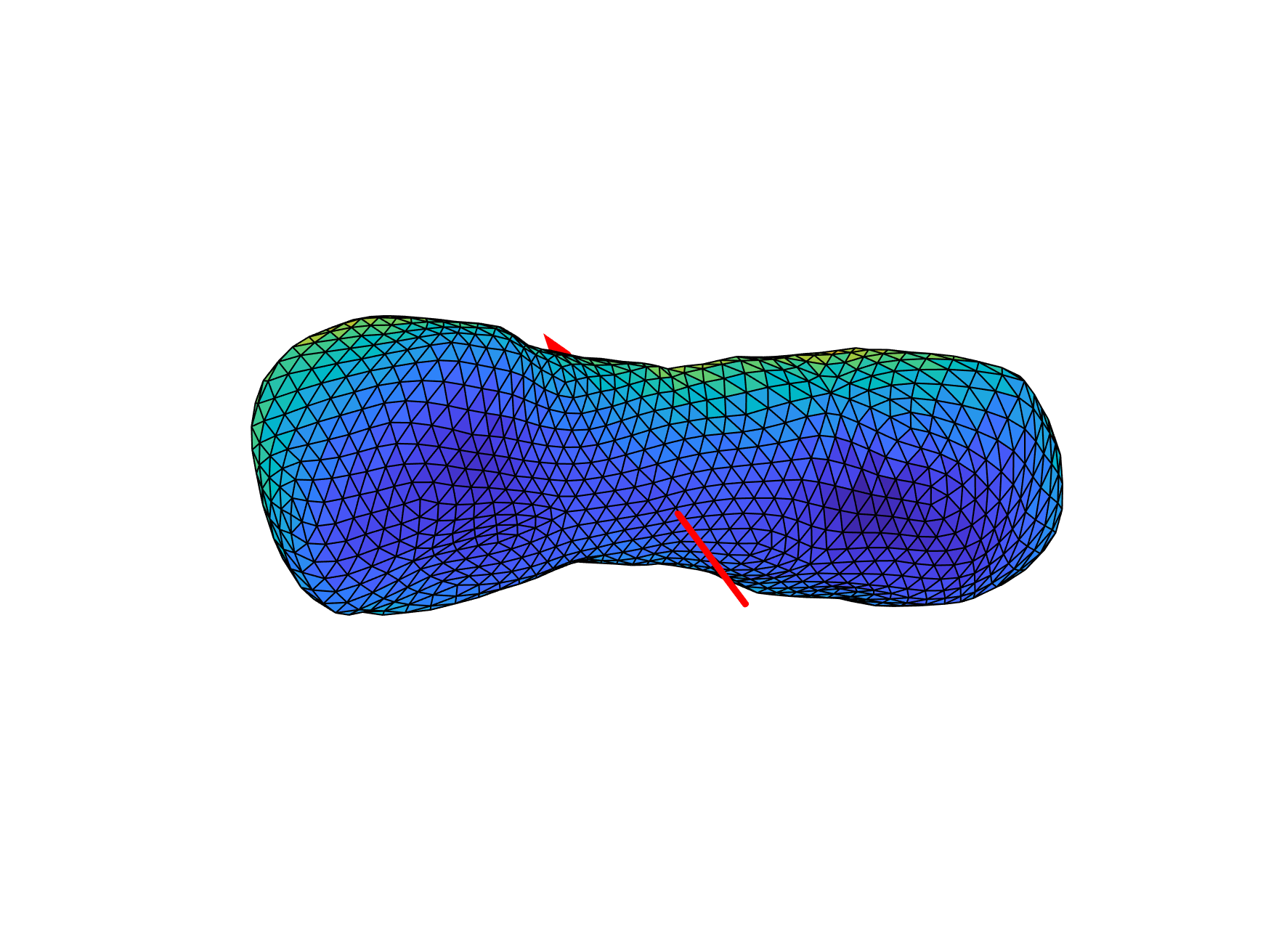}
      \centering
      \end{subfigure}
      \begin{subfigure}[b]{0.16\textwidth}
    \centering
      \includegraphics[clip=true,scale=0.25,trim=50 50 50 50]{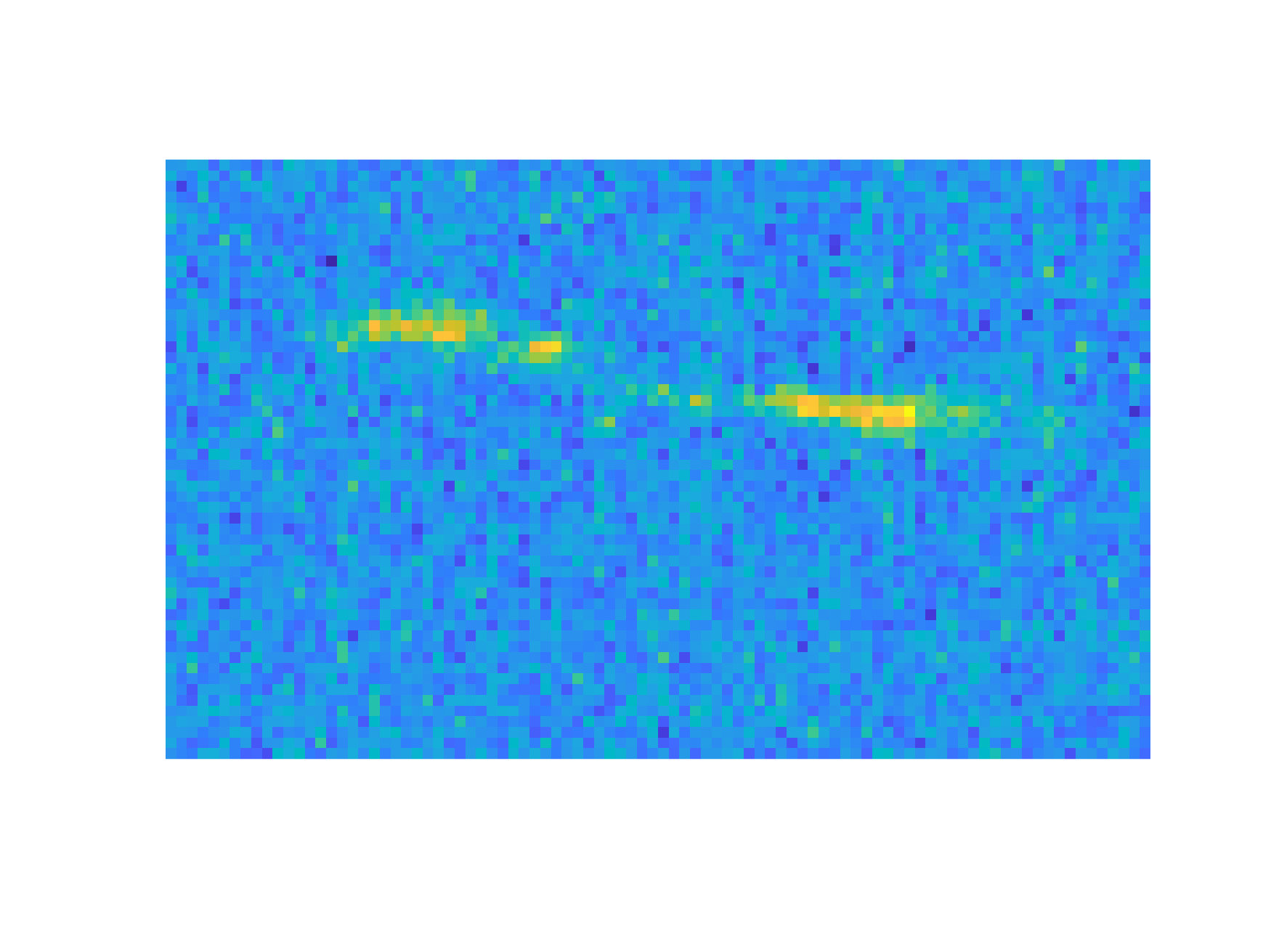}
      \end{subfigure}
      \begin{subfigure}[b]{0.16\textwidth}
      \includegraphics[clip=true,scale=0.25,trim=50 50 50 50]{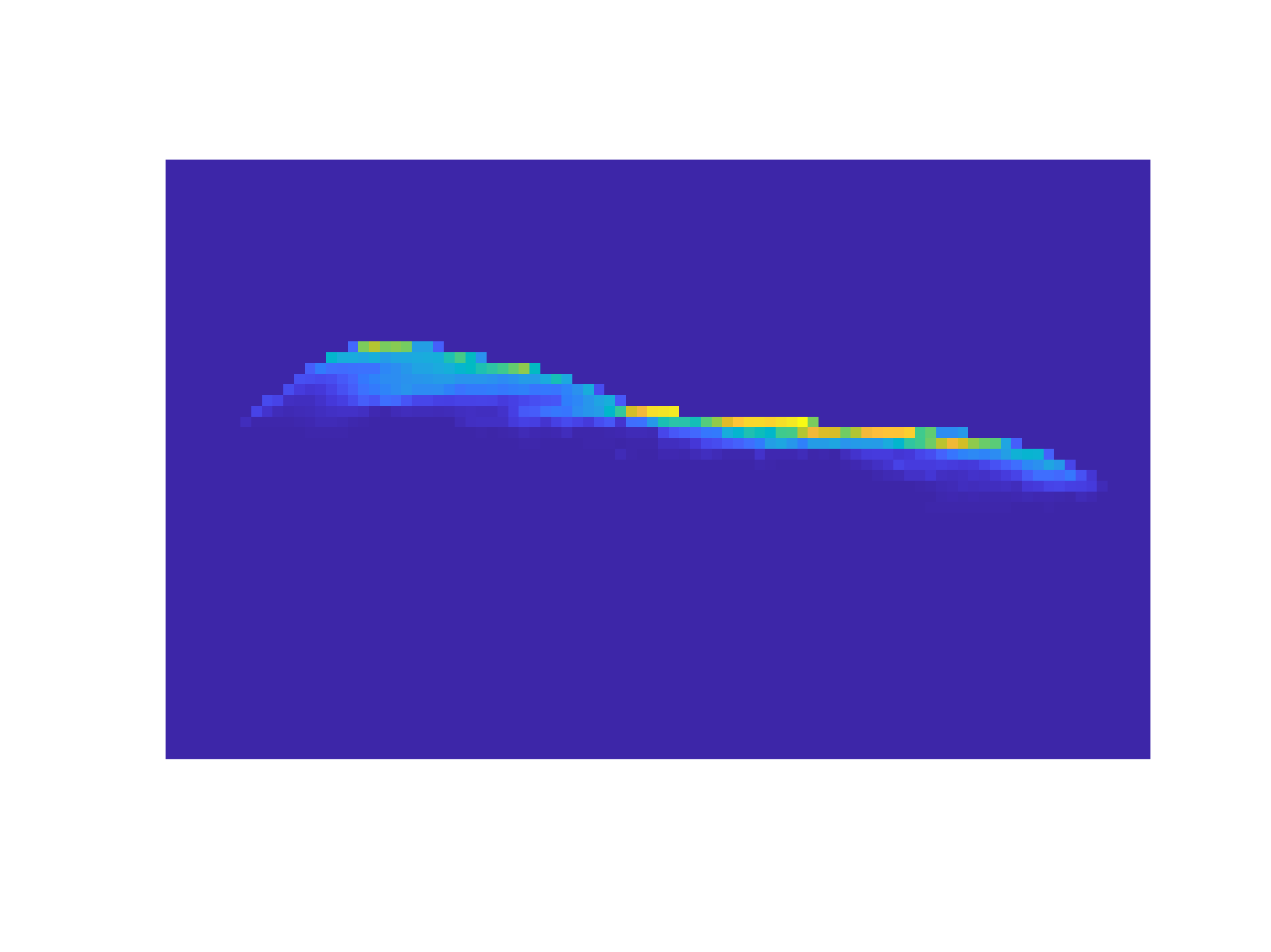}
      \centering
      \end{subfigure}
      \begin{subfigure}[b]{0.16\textwidth}
      \centering
      \includegraphics[clip=true,scale=0.2,trim=15 15 15 15]{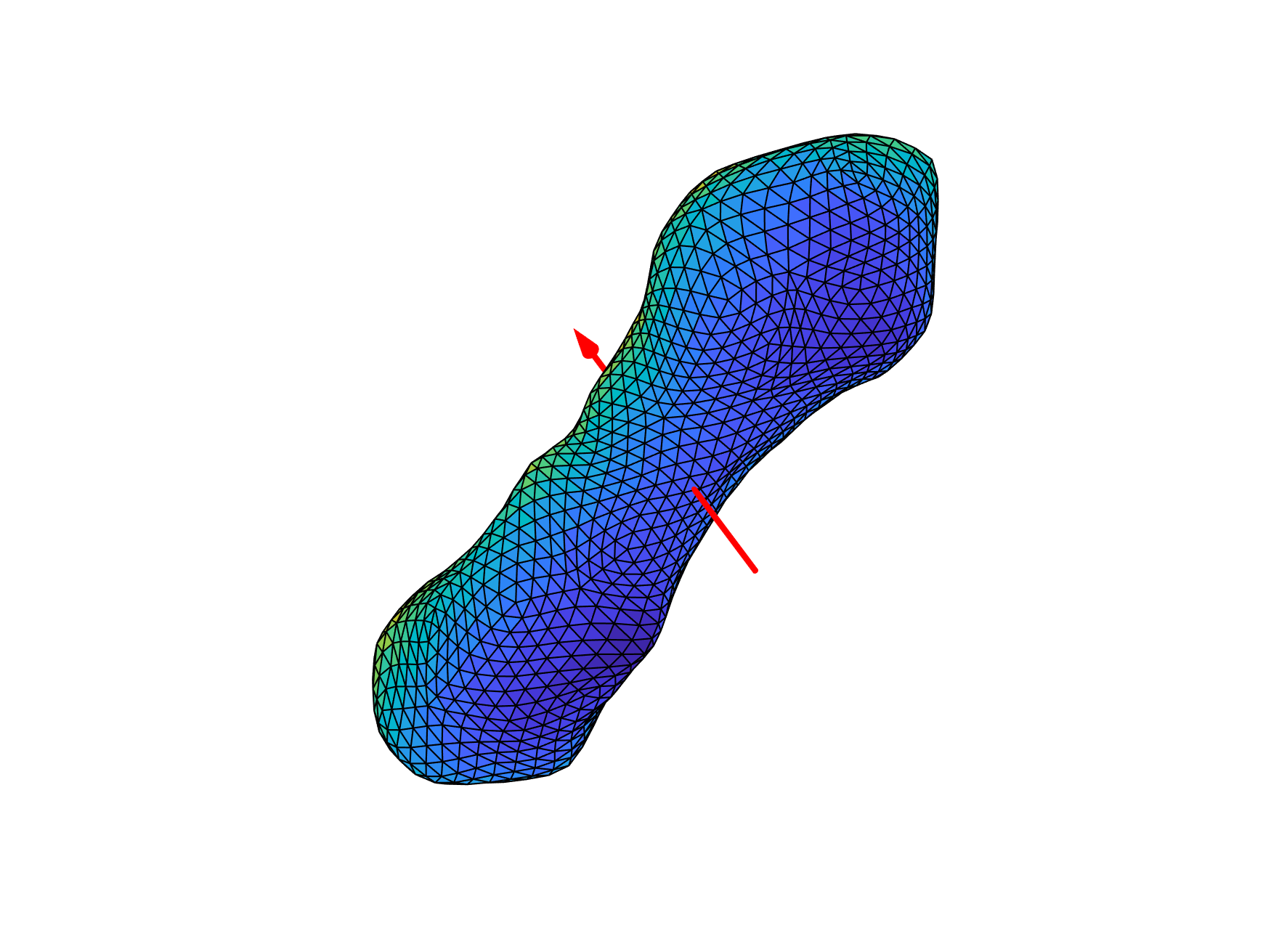}
      \end{subfigure}
            \begin{subfigure}[b]{0.16\textwidth}
    \centering
      \includegraphics[clip=true,scale=0.25,trim=50 50 50 50]{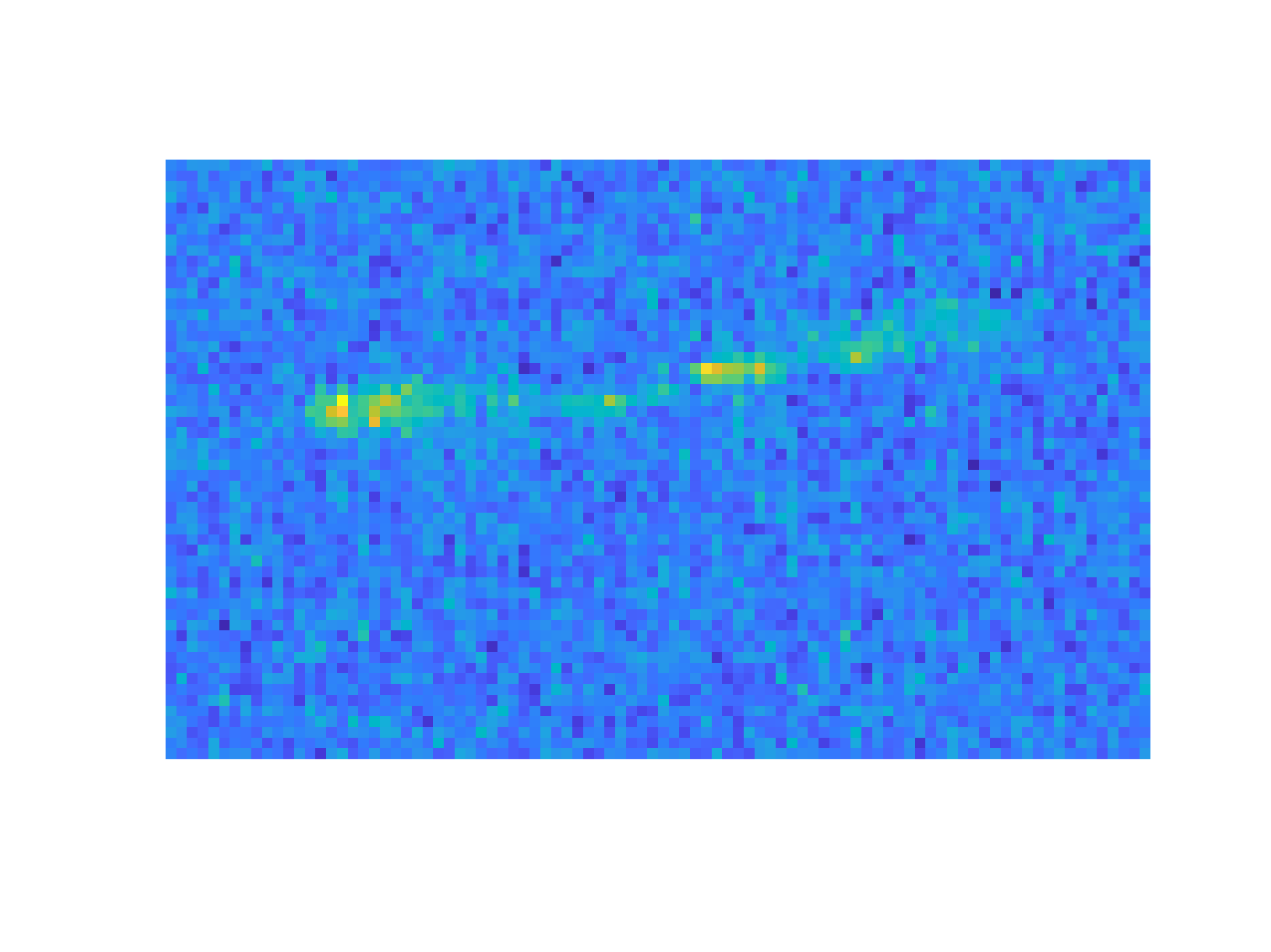}
      \end{subfigure}
      \begin{subfigure}[b]{0.16\textwidth}
      \includegraphics[clip=true,scale=0.25,trim=50 50 50 50]{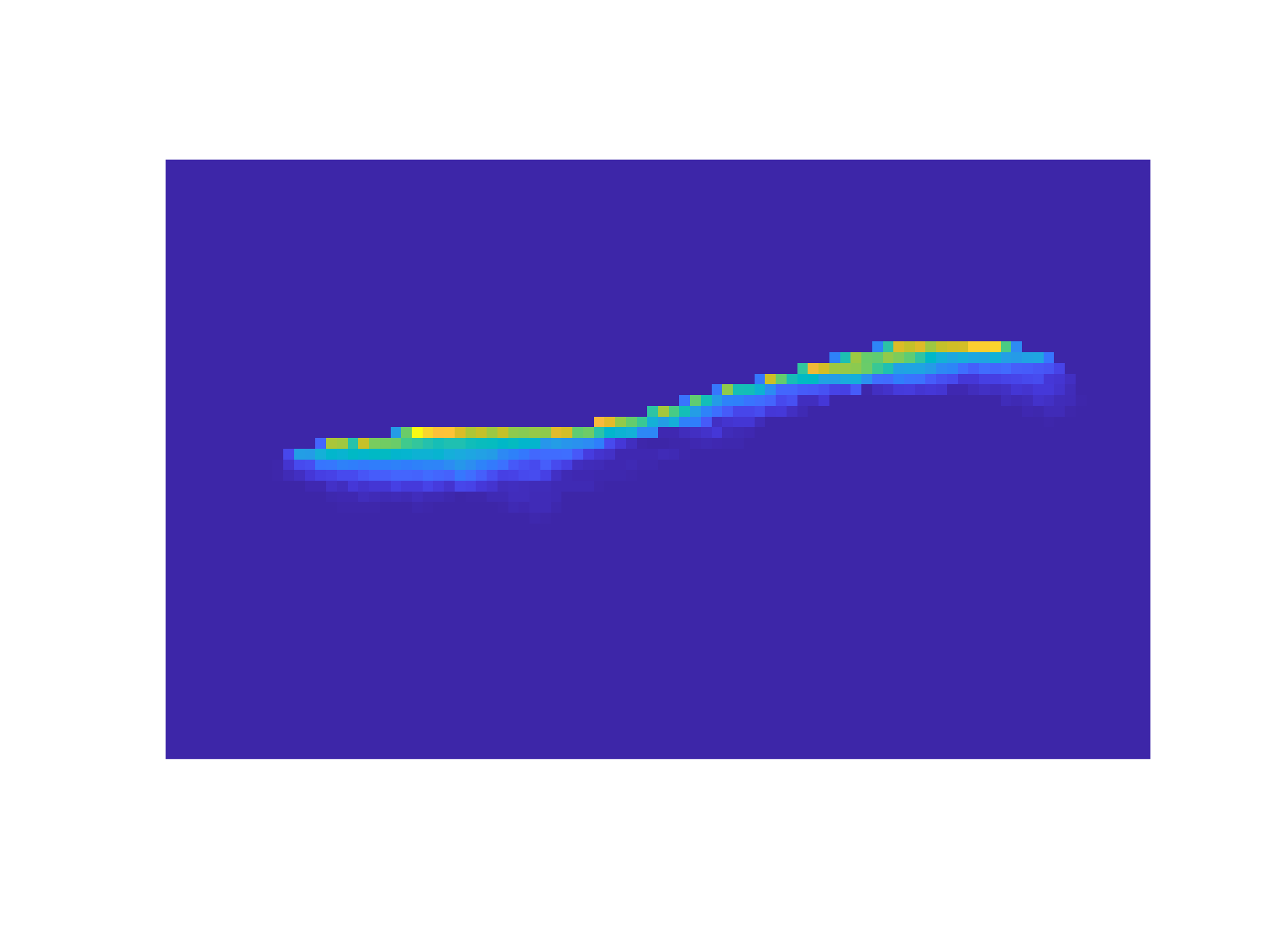}
      \centering
      \end{subfigure}
      \begin{subfigure}[b]{0.16\textwidth}
      \centering
      \includegraphics[clip=true,scale=0.2,trim=15 15 15 15]{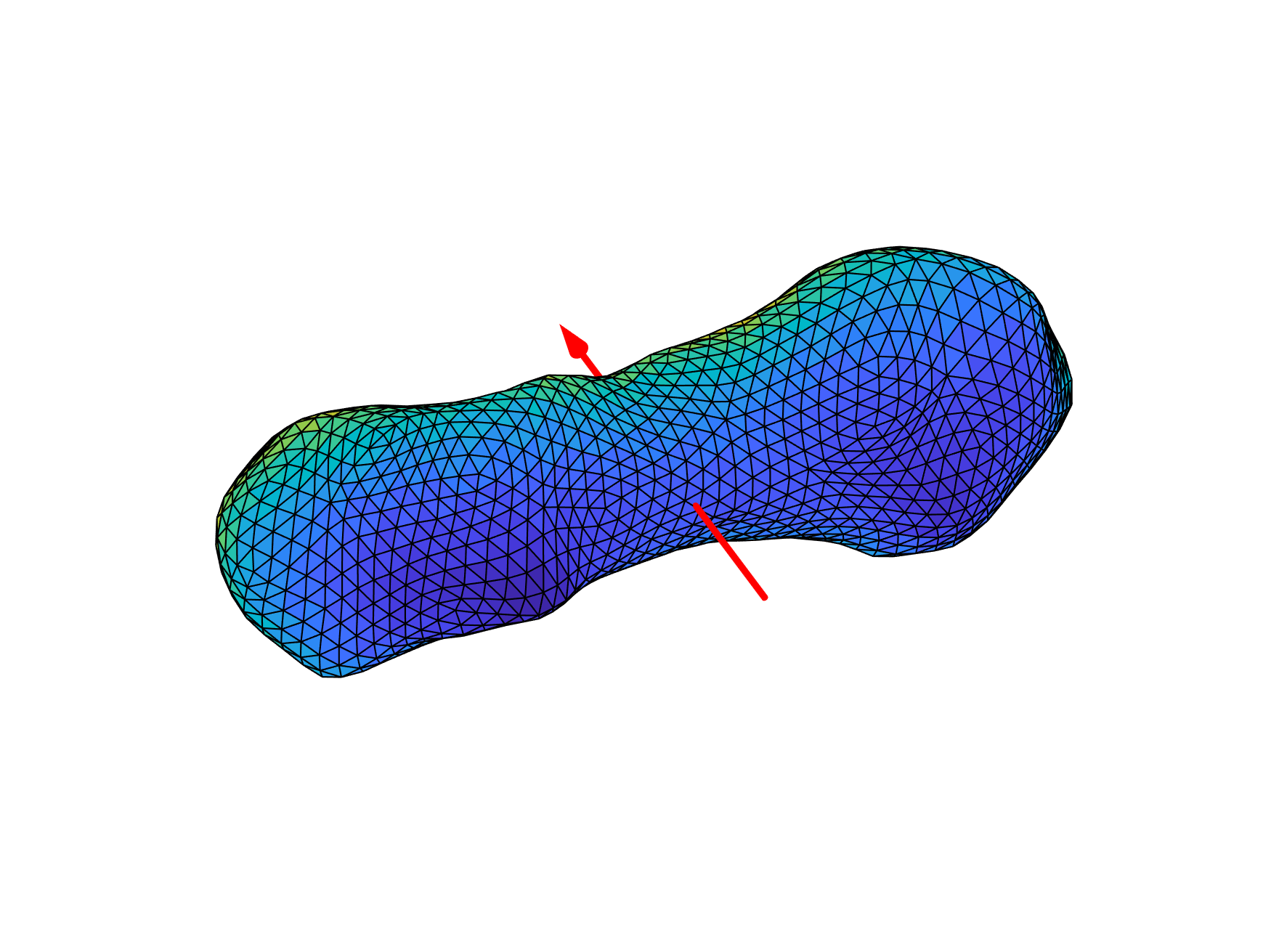}
      \end{subfigure}
\caption{\label{fig:comparisonRD}{Comparison between the delay Doppler images (right) and the synthetic images (middle) based on the best-fitting MPCD shape solution. Three images at slightly different observation times ($0.6\%$ of full rotation) overlap to simulate the effect of integration and to improve the contrast.  The third column shows the plane of the sky view of the asteroid and the direction of the rotation axis.}}
\end{figure*}
\setkeys{Gin}{draft=true}

\clearpage
\onecolumn

%Added by TeX Support
%\scriptsize{
\begin{table*}
\caption{\label{tab:ao}VLT/SPHERE disk-resolved images obtained in the \textbf{N\_R} filter by the ZIMPOL camera within the ESO's large program. For each observation, we provide the epoch, the exposure time, the airmass, the distance to the Earth $\Delta$ and the Sun $r$, the phase angle $\alpha$, and the angular diameter $D_\mathrm{a}$ of Kleopatra. 241 $s$ (176 $s$) total exposure time corresponds to 1.24\% (0.91\%) of the rotation period.}
\centering
\begin{tabular}{rr rr rrr r}
\hline 
\multicolumn{1}{c} {Date} & \multicolumn{1}{c} {UT} & \multicolumn{1}{c} {Exp} & \multicolumn{1}{c} {Airmass} & \multicolumn{1}{c} {$\Delta$} & \multicolumn{1}{c} {$r$} & \multicolumn{1}{c} {$\alpha$} & \multicolumn{1}{c} {$D_\mathrm{a}$} \\
\multicolumn{1}{c} {} & \multicolumn{1}{c} {} & \multicolumn{1}{c} {(s)} & \multicolumn{1}{c} {} & \multicolumn{1}{c} {(AU)} & \multicolumn{1}{c} {(AU)} & \multicolumn{1}{c} {(\degr)} & \multicolumn{1}{c} {(\arcsec)} \\
% &  &  &  &  &  &  &  & &  \\
\hline\hline
  2017-07-14 &      5:00:59 &  241 & 1.11 & 1.73 & 2.69 & 8.9 & 0.095  \\
  2017-07-14 &      5:05:09 &  241 & 1.11 & 1.73 & 2.69 & 8.9 & 0.095  \\
  2017-07-14 &      5:09:17 &  241 & 1.11 & 1.73 & 2.69 & 8.9 & 0.095  \\
  2017-07-14 &      5:13:26 &  241 & 1.11 & 1.73 & 2.69 & 8.9 & 0.095  \\
  2017-07-14 &      5:17:36 &  241 & 1.11 & 1.73 & 2.69 & 8.9 & 0.095  \\
  2017-07-22 &      4:18:07 &  241 & 1.11 & 1.69 & 2.66 & 7.9 & 0.097  \\
  2017-07-22 &      4:22:16 &  241 & 1.11 & 1.69 & 2.66 & 7.9 & 0.097  \\
  2017-07-22 &      4:26:27 &  241 & 1.11 & 1.69 & 2.66 & 7.9 & 0.097  \\
  2017-07-22 &      4:30:37 &  241 & 1.11 & 1.69 & 2.66 & 7.9 & 0.097  \\
  2017-07-22 &      4:34:46 &  241 & 1.11 & 1.69 & 2.66 & 7.9 & 0.097  \\
  2017-07-22 &      5:00:54 &  241 & 1.11 & 1.69 & 2.66 & 7.9 & 0.097  \\
  2017-07-22 &      5:05:06 &  241 & 1.11 & 1.69 & 2.66 & 7.9 & 0.097  \\
  2017-07-22 &      5:09:17 &  241 & 1.12 & 1.69 & 2.66 & 7.9 & 0.097  \\
  2017-07-22 &      5:13:25 &  241 & 1.12 & 1.69 & 2.66 & 7.9 & 0.097  \\
  2017-07-22 &      5:17:33 &  241 & 1.12 & 1.69 & 2.66 & 7.9 & 0.097  \\
  2017-07-27 &      4:11:03 &  241 & 1.11 & 1.68 & 2.65 & 7.9 & 0.098  \\
  2017-07-27 &      4:15:12 &  241 & 1.11 & 1.68 & 2.65 & 7.9 & 0.098  \\
  2017-07-27 &      4:19:23 &  241 & 1.11 & 1.68 & 2.65 & 7.9 & 0.098  \\
  2017-07-27 &      4:23:33 &  241 & 1.11 & 1.68 & 2.65 & 7.9 & 0.098  \\
  2017-07-27 &      4:27:42 &  241 & 1.11 & 1.68 & 2.65 & 7.9 & 0.098  \\
  2017-08-10 &      5:05:17 &  241 & 1.25 & 1.67 & 2.61 & 10.5 & 0.098  \\
  2017-08-10 &      5:09:29 &  241 & 1.26 & 1.67 & 2.61 & 10.5 & 0.098  \\
  2017-08-10 &      5:13:40 &  241 & 1.28 & 1.67 & 2.61 & 10.5 & 0.098  \\
  2017-08-10 &      5:17:48 &  241 & 1.29 & 1.67 & 2.61 & 10.5 & 0.098  \\
  2017-08-10 &      5:21:58 &  241 & 1.31 & 1.67 & 2.61 & 10.5 & 0.098  \\
  2017-08-22 &      1:42:34 &  241 & 1.10 & 1.71 & 2.58 & 13.9 & 0.096  \\
  2017-08-22 &      1:46:44 &  241 & 1.10 & 1.71 & 2.58 & 13.9 & 0.096  \\
  2017-08-22 &      1:50:54 &  241 & 1.10 & 1.71 & 2.58 & 13.9 & 0.096  \\
  2017-08-22 &      1:55:03 &  241 & 1.09 & 1.71 & 2.58 & 13.9 & 0.096  \\
  2017-08-22 &      1:59:11 &  241 & 1.09 & 1.71 & 2.58 & 13.9 & 0.096  \\
  2018-12-10 &      6:41:03 &  176 & 1.13 & 1.54 & 2.37 & 15.9 & 0.107  \\
  2018-12-10 &      6:44:11 &  176 & 1.13 & 1.54 & 2.37 & 15.9 & 0.107  \\
  2018-12-10 &      6:47:17 &  176 & 1.13 & 1.54 & 2.37 & 15.9 & 0.107  \\
  2018-12-10 &      6:50:22 &  176 & 1.13 & 1.54 & 2.37 & 15.9 & 0.107  \\
  2018-12-10 &      6:53:28 &  176 & 1.13 & 1.54 & 2.37 & 15.9 & 0.107  \\
  2018-12-19 &      6:45:02 &  176 & 1.12 & 1.51 & 2.40 & 12.9 & 0.109  \\
  2018-12-19 &      6:48:07 &  176 & 1.13 & 1.51 & 2.40 & 12.9 & 0.109  \\
  2018-12-19 &      6:51:13 &  176 & 1.13 & 1.51 & 2.40 & 12.9 & 0.109  \\
  2018-12-19 &      6:54:19 &  176 & 1.13 & 1.51 & 2.40 & 12.9 & 0.109  \\
  2018-12-19 &      6:57:24 &  176 & 1.13 & 1.51 & 2.40 & 12.9 & 0.109  \\
  2018-12-22 &      5:46:19 &  176 & 1.12 & 1.50 & 2.40 & 12.0 & 0.109  \\
  2018-12-22 &      5:49:27 &  176 & 1.12 & 1.50 & 2.40 & 12.0 & 0.109  \\
  2018-12-22 &      5:52:32 &  176 & 1.12 & 1.50 & 2.40 & 12.0 & 0.109  \\
  2018-12-22 &      5:55:38 &  176 & 1.12 & 1.50 & 2.40 & 12.0 & 0.109  \\
  2018-12-22 &      5:58:43 &  176 & 1.12 & 1.50 & 2.40 & 12.0 & 0.109  \\
  2018-12-26 &      8:08:27 &  176 & 1.37 & 1.50 & 2.41 & 10.7 & 0.109  \\
  2018-12-26 &      8:11:34 &  176 & 1.38 & 1.50 & 2.41 & 10.7 & 0.109  \\
  2018-12-26 &      8:14:41 &  176 & 1.40 & 1.50 & 2.41 & 10.7 & 0.109  \\
  2018-12-26 &      8:17:45 &  176 & 1.41 & 1.50 & 2.41 & 10.7 & 0.109  \\
  2018-12-26 &      8:20:51 &  176 & 1.43 & 1.50 & 2.41 & 10.7 & 0.109  \\
  2019-01-14 &      4:57:42 &  176 & 1.13 & 1.52 & 2.46 & 8.4 & 0.108  \\
  2019-01-14 &      5:00:49 &  176 & 1.14 & 1.52 & 2.46 & 8.4 & 0.108  \\
  2019-01-14 &      5:03:55 &  176 & 1.14 & 1.52 & 2.46 & 8.4 & 0.108  \\
  2019-01-14 &      5:07:00 &  176 & 1.14 & 1.52 & 2.46 & 8.4 & 0.108  \\
  2019-01-14 &      5:10:04 &  176 & 1.15 & 1.52 & 2.46 & 8.4 & 0.108  \\
\hline
\end{tabular}
%\tablefoot{
%    Notes...
%    }
\end{table*}

%\scriptsize{
\begin{table*}
\caption{\label{tab:aoKeck}Disk-resolved images obtained by the NIRC2 camera mounted on the Keck II telescope. For each observation, we provide the epoch, the filter, the exposure time, the airmass, the distance to the Earth $\Delta$ and the Sun $r$, the phase angle $\alpha$, the angular diameter $D_\mathrm{a}$ of Kleopatra and the program PI.}
\centering
\begin{tabular}{rr rrr rrr rr}
\hline 
\multicolumn{1}{c} {Date} & \multicolumn{1}{c} {UT} & \multicolumn{1}{c} {Filter} & \multicolumn{1}{c} {Exp} & \multicolumn{1}{c} {Airmass} & \multicolumn{1}{c} {$\Delta$} & \multicolumn{1}{c} {$r$} & \multicolumn{1}{c} {$\alpha$} & \multicolumn{1}{c} {$D_\mathrm{a}$} & \multicolumn{1}{c} {Program PI}\\
\multicolumn{1}{c} {} & \multicolumn{1}{c} {} & & \multicolumn{1}{c} {(s)} & \multicolumn{1}{c} {} & \multicolumn{1}{c} {(AU)} & \multicolumn{1}{c} {(AU)} & \multicolumn{1}{c} {(\degr)} & \multicolumn{1}{c} {(\arcsec)} & \multicolumn{1}{c} {}\\
% &  &  &  &  &  &  &  & &  \\
\hline\hline
  2002-05-07 &     10:54:36 &     H     &  60 & 1.20 & 2.45 & 3.46 & 1.8  & 0.067 &            J.L. Margot \\
  2003-07-14 &     11:11:35 &    Kp     &  10 & 1.05 & 1.70 & 2.65 & 9.8  & 0.097 &             W. Merline \\
  2008-09-19 &     06:16:59 & PK50$\_$1.5 &  60 & 1.47 & 1.24 & 2.22 & 7.5  & 0.132 &            F. Marchis \\
  2008-09-19 &     06:28:57 & PK50\_1.5 &  60 & 1.40 & 1.24 & 2.22 & 7.5  & 0.132 &            F. Marchis \\
  2008-09-19 &     11:38:20 & PK50\_1.5 &  60 & 1.18 & 1.24 & 2.22 & 7.5  & 0.132 &            F. Marchis \\
  2008-10-05 &     09:12:57 & PK50\_1.5 &  60 & 1.05 & 1.26 & 2.19 & 12.3 & 0.130 &            F. Marchis \\
  2008-10-05 &     09:48:54 & PK50\_1.5 &  60 & 1.10 & 1.26 & 2.19 & 12.3 & 0.130 &            F. Marchis \\
  2008-10-05 &     10:03:40 &     H     &  30 & 1.13 & 1.26 & 2.19 & 12.3 & 0.130 &            F. Marchis \\
  2008-10-06 &     07:18:06 & PK50\_1.5 &  60 & 1.05 & 1.26 & 2.19 & 12.6 & 0.130 &            F. Marchis \\
  2008-10-06 &     09:49:50 & PK50\_1.5 &  60 & 1.11 & 1.26 & 2.19 & 12.7 & 0.130 &            F. Marchis \\
  2008-10-09 &     05:45:41 & PK50\_1.5 &  60 & 1.23 & 1.27 & 2.18 & 13.8 & 0.129 &            F. Marchis \\
  2008-10-09 &     09:35:21 & PK50\_1.5 &  60 & 1.11 & 1.27 & 2.18 & 13.8 & 0.129 &            F. Marchis \\
  2012-08-10 &     06:14:05 &    Kp     & 0.8 & 1.14 & 2.23 & 2.92 & 16.6 & 0.074 &             W. Merline \\
  2013-08-26 &     15:38:25 &    Kp     & 0.2 & 1.00 & 1.69 & 2.11 & 28.2 & 0.097 &             W. Merline \\
\hline
\end{tabular}
%\tablefoot{
%    Notes...
%    }
\end{table*}

\onecolumn
%\scriptsize{
\begin{longtable}{rlr rrr l l}
\caption{\label{tab:lcs}Optical disk-integrated lightcurves of Kleopatra used for \adam{} shape modeling . For each lightcurve, the table gives the epoch, the number of individual measurements $N_p$, asteroid's distances to the Earth $\Delta$ and the Sun $r$, phase angle $\varphi$, photometric filter and reference.}\\
\hline 
\multicolumn{1}{c} {N} & \multicolumn{1}{c} {Epoch} & \multicolumn{1}{c} {$N_p$} & \multicolumn{1}{c} {$\Delta$} & \multicolumn{1}{c} {$r$} & \multicolumn{1}{c} {$\varphi$} & \multicolumn{1}{c} {Filter} & Reference \\
 &  &  & (AU) & (AU) & (\degr) &  &  \\
\hline\hline
\endfirsthead
\caption{continued.}\\
\hline
\multicolumn{1}{c} {N} & \multicolumn{1}{c} {Epoch} & \multicolumn{1}{c} {$N_p$} & \multicolumn{1}{c} {$\Delta$} & \multicolumn{1}{c} {$r$} & \multicolumn{1}{c} {$\varphi$} & \multicolumn{1}{c} {Filter} & Reference \\
 &  &  & (AU) & (AU) & (\degr) &  &  \\
\hline\hline
\endhead
\hline
\endfoot
\hline
     1	&  1977-01-15.9  &  277  &  1.57  &  2.51  &  8.1   &  V        &  \citet{Scaltriti1978}    \\
     2	&  1977-01-16.9  &  245  &  1.57  &  2.51  &  8.1   &  V        &  \citet{Scaltriti1978}    \\
     3	&  1980-09-06.3  &  23   &  1.24  &  2.20  &  10.3  &  C        &  \citet{Pilcher1982}       \\
     4	&  1980-09-07.3  &  27   &  1.23  &  2.20  &  9.9   &  C        &  \citet{Pilcher1982}       \\
     5	&  1980-09-09.3  &  20   &  1.23  &  2.20  &  9.2   &  C        &  \citet{Pilcher1982}       \\
     6	&  1980-09-11.1  &  11   &  1.22  &  2.19  &  8.6   &  C        &  \citet{Pilcher1982}       \\
     7	&  1980-09-14.3  &  25   &  1.21  &  2.19  &  7.8   &  C        &  \citet{Pilcher1982}       \\
     8	&  1980-09-15.6  &  41   &  1.20  &  2.19  &  7.5   &  V        &  \citet{Kennedy1982}       \\
     9	&  1980-09-18.2  &  18   &  1.20  &  2.18  &  7.1   &  C        &  \citet{Pilcher1982}       \\
    10	&  1980-09-20.3  &  14   &  1.19  &  2.18  &  7.0   &  V        &  \citet{Harris1989a}         \\
    11	&  1980-09-20.3  &  18   &  1.19  &  2.18  &  7.0   &  C        &  \citet{Pilcher1982}       \\
    12	&  1980-09-21.3  &  10   &  1.19  &  2.18  &  7.0   &  C        &  \citet{Pilcher1982}       \\
    13	&  1980-09-27.1  &  11   &  1.19  &  2.17  &  7.7   &  C        &  \citet{Pilcher1982}       \\
    14	&  1980-09-29.2  &  21   &  1.19  &  2.16  &  8.2   &  C        &  \citet{Pilcher1982}       \\
    15	&  1980-09-30.5  &  27   &  1.19  &  2.16  &  8.6   &  V        &  \citet{Kennedy1982}       \\
    16	&  1980-10-01.5  &  37   &  1.19  &  2.16  &  8.9   &  V        &  \citet{Kennedy1982}       \\
    17	&  1980-10-02.0  &  63   &  1.19  &  2.16  &  9.0   &  V        &  \citet{Zappala1983}           \\
    18	&  1980-10-02.1  &  44   &  1.19  &  2.16  &  9.1   &  V        &  \citet{Zappala1983}           \\
    19	&  1980-10-02.6  &  9    &  1.19  &  2.16  &  9.3   &  V        &  \citet{Kennedy1982}       \\
    20	&  1980-10-06.2  &  37   &  1.20  &  2.15  &  10.6  &  C        &  \citet{Pilcher1982}      \\
    21	&  1980-10-09.2  &  30   &  1.21  &  2.15  &  11.7  &  C        &  \citet{Pilcher1982}      \\
    22	&  1980-11-06.2  &  5    &  1.36  &  2.11  &  22.0  &  V        &  \citet{Harris1989a}         \\
    23	&  1980-11-07.8  &  199  &  1.38  &  2.11  &  22.4  &  V        &  \citet{Grossmann1981}          \\
    24	&  1980-11-10.2  &  24   &  1.40  &  2.11  &  23.0  &  C        &  \citet{Pilcher1982}       \\
    25	&  1981-12-01.2  &  19   &  2.48  &  2.72  &  21.2  &  V        &  \citet{Weidenschilling1987}    \\
    26	&  1981-12-02.4  &  17   &  2.47  &  2.72  &  21.2  &  V        &  \citet{Weidenschilling1987}    \\
    27	&  1982-02-13.9  &  74   &  1.96  &  2.91  &  6.4   &  V        &  \citet{Zappala1983}           \\
    28	&  1982-02-14.0  &  36   &  1.96  &  2.91  &  6.4   &  V        &  \citet{Zappala1983}           \\
    29	&  1982-02-20.3  &  30   &  1.97  &  2.93  &  6.0   &  V        &  \citet{Weidenschilling1987}    \\
    30	&  1982-03-23.2  &  162  &  2.18  &  3.00  &  12.6  &  V        &  \citet{Carlsson1983}  \\
    31	&  1982-03-26.9  &  126  &  2.22  &  3.01  &  13.5  &  V        &  \citet{Zappala1983}           \\
    32	&  1983-02-20.3  &  18   &  3.01  &  3.48  &  15.4  &  V        &  \citet{Weidenschilling1987}    \\
    33	&  1983-02-21.3  &  8    &  2.99  &  3.48  &  15.3  &  V        &  \citet{Weidenschilling1987}    \\
    34	&  1983-02-22.4  &  5    &  2.98  &  3.48  &  15.2  &  V        &  \citet{Weidenschilling1987}    \\
    35	&  1983-03-28.3  &  26   &  2.60  &  3.49  &  8.5   &  V        &  \citet{Weidenschilling1987}    \\
    36	&  1983-03-29.3  &  12   &  2.59  &  3.49  &  8.2   &  V        &  \citet{Weidenschilling1987}    \\
    37	&  1983-05-21.3  &  15   &  2.60  &  3.50  &  8.8   &  V        &  \citet{Weidenschilling1987}    \\
    38	&  1983-05-22.3  &  18   &  2.61  &  3.50  &  9.1   &  V        &  \citet{Weidenschilling1987}    \\
    39	&  1983-05-23.2  &  6    &  2.61  &  3.50  &  9.4   &  V        &  \citet{Weidenschilling1987}    \\
    40	&  1984-05-08.4  &  10   &  2.40  &  3.10  &  15.3  &  V        &  \citet{Weidenschilling1987}    \\
    41	&  1984-05-09.4  &  18   &  2.39  &  3.10  &  15.2  &  V        &  \citet{Weidenschilling1987}    \\
    42	&  1984-05-10.2  &  7    &  2.38  &  3.10  &  15.0  &  V        &  \citet{Weidenschilling1987}    \\
    43	&  1984-07-05.3  &  24   &  2.00  &  2.97  &  6.9   &  V        &  \citet{Weidenschilling1987}    \\
    44	&  1985-10-21.3  &  35   &  1.25  &  2.11  &  17.5  &  V        &  \citet{Weidenschilling1987}    \\
    45	&  1985-11-09.9  &  63   &  1.17  &  2.13  &  8.8   &  V        &  \citet{Lupishko1987}    \\
    46	&  1985-12-02.9  &  117  &  1.19  &  2.16  &  7.3   &  V        &  \citet{Dotto1992}              \\
    47	&  1985-12-08.0  &  84   &  1.22  &  2.16  &  9.4   &  V        &  \citet{Dotto1992}              \\
    48	&  1986-01-17.2  &  34   &  1.58  &  2.23  &  22.7  &  V        &  \citet{Weidenschilling1987}    \\
    49	&  1987-02-03.4  &  33   &  2.41  &  3.15  &  13.6  &  V        &  \citet{Weidenschilling1990}    \\
    50	&  1987-03-06.0  &  35   &  2.25  &  3.21  &  5.2   &  C        &  \citet{Franck1988}        \\
    51	&  1988-04-23.4  &  18   &  2.51  &  3.46  &  6.0   &  V        &  \citet{Weidenschilling1990}    \\
    52	&  1988-04-25.2  &  21   &  2.49  &  3.46  &  5.5   &  V        &  \citet{Weidenschilling1990}    \\
    53	&  1989-07-05.2  &  48   &  1.77  &  2.68  &  12.1  &  VB       &  \citet{Hutton1990}              \\
    54	&  1994-09-05.1  &  16   &  1.26  &  2.23  &  9.7   &  R        &  \citet{Fauvaud2001}        \\
    55	&  1994-09-06.1  &  15   &  1.25  &  2.22  &  9.4   &  R        &  \citet{Fauvaud2001}           \\
    56	&  1994-09-07.2  &  50   &  1.25  &  2.22  &  9.1   &  R        &  \citet{Fauvaud2001}           \\
    57	&  1994-09-08.1  &  163  &  1.25  &  2.22  &  8.8   &  R        &  \citet{Fauvaud2001}           \\
    58	&  2006-04-02.3  &  134  &  2.42  &  3.42  &  2.1   &  R        &  \citet{Warner2006c}                \\
    59	&  2006-04-03.2  &  149  &  2.42  &  3.42  &  2.1   &  R        &  \citet{Warner2006c}                \\
    60	&  2008-07-26    &  56   &  1.59  &  2.33  &  21.0  &  C    &  \citet{Grice2017}                     \\
    61	&  2008-07-27    &  61   &  1.58  &  2.33  &  20.8  &  C    &  \citet{Grice2017}                     \\
    62	&  2008-07-28    &  65   &  1.57  &  2.32  &  20.6  &  C    &  \citet{Grice2017}                     \\
    63	&  2008-07-29    &  53   &  1.56  &  2.32  &  20.4  &  C    &  \citet{Grice2017}                     \\
    64	&  2008-07-30    &  61   &  1.55  &  2.32  &  20.1  &  C    &  \citet{Grice2017}                     \\
    65	&  2008-08-01    &  83   &  1.53  &  2.31  &  19.7  &  C    &  \citet{Grice2017}                     \\
    66	&  2008-08-02    &  37   &  1.52  &  2.31  &  19.4  &  C    &  \citet{Grice2017}                     \\
    67	&  2008-08-04    &  41   &  1.50  &  2.31  &  18.9  &  C    &  \citet{Grice2017}                     \\
    68	&  2008-08-05    &  49   &  1.49  &  2.31  &  18.7  &  C    &  \citet{Grice2017}                     \\
    69	&  2008-08-06    &  59   &  1.48  &  2.30  &  18.4  &  C    &  \citet{Grice2017}                     \\
    70	&  2008-08-07    &  38   &  1.47  &  2.30  &  18.1  &  C    &  \citet{Grice2017}                     \\
    71	&  2008-08-09    &  69   &  1.45  &  2.30  &  17.6  &  C    &  \citet{Grice2017}                     \\
    72	&  2008-08-10    &  61   &  1.44  &  2.30  &  17.3  &  C    &  \citet{Grice2017}                     \\
    73	&  2008-08-11    &  65   &  1.43  &  2.29  &  17.0  &  C    &  \citet{Grice2017}                     \\
    74	&  2008-09-23    &  39   &  1.24  &  2.21  &  8.4   &  C    &  \citet{Grice2017}                     \\
    75	&  2008-09-23    &  40   &  1.24  &  2.21  &  8.4   &  C    &  \citet{Grice2017}                     \\
    76	&  2008-10-03    &  67   &  1.25  &  2.19  &  11.7  &  C    &  \citet{Grice2017}                     \\
    77	&  2008-10-04    &  70   &  1.25  &  2.19  &  12.1  &  C    &  \citet{Grice2017}                     \\
    78	&  2008-10-05    &  37   &  1.26  &  2.19  &  12.5  &  C    &  \citet{Grice2017}                     \\
    79	&  2008-10-06    &  48   &  1.26  &  2.19  &  12.9  &  C    &  \citet{Grice2017}                     \\
    80	&  2008-10-07    &  64   &  1.26  &  2.19  &  13.2  &  C    &  \citet{Grice2017}                     \\
    81	&  2008-10-08    &  64   &  1.27  &  2.19  &  13.6  &  C    &  \citet{Grice2017}                     \\
    82	&  2008-10-14    &  47   &  1.29  &  2.18  &  15.9  &  C    &  \citet{Grice2017}                     \\
    83	&  2008-10-14    &  124  &  1.29  &  2.18  &  15.9  &  C    &  \citet{Grice2017}                     \\
    84	&  2008-10-15    &  38   &  1.30  &  2.17  &  16.3  &  C    &  \citet{Grice2017}                     \\
    85	&  2008-10-16    &  96   &  1.30  &  2.17  &  16.7  &  C    &  \citet{Grice2017}                     \\
    86	&  2008-10-17    &  41   &  1.31  &  2.17  &  17.0  &  C    &  \citet{Grice2017}                     \\
    87	&  2008-10-18    &  34   &  1.31  &  2.17  &  17.4  &  C    &  \citet{Grice2017}                     \\
    88	&  2008-10-20    &  92   &  1.33  &  2.17  &  18.1  &  C    &  \citet{Grice2017}                     \\
    89	&  2008-10-23    &  59   &  1.35  &  2.16  &  19.1  &  C    &  \citet{Grice2017}                     \\
    90	&  2008-10-24    &  28   &  1.35  &  2.16  &  19.4  &  C    &  \citet{Grice2017}                     \\
    91	&  2008-10-24    &  76   &  1.35  &  2.16  &  19.4  &  C    &  \citet{Grice2017}                     \\
    92	&  2008-10-25    &  30   &  1.36  &  2.16  &  19.8  &  C    &  \citet{Grice2017}                     \\
    93	&  2008-10-26    &  44   &  1.37  &  2.16  &  20.1  &  C    &  \citet{Grice2017}                     \\
    94	&  2008-10-28    &  29   &  1.38  &  2.16  &  20.7  &  C    &  \citet{Grice2017}                     \\
    95	&  2008-10-31    &  35   &  1.40  &  2.15  &  21.6  &  C    &  \citet{Grice2017}                     \\
    96	&  2008-11-01    &  40   &  1.41  &  2.15  &  21.8  &  C    &  \citet{Grice2017}                     \\
    97	&  2008-11-03    &  37   &  1.43  &  2.15  &  22.4  &  C    &  \citet{Grice2017}                     \\
    98	&  2008-11-04    &  28   &  1.44  &  2.15  &  22.6  &  C    &  \citet{Grice2017}                     \\
    99	&  2008-11-12    &  55   &  1.51  &  2.14  &  24.5  &  C    &  \citet{Grice2017}                     \\
   100	&  2008-11-12    &  78   &  1.51  &  2.14  &  24.5  &  C    &  \citet{Grice2017}                     \\
   101	&  2008-11-14    &  38   &  1.52  &  2.14  &  24.8  &  C    &  \citet{Grice2017}                     \\
   102	&  2009-12-30    &  71   &  2.10  &  2.76  &  17.4  &  C    &  \citet{Grice2017}                     \\
   103	&  2010-01-05    &  80   &  2.05  &  2.78  &  16.0  &  C    &  \citet{Grice2017}                     \\
   104	&  2010-01-07    &  31   &  2.04  &  2.78  &  15.5  &  C    &  \citet{Grice2017}                     \\
   105	&  2010-01-10    &  25   &  2.02  &  2.79  &  14.7  &  C    &  \citet{Grice2017}                     \\
   106	&  2010-01-11    &  60   &  2.01  &  2.79  &  14.5  &  C    &  \citet{Grice2017}                     \\
   107	&  2010-01-12    &  27   &  2.00  &  2.79  &  14.2  &  C    &  \citet{Grice2017}                     \\
   108	&  2010-01-13    &  41   &  2.00  &  2.80  &  13.9  &  C    &  \citet{Grice2017}                     \\
   109	&  2010-01-14    &  26   &  1.99  &  2.80  &  13.7  &  C    &  \citet{Grice2017}                     \\
   110	&  2010-01-15    &  47   &  1.99  &  2.80  &  13.4  &  C    &  \citet{Grice2017}                     \\
   111	&  2010-01-28    &  60   &  1.93  &  2.83  &  9.6   &  C    &  \citet{Grice2017}                     \\
   112	&  2010-02-02    &  122  &  1.92  &  2.85  &  8.3   &  C    &  \citet{Grice2017}                     \\
   113	&  2010-02-03    &  55   &  1.92  &  2.85  &  8.0   &  C    &  \citet{Grice2017}                     \\
   114	&  2010-02-04    &  49   &  1.92  &  2.85  &  7.7   &  C    &  \citet{Grice2017}                     \\
   115	&  2010-02-04    &  78   &  1.92  &  2.85  &  7.8   &  C    &  \citet{Grice2017}                     \\
   116	&  2010-02-05    &  45   &  1.92  &  2.86  &  7.5   &  C    &  \citet{Grice2017}                     \\
   117	&  2010-02-05    &  46   &  1.92  &  2.85  &  7.5   &  C    &  \citet{Grice2017}                     \\
   118	&  2010-02-06    &  29   &  1.92  &  2.86  &  7.3   &  C    &  \citet{Grice2017}                     \\
   119	&  2010-02-06    &  34   &  1.92  &  2.86  &  7.3   &  C    &  \citet{Grice2017}                     \\
   120	&  2010-02-07    &  35   &  1.92  &  2.86  &  7.1   &  C    &  \citet{Grice2017}                     \\
   121	&  2010-02-08    &  22   &  1.92  &  2.86  &  6.9   &  C    &  \citet{Grice2017}                     \\
   122	&  2010-02-08    &  25   &  1.92  &  2.86  &  6.9   &  C    &  \citet{Grice2017}                     \\
   123	&  2010-02-09    &  48   &  1.92  &  2.87  &  6.7   &  C    &  \citet{Grice2017}                     \\
   124	&  2010-02-10    &  28   &  1.92  &  2.87  &  6.6   &  C    &  \citet{Grice2017}                     \\
   125	&  2010-02-11    &  55   &  1.92  &  2.87  &  6.5   &  C    &  \citet{Grice2017}                     \\
   126	&  2010-02-12    &  58   &  1.92  &  2.87  &  6.4   &  C    &  \citet{Grice2017}                     \\
   127	&  2010-02-14    &  58   &  1.92  &  2.88  &  6.3   &  C    &  \citet{Grice2017}                     \\
   128	&  2010-02-15    &  37   &  1.93  &  2.88  &  6.2   &  C    &  \citet{Grice2017}                     \\
   129	&  2010-02-15    &  38   &  1.93  &  2.88  &  6.2   &  C    &  \citet{Grice2017}                     \\
   130	&  2010-02-18    &  41   &  1.93  &  2.89  &  6.3   &  C    &  \citet{Grice2017}                     \\
   131	&  2010-02-19    &  67   &  1.94  &  2.89  &  6.4   &  C    &  \citet{Grice2017}                     \\
   132	&  2010-02-20    &  30   &  1.94  &  2.89  &  6.5   &  C    &  \citet{Grice2017}                     \\
   133	&  2010-02-23    &  28   &  1.95  &  2.90  &  6.9   &  C    &  \citet{Grice2017}                     \\
   134	&  2010-02-23    &  34   &  1.95  &  2.90  &  6.9   &  C    &  \citet{Grice2017}                     \\
   135	&  2010-02-24    &  89   &  1.96  &  2.90  &  7.1   &  C    &  \citet{Grice2017}                     \\
   136	&  2010-02-24    &  90   &  1.96  &  2.90  &  7.1   &  C    &  \citet{Grice2017}                     \\
   137	&  2010-03-01    &  32   &  1.98  &  2.91  &  8.1   &  C    &  \citet{Grice2017}                     \\
   138	&  2010-03-02    &  17   &  1.99  &  2.92  &  8.4   &  C    &  \citet{Grice2017}                     \\
   139	&  2010-03-02    &  23   &  1.99  &  2.92  &  8.4   &  C    &  \citet{Grice2017}                     \\
   140	&  2010-03-02    &  23   &  1.99  &  2.92  &  8.4   &  C    &  \citet{Grice2017}                     \\
   141	&  2010-03-03    &  44   &  2.00  &  2.92  &  8.6   &  C    &  \citet{Grice2017}                     \\
   142	&  2010-03-03    &  46   &  2.00  &  2.92  &  8.6   &  C    &  \citet{Grice2017}                     \\
   143	&  2010-03-03    &  53   &  2.00  &  2.92  &  8.6   &  C    &  \citet{Grice2017}                     \\
   144	&  2010-03-10    &  45   &  2.05  &  2.94  &  10.5  &  C    &  \citet{Grice2017}                     \\
   145	&  2010-03-13    &  36   &  2.07  &  2.94  &  11.2  &  C    &  \citet{Grice2017}                     \\
   146	&  2010-03-13    &  48   &  2.07  &  2.94  &  11.3  &  C    &  \citet{Grice2017}                     \\
   147	&  2010-03-14    &  54   &  2.08  &  2.94  &  11.5  &  C    &  \citet{Grice2017}                     \\
   148	&  2010-03-14    &  71   &  2.08  &  2.94  &  11.5  &  C    &  \citet{Grice2017}                     \\
   149	&  2010-03-15    &  32   &  2.09  &  2.95  &  11.7  &  C    &  \citet{Grice2017}                     \\
   150	&  2010-03-16    &  35   &  2.10  &  2.95  &  12.0  &  C    &  \citet{Grice2017}                     \\
   151	&  2010-03-17    &  53   &  2.11  &  2.95  &  12.3  &  C    &  \citet{Grice2017}                     \\
   152	&  2010-03-18    &  24   &  2.12  &  2.95  &  12.5  &  C    &  \citet{Grice2017}                     \\
   153	&  2010-03-18    &  43   &  2.12  &  2.95  &  12.5  &  C    &  \citet{Grice2017}                     \\
   154	&  2010-03-19    &  15   &  2.13  &  2.96  &  12.8  &  C    &  \citet{Grice2017}                     \\
   155	&  2010-03-20    &  50   &  2.14  &  2.96  &  13.0  &  C    &  \citet{Grice2017}                     \\
   156	&  2010-03-20    &  96   &  2.14  &  2.96  &  13.0  &  C    &  \citet{Grice2017}                     \\
   157	&  2010-03-21    &  33   &  2.15  &  2.96  &  13.2  &  C    &  \citet{Grice2017}                     \\
   158	&  2010-03-21    &  36   &  2.15  &  2.96  &  13.3  &  C    &  \citet{Grice2017}                     \\
   159	&  2010-03-22    &  23   &  2.16  &  2.96  &  13.5  &  C    &  \citet{Grice2017}                     \\
   160	&  2010-03-22    &  40   &  2.16  &  2.96  &  13.5  &  C    &  \citet{Grice2017}                     \\
   161	&  2010-03-23    &  64   &  2.18  &  2.97  &  13.7  &  C    &  \citet{Grice2017}                     \\
   162	&  2010-03-24    &  26   &  2.19  &  2.97  &  13.9  &  C    &  \citet{Grice2017}                     \\
   163	&  2010-03-24    &  62   &  2.19  &  2.97  &  14.0  &  C    &  \citet{Grice2017}                     \\
   164	&  2010-03-29    &  55   &  2.25  &  2.98  &  15.0  &  C    &  \citet{Grice2017}                     \\
   165	&  2010-03-29    &  116  &  2.24  &  2.98  &  15.0  &  C    &  \citet{Grice2017}                     \\
   166	&  2010-03-30    &  50   &  2.26  &  2.98  &  15.2  &  C    &  \citet{Grice2017}                     \\
   167	&  2010-03-30    &  70   &  2.26  &  2.98  &  15.2  &  C    &  \citet{Grice2017}                     \\
   168	&  2010-03-31    &  61   &  2.27  &  2.98  &  15.4  &  C    &  \citet{Grice2017}                     \\
   169	&  2010-04-01    &  67   &  2.28  &  2.99  &  15.6  &  C    &  \citet{Grice2017}                     \\
   170	&  2010-04-04    &  41   &  2.32  &  2.99  &  16.1  &  C    &  \citet{Grice2017}                     \\
   171	&  2010-04-06    &  24   &  2.35  &  3.00  &  16.5  &  C    &  \citet{Grice2017}                     \\
   172	&  2010-04-07    &  40   &  2.36  &  3.00  &  16.6  &  C    &  \citet{Grice2017}                     \\
   173	&  2010-04-08    &  37   &  2.37  &  3.00  &  16.8  &  C    &  \citet{Grice2017}                     \\
   174	&  2010-04-10    &  39   &  2.40  &  3.01  &  17.1  &  C    &  \citet{Grice2017}                     \\
   175	&  2010-04-12    &  35   &  2.43  &  3.01  &  17.4  &  C    &  \citet{Grice2017}                     \\
   176	&  2011-02-20    &  9    &  2.96  &  3.47  &  15.2  &  C    &  \citet{Grice2017}                     \\
   177	&  2013-10-23    &  200  &  1.18  &  2.11  &  13.1  &  C    &  \citet{Grice2017}                     \\
   178	&  2013-10-28    &  174  &  1.16  &  2.11  &  10.7  &  C    &  \citet{Grice2017}                     \\
   179	&  2013-10-30    &  115  &  1.15  &  2.11  &  9.7   &  C    &  \citet{Grice2017}                     \\
   180	&  2015-04-15.0  &  245  &  2.41  &  3.25  &  11.4  &  R    &  \textbf{This work} \\ %St\'{e}phane Fauvaud               \\
   \hline
\end{longtable}
%\tablefoot{
%     Gaia-GOSA (Gaia-Ground-based Observational Service for Asteroids, \url{www.gaiagosa.eu}).
%    }

\onecolumn
%\twocolumn
%\begin{multicols}{2}
%\shortlipsum
%\medskip
%\scriptsize{
%\longtab{2}{
%\begin{tabular}{r@{\,\,\,}l lc| r@{\,\,\,}l lc} \hline
% \multicolumn{2}{c} {Asteroid} & Date & Observer &  \multicolumn{2}{c}{Asteroid} & Date & Observer\\ \hline\hline
\begin{longtable}{l}
\caption{\label{tab:occ}Individual observers that participated in the stellar occultation campaigns targeting asteroid Kleopatra \citep{Herald2020}.}\\
\hline
Observer \\ \hline\hline

\endfirsthead
\caption{continued.}\\

\hline
 Observer \\ \hline\hline
\endhead
\hline
\endfoot
\multicolumn{1}{c} {\textbf{1980-10-10}} \\
S. Krysko, Linbrook, Alberta        \\
G. Stokes, Richland, WA              \\
Beals/Belcher/Loehde, Cooking Lake   \\
D. Scarlett, Kaslo, B.C.            \\
Jones/Bowen, Castlegar, B.C.         \\
D. Hube, Devon, U. of Alberta, AL   \\
G. Fouts, Goldendale, WA            \\
Mitchell, Ellensburg, WA            \\
E. Mannery, Ellensburg, WA          \\
\multicolumn{1}{c} {\textbf{1991-01-19}} \\
D. Dunham/W. Warren, Maryland, NJ   \\
Jeff Guerber, Greenbelt, MD         \\
C. Aikman/M. Fletcher, Victoria, BC \\
Robert J. Mordic, Richmond Hts, OH  \\
Samuel Storch, Jones Beach, NY      \\
Jim Pyral, North Bend, WA           \\
Edwin Lurcott, West Chester, PA     \\
M. Henry, Canton, OH                \\
James H. Fox, Afton, MN             \\
Robert Bolster, Alexandria, VA      \\
Dan Grieser, Columbus, OH           \\
\multicolumn{1}{c} {\textbf{2009-12-24}} \\
D. Dunham et al, Piedra, AZ             \\
J. Ray, Glendale, AZ                    \\
R. Peterson, Phoenix, AZ                \\
R. Peterson, Scottsdale, AZ             \\
P. Maley, Sun Lakes, AZ                 \\
G. Rattley, Gilbert, AZ                 \\
P. Maley, Santan, AZ                    \\
S. Degenhardt, Quijotoa, AZ             \\
L. Martinez, Casa Grande, AZ            \\
S. Degenhardt, Gu Oldak, AZ             \\
S. Degenhardt, Sells, AZ                \\
S. Degenhardt, Ali Chukson, AZ          \\
S. Degenhardt, Schuchk, AZ              \\
S. Degenhardt, Three Points, AZ         \\
J. Stamm, Oro Valley, AZ                \\
\multicolumn{1}{c} {\textbf{2015-03-12}} \\
Henk Bulder, NL                       \\
Friedhelm Dorst, DE                   \\
Oliver Kloes, DE                      \\
Jan-Maarten Winkel, NL                \\
Otto Farago, DE                       \\
Vasilis Metallinos, GR                \\
Harrie Rutten, NL                     \\
Henk De Groot, NL                     \\
Bernd Gaehrken, DE                    \\
Hans Kostense, NL                     \\
D. Fischer, H.G. Purucker, R. Stoyan, DE    \\
Eberhard Bredner, FR                  \\
Andre Mueller, DE                     \\
Lex Blommers, NL                      \\
K. Moddemeijer, P. Bastiaansen, W. Nobel, NL     \\
Christof Sauter, CH                   \\
Maxime Devogele, BE                   \\
Mike Kohl, CH                         \\
Jose De Queiroz, CH                   \\
Karl-Ludwig Bath, DE                  \\
Martin Federspiel, DE                 \\
Fernand Emering, LU                   \\
F. Van Den Abbeel, BE                 \\
Rene Bourtembourg, BE                 \\
Jonas Schenker, CH                    \\
J. Lecacheux, E. Meza, FR                \\
Stefano Sposetti, CH                  \\
Roberto Di Luca, IT                   \\
T. Pauwels, P. De Cat, BE              \\
C. Demeautis. D. Matter, FR              \\
Stefano Sposetti, CH                  \\
Andrea Manna, CH                      \\
Alberto Ossola, CH                    \\
Carlo Gualdoni, IT                    \\
Roland Decellier, BE                  \\
Fabrizio Ciabattari, IT               \\
Mauro Bachini, IT                     \\
Giancarlo Bonatti, IT                 \\
Alex Pratt, UK                        \\
Gilles Sautot, FR                     \\
Roland Boninsegna, BE                 \\
Fausto Delucchi, CH                   \\
Martin Federspiel, DE                 \\
G. Sautot, E. Vauthrin, FR               \\
Olivier Dechambre, FR                 \\
Jerome Berthier, FR                   \\
Frederic Vachier, FR                  \\
B. Carry, M. Pajuelo, FR               \\
Joan Rovira, ES                       \\
\multicolumn{1}{c} {\textbf{2016-04-05}} \\
B. Dunford, Naperville, IL            \\ 
A. Olsen, Urbana, IL                   \\
D. Dunham/J. Dunham, Yemasse, SC        \\
D. Dunham/J. Dunham, Coosawhatchie, SC  \\
D. Dunham/J. Dunham, Hardeesville, SC   \\
D. Dunham/J. Dunham, Savannah, GA       \\
D. Dunham/J. Dunham, Midway, GA         \\
D. Dunham/J. Dunham, South Newport, GA  \\
D. Dunham/J. Dunham, Darien, GA         \\
N. Smith, Trenton, GA                  \\
R. Venable, Yonkers, GA                \\
R. Venable, Hawkinsville, GA           \\
S. Messner, Moravia, IA                \\
S. Messner, Iconium, IA                \\
R. Venable, Oakfield, GA               \\
R. Venable, Newton, GA   \\
\hline
 \end{longtable}
%\tablefoot{
%\tablefoottext{a}{On line at \texttt{http://www.david-higgins.com/Astronomy/asteroid/lightcurves.htm}} %Higgins
%}
%}
%}
%\end{multicols}

\end{appendix}

\end{document}